\documentclass[traditabstract]{aa}
\usepackage{graphicx,amsmath}
\usepackage{epsf}
\usepackage{txfonts}
\usepackage{natbib}
\usepackage{longtable}
\usepackage{upgreek}
\usepackage{lscape}
\usepackage[usenames]{color}
\definecolor{Blue}{rgb}{0.1,0,0.55} 
\definecolor{Red}{rgb}{1.0,0.1,0.1} 
\usepackage[colorlinks=true,citecolor=blue]{hyperref}
\usepackage{fixltx2e}

\bibpunct{(}{)}{;}{a}{}{,} 

\def\setsymbol#1#2{\expandafter\def\csname #1\endcsname{#2}}
\def\getsymbol#1{\csname #1\endcsname}

\def\Planck{\textit{Planck}}

\newbox\tablebox    \newdimen\tablewidth
\def\leaderfil{\leaders\hbox to 5pt{\hss.\hss}\hfil}
\def\tablenote#1 #2\par{\begingroup \parindent=0.8em
    \abovedisplayshortskip=0pt\belowdisplayshortskip=0pt
    \noindent
    $$\hss\vbox{\hsize\tablewidth \hangindent=\parindent \hangafter=1 \noindent
    \hbox to \parindent{$^#1$\hss}\strut#2\strut\par}\hss$$
    \endgroup}

\def\L2{\ifmmode L_2\else $L_2$\fi}

\def\DeltaT{\ifmmode \Delta T\else $\Delta T$\fi}
\def\deltat{\ifmmode \Delta t\else $\Delta t$\fi}
\def\fknee{\ifmmode f_{\rm knee}\else $f_{\rm knee}$\fi}
\def\Fmax{\ifmmode F_{\rm max}\else $F_{\rm max}$\fi}
\def\solar{\ifmmode{\rm M}_{\mathord\odot}\else${\rm M}_{\mathord\odot}$\fi}
\def\Msolar{\ifmmode{\rm M}_{\mathord\odot}\else${\rm M}_{\mathord\odot}$\fi}
\def\Lsolar{\ifmmode{\rm L}_{\mathord\odot}\else${\rm L}_{\mathord\odot}$\fi}

\def\inv{\ifmmode^{-1}\else$^{-1}$\fi}
\def\mo{\ifmmode^{-1}\else$^{-1}$\fi}
\def\sup#1{\ifmmode ^{\rm #1}\else $^{\rm #1}$\fi}
\def\expo#1{\ifmmode \times 10^{#1}\else $\times 10^{#1}$\fi}
\def\,{\thinspace}
\def\lsim{\mathrel{\raise .4ex\hbox{\rlap{$<$}\lower 1.2ex\hbox{$\sim$}}}}
\def\gsim{\mathrel{\raise .4ex\hbox{\rlap{$>$}\lower 1.2ex\hbox{$\sim$}}}}

\def\simprop{\mathrel{\raise .4ex\hbox{\rlap{$\propto$}\lower 1.2ex\hbox{$\sim$}}}}
\def\deg{\ifmmode^\circ\else$^\circ$\fi}
\def\pdeg{\ifmmode $\setbox0=\hbox{$^{\circ}$}\rlap{\hskip.11\wd0 .}$^{\circ}
          \else \setbox0=\hbox{$^{\circ}$}\rlap{\hskip.11\wd0 .}$^{\circ}$\fi}
\def\arcs{\ifmmode {^{\scriptstyle\prime\prime}}
          \else $^{\scriptstyle\prime\prime}$\fi}
\def\arcm{\ifmmode {^{\scriptstyle\prime}}
          \else $^{\scriptstyle\prime}$\fi}
\newdimen\sa  \newdimen\sb
\def\parcs{\sa=.07em \sb=.03em
     \ifmmode \hbox{\rlap{.}}^{\scriptstyle\prime\kern -\sb\prime}\hbox{\kern -\sa}
     \else \rlap{.}$^{\scriptstyle\prime\kern -\sb\prime}$\kern -\sa\fi}
\def\parcm{\sa=.08em \sb=.03em
     \ifmmode \hbox{\rlap{.}\kern\sa}^{\scriptstyle\prime}\hbox{\kern-\sb}
     \else \rlap{.}\kern\sa$^{\scriptstyle\prime}$\kern-\sb\fi}
\def\ra[#1 #2 #3.#4]{#1\sup{h}#2\sup{m}#3\sup{s}\llap.#4}
\def\dec[#1 #2 #3.#4]{#1\deg#2\arcm#3\arcs\llap.#4}
\def\deco[#1 #2 #3]{#1\deg#2\arcm#3\arcs}
\def\rra[#1 #2]{#1\sup{h}#2\sup{m}}

\def\dots{\relax\ifmmode \ldots\else $\ldots$\fi}
\def\WHzsr{\ifmmode $W\,Hz\mo\,sr\mo$\else W\,Hz\mo\,sr\mo\fi}
\def\mHz{\ifmmode $\,mHz$\else \,mHz\fi}
\def\GHz{\ifmmode $\,GHz$\else \,GHz\fi}
\def\mKs{\ifmmode $\,mK\,s$^{1/2}\else \,mK\,s$^{1/2}$\fi}
\def\muKs{\ifmmode \,\mu$K\,s$^{1/2}\else \,$\mu$K\,s$^{1/2}$\fi}
\def\muKRJs{\ifmmode \,\mu$K$_{\rm RJ}$\,s$^{1/2}\else \,$\mu$K$_{\rm RJ}$\,s$^{1/2}$\fi}
\def\muKHz{\ifmmode \,\mu$K\,Hz$^{-1/2}\else \,$\mu$K\,Hz$^{-1/2}$\fi}
\def\MJysr{\ifmmode \,$MJy\,sr\mo$\else \,MJy\,sr\mo\fi}
\def\MJysrmK{\ifmmode \,$MJy\,sr\mo$\,mK$_{\rm CMB}\mo\else \,MJy\,sr\mo\,mK$_{\rm CMB}\mo$\fi}
\def\microns{\ifmmode \,\mu$m$\else \,$\mu$m\fi}

\def\muK{\ifmmode \,\mu$K$\else \,$\mu$\hbox{K}\fi}
\def\microK{\ifmmode \,\mu$K$\else \,$\mu$\hbox{K}\fi}
\def\muW{\ifmmode \,\mu$W$\else \,$\mu$\hbox{W}\fi}
\def\kms{\ifmmode $\,km\,s$^{-1}\else \,km\,s$^{-1}$\fi}
\def\kmsMpc{\ifmmode $\,\kms\,Mpc\mo$\else \,\kms\,Mpc\mo\fi}
\setsymbol{LFI:center:frequency:70GHz:units}{70.3\,GHz}
\setsymbol{LFI:center:frequency:44GHz:units}{44.1\,GHz}
\setsymbol{LFI:center:frequency:30GHz:units}{28.5\,GHz}

\setsymbol{LFI:center:frequency:70GHz}{70.3}
\setsymbol{LFI:center:frequency:44GHz}{44.1}
\setsymbol{LFI:center:frequency:30GHz}{28.5}

\setsymbol{LFI:center:frequency:LFI18:Rad:M:units}{71.7\GHz}
\setsymbol{LFI:center:frequency:LFI19:Rad:M:units}{67.5\GHz}
\setsymbol{LFI:center:frequency:LFI20:Rad:M:units}{69.2\GHz}
\setsymbol{LFI:center:frequency:LFI21:Rad:M:units}{70.4\GHz}
\setsymbol{LFI:center:frequency:LFI22:Rad:M:units}{71.5\GHz}
\setsymbol{LFI:center:frequency:LFI23:Rad:M:units}{70.8\GHz}
\setsymbol{LFI:center:frequency:LFI24:Rad:M:units}{44.4\GHz}
\setsymbol{LFI:center:frequency:LFI25:Rad:M:units}{44.0\GHz}
\setsymbol{LFI:center:frequency:LFI26:Rad:M:units}{43.9\GHz}
\setsymbol{LFI:center:frequency:LFI27:Rad:M:units}{28.3\GHz}
\setsymbol{LFI:center:frequency:LFI28:Rad:M:units}{28.8\GHz}
\setsymbol{LFI:center:frequency:LFI18:Rad:S:units}{70.1\GHz}
\setsymbol{LFI:center:frequency:LFI19:Rad:S:units}{69.6\GHz}
\setsymbol{LFI:center:frequency:LFI20:Rad:S:units}{69.5\GHz}
\setsymbol{LFI:center:frequency:LFI21:Rad:S:units}{69.5\GHz}
\setsymbol{LFI:center:frequency:LFI22:Rad:S:units}{72.8\GHz}
\setsymbol{LFI:center:frequency:LFI23:Rad:S:units}{71.3\GHz}
\setsymbol{LFI:center:frequency:LFI24:Rad:S:units}{44.1\GHz}
\setsymbol{LFI:center:frequency:LFI25:Rad:S:units}{44.1\GHz}
\setsymbol{LFI:center:frequency:LFI26:Rad:S:units}{44.1\GHz}
\setsymbol{LFI:center:frequency:LFI27:Rad:S:units}{28.5\GHz}
\setsymbol{LFI:center:frequency:LFI28:Rad:S:units}{28.2\GHz}

\setsymbol{LFI:center:frequency:LFI18:Rad:M}{71.7}
\setsymbol{LFI:center:frequency:LFI19:Rad:M}{67.5}
\setsymbol{LFI:center:frequency:LFI20:Rad:M}{69.2}
\setsymbol{LFI:center:frequency:LFI21:Rad:M}{70.4}
\setsymbol{LFI:center:frequency:LFI22:Rad:M}{71.5}
\setsymbol{LFI:center:frequency:LFI23:Rad:M}{70.8}
\setsymbol{LFI:center:frequency:LFI24:Rad:M}{44.4}
\setsymbol{LFI:center:frequency:LFI25:Rad:M}{44.0}
\setsymbol{LFI:center:frequency:LFI26:Rad:M}{43.9}
\setsymbol{LFI:center:frequency:LFI27:Rad:M}{28.3}
\setsymbol{LFI:center:frequency:LFI28:Rad:M}{28.8}
\setsymbol{LFI:center:frequency:LFI18:Rad:S}{70.1}
\setsymbol{LFI:center:frequency:LFI19:Rad:S}{69.6}
\setsymbol{LFI:center:frequency:LFI20:Rad:S}{69.5}
\setsymbol{LFI:center:frequency:LFI21:Rad:S}{69.5}
\setsymbol{LFI:center:frequency:LFI22:Rad:S}{72.8}
\setsymbol{LFI:center:frequency:LFI23:Rad:S}{71.3}
\setsymbol{LFI:center:frequency:LFI24:Rad:S}{44.1}
\setsymbol{LFI:center:frequency:LFI25:Rad:S}{44.1}
\setsymbol{LFI:center:frequency:LFI26:Rad:S}{44.1}
\setsymbol{LFI:center:frequency:LFI27:Rad:S}{28.5}
\setsymbol{LFI:center:frequency:LFI28:Rad:S}{28.2}

\setsymbol{LFI:white:noise:sensitivity:70GHz:units}{134.7\muKs}
\setsymbol{LFI:white:noise:sensitivity:44GHz:units}{164.7\muKs}
\setsymbol{LFI:white:noise:sensitivity:30GHz:units}{143.4\muKs}

\setsymbol{LFI:white:noise:sensitivity:70GHz}{134.7}
\setsymbol{LFI:white:noise:sensitivity:44GHz}{164.7}
\setsymbol{LFI:white:noise:sensitivity:30GHz}{143.4}

\setsymbol{LFI:white:noise:sensitivity:LFI18:Rad:M:units}{512.0\muKs}
\setsymbol{LFI:white:noise:sensitivity:LFI19:Rad:M:units}{581.4\muKs}
\setsymbol{LFI:white:noise:sensitivity:LFI20:Rad:M:units}{590.8\muKs}
\setsymbol{LFI:white:noise:sensitivity:LFI21:Rad:M:units}{455.2\muKs}
\setsymbol{LFI:white:noise:sensitivity:LFI22:Rad:M:units}{492.0\muKs}
\setsymbol{LFI:white:noise:sensitivity:LFI23:Rad:M:units}{507.7\muKs}
\setsymbol{LFI:white:noise:sensitivity:LFI24:Rad:M:units}{462.2\muKs}
\setsymbol{LFI:white:noise:sensitivity:LFI25:Rad:M:units}{413.6\muKs}
\setsymbol{LFI:white:noise:sensitivity:LFI26:Rad:M:units}{478.6\muKs}
\setsymbol{LFI:white:noise:sensitivity:LFI27:Rad:M:units}{277.7\muKs}
\setsymbol{LFI:white:noise:sensitivity:LFI28:Rad:M:units}{312.3\muKs}
\setsymbol{LFI:white:noise:sensitivity:LFI18:Rad:S:units}{465.7\muKs}
\setsymbol{LFI:white:noise:sensitivity:LFI19:Rad:S:units}{555.6\muKs}
\setsymbol{LFI:white:noise:sensitivity:LFI20:Rad:S:units}{623.2\muKs}
\setsymbol{LFI:white:noise:sensitivity:LFI21:Rad:S:units}{564.1\muKs}
\setsymbol{LFI:white:noise:sensitivity:LFI22:Rad:S:units}{534.4\muKs}
\setsymbol{LFI:white:noise:sensitivity:LFI23:Rad:S:units}{542.4\muKs}
\setsymbol{LFI:white:noise:sensitivity:LFI24:Rad:S:units}{399.2\muKs}
\setsymbol{LFI:white:noise:sensitivity:LFI25:Rad:S:units}{392.6\muKs}
\setsymbol{LFI:white:noise:sensitivity:LFI26:Rad:S:units}{418.6\muKs}
\setsymbol{LFI:white:noise:sensitivity:LFI27:Rad:S:units}{302.9\muKs}
\setsymbol{LFI:white:noise:sensitivity:LFI28:Rad:S:units}{285.3\muKs}

\setsymbol{LFI:white:noise:sensitivity:LFI18:Rad:M}{512.0}
\setsymbol{LFI:white:noise:sensitivity:LFI19:Rad:M}{581.4}
\setsymbol{LFI:white:noise:sensitivity:LFI20:Rad:M}{590.8}
\setsymbol{LFI:white:noise:sensitivity:LFI21:Rad:M}{455.2}
\setsymbol{LFI:white:noise:sensitivity:LFI22:Rad:M}{492.0}
\setsymbol{LFI:white:noise:sensitivity:LFI23:Rad:M}{507.7}
\setsymbol{LFI:white:noise:sensitivity:LFI24:Rad:M}{462.2}
\setsymbol{LFI:white:noise:sensitivity:LFI25:Rad:M}{413.6}
\setsymbol{LFI:white:noise:sensitivity:LFI26:Rad:M}{478.6}
\setsymbol{LFI:white:noise:sensitivity:LFI27:Rad:M}{277.7}
\setsymbol{LFI:white:noise:sensitivity:LFI28:Rad:M}{312.3}
\setsymbol{LFI:white:noise:sensitivity:LFI18:Rad:S}{465.7}
\setsymbol{LFI:white:noise:sensitivity:LFI19:Rad:S}{555.6}
\setsymbol{LFI:white:noise:sensitivity:LFI20:Rad:S}{623.2}
\setsymbol{LFI:white:noise:sensitivity:LFI21:Rad:S}{564.1}
\setsymbol{LFI:white:noise:sensitivity:LFI22:Rad:S}{534.4}
\setsymbol{LFI:white:noise:sensitivity:LFI23:Rad:S}{542.4}
\setsymbol{LFI:white:noise:sensitivity:LFI24:Rad:S}{399.2}
\setsymbol{LFI:white:noise:sensitivity:LFI25:Rad:S}{392.6}
\setsymbol{LFI:white:noise:sensitivity:LFI26:Rad:S}{418.6}
\setsymbol{LFI:white:noise:sensitivity:LFI27:Rad:S}{302.9}
\setsymbol{LFI:white:noise:sensitivity:LFI28:Rad:S}{285.3}

\setsymbol{LFI:knee:frequency:70GHz:units}{29.5\mHz}
\setsymbol{LFI:knee:frequency:44GHz:units}{56.2\mHz}
\setsymbol{LFI:knee:frequency:30GHz:units}{113.7\mHz}

\setsymbol{LFI:knee:frequency:70GHz}{29.5}
\setsymbol{LFI:knee:frequency:44GHz}{56.2}
\setsymbol{LFI:knee:frequency:30GHz}{113.7}

\setsymbol{LFI:knee:frequency:LFI18:Rad:M:units}{16.3\mHz}
\setsymbol{LFI:knee:frequency:LFI19:Rad:M:units}{15.1\mHz}
\setsymbol{LFI:knee:frequency:LFI20:Rad:M:units}{18.7\mHz}
\setsymbol{LFI:knee:frequency:LFI21:Rad:M:units}{37.2\mHz}
\setsymbol{LFI:knee:frequency:LFI22:Rad:M:units}{12.7\mHz}
\setsymbol{LFI:knee:frequency:LFI23:Rad:M:units}{34.6\mHz}
\setsymbol{LFI:knee:frequency:LFI24:Rad:M:units}{46.2\mHz}
\setsymbol{LFI:knee:frequency:LFI25:Rad:M:units}{24.9\mHz}
\setsymbol{LFI:knee:frequency:LFI26:Rad:M:units}{67.6\mHz}
\setsymbol{LFI:knee:frequency:LFI27:Rad:M:units}{187.4\mHz}
\setsymbol{LFI:knee:frequency:LFI28:Rad:M:units}{122.2\mHz}
\setsymbol{LFI:knee:frequency:LFI18:Rad:S:units}{17.7\mHz}
\setsymbol{LFI:knee:frequency:LFI19:Rad:S:units}{22.0\mHz}
\setsymbol{LFI:knee:frequency:LFI20:Rad:S:units}{8.7\mHz}
\setsymbol{LFI:knee:frequency:LFI21:Rad:S:units}{25.9\mHz}
\setsymbol{LFI:knee:frequency:LFI22:Rad:S:units}{15.8\mHz}
\setsymbol{LFI:knee:frequency:LFI23:Rad:S:units}{129.8\mHz}
\setsymbol{LFI:knee:frequency:LFI24:Rad:S:units}{100.9\mHz}
\setsymbol{LFI:knee:frequency:LFI25:Rad:S:units}{38.9\mHz}
\setsymbol{LFI:knee:frequency:LFI26:Rad:S:units}{58.9\mHz}
\setsymbol{LFI:knee:frequency:LFI27:Rad:S:units}{104.4\mHz}
\setsymbol{LFI:knee:frequency:LFI28:Rad:S:units}{40.7\mHz}

\setsymbol{LFI:knee:frequency:LFI18:Rad:M}{16.3}
\setsymbol{LFI:knee:frequency:LFI19:Rad:M}{15.1}
\setsymbol{LFI:knee:frequency:LFI20:Rad:M}{18.7}
\setsymbol{LFI:knee:frequency:LFI21:Rad:M}{37.2}
\setsymbol{LFI:knee:frequency:LFI22:Rad:M}{12.7}
\setsymbol{LFI:knee:frequency:LFI23:Rad:M}{34.6}
\setsymbol{LFI:knee:frequency:LFI24:Rad:M}{46.2}
\setsymbol{LFI:knee:frequency:LFI25:Rad:M}{24.9}
\setsymbol{LFI:knee:frequency:LFI26:Rad:M}{67.6}
\setsymbol{LFI:knee:frequency:LFI27:Rad:M}{187.4}
\setsymbol{LFI:knee:frequency:LFI28:Rad:M}{122.2}
\setsymbol{LFI:knee:frequency:LFI18:Rad:S}{17.7}
\setsymbol{LFI:knee:frequency:LFI19:Rad:S}{22.0}
\setsymbol{LFI:knee:frequency:LFI20:Rad:S}{8.7}
\setsymbol{LFI:knee:frequency:LFI21:Rad:S}{25.9}
\setsymbol{LFI:knee:frequency:LFI22:Rad:S}{15.8}
\setsymbol{LFI:knee:frequency:LFI23:Rad:S}{129.8}
\setsymbol{LFI:knee:frequency:LFI24:Rad:S}{100.9}
\setsymbol{LFI:knee:frequency:LFI25:Rad:S}{38.9}
\setsymbol{LFI:knee:frequency:LFI26:Rad:S}{58.9}
\setsymbol{LFI:knee:frequency:LFI27:Rad:S}{104.4}
\setsymbol{LFI:knee:frequency:LFI28:Rad:S}{40.7}

\setsymbol{LFI:slope:70GHz:units}{$-1.03$\mHz}
\setsymbol{LFI:slope:44GHz:units}{$-0.89$\mHz}
\setsymbol{LFI:slope:30GHz:units}{$-0.87$\mHz}

\setsymbol{LFI:slope:70GHz}{$-1.03$}
\setsymbol{LFI:slope:44GHz}{$-0.89$}
\setsymbol{LFI:slope:30GHz}{$-0.87$}

\setsymbol{LFI:slope:LFI18:Rad:M:units}{$-1.04$\mHz}
\setsymbol{LFI:slope:LFI19:Rad:M:units}{$-1.09$\mHz}
\setsymbol{LFI:slope:LFI20:Rad:M:units}{$-0.69$\mHz}
\setsymbol{LFI:slope:LFI21:Rad:M:units}{$-1.56$\mHz}
\setsymbol{LFI:slope:LFI22:Rad:M:units}{$-1.01$\mHz}
\setsymbol{LFI:slope:LFI23:Rad:M:units}{$-0.96$\mHz}
\setsymbol{LFI:slope:LFI24:Rad:M:units}{$-0.83$\mHz}
\setsymbol{LFI:slope:LFI25:Rad:M:units}{$-0.91$\mHz}
\setsymbol{LFI:slope:LFI26:Rad:M:units}{$-0.95$\mHz}
\setsymbol{LFI:slope:LFI27:Rad:M:units}{$-0.87$\mHz}
\setsymbol{LFI:slope:LFI28:Rad:M:units}{$-0.88$\mHz}
\setsymbol{LFI:slope:LFI18:Rad:S:units}{$-1.15$\mHz}
\setsymbol{LFI:slope:LFI19:Rad:S:units}{$-1.00$\mHz}
\setsymbol{LFI:slope:LFI20:Rad:S:units}{$-0.95$\mHz}
\setsymbol{LFI:slope:LFI21:Rad:S:units}{$-0.92$\mHz}
\setsymbol{LFI:slope:LFI22:Rad:S:units}{$-1.01$\mHz}
\setsymbol{LFI:slope:LFI23:Rad:S:units}{$-0.95$\mHz}
\setsymbol{LFI:slope:LFI24:Rad:S:units}{$-0.73$\mHz}
\setsymbol{LFI:slope:LFI25:Rad:S:units}{$-1.16$\mHz}
\setsymbol{LFI:slope:LFI26:Rad:S:units}{$-0.79$\mHz}
\setsymbol{LFI:slope:LFI27:Rad:S:units}{$-0.82$\mHz}
\setsymbol{LFI:slope:LFI28:Rad:S:units}{$-0.91$\mHz}

\setsymbol{LFI:slope:LFI18:Rad:M}{$-1.04$}
\setsymbol{LFI:slope:LFI19:Rad:M}{$-1.09$}
\setsymbol{LFI:slope:LFI20:Rad:M}{$-0.69$}
\setsymbol{LFI:slope:LFI21:Rad:M}{$-1.56$}
\setsymbol{LFI:slope:LFI22:Rad:M}{$-1.01$}
\setsymbol{LFI:slope:LFI23:Rad:M}{$-0.96$}
\setsymbol{LFI:slope:LFI24:Rad:M}{$-0.83$}
\setsymbol{LFI:slope:LFI25:Rad:M}{$-0.91$}
\setsymbol{LFI:slope:LFI26:Rad:M}{$-0.95$}
\setsymbol{LFI:slope:LFI27:Rad:M}{$-0.87$}
\setsymbol{LFI:slope:LFI28:Rad:M}{$-0.88$}
\setsymbol{LFI:slope:LFI18:Rad:S}{$-1.15$}
\setsymbol{LFI:slope:LFI19:Rad:S}{$-1.00$}
\setsymbol{LFI:slope:LFI20:Rad:S}{$-0.95$}
\setsymbol{LFI:slope:LFI21:Rad:S}{$-0.92$}
\setsymbol{LFI:slope:LFI22:Rad:S}{$-1.01$}
\setsymbol{LFI:slope:LFI23:Rad:S}{$-0.95$}
\setsymbol{LFI:slope:LFI24:Rad:S}{$-0.73$}
\setsymbol{LFI:slope:LFI25:Rad:S}{$-1.16$}
\setsymbol{LFI:slope:LFI26:Rad:S}{$-0.79$}
\setsymbol{LFI:slope:LFI27:Rad:S}{$-0.82$}
\setsymbol{LFI:slope:LFI28:Rad:S}{$-0.91$}

\setsymbol{LFI:FWHM:70GHz:units}{13\parcm01}
\setsymbol{LFI:FWHM:44GHz:units}{27\parcm92}
\setsymbol{LFI:FWHM:30GHz:units}{32\parcm65}

\setsymbol{LFI:FWHM:70GHz}{13.01}
\setsymbol{LFI:FWHM:44GHz}{27.92}
\setsymbol{LFI:FWHM:30GHz}{32.65}

\setsymbol{LFI:FWHM:LFI18:units}{13\parcm39}
\setsymbol{LFI:FWHM:LFI19:units}{13\parcm01}
\setsymbol{LFI:FWHM:LFI20:units}{12\parcm75}
\setsymbol{LFI:FWHM:LFI21:units}{12\parcm74}
\setsymbol{LFI:FWHM:LFI22:units}{12\parcm87}
\setsymbol{LFI:FWHM:LFI23:units}{13\parcm27}
\setsymbol{LFI:FWHM:LFI24:units}{22\parcm98}
\setsymbol{LFI:FWHM:LFI25:units}{30\parcm46}
\setsymbol{LFI:FWHM:LFI26:units}{30\parcm31}
\setsymbol{LFI:FWHM:LFI27:units}{32\parcm65}
\setsymbol{LFI:FWHM:LFI28:units}{32\parcm66}

\setsymbol{LFI:FWHM:LFI18}{13.39}
\setsymbol{LFI:FWHM:LFI19}{13.01}
\setsymbol{LFI:FWHM:LFI20}{12.75}
\setsymbol{LFI:FWHM:LFI21}{12.74}
\setsymbol{LFI:FWHM:LFI22}{12.87}
\setsymbol{LFI:FWHM:LFI23}{13.27}
\setsymbol{LFI:FWHM:LFI24}{22.98}
\setsymbol{LFI:FWHM:LFI25}{30.46}
\setsymbol{LFI:FWHM:LFI26}{30.31}
\setsymbol{LFI:FWHM:LFI27}{32.65}
\setsymbol{LFI:FWHM:LFI28}{32.66}

\setsymbol{LFI:FWHM:uncertainty:LFI18:units}{0.170\arcm}
\setsymbol{LFI:FWHM:uncertainty:LFI19:units}{0.174\arcm}
\setsymbol{LFI:FWHM:uncertainty:LFI20:units}{0.170\arcm}
\setsymbol{LFI:FWHM:uncertainty:LFI21:units}{0.156\arcm}
\setsymbol{LFI:FWHM:uncertainty:LFI22:units}{0.164\arcm}
\setsymbol{LFI:FWHM:uncertainty:LFI23:units}{0.171\arcm}
\setsymbol{LFI:FWHM:uncertainty:LFI24:units}{0.652\arcm}
\setsymbol{LFI:FWHM:uncertainty:LFI25:units}{1.075\arcm}
\setsymbol{LFI:FWHM:uncertainty:LFI26:units}{1.131\arcm}
\setsymbol{LFI:FWHM:uncertainty:LFI27:units}{1.266\arcm}
\setsymbol{LFI:FWHM:uncertainty:LFI28:units}{1.287\arcm}

\setsymbol{LFI:FWHM:uncertainty:LFI18}{0.170}
\setsymbol{LFI:FWHM:uncertainty:LFI19}{0.174}
\setsymbol{LFI:FWHM:uncertainty:LFI20}{0.170}
\setsymbol{LFI:FWHM:uncertainty:LFI21}{0.156}
\setsymbol{LFI:FWHM:uncertainty:LFI22}{0.164}
\setsymbol{LFI:FWHM:uncertainty:LFI23}{0.171}
\setsymbol{LFI:FWHM:uncertainty:LFI24}{0.652}
\setsymbol{LFI:FWHM:uncertainty:LFI25}{1.075}
\setsymbol{LFI:FWHM:uncertainty:LFI26}{1.131}
\setsymbol{LFI:FWHM:uncertainty:LFI27}{1.266}
\setsymbol{LFI:FWHM:uncertainty:LFI28}{1.287}

\setsymbol{HFI:center:frequency:100GHz:units}{100\,GHz}
\setsymbol{HFI:center:frequency:143GHz:units}{143\,GHz}
\setsymbol{HFI:center:frequency:217GHz:units}{217\,GHz}
\setsymbol{HFI:center:frequency:353GHz:units}{353\,GHz}
\setsymbol{HFI:center:frequency:545GHz:units}{545\,GHz}
\setsymbol{HFI:center:frequency:857GHz:units}{857\,GHz}

\setsymbol{HFI:center:frequency:100GHz}{100}
\setsymbol{HFI:center:frequency:143GHz}{143}
\setsymbol{HFI:center:frequency:217GHz}{217}
\setsymbol{HFI:center:frequency:353GHz}{353}
\setsymbol{HFI:center:frequency:545GHz}{545}
\setsymbol{HFI:center:frequency:857GHz}{857}

\setsymbol{HFI:Ndetectors:100GHz}{8}
\setsymbol{HFI:Ndetectors:143GHz}{11}
\setsymbol{HFI:Ndetectors:217GHz}{12}
\setsymbol{HFI:Ndetectors:353GHz}{12}
\setsymbol{HFI:Ndetectors:545GHz}{3}
\setsymbol{HFI:Ndetectors:857GHz}{4}

\setsymbol{HFI:FWHM:Maps:100GHz:units}{9\parcm88}
\setsymbol{HFI:FWHM:Maps:143GHz:units}{7\parcm18}
\setsymbol{HFI:FWHM:Maps:217GHz:units}{4\parcm87}
\setsymbol{HFI:FWHM:Maps:353GHz:units}{4\parcm65}
\setsymbol{HFI:FWHM:Maps:545GHz:units}{4\parcm72}
\setsymbol{HFI:FWHM:Maps:857GHz:units}{4\parcm39}
\setsymbol{HFI:FWHM:Maps:100GHz}{9.88}
\setsymbol{HFI:FWHM:Maps:143GHz}{7.18}
\setsymbol{HFI:FWHM:Maps:217GHz}{4.87}
\setsymbol{HFI:FWHM:Maps:353GHz}{4.65}
\setsymbol{HFI:FWHM:Maps:545GHz}{4.72}
\setsymbol{HFI:FWHM:Maps:857GHz}{4.39}

\setsymbol{HFI:beam:ellipticity:Maps:100GHz}{1.15}
\setsymbol{HFI:beam:ellipticity:Maps:143GHz}{1.01}
\setsymbol{HFI:beam:ellipticity:Maps:217GHz}{1.06}
\setsymbol{HFI:beam:ellipticity:Maps:353GHz}{1.05}
\setsymbol{HFI:beam:ellipticity:Maps:545GHz}{1.14}
\setsymbol{HFI:beam:ellipticity:Maps:857GHz}{1.19}

\setsymbol{HFI:FWHM:Mars:100GHz:units}{9\parcm37}
\setsymbol{HFI:FWHM:Mars:143GHz:units}{7\parcm04}
\setsymbol{HFI:FWHM:Mars:217GHz:units}{4\parcm68}
\setsymbol{HFI:FWHM:Mars:353GHz:units}{4\parcm43}
\setsymbol{HFI:FWHM:Mars:545GHz:units}{3\parcm80}
\setsymbol{HFI:FWHM:Mars:857GHz:units}{3\parcm67}

\setsymbol{HFI:FWHM:Mars:100GHz}{9.37}
\setsymbol{HFI:FWHM:Mars:143GHz}{7.04}
\setsymbol{HFI:FWHM:Mars:217GHz}{4.68}
\setsymbol{HFI:FWHM:Mars:353GHz}{4.43}
\setsymbol{HFI:FWHM:Mars:545GHz}{3.80}
\setsymbol{HFI:FWHM:Mars:857GHz}{3.67}

\setsymbol{HFI:beam:ellipticity:Mars:100GHz}{1.18}
\setsymbol{HFI:beam:ellipticity:Mars:143GHz}{1.03}
\setsymbol{HFI:beam:ellipticity:Mars:217GHz}{1.14}
\setsymbol{HFI:beam:ellipticity:Mars:353GHz}{1.09}
\setsymbol{HFI:beam:ellipticity:Mars:545GHz}{1.25}
\setsymbol{HFI:beam:ellipticity:Mars:857GHz}{1.03}

\setsymbol{HFI:CMB:relative:calibration:100GHz}{$\lsim 1\%$}
\setsymbol{HFI:CMB:relative:calibration:143GHz}{$\lsim 1\%$}
\setsymbol{HFI:CMB:relative:calibration:217GHz}{$\lsim 1\%$}
\setsymbol{HFI:CMB:relative:calibration:353GHz}{$\lsim 1\%$}
\setsymbol{HFI:CMB:relative:calibration:545GHz}{}
\setsymbol{HFI:CMB:relative:calibration:857GHz}{}

\setsymbol{HFI:CMB:absolute:calibration:100GHz}{$\lsim 2\%$}
\setsymbol{HFI:CMB:absolute:calibration:143GHz}{$\lsim 2\%$}
\setsymbol{HFI:CMB:absolute:calibration:217GHz}{$\lsim 2\%$}
\setsymbol{HFI:CMB:absolute:calibration:353GHz}{$\lsim 2\%$}
\setsymbol{HFI:CMB:absolute:calibration:545GHz}{}
\setsymbol{HFI:CMB:absolute:calibration:857GHz}{}

\setsymbol{HFI:FIRAS:gain:calibration:accuracy:statistical:100GHz}{}
\setsymbol{HFI:FIRAS:gain:calibration:accuracy:statistical:143GHz}{}
\setsymbol{HFI:FIRAS:gain:calibration:accuracy:statistical:217GHz}{}
\setsymbol{HFI:FIRAS:gain:calibration:accuracy:statistical:353GHz}{2.5\%}
\setsymbol{HFI:FIRAS:gain:calibration:accuracy:statistical:545GHz}{1\%}
\setsymbol{HFI:FIRAS:gain:calibration:accuracy:statistical:857GHz}{0.5\%}

\setsymbol{HFI:FIRAS:gain:calibration:accuracy:systematic:100GHz}{}
\setsymbol{HFI:FIRAS:gain:calibration:accuracy:systematic:143GHz}{}
\setsymbol{HFI:FIRAS:gain:calibration:accuracy:systematic:217GHz}{}
\setsymbol{HFI:FIRAS:gain:calibration:accuracy:systematic:353GHz}{}
\setsymbol{HFI:FIRAS:gain:calibration:accuracy:systematic:545GHz}{7\%}
\setsymbol{HFI:FIRAS:gain:calibration:accuracy:systematic:857GHz}{7\%}

\setsymbol{HFI:FIRAS:zero:point:accuracy:100GHz:units}{0.8\MJysr}
\setsymbol{HFI:FIRAS:zero:point:accuracy:143GHz:units}{}
\setsymbol{HFI:FIRAS:zero:point:accuracy:217GHz:units}{}
\setsymbol{HFI:FIRAS:zero:point:accuracy:353GHz:units}{1.4\MJysr}
\setsymbol{HFI:FIRAS:zero:point:accuracy:545GHz:units}{2.2\MJysr}
\setsymbol{HFI:FIRAS:zero:point:accuracy:857GHz:units}{1.7\MJysr}

\setsymbol{HFI:FIRAS:zero:point:accuracy:100GHz}{0.8}
\setsymbol{HFI:FIRAS:zero:point:accuracy:143GHz}{}
\setsymbol{HFI:FIRAS:zero:point:accuracy:217GHz}{}
\setsymbol{HFI:FIRAS:zero:point:accuracy:353GHz}{1.4}
\setsymbol{HFI:FIRAS:zero:point:accuracy:545GHz}{2.2}
\setsymbol{HFI:FIRAS:zero:point:accuracy:857GHz}{1.7}

\setsymbol{HFI:unit:conversion:100GHz:units}{0.2415\MJysrmK}
\setsymbol{HFI:unit:conversion:143GHz:units}{0.3694\MJysrmK}
\setsymbol{HFI:unit:conversion:217GHz:units}{0.4811\MJysrmK}
\setsymbol{HFI:unit:conversion:353GHz:units}{0.2883\MJysrmK}
\setsymbol{HFI:unit:conversion:545GHz:units}{0.05826\MJysrmK}
\setsymbol{HFI:unit:conversion:857GHz:units}{0.002238\MJysrmK}

\setsymbol{HFI:unit:conversion:100GHz}{0.2415}
\setsymbol{HFI:unit:conversion:143GHz}{0.3694}
\setsymbol{HFI:unit:conversion:217GHz}{0.4811}
\setsymbol{HFI:unit:conversion:353GHz}{0.2883}
\setsymbol{HFI:unit:conversion:545GHz}{0.05826}
\setsymbol{HFI:unit:conversion:857GHz}{0.002238}

\setsymbol{HFI:colour:correction:alpha=-2:V1.01:100GHz}{0.9893}
\setsymbol{HFI:colour:correction:alpha=-2:V1.01:143GHz}{0.9759}
\setsymbol{HFI:colour:correction:alpha=-2:V1.01:217GHz}{1.0007}
\setsymbol{HFI:colour:correction:alpha=-2:V1.01:353GHz}{1.0028}
\setsymbol{HFI:colour:correction:alpha=-2:V1.01:545GHz}{1.0019}
\setsymbol{HFI:colour:correction:alpha=-2:V1.01:857GHz}{0.9889}

\setsymbol{HFI:colour:correction:alpha=0:V1.01:100GHz}{1.0008}
\setsymbol{HFI:colour:correction:alpha=0:V1.01:143GHz}{1.0148}
\setsymbol{HFI:colour:correction:alpha=0:V1.01:217GHz}{0.9909}
\setsymbol{HFI:colour:correction:alpha=0:V1.01:353GHz}{0.9888}
\setsymbol{HFI:colour:correction:alpha=0:V1.01:545GHz}{0.9878}
\setsymbol{HFI:colour:correction:alpha=0:V1.01:857GHz}{1.0014}

\newcommand{\wmean}[1]{\left\langle #1\right\rangle_{\rm w}}
\newcommand{\WMAP}{\textit{WMAP}}
\renewcommand{\S}{{Sect.}}

\begin{document}

\title{Long-term variability of extragalactic radio sources in the \Planck\ Early Release Compact Source Catalogue}
\titlerunning{Long-term variability of extragalactic radio sources in the \Planck\ ERCSC}

\author{
X.~Chen\inst{1} 
\and
J.~P.~Rachen\inst{2, 8} 
\and 
M.~L\'{o}pez-Caniego\inst{3}
\and 
C.~Dickinson\inst{4}
\and 
T.~J.~Pearson\inst{1, 5}
\and
L.~Fuhrmann\inst{6}
\and
T.~P.~Krichbaum\inst{6}
\and 
B.~Partridge\inst{7}
}
\institute{\small
Infrared Processing and Analysis Center, California Institute of Technology,
Pasadena, CA 91125, U.S.A.
\and
Max-Planck-Institut f\"{u}r Astrophysik, Karl-Schwarzschild-Str. 1, 85741
Garching, Germany
\and 
Instituto de F\'{\i}sica de Cantabria (CSIC-Universidad de Cantabria), Avda. de
los Castros s/n, Santander, Spain
\and 
Jodrell Bank Centre for Astrophysics, Alan Turing Building, School of Physics
and Astronomy, The University of Manchester, Oxford Road, Manchester, M13 9PL,
U.K.
\and
California Institute of Technology, Pasadena, California, U.S.A.
\and
Max-Planck-Institut f\"{u}r Radioastronomie, Auf dem H\"{u}gel 69, 53121 Bonn,
Germany
\and
Haverford College Astronomy Department, 370 Lancaster Avenue, Haverford,
Pennsylvania, U.S.A.
\and
Department of Astrophysics / IMAPP, Radboud University Nijmegen, P.O. Box 9010,
6500
GL Nijmegen, The Netherlands 
}

\abstract{Combining measurements taken using the \textit{Wilkinson
    Microwave Anisotropy Probe} (\WMAP) from 2001 to 2008 with
  measurements taken using \Planck\ from 2009 to 2010, we investigate
  the long-term flux density variability of extragalactic radio
  sources selected from the \Planck\ Early Release Compact Source
  Catalogue. The single-year, single-frequency \WMAP\ maps are used to
  estimate yearly-averaged flux densities of the sources in the four
  \WMAP\ bands: Ka (33\,GHz), Q (41\,GHz), V (61\,GHz), and W
  (94\,GHz). We identify 82, 67, 32, and 15 sources respectively as variable at greater
than
$99\%$ confidence level in these four bands. The amplitudes of
variation are comparable between bands, and are not correlated with either
the flux densities or the spectral indices of the sources. The number counts of \WMAP\ Ka-band sources are
stable from year to year despite the fluctuation caused by individual source variability. Most of
our sources show strong correlation in variability between bands. Almost all the
sources that show variability are blazars. We have attempted to fit two simple,
four-parameter models to the time-series of 32 sources showing correlated
variability at multiple frequencies -- a long-term
flaring model and a rotating-jet model. We find that 19 sources
(60\%) can be fit with the simple rotating-jet model, and ten of
these also fit the simple
long-term flaring model. The remaining 13 sources (40\%) show more complex
variability behaviour that is not consistent with either model. Extended
radio galaxies in our sample show no sign of variability, as expected, with the
exception of Pictor A for which we report evidence for a millimetre flare
lasting between 2002 and 2010. }   

\keywords{surveys: radio sources -- radio continuum: galaxies -- BL Lacertae objects: general --
quasars: general -- galaxies: individual (Pic\,A)}

\maketitle   
   
\section{Introduction}
\label{sec:introduction}

Compact radio sources are the most significant contaminant at
small angular scales in high-precision cosmic microwave background
(CMB) experiments such as the \textit{Wilkinson Microwave Anisotropy
  Probe} (\WMAP, \citealt{Bennett2003a}) and the \Planck\footnote{\Planck\
(\url{http://www.esa.int/Planck}) is a project of the European Space Agency (ESA)
with instruments provided by two scientific consortia funded by ESA member
states (in particular the lead countries France and Italy), with contributions
from NASA (USA) and telescope reflectors provided by a collaboration between ESA
and a scientific consortium led and funded by Denmark.}  satellite
\citep{planck2011-1.1}.  In the frequency range 20--100\,GHz, which
includes all the \WMAP\ bands and the lower-frequency \Planck\
channels, the radio source population is dominated by compact,
flat-spectrum sources \citep{Bennett2003b, Hinshaw2007, Wright2009,
planck2011-6.1}, many of which are variable on time-scales ranging from
days to years \citep{planck2011-6.2, planck2011-6.3a}. The \Planck\ Early
Release Compact Source Catalogue (ERCSC, \citealt{planck2011-1.10}) lists bright sources detected by the \Planck\ satellite during the first 1.6 sky surveys at all nine frequency channels. For
some of the extragalactic radio sources, the spectral shape at \Planck\ 
frequencies suggests that they have been caught in a bright, flaring stage
\citep{planck2011-6.2}. Knowledge of variability is thus essential in order to
correctly account for the contamination of the CMB measurements by radio
sources.

Most variable extragalactic sources are active galactic nuclei (AGN),
and the most extreme variability is seen in blazars. 
Relativistically beamed, non-thermal emission dominates
the spectral energy distribution (SED) of blazars over other
components such as dust emission from the galaxy or
unbeamed synchrotron emission from radio galaxy lobes.
Relativistically beamed sources are
prone to variability for two reasons: (a)~the emission is strongly
Doppler boosted, so small changes of the Doppler factor caused by
variations of the bulk Lorentz factor or the viewing angle can lead to
strong variability; (b)~in a relativistic jet, any small perturbation at
the origin of the jet can cause the formation of shocks
\citep[e.g.,][]{Begelman1984}. Variability caused by changes in Doppler
boosting, also called \textit{geometric variability}, changes the spectrum only slowly if at all \citep[e.g.,][]{Begelman1980,
Camenzind1992, Steffen1995, Villata1999}.
Flares produced by shocks in the plasma flow can
lead to strongly frequency-dependent variability on shorter time scales,
while also contribute to largely achromatic variability on longer time scales \citep[e.g.,][]{Marscher1985, Valtaoja1988}.

Studies of radio source variability in the radio-to-millimetre regime on time-scales up to several decades have been facilitated by monitoring programs targeting medium or
large source samples selected by known variability \citep{Aller1996,
Tornikoski1996, Aller1999, Stevens1994}, radio spectrum
\citep{Valtaoja1992, Terasranta1998, Nieppola2007}, or gamma-ray brightness
\citep{Fuhrmann2013}. Discussions of blazar variability have mostly been focused
on understanding the characteristics of flares,
which are observed on time scales ranging from one day or less \citep{Wagner1995} to weeks 
or months \citep[e.g.,][]{Angelakis2012} to several years \citep{Hovatta2009}, and their interpretation in
the framework of jet models \citep[e.g.,][]{Valtaoja1999, Lahteenmaki1999a,
Lahteenmaki1999b, Turler2000, Ciaramella2004, Hovatta2009, Angelakis2012}. 
While it is generally accepted that shock-produced flares dominate the
variability on very short time scales (days to weeks), geometric
variability should have an effect
on longer time scales because rotation or precession, as suggested by VLBI observations of many jets
(e.g., \citealt{Kellermann2004}), would inevitably lead to changes in Doppler
boosting of the radiative zones in the jet. Indeed, \citet{Bach2006} have modelled
optical and radio time-series of blazars observed in the monitoring campaigns of
the Whole Earth Blazar Telescope \citep[WEBT,][]{Villata2002,Villata2004} in the
framework of inhomogeneous, rotating helical jets \citep{Villata1998,
Villata1999, Ostorero2004}, and have  suggested that all variability down to time scales of
${\sim}\,100\,$days can be explained by geometric effects alone.

In this paper, we report results on long-term
flux density variations of a sample of sources selected from the \Planck\ ERCSC.
We primarily use data collected by \WMAP\ and \Planck. 
\WMAP\ started its observations at the second Sun-Earth Lagrange
point (L2) in August 2001, and ended the collection of science data on 19 August 2010. \Planck\ started its observation at L2
on 13 August 2009 and is still taking data. Both telescopes take about 6 months to complete a single sky survey.
Since the lower four frequency bands of \Planck\ overlap with the upper four frequency bands
of \WMAP, and both telescopes have the powerful ability to make near-simultaneous 
measurements at multiple frequency bands, combining the observations from the two experiments 
across an interval of nearly ten years provides the best opportunity to track the long-term variations
of radio sources at frequencies between 30 and 100\,GHz. Table~\ref{tab:basics} lists the central frequencies
and beam-widths of the \WMAP\ and \Planck\ bands that are of interest here.  

We describe the selection of our sample from the \Planck\
ERCSC in Sect.~\ref{sec:sample}. In Sect.~\ref{sec:data}  we introduce the
\WMAP\ data used in our analysis and
the method of estimating source flux densities from the \WMAP\ maps. 
Source variability is quantified and studied statistically in
Sect.~\ref{sec:stats}. In Sect.~\ref{sec:interpretation}, we
analyse the long-term averaged variability patterns of strongly variable blazars
in our sample. We then discuss the interesting case of long-term variability in
unbeamed radio sources in Sect.~\ref{sec:unbeamed}. The results are put into
perspective with other studies in Sect.~\ref{sec:discussion}, and we conclude
in Sect.~\ref{sec:conclusions}.

\begin{table*}
\begin{center}
\caption{Characteristics of relevant \WMAP\ and \Planck\ bands in this study.\label{tab:basics}}
\begin{tabular}{ccccccc}
\hline
\hline
\noalign{\vskip2pt}
\multicolumn{3}{c}{\WMAP} & & \multicolumn{3}{c}{\Planck} \\
\noalign{\vskip2pt}
\multispan{3}\hrulefill & & \multispan{3}\hrulefill \cr 
\noalign{\vskip3pt}
Band & Center Frequency [GHz] & Beam FWHM [arcmin] & & Band & Center Frequency [GHz] & Beam FWHM [arcmin]  \\
\noalign{\vskip 3pt\hrule\vskip 5pt}
K & 22.8 & 49.2 &  &  &  &  \\
Ka  & 33.0 & 37.2 & & \phantom{0}30 & \phantom{0}28.5 & 32.65 \\
Q  & 40.7 & 29.4 &  & \phantom{0}44 & \phantom{0}44.1 & 27.92 \\
V  & 60.8 & 19.8 & & \phantom{0}70 & \phantom{0}70.3 & 13.01 \\
W  & 93.5 & 12.6 & & 100 & 100.0 & \phantom{0}9.37 \\
\noalign{\vskip3pt\hrule}
\end{tabular}
\tablefoot{Values for \WMAP\ are from \citet{Bennett2003a} and those for \Planck\ are from \citet{planck2011-1.1}.}
\end{center}
\end{table*}

\section{Sample Selection and \Planck\ Data}
\label{sec:sample}

\begin{figure*}
\centering
\includegraphics[width=0.54\textwidth, angle = 90]{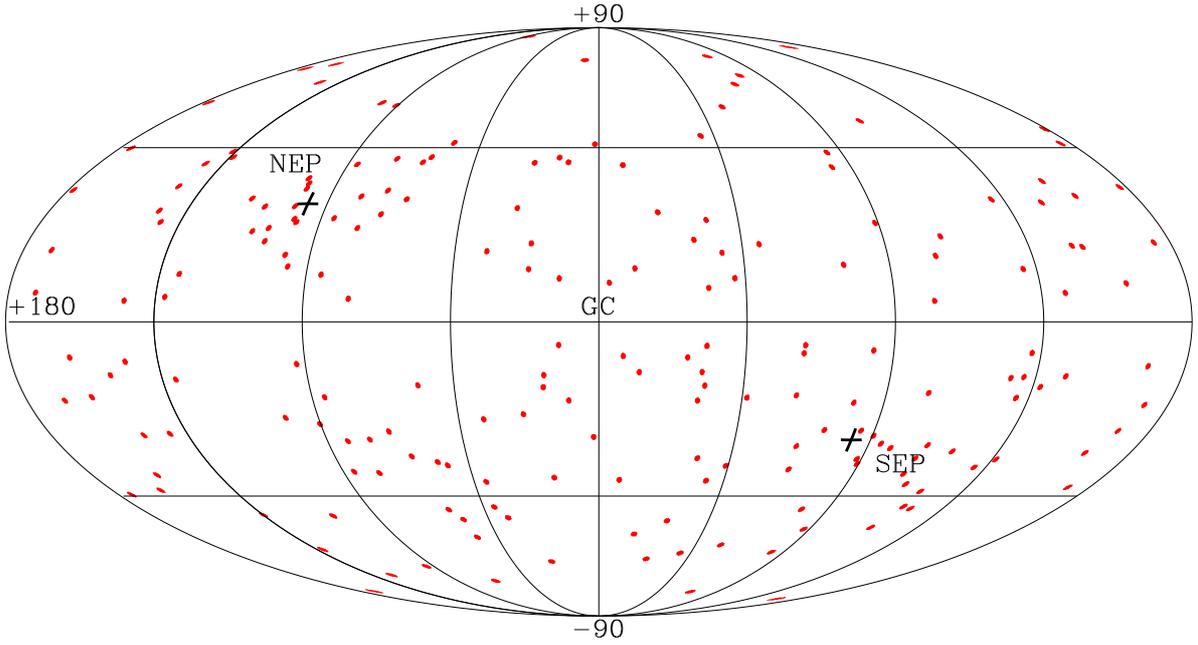}
\caption{Map in Galactic coordinates showing the locations of the 198 extragalactic sources selected from the \Planck\ ERCSC. The north ecliptic pole (NEP) and the south ecliptic pole (SEP) are indicated by crosses. Source densities are higher near the ecliptic poles, where \Planck\ has longer integration time.  \label{fig:skymap}}
\end{figure*}

The \Planck\ Early Release Compact Source Catalogue \citep{planck2011-1.10} is a catalogue
of highly reliable sources (with $\ge 90\%$ cumulative reliability), both Galactic and extragalactic, detected over the
entire sky in the first 1.6 \Planck\ sky surveys. This catalogue provides separate lists of
sources detected in each of the nine \Planck\
frequency channels (30 to 857\,GHz). We restricted our sample
selection to the \Planck\ 30, 44, 70, and 100\,GHz channels that overlap with the \WMAP\ Ka to W bands, and excluded sources close to the Galactic plane ($|b| < 5\degr$). We first collated the lists at these four frequencies and constructed a
``band-merged'' catalogue listing the sources detected in all four channels. The band-merging process was based on positions only. No brightness information was used, to avoid biasing
towards or against any class of sources. 
The source list at each frequency was used as the {\it seed} catalogue and counterparts were sought in the {\it candidate} catalogues, i.e., the source lists at the other frequencies. The matching radius was always defined as half of the 
full width at half maximum (FWHM) of whichever band (seed or candidate) has lower
resolution. A confusion check was then
performed to ensure a one-to-one reciprocal relation between matches across all four
bands. Due to the change in resolution, there was one case in which the 44\,GHz source could be matched with two sources at
100\,GHz; we excluded this source. We then cross-checked this list with both the
NASA/IPAC Extragalactic Database (NED\footnote{\url{http://ned.ipac.caltech.edu/}}) and the SIMBAD Astronomical Database\footnote{\url{http://simbad.u-strasbg.fr/simbad/}.} to
remove any Galactic objects (\ion{H}{ii} regions, planetary nebulae,
supernova remnants, etc.). This process yielded a final list of 198 extragalactic
sources, including 135 flat-spectrum radio quasars (FSRQs), 26 BL Lacertae
objects (BLLs), 28 non-blazar active galaxies (AGNs, including many extended radio galaxies), 1 starburst galaxy (SBG),
1 normal galaxy (GAL), and 7 radio sources of uncertain type. The classification
of the active galaxies was taken from the Candidate
Gamma-Ray Blazar Survey (CGRaBS, \citealt{Healey2008}), the third edition of the
Roma-BZCAT blazar catalogue \citep{Massaro2011}, and the twelfth edition of the
Catalogue of Quasars and Active Nuclei \citep{Veron2006}. Fig.~\ref{fig:skymap}
shows the distribution of the sources on the sky.
We emphasize that there is no guarantee that the sample is complete. 
In addition, requiring that a source be seen in all the \Planck\ bands between 30 and 100\,GHz will inevitably favour flat-spectrum radio sources.

\begin{figure}
\centering
\includegraphics[width=0.45\textwidth]{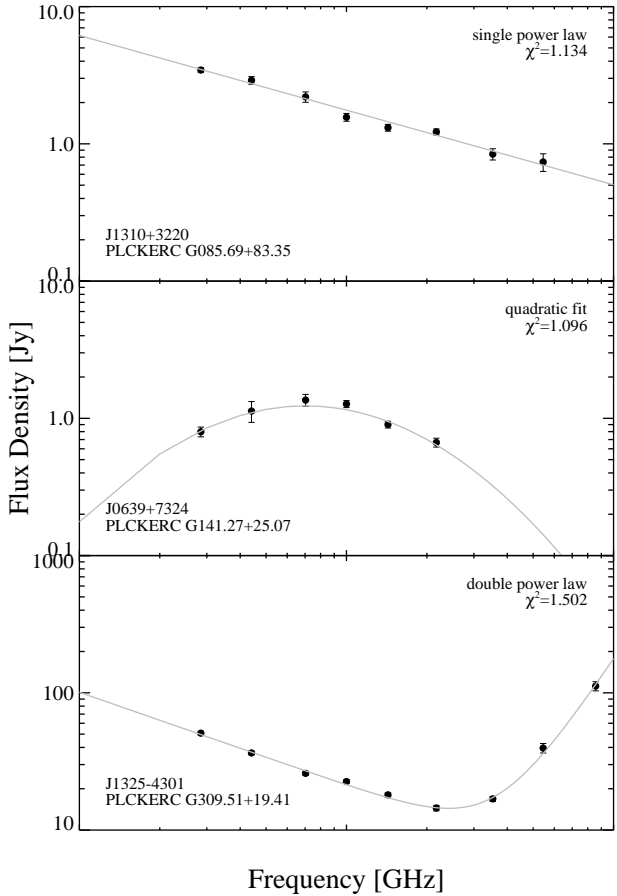}
\caption{Examples of source SEDs and fitted models (grey solid line). The reduced $\chi^2$ values are indicated. \label{fig:sed}}
\end{figure}

The calibration of the \Planck\ 30--100\,GHz maps is currently based on the CMB dipole \citep{planck2011-1.6,
planck2011-1.7}. Since the ERCSC flux densities were calculated directly
from the maps and are therefore only correct if the source has a CMB spectrum, 
colour corrections are needed for sources with different spectra. To determine the spectra, we 
looked for the counterparts of our sources in the higher-frequency (143--857\,GHz) ERCSC source lists.
We then fit the SED of each source over all the frequencies at which it was detected,
using three simple models, when there were enough data points for fitting:

\noindent i) single power law 
\begin{equation}
\log S = p_0 + p_1 \log\nu \,; 
\label{eq:sp}
\end{equation}
ii) quadratic fit
\begin{equation}
\log S = p_0 + p_1 \log\nu + p_2 (\log\nu)^2 \,; 
\end{equation}
iii) double power law 
\begin{equation}
S = p_0 (\nu/\nu_0)^{p_1} + p_2 (\nu/\nu_0)^{p_3} \,.
\end{equation}
Here $p_i \, (i\,{=}\,0, 1, 2, 3)$ denote parameters that are different in each
model.
Since the \Planck\ 100\,GHz data can be significantly contaminated by the CO ($J$=1$\rightarrow$0) line at 115\,GHz \citep{planck2011-1.7}, we did not include the 100\,GHz
flux densities in the SED fitting. A model fit was accepted when the $\chi^2$ values indicated that the model was compatible with 95\% of the data. In cases when more than one model was acceptable, we always selected the
model with fewest parameters unless an additional parameter significantly
improved the value of the reduced $\chi^2$ (as described in
\citealt{Bevington2003}). A few examples of the source SEDs
and our fitted models are given in Fig.~\ref{fig:sed}. For sources with a
good fit (144 out of the 198 sources), we used the model to estimate
the spectral index $\alpha~(S \propto \nu^{\,\alpha})$ at the \Planck\ central
frequencies. For sources that could not be reasonably fit by any of the models, flux densities in the adjacent bands were used to estimate the spectral
indices.  We then applied the corresponding colour corrections given in
\citet{planck2011-1.6} and \citet{planck2011-1.7}. 
These range from almost no change to 5.7$\%$, depending on frequency and source spectral shape.
The absolute calibration errors are at the 1$\%$ level for the \Planck\ 30--70\,GHz channels, and 2$\%$ level for the 100\,GHz channel. 
We conservatively assumed an overall calibration uncertainty of 3$\%$ and added it in quadrature with the ERCSC flux density errors prior to modelling  the spectra.

\section{\WMAP\ Data}
\label{sec:data}

The \WMAP\ individual-year maps from 2001 to 2008 and the flux measurements of our
sources from these maps constitute the main dataset of this paper, which is described
in  this section. To support our analysis, we have also used some ground-based data from the Effelsberg 100-m telescope, the IRAM 30-m
telescope, and the Mets\"{a}hovi 14-m telescope. We defer the description of
these data to the sections where they are used. 

\subsection{\WMAP\ maps and photometry}

The high-resolution ($N_{\rm side} = 1024$ in the HEALPix scheme, \citealt{Gorski2005}) \WMAP\
single-year, single differencing-assembly (DA) maps\footnote{All the maps are
available from NASA's Legacy Archive for Microwave Background Data Analysis
(LAMBDA) at \url{http://lambda.gsfc.nasa.gov/}.} from 2001 to 2008 are the basis of
this study.  Since the ERCSC was derived from single-frequency \Planck\ maps, we
performed a weighted, pixel-by-pixel, mean of the single DA maps from each year to
generate single-year maps at each  \WMAP\ frequency. The number of observations per pixel was used 
as the weight. The average temperature outside the \WMAP\ seven-year KQ85
mask \citep{Gold2011} was subtracted from the single DA maps before the
weighted mean was calculated.

For consistency with the ERCSC flux density measurements, we used
aperture photometry to obtain flux densities for the \Planck-detected sources from the
\WMAP\ maps. The source position was taken from the ERCSC 100\,GHz catalogue as this band has the highest resolution of the four bands considered here. We measured the flux density in a circular aperture with
radius equal to 1.1 times the nominal FWHM of the \WMAP\ beam at each
frequency, and subtracted an estimate of the local background, computed
as the median intensity within a concentric annulus. The size of the aperture is chosen to include $\gtrsim 95\%$ of the source power in the K- to V-bands, which was calculated from the \WMAP\ seven-year beam radial profile\footnote{Also available from NASA's LAMBDA site.}. Due to a significant shoulder in the W-band beam profile at 0\fdg2--0\fdg5 \citep{page2003a}, this aperture only includes $\sim 85\%$ of the source flux in the W-band. To have $\gtrsim 95\%$ of the source power in the aperture, we would have to adopt an aperture radius of 2$\times$FWHM in this band, which would then include too much noise from the background. We therefore decided to stay with an aperture radius of 1.1$\times$FWHM for the W-band, and correct for the flux loss using Monte Carlo simulations (see the next paragraph). Similarly, we 
calculated at each band the radial distance from the center of the beam
that includes 97.5$\%$ of the source power ($\sim$1.3, 1.3, 1.4,
1.7, or 2.4 times the corresponding beam FWHM at K, Ka, Q, V, or W
band), and used this for the inner radius of the annulus to avoid
over-subtraction of the source. We then chose the outer radius to
make the area of the annulus approximately the same as
the aperture. 
 
For studying variability, it is important to have unbiased flux
density measurements and reliable estimates of their uncertainties.
Since we do not know whether the
aperture is perfectly centered on each source and we do not have
precise estimates of the background, we estimated aperture corrections
using a Monte Carlo method. We injected fake point sources into the
\WMAP\ maps at random positions and compared the extracted flux
densities with the input. The flux densities of the artificial sources
had a flat $d \log N/d\log S$ distribution (equal numbers of sources per decade of $S$) in the range $100\,{\rm mJy} < S
< 200\,{\rm Jy}$. The sources were first injected into single pixels
in an empty $N_{\rm side} = 4096$ map, which was then smoothed with
the {\it WMAP} beam profile appropriate for the chosen frequency and
degraded to $N_{\rm side} = 1024$ before being added to the \WMAP\
single-year maps at that frequency. We then performed the same
aperture photometry at the input source locations, excluding those
at $|b| < 5\degr$ and those within 2\degr\ of other fake
sources or sources in the \WMAP\ seven-year source catalogue
 \citep{Gold2011}. To minimize collisions of fake sources, we injected 1000 sources in each map and iterated five times to 
get a sample of ${\sim}\,5000$ fake measurements. We used the distribution of the ratio of extracted to input flux density to estimate a bias correction factor, which was  typically ${\sim}\,10$--$20\%$ except at the W-band. 

We estimated the uncertainties in the flux densities obtained from the aperture
  photometry by error
  propagation in the formula for the aperture flux density, assuming
  white uncorrelated noise. This assumption is justified for instrumental noise, but not for noise in the background. We therefore used Monte Carlo simulations to test our error estimates, by injecting 1000 sources with the same flux density (from 1\,Jy to 20\,Jy in 1\,Jy steps) in each map, and using the
    dispersion of the recovered flux densities as an indicator of the
  ``true" uncertainty. We found that the flux density uncertainties
  from the white noise model were always less than those returned from Monte Carlo
 simulations, as expected. However, when repeating the same process on CMB-subtracted maps
  (subtracting the \WMAP\ seven-year internal linear combination CMB map at $|b| >
  30\degr$), the simulation results are in good agreement with the white-noise estimates, showing that the correlated noise is mostly CMB. Since CMB noise is
  constant from year to year and therefore will not affect the
  variability analysis, we decided to use the flux density uncertainty
  returned directly from the white-noise model in our study.  

The
  absolute calibration of \WMAP\ data is based on the CMB monopole temperature and the
  velocity-dependent dipole resulting from \WMAP's orbit about the
  solar system barycentre, and has an absolute
  calibration uncertainty of $0.2\%$ \citep{Jarosik2011}. The aperture correction error measured
  from simulations varied between $0.4\%$ and $1.3\%$ in the K to W bands. We
  therefore assumed an overall $2\%$ error from aperture correction and
  calibration, and added it in quadrature with the flux density uncertainty to constitute the final error term for the flux density of
  each source.

We fit the SED of each source at each year with a single power law
(Eq.~\ref{eq:sp}), and applied colour corrections to the central frequencies,
following the recipe of \cite{Jarosik2003}. The \WMAP\ single-year maps are generally much noisier than the \Planck\ maps,
especially at the V and W bands where even the seven-year co-added maps are 
less sensitive than the one-year \Planck\ maps \citep[see][Fig.~5]{planck2011-1.10}. As a
result, the flux density estimates can sometimes be negative for faint
sources at certain frequencies in some years. In the SED fitting, we only used the good quality (defined as $S > 3\sigma$) flux density measurements. 
A 3$\sigma$ threshold ($\sim 99.7\%$ probability that the signal is truly positive assuming the background is Gaussian) is adopted and considered sufficient in our case as we look for a positive signal at the location of a known source. We do, however, caution that in non-Gaussian backgrounds a 3$\sigma$ threshold raises the risk of deriving incorrect flux density and error estimates.
Although the \WMAP\ K-band (23\,GHz) flux densities were not used in the variability study
because there is no corresponding channel in \Planck, we included them in the SED
fitting to obtain more accurate spectral indices.

The \WMAP\ single-year flux densities for our sources, together with
the \Planck\ ERCSC flux densities extrapolated to the \WMAP\ centre
frequencies (using the SED models described in \S\,\ref{sec:sample}),
are listed in Table~\ref{tab:sample}. These flux density values are  used to construct light curves for all the sources in our sample. 
The full set of light curves is available in the online supplementary material, arranged in ascending order of Right Ascension. 
Only light curves that have four or more good quality
measurements in the seven years of \WMAP\ data are plotted (see
\S\,\ref{sec:stats} for detailed discussion). For sources that are specifically discussed in the paper, individual light curves are presented where they are discussed. 
We show all the flux density measurements with $S > 2\sigma$, with dash-dot lines indicating
$3\sigma$ noise level at each frequency. Since the flux measurements here are generally averages over a fraction of a year (see \S~\ref{sec:scan}), all the data points are plotted at a nominal epoch in February of each year, as both \WMAP\ and \Planck\ single year observations start in mid-August. 
The effective epoch of each  observation may differ by a few months from the nominal epoch.

\begin{table*}[p]
\centering
\rotatebox{90}{\begin{minipage}{\textheight}
\caption{Flux densities at \WMAP\ Ka, Q, V, and W band for the 198 extragalactic radio sources selected from the ERCSC (Example, full table can be retrieved online).}
\label{tab:sample}
\centering
\begin{tabular}{cccccrrrrrrrrrc}
\hline\hline   
\noalign{\vskip3pt}              
\footnotesize
Name\tablefootmark{a} & GLON & GLAT & Type & Band & \multicolumn{1}{c}{WMAP1} & \multicolumn{1}{c}{WMAP2} & \multicolumn{1}{c}{WMAP3} & \multicolumn{1}{c}{WMAP4} & \multicolumn{1}{c}{WMAP5} & \multicolumn{1}{c}{WMAP6} & \multicolumn{1}{c}{WMAP7} & \multicolumn{1}{c}{ERCSC\tablefootmark{b}}  & \multicolumn{1}{c}{$V_{\rm rms}$\tablefootmark{c}}  & Notes\tablefootmark{d}  \\
 & [deg] & [deg] &  &  & \multicolumn{1}{c}{[Jy]} & \multicolumn{1}{c}{[Jy]} & \multicolumn{1}{c}{[Jy]} & \multicolumn{1}{c}{[Jy]} & \multicolumn{1}{c}{[Jy]} & \multicolumn{1}{c}{[Jy]} & \multicolumn{1}{c}{[Jy]} & \multicolumn{1}{c}{[Jy]} & &  \\
\noalign{\vskip3pt\hrule\vskip3pt}                              
    J0006$-$0623 &  \phantom{0}93.503 &  $-$66.636 &   BLL      & Ka &       1.44$\pm$0.33  &       1.88$\pm$0.30  &       1.86$\pm$0.29  &       1.54$\pm$0.29  &       1.26$\pm$0.29  &       1.05$\pm$0.30  &       0.77$\pm$0.30  &       1.46$\pm$0.08  &  17.24 &   \\
    (0003$-$066) & & & &  Q &       1.69$\pm$0.37  &       1.32$\pm$0.34  &       2.31$\pm$0.34  &       1.74$\pm$0.35  &       1.50$\pm$0.35  &       0.67$\pm$0.41  &       1.20$\pm$0.36  &       1.43$\pm$0.07  &  10.96 &   \\
  &  &  &  &  V &       1.72$\pm$0.58  &       1.23$\pm$0.54  &       1.19$\pm$0.63  &       1.14$\pm$0.56  &       1.49$\pm$0.59  &       0.90$\pm$0.56  &       1.27$\pm$0.63  &       1.35$\pm$0.09  &    &   \\
  &  &  &  &  W &      $-$0.04$\pm$0.93  &       0.45$\pm$0.84  &       0.48$\pm$0.91  &       1.17$\pm$0.85  &       3.95$\pm$1.09  &      $-$0.37$\pm$0.72  &       1.25$\pm$0.84  &       1.22$\pm$0.08  &   &   \\
    J0010+1058 &  106.981 &  $-$50.622 &   AGN      & Ka &       0.65$\pm$0.29  &       1.31$\pm$0.30  &       1.60$\pm$0.32  &       2.09$\pm$0.30  &       0.90$\pm$0.29  &       0.92$\pm$0.30  &       0.20$\pm$0.30  &       2.51$\pm$0.10  &  29.68 & v \\
    (Mrk 1501) & & & &  Q &       0.38$\pm$0.37  &       0.70$\pm$0.37  &       1.54$\pm$0.34  &       2.11$\pm$0.37  &       1.04$\pm$0.36  &       0.53$\pm$0.37  &       1.75$\pm$0.36  &       2.68$\pm$0.16  &    &   \\
  &  &  &  &  V &      $-$0.13$\pm$0.63  &       0.75$\pm$0.56  &       1.73$\pm$0.63  &       0.75$\pm$0.60  &       0.90$\pm$0.61  &       0.54$\pm$0.64  &       1.93$\pm$0.60  &       2.27$\pm$0.17  &    &   \\
  &  &  &  &  W &       0.69$\pm$0.99  &       1.07$\pm$0.97  &       2.20$\pm$1.12  &       3.54$\pm$0.75  &       2.01$\pm$0.93  &      $-$1.11$\pm$0.91  &       0.98$\pm$1.01  &       1.89$\pm$0.10  &   &   \\
    J0050$-$5738 &  303.285 &  $-$59.481 &   FSRQ     & Ka &       0.97$\pm$0.23  &       0.94$\pm$0.22  &       0.92$\pm$0.23  &       1.61$\pm$0.22  &       1.38$\pm$0.23  &       1.64$\pm$0.22  &       1.55$\pm$0.23  &       1.33$\pm$0.09  &  18.85 & v \\
    (0047$-$579) & & & &  Q &       1.30$\pm$0.27  &       1.21$\pm$0.25  &       1.77$\pm$0.26  &       1.62$\pm$0.27  &       1.27$\pm$0.26  &       1.14$\pm$0.27  &       1.45$\pm$0.27  &       1.36$\pm$0.18  &  $-$9.43 &   \\
  &  &  &  &  V &       1.46$\pm$0.42  &       1.03$\pm$0.45  &       1.21$\pm$0.45  &       1.40$\pm$0.46  &       0.57$\pm$0.43  &       0.42$\pm$0.44  &       2.08$\pm$0.45  &       1.25$\pm$0.15  &    &   \\
  &  &  &  &  W &       0.24$\pm$0.72  &       1.53$\pm$0.73  &       0.94$\pm$0.57  &       1.71$\pm$0.81  &       1.24$\pm$0.69  &       0.52$\pm$0.68  &       0.04$\pm$0.68  &       0.81$\pm$0.08  &   &   \\
    J0051$-$0650 &  122.730 &  $-$69.711 &   FSRQ     & Ka &       1.53$\pm$0.30  &       1.11$\pm$0.30  &       1.44$\pm$0.31  &       1.36$\pm$0.30  &       1.16$\pm$0.29  &       1.24$\pm$0.30  &       2.01$\pm$0.30  &       1.38$\pm$0.08  &   5.43 &   \\
    (0048$-$071) & & & &  Q &       1.70$\pm$0.38  &       0.77$\pm$0.36  &       1.06$\pm$0.37  &       1.17$\pm$0.38  &      $-$0.41$\pm$0.38  &       0.76$\pm$0.38  &       1.38$\pm$0.35  &       1.49$\pm$0.09  &    &   \\
  &  &  &  &  V &       2.25$\pm$0.60  &       1.25$\pm$0.63  &       1.58$\pm$0.57  &       0.44$\pm$0.58  &       2.32$\pm$0.59  &       2.62$\pm$0.60  &       3.78$\pm$0.68  &       1.55$\pm$0.11  &  12.76 &   \\
  &  &  &  &  W &       2.09$\pm$0.90  &       0.66$\pm$0.94  &       1.94$\pm$0.90  &       2.70$\pm$0.83  &       1.94$\pm$0.85  &      $-$0.89$\pm$0.92  &       2.05$\pm$1.16  &       1.39$\pm$0.09  &    &   \\
    J0106$-$4034 &  290.673 &  $-$76.187 &   FSRQ     & Ka &       2.37$\pm$0.22  &       1.42$\pm$0.22  &       1.90$\pm$0.22  &       2.64$\pm$0.22  &       3.77$\pm$0.22  &       3.51$\pm$0.22  &       3.40$\pm$0.22  &       1.30$\pm$0.07  &  31.75 & V \\
    (0104$-$408) & & & &  Q &       2.34$\pm$0.28  &       1.67$\pm$0.28  &       2.10$\pm$0.29  &       2.60$\pm$0.28  &       3.55$\pm$0.28  &       3.39$\pm$0.28  &       3.17$\pm$0.29  &       1.33$\pm$0.08  &  24.00 & V \\
  &  &  &  &  V &       1.47$\pm$0.45  &       1.78$\pm$0.47  &       2.24$\pm$0.49  &       1.26$\pm$0.49  &       3.39$\pm$0.44  &       3.50$\pm$0.46  &       2.47$\pm$0.45  &       1.23$\pm$0.11  &  30.58 & V \\
  &  &  &  &  W &       3.27$\pm$0.60  &       1.57$\pm$0.62  &       1.57$\pm$0.75  &       3.22$\pm$0.67  &       2.74$\pm$0.63  &       2.67$\pm$0.73  &       2.51$\pm$0.76  &       0.97$\pm$0.08  & $-$15.32 &   \\
    J0108+0135 &  131.817 &  $-$60.985 &   FSRQ     & Ka &       2.33$\pm$0.30  &       3.32$\pm$0.31  &       2.43$\pm$0.30  &       1.12$\pm$0.31  &       1.45$\pm$0.32  &       1.05$\pm$0.30  &       1.56$\pm$0.31  &       3.13$\pm$0.10  &  40.77 & V \\
    (0106+013) & & & &  Q &       2.05$\pm$0.36  &       3.06$\pm$0.37  &       1.39$\pm$0.36  &       1.87$\pm$0.36  &       0.21$\pm$0.34  &       0.47$\pm$0.37  &       0.09$\pm$0.39  &       2.85$\pm$0.08  &  37.99 & V \\
  &  &  &  &  V &       2.13$\pm$0.56  &       2.36$\pm$0.62  &       2.31$\pm$0.61  &       1.64$\pm$0.63  &       0.46$\pm$0.63  &       0.67$\pm$0.62  &       1.68$\pm$0.57  &       2.38$\pm$0.06  &    &   \\
  &  &  &  &  W &      $-$1.17$\pm$1.15  &       1.89$\pm$1.06  &      $-$1.34$\pm$1.17  &      $-$1.63$\pm$1.20  &      $-$0.34$\pm$0.73  &       1.49$\pm$0.91  &       0.07$\pm$1.20  &       1.97$\pm$0.05  &    &   \\
    J0121+1149 &  134.577 &  $-$50.354 &   FSRQ     & Ka &       0.94$\pm$0.32  &       0.82$\pm$0.31  &       0.66$\pm$0.32  &       0.99$\pm$0.31  &       0.39$\pm$0.33  &       1.77$\pm$0.31  &       1.81$\pm$0.31  &       1.75$\pm$0.08  &   &   \\
    (0119+115) & & & &  Q &       0.38$\pm$0.39  &       0.92$\pm$0.36  &       1.03$\pm$0.36  &       0.83$\pm$0.37  &       0.81$\pm$0.35  &       1.57$\pm$0.35  &       2.65$\pm$0.37  &       1.56$\pm$0.07  &   &   \\
  &  &  &  &  V &       0.74$\pm$0.63  &      $-$0.16$\pm$0.60  &       0.95$\pm$0.66  &       0.63$\pm$0.63  &       1.14$\pm$0.59  &       1.03$\pm$0.59  &       0.30$\pm$0.58  &       1.26$\pm$0.05  &   &   \\
  &  &  &  &  W &       0.65$\pm$0.81  &      $-$0.79$\pm$0.94  &       0.15$\pm$0.89  &      $-$1.63$\pm$0.89  &       0.27$\pm$0.77  &      $-$0.40$\pm$0.93  &       0.84$\pm$1.07  &       1.00$\pm$0.05  &   &   \\
    J0132$-$1655 &  168.145 &  $-$76.020 &   FSRQ     & Ka &       2.10$\pm$0.28  &       2.56$\pm$0.30  &       2.38$\pm$0.30  &       2.47$\pm$0.30  &       2.18$\pm$0.29  &       1.82$\pm$0.29  &       1.87$\pm$0.29  &       1.71$\pm$0.08  &  $-$3.11 &   \\
    (0130$-$171) & & & &  Q &       1.66$\pm$0.33  &       2.09$\pm$0.33  &       2.20$\pm$0.36  &       1.76$\pm$0.36  &       2.16$\pm$0.35  &       1.54$\pm$0.34  &       1.29$\pm$0.34  &       1.53$\pm$0.06  &   0.73 &   \\
  &  &  &  &  V &       0.95$\pm$0.51  &       2.71$\pm$0.55  &       2.28$\pm$0.54  &       2.77$\pm$0.56  &       1.60$\pm$0.58  &       2.21$\pm$0.58  &       1.24$\pm$0.52  &       1.24$\pm$0.04  &  $-$7.05 &   \\
  &  &  &  &  W &       1.56$\pm$0.84  &       4.50$\pm$0.77  &       0.38$\pm$0.78  &       1.54$\pm$0.98  &       0.73$\pm$1.01  &       1.49$\pm$0.75  &       2.62$\pm$0.89  &       0.99$\pm$0.04  &   &   \\
    J0136+4751 &  130.788 &  $-$14.316 &   FSRQ     & Ka &       4.33$\pm$0.28  &       4.86$\pm$0.28  &       4.61$\pm$0.28  &       4.13$\pm$0.29  &       3.50$\pm$0.28  &       3.66$\pm$0.29  &       4.32$\pm$0.29  &       4.68$\pm$0.12  &   9.59 & V \\
    (0133+476) & & & &  Q &       5.02$\pm$0.37  &       5.42$\pm$0.38  &       4.86$\pm$0.36  &       4.24$\pm$0.39  &       2.42$\pm$0.35  &       4.10$\pm$0.36  &       4.01$\pm$0.36  &       4.42$\pm$0.11  &  21.99 & V \\
  &  &  &  &  V &       4.15$\pm$0.58  &       6.32$\pm$0.56  &       3.89$\pm$0.57  &       3.39$\pm$0.60  &       2.98$\pm$0.56  &       3.44$\pm$0.61  &       3.33$\pm$0.63  &       3.92$\pm$0.12  &  25.29 & V \\
  &  &  &  &  W &       3.15$\pm$0.90  &       2.49$\pm$0.69  &       1.73$\pm$0.85  &       1.80$\pm$0.93  &       0.07$\pm$0.95  &       0.66$\pm$1.09  &       3.56$\pm$0.90  &       3.40$\pm$0.11  &    &   \\
    J0137$-$2431 &  201.403 &  $-$79.286 &   FSRQ     & Ka &       0.53$\pm$0.28  &       1.44$\pm$0.26  &       1.43$\pm$0.28  &       1.07$\pm$0.24  &       1.23$\pm$0.27  &       1.05$\pm$0.27  &       0.57$\pm$0.27  &       2.45$\pm$0.09  &  $-$8.72 &   \\
    (0135$-$247) & & & &  Q &       0.82$\pm$0.33  &       1.72$\pm$0.32  &       1.37$\pm$0.31  &       1.64$\pm$0.31  &       1.27$\pm$0.33  &       1.20$\pm$0.33  &       1.41$\pm$0.33  &       2.49$\pm$0.15  & $-$14.46 &   \\
  &  &  &  &  V &       0.80$\pm$0.53  &       1.12$\pm$0.50  &       3.21$\pm$0.56  &       1.72$\pm$0.52  &       1.52$\pm$0.50  &       0.98$\pm$0.52  &       1.69$\pm$0.54  &       2.73$\pm$0.14  &  14.00 &   \\
  &  &  &  &  W &      $-$1.96$\pm$0.81  &       2.39$\pm$0.82  &       2.73$\pm$1.01  &       1.47$\pm$0.81  &       1.30$\pm$0.81  &      $-$1.11$\pm$0.75  &      $-$0.23$\pm$0.83  &       2.25$\pm$0.09  &    &   \\
\noalign{\vskip3pt\hrule}                              
\end{tabular}
\tablefoot{
\tablefoottext{a}{The J2000 name for each source is given. For easy reference, we have also included in parentheses the commonly known name or its B1950 name.}
\tablefoottext{b}{These are the ERCSC flux densities extrapolated to the center frequencies of \WMAP\ Ka, Q, V, and W-band.}
\tablefoottext{c}{This field is left blank when a timeline is not included in the variability analysis. Negative values indicate that the flux density fluctuation level is less than the noise level (see \S\,\ref{sec:var}).}
\tablefoottext{d}{``V'' denotes strongly variable sources (${>}\,99\%$ confidence), and ``v'' denotes moderately variable sources (${>}\,95\%$ confidence) at a given frequency.}
}
\end{minipage}}
\end{table*}

\subsection{Difference between \WMAP\ and \Planck\ flux density measurements}
\label{sec:scan}

\WMAP\ and \Planck\ employ different scanning strategies, and when
comparing flux density measurements from the two telescopes it is
important to understand how they are affected by the scanning strategy.

The \WMAP\ satellite is a differential experiment, and its
optical system consists of two back-to-back telescopes that measure the
difference in temperature between two points separated by 141$\deg$ on the sky. The satellite spins around its
symmetry axis once every 2.2\,min while the spin axis precesses at 22.5$\deg$\,h$^{-1}$
about the Sun--\WMAP\ line. Since each telescope line of sight is 
${\sim}\,70\deg$ off the symmetry axis, the combined motion of spin and precession
causes the observing beams to fill an annulus centered on the local solar vector,
with inner and outer radii of ${\sim}\,48\deg$ and ${\sim}\,93\deg$, respectively.\footnote{See the \WMAP\ mission site (\url{http://wmap.gsfc.nasa.gov/}) for more details of scanning strategy.}
Although this complex scanning strategy provides many repeated
observations (\WMAP\ observes more than 30$\%$ of the sky each day), the coverage is not uniform. For
example, in one year, a source on the ecliptic plane can be seen
continuously for 1.5 months, then off for 3 months, then on for another 1.5
months, then off for 6 months; while a source at 48$\deg$ ecliptic latitude or
higher can be seen for 6 months straight, and then not seen for another 6
months. Sources close to the ecliptic poles are seen continuously throughout the
year. Therefore, flux densities estimated from  a single-year map are averages 
over a different fraction of the year for each source.

The \Planck\ satellite
spins around its axis at a rate of one rotation per minute. The focal plane is
oriented at an angle of 85$\deg$ to the spin axis,  which tracks the direction
of the Sun at ${\sim}\, 1\deg$ per day \citep{planck2011-1.1}. This scan pattern also results in
inhomogeneous integration time, and areas
near the ecliptic poles are observed with greater depth than the rest of the sky. Most sources are seen for a few days every six months. Since the data
included in the ERCSC amount to 1.6 full-sky surveys \citep{planck2011-1.10}, some of
the sources in our sample have been covered twice with a time separation of
${\sim}\,6$ months. Therefore, with the exception of sources close to the ecliptic poles, the flux densities given in  the ERCSC are mostly
from a single snapshot or an average of two passes, rather than an average over a substantial
fraction of a year.

\begin{figure}[t]
\centering
\includegraphics[width=0.45\textwidth]{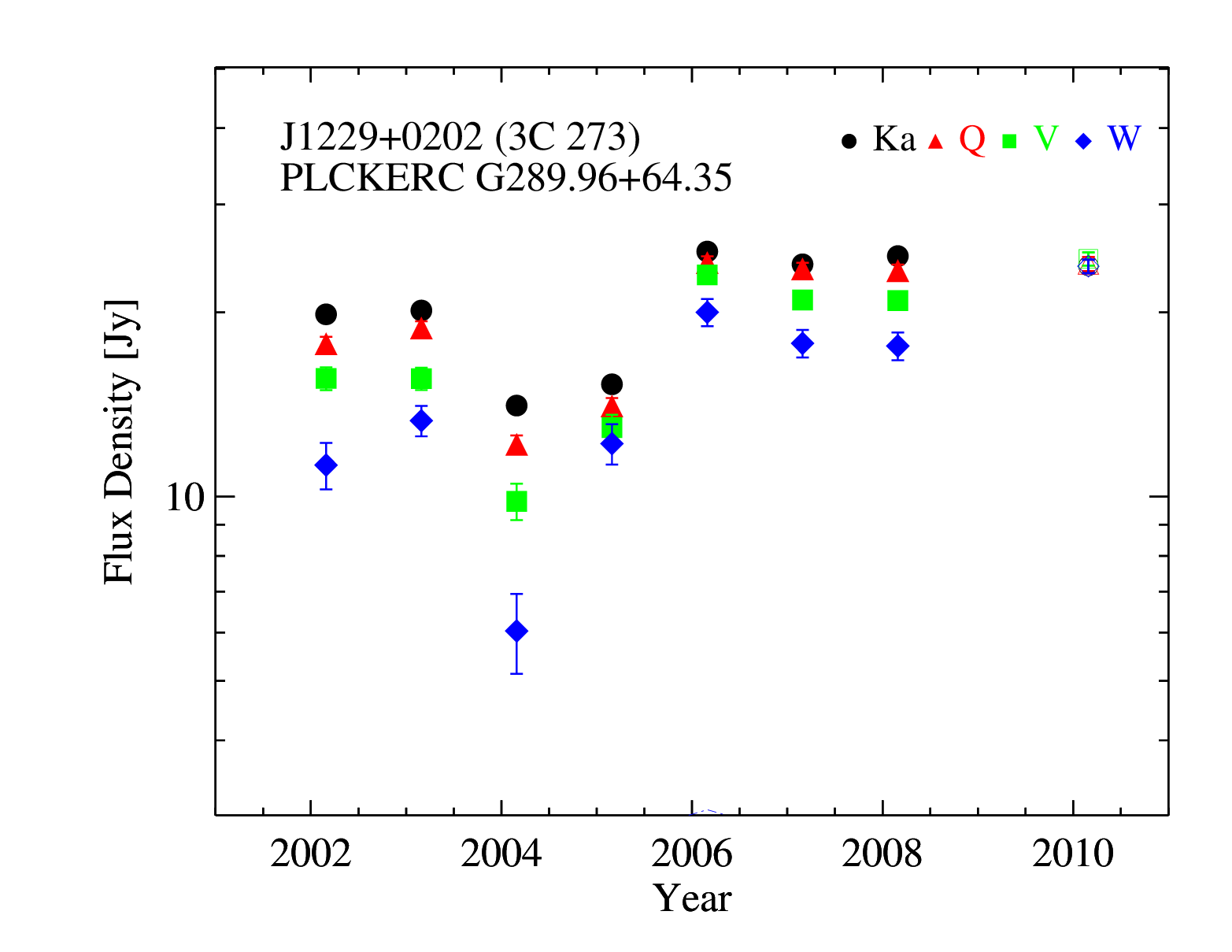}
\includegraphics[width=0.45\textwidth]{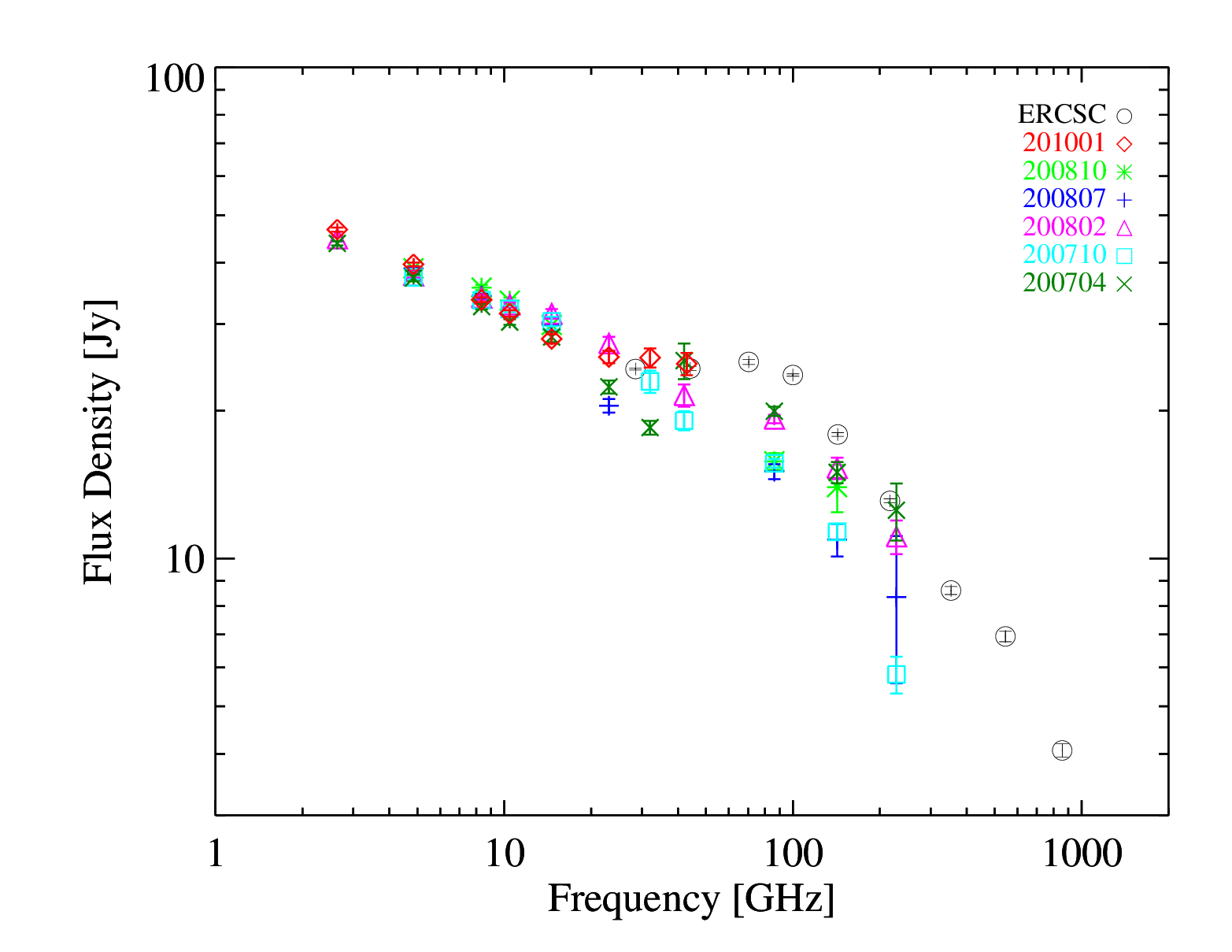}
\\
\caption{{\it Top}: \WMAP/\Planck\ light curves of 3C\,273. The \WMAP\ flux densities  are shown by filled symbols and the \Planck\ flux densities are shown by open symbols. 
{\it Bottom}: spectra of 3C\,273 from \Planck, Effelsberg, and IRAM observations at several epochs. The 201001 Effelsberg 100-m observation is nearly simultaneous with the \Planck\ observation, and confirms the flare seen by \Planck.
\label{fig:3c273}}
\end{figure}

Because of these differences, we treat the \WMAP\ single-year flux densities and the ERCSC
flux densities differently in our analysis.  To avoid
systematic errors, we include only the
\WMAP\ flux densities in the statistical studies of the source variability. The ERCSC flux densities, extrapolated to the \WMAP\
frequencies, while helpful in confirming the long-term trend
for sources exhibiting long-term variability, are more sensitive to short-term variability (e.g., flares).  
An example is given in Fig.~\ref{fig:3c273}, where the top panel shows that the
highly variable source 3C\,273 maintains a steep spectrum in the
\WMAP\ year~1 to year~7 observations, but its spectrum suddenly becomes flat in the ERCSC.
This is because this source was undergoing a strong flare during the time \Planck\
observed it (1--10 January 2010). Ground-based observations with the Effelsberg 100-m and IRAM 30-m
telescopes (obtained within the framework of the F-GAMMA program;
\citealt{Fuhrmann2013}) show that this source has undergone a few flares in the
years 2007--2008 while generally preserving a steep spectrum
(Fig.~\ref{fig:3c273}, bottom panel). The January 2010 observation at the 
Effelsberg telescope was obtained near-simultaneously with the \Planck\ data,
and confirms the flaring activity captured by \Planck.

\section{Flux Density Variability}
\label{sec:stats}

In this section we quantify and discuss the variability properties of the
sources in our sample. We first define and discuss different measures of
variability in accordance with those used in the literature, then investigate
possible effects of variability on source number counts, and finally discuss
correlation between variability found in different bands. In all steps of our
analysis, we define incremental quality cuts to select sources which are
included in the analysis. These cuts are chosen in order to maximize the number
of sources included while keeping the results robust. We
discuss qualitative statistical trends found in our sample, but emphasize that
no quantitative significance levels can be assigned to them owing to the incompleteness of the sample.

\subsection{Quantifying Variability}
\label{sec:var}

We define the fractional variability index, $V_{\rm rms}$, by
\begin{equation}\label{eq:Vrms}
V_{\rm rms} = \frac{\wmean{\sigma}}{\wmean{S}} 
	      \sqrt{\frac{\chi^2}{N-1} - 1} \;\equiv\;
\frac{\wmean{\sigma}}{\wmean{S}} X_{\rm rms} \quad,
\end{equation}
where 
\begin{equation}
\wmean{S} = \frac{\sum_i S_i/\sigma_i^2}{\sum_i 1/\sigma_i^2}
\end{equation}
is the mean of the $N$ individual flux measurements $S_i$, weighted by their
individual inverse variances corresponding to the measurement error $\sigma_i$,
and 
\begin{equation}
\wmean{\sigma} = \sqrt{\frac{N}{\sum_i 1/\sigma_i^2}} 
\end{equation}
is the average error of the measurements contributing to the weighted
mean. The quantity $\chi^2$ is defined as
\begin{equation}
\label{eq:chi2}
 \chi^2 =  \sum_{i=1}^N \frac{(S_i - \bar S)^2}{\sigma_i^2}\quad,
\end{equation}
where
$\bar S = \wmean{S}$, and $\chi^2/(N-1)$ is the reduced $\chi^2$ of
the sample for $N -1$ degrees of freedom (assuming Gaussian errors). The term $X_{\rm rms}$ gives the variability amplitude in units of the average measurement error.
As explained in Appendix
\ref{app:var}, our $V_{\rm rms}$ is equivalent to the quantity used 
by \cite{Sadler2006}, but generalized for the case of small sample size and
unequal measurement errors. 

As defined in Eq.~\ref{eq:Vrms},  $V_{\rm rms}$ naturally
connects the variability amplitude of a source with the
statistical significance of its being variable (see 
Appendix~\ref{app:limit}). Following \citet{Bolton2006} and \citet{Franzen2009},
$\chi^2$ is used to test the null hypothesis that each source is non-variable
(i.e., having constant flux density) at each frequency. We classify a source as
\textit{strongly variable} when the $\chi^2$ value indicates that the
probability of the null hypothesis is less than 1$\%$ (i.e., source is variable
at ${>}\, 99\%$ confidence level). 
For $N = 7$ (corresponding to seven years of \WMAP\
measurements), we have $N-1 = 6$ degrees of freedom, which requires $\chi^2 >
16.8$ or  $V_{\rm rms} > 1.34 \wmean{\sigma}/\wmean{S}$ in order to claim a
source \textit{strongly variable}. Additionally, we use $V_{\rm rms} > \wmean{\sigma}/\wmean{S}$ (or $X_{\rm rms} > 1$), corresponding to a confidence level of ${\sim}\, 95\%$, to classify a
source as \textit{moderately variable}. Other sources are 
not considered to be significantly variable.
Since the sampling of \WMAP\ and \Planck\ are different (see
\S\,\ref{sec:scan}), we have used only the \WMAP\ flux density measurements in
the calculations here to avoid systematic errors. 

For many sources, especially at higher frequencies, we do not have good quality
(i.e., $S_i > 3\sigma_i$) measurements for all years. 
We require at least four good data points (i.e., $N_{3\sigma} \geq 4$) at a
given frequency to include a timeline  in our variability analysis, and use $S_i = 3\sigma_i$ 
as fiducial data points to fill in the years without good measurements.  
As discussed in Appendix~\ref{app:limit}, this leads to true lower limits
for $\chi^2$ and $V_{\rm rms}$.
This approach also ensures that
we have no variation in the sampling frequency of our data and formally
have $N = 7$ in all cases. We include all the available $V_{\rm rms}$
values in Table~\ref{tab:sample} (expressed as a percentage), and flag strongly variable sources with a
``V'' and moderately variable sources with a ``v'' in the notes column.

\begin{table*}
\begin{center}
\caption{Variable Sources in our sample (198 sources in total).\label{tab:var}}
\begin{tabular}{ccccccc}
\noalign{\hrule\vskip2pt\hrule\vskip2pt}
& &\multispan2{\hfil Strongly variable\tablefootmark{a}\hfil}&&\multispan2{\hfil Strongly + Moderately variable\tablefootmark{b}\hfil}\\[-2pt]
&&\multispan{2}\hrulefill & & \multispan{2}\hrulefill \\
Band & Number & Number & Median $V_{\rm rms}$ && Number & Median $V_{\rm rms}$ \\
 & ($N_{3\sigma} \geq 4$)  &  & $[\%]$ &&  & $[\%]$ \\
 \hline
Ka  & 165 &  82 (49.7$\%$)  & 26.9 && 97 (58.8$\%$) & 25.1 \\
Q &  157 & 67 (42.7$\%$) & 27.2  && 84 (53.5$\%$) & 25.4 \\
V  & 106 & 32 (30.2$\%$) & 31.4 && 42 (39.6$\%$) & 29.6 \\
W  & 51 & 15 (29.4$\%$) & 33.4 && 20 (39.2$\%$) & 33.0  \\
\hline
\end{tabular}
\tablefoot{
\tablefoottext{a}{These are sources that are variable at ${>}\, 99\%$ confidence level.}
\tablefoottext{b}{These are sources that are variable at ${>}\,95\%$ confidence level, including both strongly variable and moderately variable sources as defined in \S\,\ref{sec:var}.}
}
\end{center}
\end{table*}

Table~\ref{tab:var} summarizes, for each band, the total number of sources qualified for variability analysis, the number (and fraction) of strongly variable sources with their median fractional variability index $V_{\rm rms}$ (expressed as a percentage), and the number (and fraction) of all the variable sources (both strongly and moderately variable) with their median $V_{\rm rms}$.
The percentage of variable sources decreases with increasing frequency, while the level of variability, indicated by median $V_{\rm rms}$, increases with frequency. 
A similar trend has been reported by \citet{Franzen2009} who suggested that it arises because sources tend to have larger flux density uncertainties and smaller flux densities at higher frequencies. 
Therefore only the
sources with the strongest fractional variability can still be found to be variable at high significance at these frequencies.
Our definition of $V_{\rm rms}$ makes this explicit: 
$\langle V_{\rm rms}\rangle \propto \langle\,\wmean{\sigma}/\wmean{S}\rangle$ when calculated on a set of sources limited by
variability significance (i.e., with a fixed threshold on $\chi^2$ and hence $X_{\rm rms}$). Since the flux density errors tend to increase and flux densities tend to decrease with frequency in our sample, $\langle V_{\rm rms}\rangle$ will inevitably increase with frequency even when no true correlation of physical variability and frequency is present. 
To further test our hypothesis, we selected a subset of sources in our sample that are limited by a minimum $\check V_{\rm rms}$ instead of $X_{\rm rms}$. We define $\check V_{\rm rms}$ so that all the sources (across all the frequencies) with $V_{\rm rms} \ge \check V_{\rm rms}$ would have  $X_{\rm rms}>1$.  
We find $\check V_{\rm rms} = 22.7\%$ in our sample, which leaves us a subsample of 56, 50, 31, and 20 sources at Ka, Q, V, and W bands, respectively. The median variability indices for this subsample are found to be comparable in all four
bands: 30.5$\%$ (Ka); 33.0$\%$ (Q); 31.4$\%$ (V); and 33.0$\%$ (W). Although
our sample is not statistically complete, this confirms
qualitatively the proposed explanation of \citet{Franzen2009} that the trend of increasing
variability with frequency may be caused entirely by systematics in source selection,
and suggests that on the time scales and the frequency range we are
probing, there is no significant increase of physical variability with
frequency. Since different physical processes may dominate
variability at short and long time scales (see \S\,\ref{sec:interpretation}),
our finding is not necessarily inconsistent with the conflicting results found on shorter time
scales (e.g., \citealt{Angelakis2012}).

\begin{figure}
\centering
\includegraphics[width=0.4\textwidth]{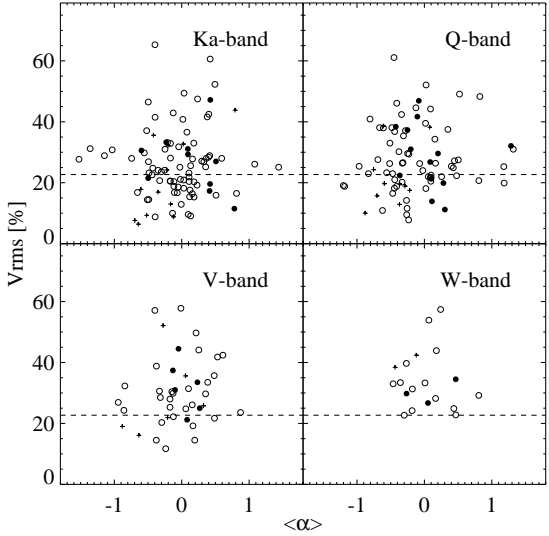}
\caption{Dependence of fractional variability index $V_{\rm rms}$ on mean spectral index
$\langle\alpha\rangle$ in Ka, Q, V, and W bands for variable sources. BLLs
are shown by filled circles, FSRQs by open circles, and all the other sources by
plus signs. The dashed lines indicate the level of $\check V_{\rm rms} = 22.7\%$.
\label{fig:varspec}}
\end{figure}

\begin{figure}
\centering
\includegraphics[width=0.4\textwidth]{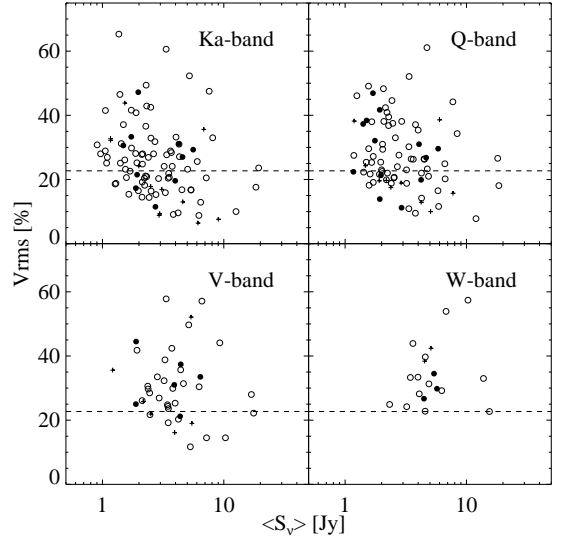}
\caption{Dependence of fractional variability index $V_{\rm rms}$ on mean flux density
$\langle S_\nu\rangle$ in Ka, Q, V, and W bands for variable sources. BLLs
are shown by filled circles, FSRQs by open circles, and all the other sources by
plus signs. The dashed lines indicate the level of $\check V_{\rm rms}$.
The apparent trend of increasing variability with decreasing flux density is
likely an artefact arising from the application of a significance cut in the
presence of large measurement errors.
\label{fig:varflux}}
\end{figure}

We also tested for possible dependences of variability index on source
spectral index and total flux density. For all the variable sources, we calculated the
two-point spectral indices (i.e., $\alpha_{\rm K}^{\rm Ka}$, $\alpha_{\rm Ka}^{\rm Q}$,
$\alpha_{\rm Q}^{\rm V}$, and $\alpha_{\rm V}^{\rm W}$) for the years when good quality data
points were available. Fig.~\ref{fig:varspec} plots the fractional variability index  of each source against the time-averaged spectral index
 in each band. We use
$\alpha_{\rm K}^{\rm Ka}$ to approximate the spectral index at Ka-band,
$\alpha_{\rm Ka}^{\rm Q}$ for Q-band, $\alpha_{\rm Q}^{\rm V}$ for V-band, and $\alpha_{\rm V}^{\rm W}$
for W-band. 
The variable sources are clearly dominated by flat-spectrum
sources. Unlike \citet{Bolton2006} and \citet{Franzen2009}, we do not find that sources with rising spectrum are more
variable. We have
also plotted the variability indices of  all variable sources against their time-averaged flux densities 
 at each frequency (Fig.~\ref{fig:varflux}). 
The apparent tendency for fainter sources to be more variable is not a physical trend, and is caused by the fact that fainter sources have larger relative flux density errors, and thus have a lower $V_{\rm rms}$ when applying a significance cut (e.g.,  $X_{\rm rms}>1$). As can be seen from Fig.~\ref{fig:varflux}, when only sources with a fractional variability above $\check V_{\rm rms}$ are considered, no significant correlation is found between the amplitude of variability and flux density.

\begin{figure}
\centering
\includegraphics[width=0.42\textwidth]{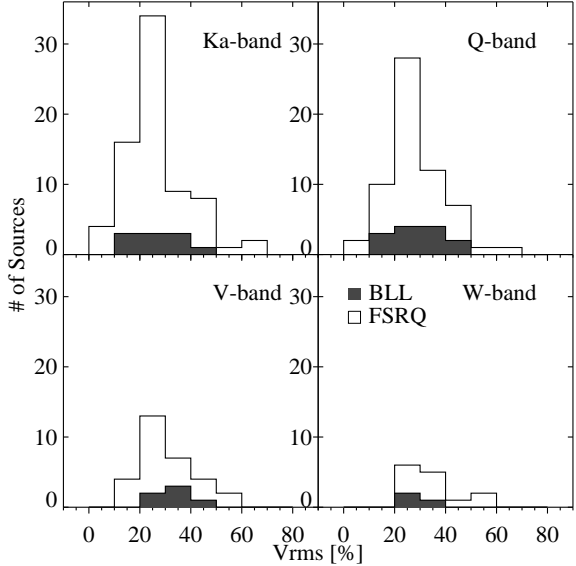}
\caption{Histograms of fractional variability index for variable
sources;  FSRQs are shown in white and
BLLs in grey.\label{fig:vartype}}
\end{figure}

As mentioned in \S\,\ref{sec:sample}, most of the sources in our sample
can be classified as either luminous broad-line FSRQs or
continuum-dominated BLLs. In Fig.~\ref{fig:vartype} we plot the
distribution of fractional variability index for the sources that are classified
as variable in each band. There is no evidence that one class is more
variable than the other.

\subsection{Modification of source number counts due to variability?}

\begin{figure}
\centering
\includegraphics[width=0.47\textwidth]{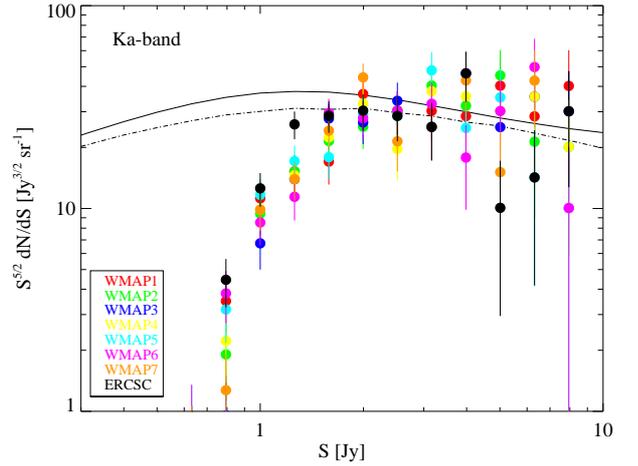}
\caption{Euclidean normalized differential number counts at WMAP Ka-band (33\,GHz) in each year. The solid curve shows the total number counts of extragalactic radio sources predicted by the \citet{deZotti2005} evolution model, and the dashed line represents the total number counts from the \citet{Tucci2011} model. 
\label{fig:numcnt_Ka}}
\end{figure}

The number of sources with good quality ($S > 3\sigma$) data across all seven years of \WMAP\ observations is greatest at the \WMAP\ Ka-band (33\,GHz), with ${\sim}\, 168\pm12$ sources in each year. The number decreases with frequency because most sources have flat or falling spectra, and flux uncertainties are also higher at higher frequencies due to the sensitivities of the maps. We therefore study the influence of source variability on number counts only in the Ka-band. In Fig.~\ref{fig:numcnt_Ka}, we plot the differential number counts of the whole sample at the Ka-band in each year, including the ERCSC observation period. The ERCSC data points displayed here are again the ERCSC 30\,GHz flux densities extrapolated to the Ka-band central frequency. Although they vary from year to year, our source counts are largely consistent with the prediction of extragalactic radio source evolutionary models from \citet{deZotti2005} and \citet{Tucci2011}. The departure of the data from the models suggests 
that our sample is incomplete below 2\,Jy. We thus model the source count distribution $dN/dS$ in Ka band as a power law $\kappa (S/{\rm Jy})^{\beta}$ in each year, with a flux density cut of 2\,Jy, using least-squares. The values we obtain are presented in Table~\ref{tab:numcnt}. We apply a completeness correction to account for the Galactic cut  of $|b| > 5\deg$ that we used in selecting our sample. The slopes are close to Euclidean. It is clear that although the number counts fluctuate from year to year, they are consistent within the uncertainty. We therefore conclude that individual source variability does not have a significant effect on the source number counts, i.e., the counts are statistically stable, and so the contamination by unresolved sources in the CMB power spectrum that are estimated from the counts will also be stable.

\begin{table}
\begin{center}
\caption{WMAP Ka-band source counts in the flux density range 2--20\,Jy.\label{tab:numcnt}}
\begin{tabular}{ccccc}
\noalign{\hrule\vskip2pt\hrule\vskip3pt}                              
Year & $N_{\rm src}$ & $\kappa$ $[{\rm Jy}^{-1} \,{\rm sr}^{-1}]$ & $\beta$  \\
\noalign{\vskip3pt\hrule\vskip3pt}                              
WMAP1 & 86 & 41$\pm$12 & $-$2.7$\pm$0.2  \\
WMAP2 & 78 & 29$\pm$\phantom{0}9 & $-$2.5$\pm$0.2 \\
WMAP3 & 81 &  25$\pm$\phantom{0}8 & $-$2.4$\pm$0.2 \\
WMAP4 & 79 & 25$\pm$\phantom{0}8 & $-$2.4$\pm$0.2 \\
WMAP5 & 78 & 37$\pm$12 & $-$2.7$\pm$0.3 \\
WMAP6 & 73 & 36$\pm$12 & $-$2.8$\pm$0.3 \\
WMAP7 & 84 & 36$\pm$14 & $-$2.7$\pm$0.3 \\
ERCSC\tablefootmark{a} & 75 & 43$\pm$16 & $-$2.9$\pm$0.3 \\
\noalign{\vskip3pt\hrule}                              
\end{tabular}
\tablefoot{
\tablefoottext{a}{Based on ERCSC 30\,GHz flux densities extrapolated to the centre frequency of \WMAP\ Ka-band.}
}
\end{center}
\end{table}

\subsection{Correlation Analysis}
\label{sec:cor}

As mentioned in \S\,\ref{sec:introduction} (see also \S\,\ref{sec:interpretation}) geometric variability or low-peaked flares are likely responsible for the variability at the time scales and frequencies probed in our study. Both physical processes predict that the average radio--millimetre spectral shape should remain nearly unchanged during flux
variations.
We thus expect  flux density variations in different bands to be directly correlated. To quantify
this correlation,  we calculate the Pearson product correlation coefficient, $r$, between bands for each source.
We note that this method quantifies the instantaneous correlation between flux
density changes, but is less sensitive to correlated changes with time lags between bands.
The existence of significant time lags would reduce the
correlation coefficient in our analysis. However, it is our intention in this study 
to distinguish between long-term (${\gsim}\,3$\,yr), frequency
independent patterns, and short-term (${\lsim}\,1$\,yr), high-peaked flares (see \S\,\ref{sec:blazars}).
We do not employ more sophisticated methods such as discrete correlation function analysis to quantify
possible time lags, as this is not meaningful for a sample containing only seven
data points in each timeline. 

We analyse the correlations between Ka-band and each of the other bands (Q, V, and W). 
We restrict our analysis to sources which have been classified as strongly
variable in the Ka-band, and satisfy the requirement of $N_{3\sigma} \geq 4$ and $N_{2\sigma}=7$ in each band; we do not require that the source to be classified as variable in the Q, V, or W bands. 
We include all data points with $S>2\sigma$ here to ensure
that we have no missing data points in the time lines. Unlike in \S\,\ref{sec:var}, we cannot use upper limits
for flux densities here as this would not lead to any useful limits on $r$, and the
more restrictive requirement $N_{3\sigma}=7$ would leave us with too few time lines to analyse. 
We regard the variability between two bands as
``strongly correlated'' if $r>0.875$, which corresponds to a probability
of less than $\sim 1\%$ 
that the null hypothesis of an uncorrelated variability is true
(see Appendix \ref{app:cor}). 
We regard a value $r<0.3$ as indication of no correlation, 
i.e., a $50\%$ probability to \textit{reject} perfect correlation. Sources with 
$0.3 < r < 0.875$ are considered ``moderately'' correlated.

Our criteria select $55$, $39$, and $21$ sources in the Q, V, and W band,
respectively, for correlation analysis. Of these sources, 
$58\%$, $46\%$, and $28\%$  show strong 
correlation with the Ka band, and $40\%$, $49\%$, and $57\%$ show moderate correlation. The majority of our sources show correlated
variability between Ka and Q band, which confirms a similar finding by
\citet{Franzen2009} between 15 and 32\,GHz. 
Although there may be a moderate trend towards less correlated
variability at higher frequencies, this should not be overstated as it may be
due to the larger error bars affecting 
these bands. In general we find that very few variable sources show no
direct correlation between Ka and higher-frequency bands. This result may
suggest that variability in these frequencies on long time scales is 
less frequency-dependent than on short time scales (e.g.,
\citealt{Angelakis2012}).

\section{Variability patterns in blazars}
\label{sec:interpretation}

In this section we describe the variability patterns expected in simple models motivated
by current understanding of AGNs and extragalactic jets, and search for the signatures of these patterns in the time series of sources in our sample.

\subsection{Physics of blazars and models for variability}
\label{sec:blazars}

The term \textit{blazar} was introduced in the late 1970s to denote the
combined class of optical violent variable (OVV) quasars and BL Lac objects
\citep{Angel1980}. In the radio-to-millimetre range, blazars show featureless, often highly
polarized, continuum spectra
which are attributed to
partially self-absorbed synchrotron emission from relativistically boosted,
compact jets emerging from AGN at a small angle $\theta$ to the line of
sight \citep[e.g.,][]{Begelman1984}. If $\Gamma = 1/\sqrt{1-\beta^2} \gg 1$ is the bulk
Lorentz factor of the relativistic jet plasma, the Doppler boosting factor $D =
[\Gamma (1 - \beta\,\cos\theta)]^{-1}$
is ${\sim}\,\Gamma$ for $\theta \lsim \Gamma^{-1}$. This causes the flux density
of
the synchrotron emission to dominate over unboosted thermal components of the
AGN, as $S_\nu\propto D^{3+\alpha}$. 
The Doppler factor, bulk Lorentz factor and inclination angle of the jets
can be estimated by several methods. Firstly, by comparing number counts of
blazars with those of unboosted radio galaxies \citep{Urry1995}. Secondly,  from
superluminal apparent speeds of VLBI components
\citep[e.g.,][]{Blandford1977,Ghisellini1993,Zensus1997, Kellermann2004}. And
thirdly,  by exploiting physical constraints on models of radiative emission 
\citep{Ghisellini1998, Lahteenmaki1999a, Lahteenmaki1999b, Hovatta2009}. All
these methods allow a consistent estimate of $D \sim \Gamma\sim 10$ and
$\theta\sim 5^\circ$. Estimates of individual Doppler factors and viewing
angles, however, are too model-dependent to be considered reliable
\citep[e.g.,][]{Hovatta2009}.

Blazars are strongly variable down to time scales of ${\sim}\, 1$
day or less. Such rapid variability is usually attributed to the presence of shocks in the jets, which
are expected to arise from instabilities inside the jet or from
interactions of the jet with an external medium \citep{Begelman1984}.
Several phenomenological approaches to the formation and evolution of
shocks in jets have been suggested, ranging from persistent moving or standing
shock waves \citep{Marscher1985, Marscher2008} to rapid sequences of shocks
formed over a large range of scales in the jet \citep{Spada2001, Guetta2004,
Rachen2010}. Shocks in AGN jets can also  provide the sites for efficient
particle acceleration \citep[e.g.,][]{Biermann1987} needed to produce the observed high
energy gamma-ray emission \citep[see, e.g.,][]{Boettcher2012}.

Another type of
variability, usually called \textit{geometric variability}, is expected in
a number of phenomenological models describing helical or precessing motion in
jets \citep{Camenzind1992, Steffen1995, Valtaoja1989, Wagner1995,
Villata1999, Ostorero2004, Valtonen2009, Valtonen2011}. Because of the strong dependence of the beamed flux on the 
Doppler factor ($\propto D^{3+\alpha}$),  small changes of the inclination
angle $\theta$ or the bulk Lorentz factor $\Gamma$ of the jet will induce large
changes in the observed flux. For example, in a source with $\Gamma=10$, $\theta
= 5^\circ$, and $\alpha=1$, a change of inclination angle by one degree towards
the observer increases the measured flux by a factor of two.

These two types of variations are expected to affect the spectrum of the source differently:
while shocks naturally produce
spectral changes by injecting a new population of energetic particles,
geometric effects should not have a large effect on the emitted spectrum. Detailed
modelling, however, shows that this simple criterion has to be used with care.
For example, the strongly inverted ``edges'' often seen in the radio--millimetre
spectra of variable blazars \citep{Valtaoja1988, Brown1989, Krichbaum2008,
planck2011-6.2, planck2011-6.3a} can be explained both by synchrotron
self-absorbed emission from a compact, shocked region with
temporarily enhanced emission compared to the surrounding jet
\citep[e.g.,][]{Valtaoja1988, Brown1989, Turler2000, Rachen2010}, and by
decreasing inclination angle in a steady but
rotating helical jet \citep{Ostorero2004}. The same ambiguity affects the
interpretation of helically moving VLBI components \citep{Savolainen2002,
Bach2006, Rastorgueva2009, Jorstad2010}. On the other hand, achromatic
variability, canonically expected from geometric effects, can also be explained
by shock-induced flares if they are observed only above their
peak-frequency \citep[so-called \textit{low-peaked flares},][]{Valtaoja1988}. 
Thus, although the overall properties of
variability from shocks and from geometric effects can be quite different, the spectral behavior during flares may not be a reliable way to distinguish the processes.

\subsection{Short-term vs. long-term variability}
\label{sec:longtermvar}

A major difference
between geometric and shock induced variability is the observed time scale. In a
relativistically boosted emitter, all \textit{intrinsic} time scales of
variability are compressed by the Doppler factor,  $\tau_{\rm obs} = \tau_{\rm
int}/D \ll \tau_{\rm int}$. This applies in particular to the time scales
connected to the formation and propagation of shocks in the jet, and allows
emission regions of sub-parsec size to produce variability on observed time
scales down to a few days. In contrast, the physical time scale of geometric
changes are not Doppler compressed, as the rotation is transverse to the jet
propagation ($\tau_{\rm obs} \sim \tau_{\rm int}$). If indeed \textit{both}
processes contribute to variability with comparable \textit{intrinsic} time
scale and amplitude, geometric variability should then dominate the amplitude
on long time scales. We illustrate below how this affects the patterns observed in long-term averaged data.

Let ${\cal G}(t)$ and ${\cal F}(t,\nu)$ denote the time structure of geometric
and shock induced (flare) variability respectively, with the total flux of the
source being $S(t,\nu) = {\cal G}(t)\,{\cal F}(t,\nu)$. The \WMAP\ data used in
this paper provide regularly sampled 1-year averages, i.e.,
\begin{equation}\label{eq:lowpassfilter}
S_{\it WMAP}(t_i,\nu_b) = \int_{t_{i-1}}^{t_i} dt\,{\cal G}(t)\,{\cal F}(t,\nu_b)\,w(t_i-t) \,,
\end{equation}
where $t_i$ is the end-time of operation year $i$, $\nu_b$ is the central frequency of the appropriate \WMAP\ band,
and $w$ is the exposure function of \WMAP\ in this band (which we assume is approximately the same in
every operation year). 
Eq.~\ref{eq:lowpassfilter} describes a {low-pass filter}, suppressing all fluctuations in $S(t,\nu_b)$ on time scales $\tau \ll\Delta t \equiv t_i-t_{i-1} \sim 1$\,yr. This is typically the case for the flare process ${\cal F}(t,\nu_b)$, but not for the geometric process ${\cal G}(t)$, as its characteristic time scale is not Doppler compressed. Therefore, although large amplitudes of ${\cal F}(t,\nu_b)$ may cause an irregular structure in $S(t,\nu_b)$, characteristic patterns of ${\cal G}(t)$ may become more evident when using averaged data. We note that these considerations apply to the \WMAP\ data, but not to the \Planck\ ERCSC data in general. 
Most ERCSC flux densities represent averages of one or two snapshots, except for a few sources near the ecliptic poles which were observed more often. The flux densities thus retain a significant fraction of the short-term variability amplitude. We therefore discuss the \WMAP\ and ERCSC data separately.

\subsection{Test for variability patterns}

Geometric variability is not bound to specific simple patterns, as the direction
of jet emission can change in potentially chaotic ways. However, a
major motivation for proposing geometric variability is
the possible connection to helical jet structures shown by VLBI monitoring
(see \S\,\ref{sec:helical}), which are thought to originate in a rotating jet base. As
discussed in Appendix~\ref{app:pattern}, a simple rotating relativistic
jet will produce a flux variability pattern of the form
\begin{equation}\label{eq:rotpattern}
S_{\rm rot}(t) = \frac{S_0}{\big( 1 + A \cos[\omega(t-t_0)]\big)^{3+\alpha}} \,.
\end{equation}
Assuming the amplitude $A$ and frequency $\omega$ of the oscillation are
time-independent, this model involves four free parameters
(the spectral index $\alpha$ is determined from the
data). 

\defcitealias{Valtaoja1999}{VLTL}

A similarly simple phenomenological description of the time evolution of
flares caused by shocks
has been introduced by \citet[hereafter VLTL]{Valtaoja1999}, based on observed flare properties
and a generalization of the ``shock in jet'' model \citep{Marscher1985,
Valtaoja1988}. It has been successfully applied to decompose single-frequency
 blazar light-curves  into a constant background and a number
of separate exponential flares (see also \citealt{Nieppola2009}).
In the original VLTL model, a flare is described by an
exponential rise with time constant
$\tau_{\rm r}$ followed by an exponential decay with time constant $\tau_{\rm f}
\approx 1.3\tau_{\rm r}$.
The time scales of such flares can vary between a few months and
${\sim}\,10$\,yr, 
with a median of ${\sim}\,3$\,yr \citep{Nieppola2009}, which is
comparable to the time scales probed in this paper. This motivates 
a four-parameter \textit{VLTL pattern} as a heuristic description
of a single flare or activity phase:
\begin{equation}\label{eq:vltpattern}
S_{\rm VLTL}(t) = S_0 + \Delta S_{\rm max}\; \left\{\begin{array}{ll}
 e^{(t - t_{\rm max})/\tau}    & \quad t \le t_{\rm max} \\
 e^{(t_{\rm max} - t)/1.3\tau} & \quad t > t_{\rm max}  \,.\\
\end{array}\right. 
\end{equation}

Both models are clearly simplified. Realistic models of helical
jets involve complex dependencies on more parameters \citep{Camenzind1992,
Steffen1995, Villata1999}, and may potentially include static inhomgenities in
the jets \citep{Ostorero2004}. For the VLTL model, \citet{Nieppola2009} showed
that the relation $\tau_{\rm f}\approx 1.3\tau_{\rm r}$ holds only as a gross
average, while the individual flare decay-to-rise time scale ratio varies
between $0.3$ and $5$. Moreover, it is likely that more than one
flare appears during the decade probed by our data. We therefore
emphasize that our quest is not to provide fits to the data with realistic
models, but to test how well our data can be represented by the 
simplified models discussed here.

\begin{figure*}
\centering
 \begin{tabular}{ccc}
\includegraphics[width=0.315\textwidth]{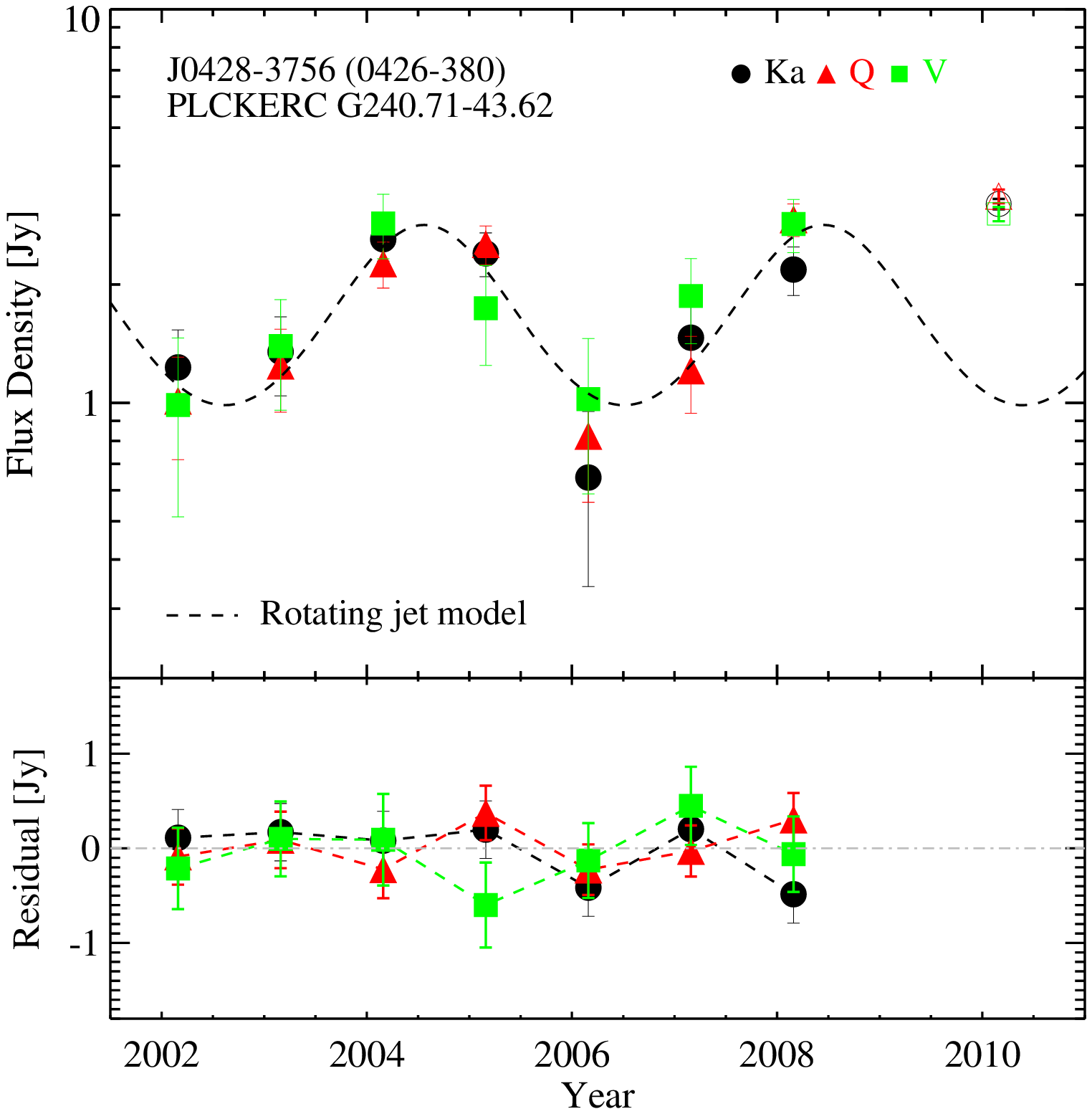} &
\includegraphics[width=0.315\textwidth]{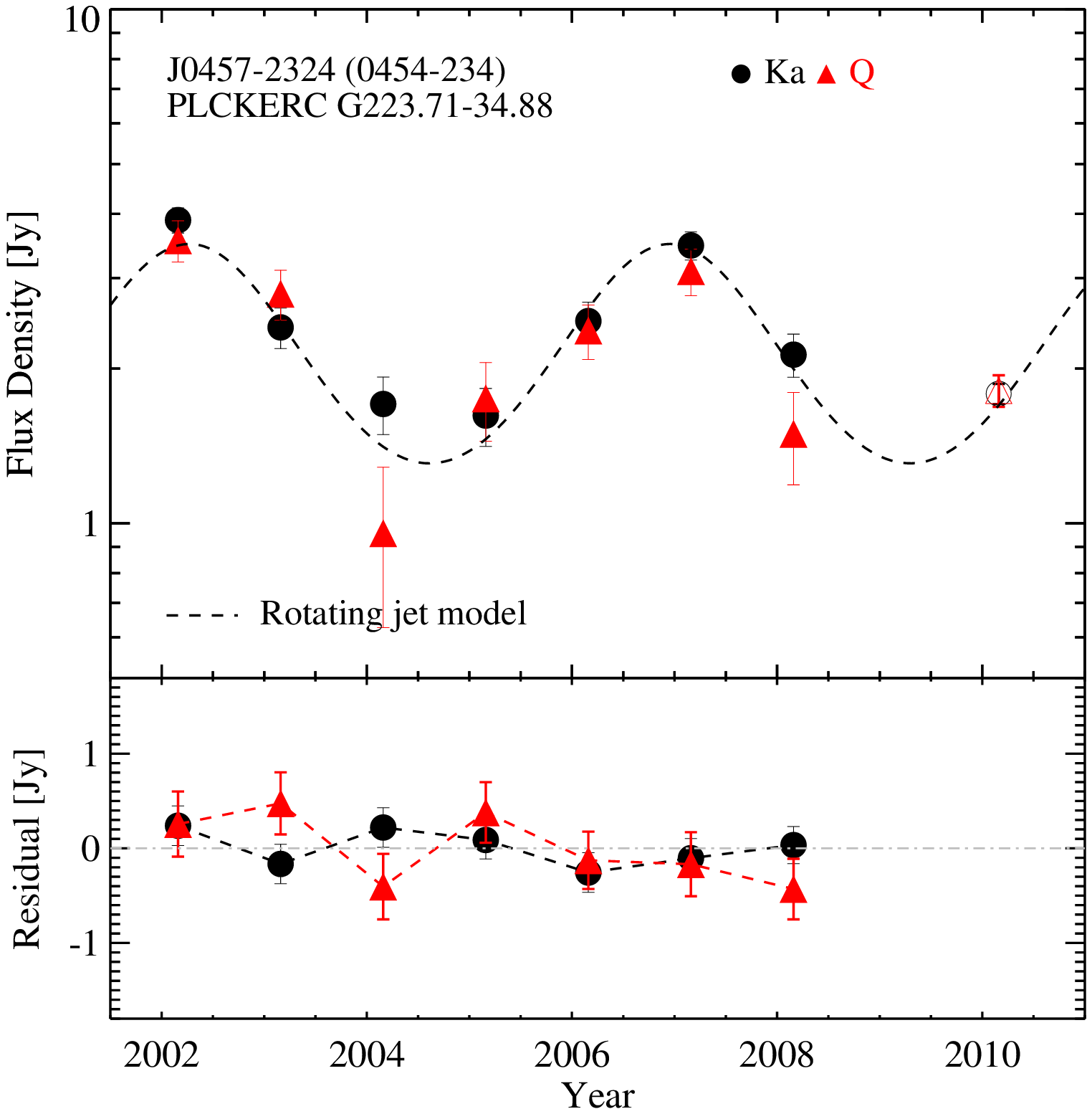} &
\includegraphics[width=0.315\textwidth]{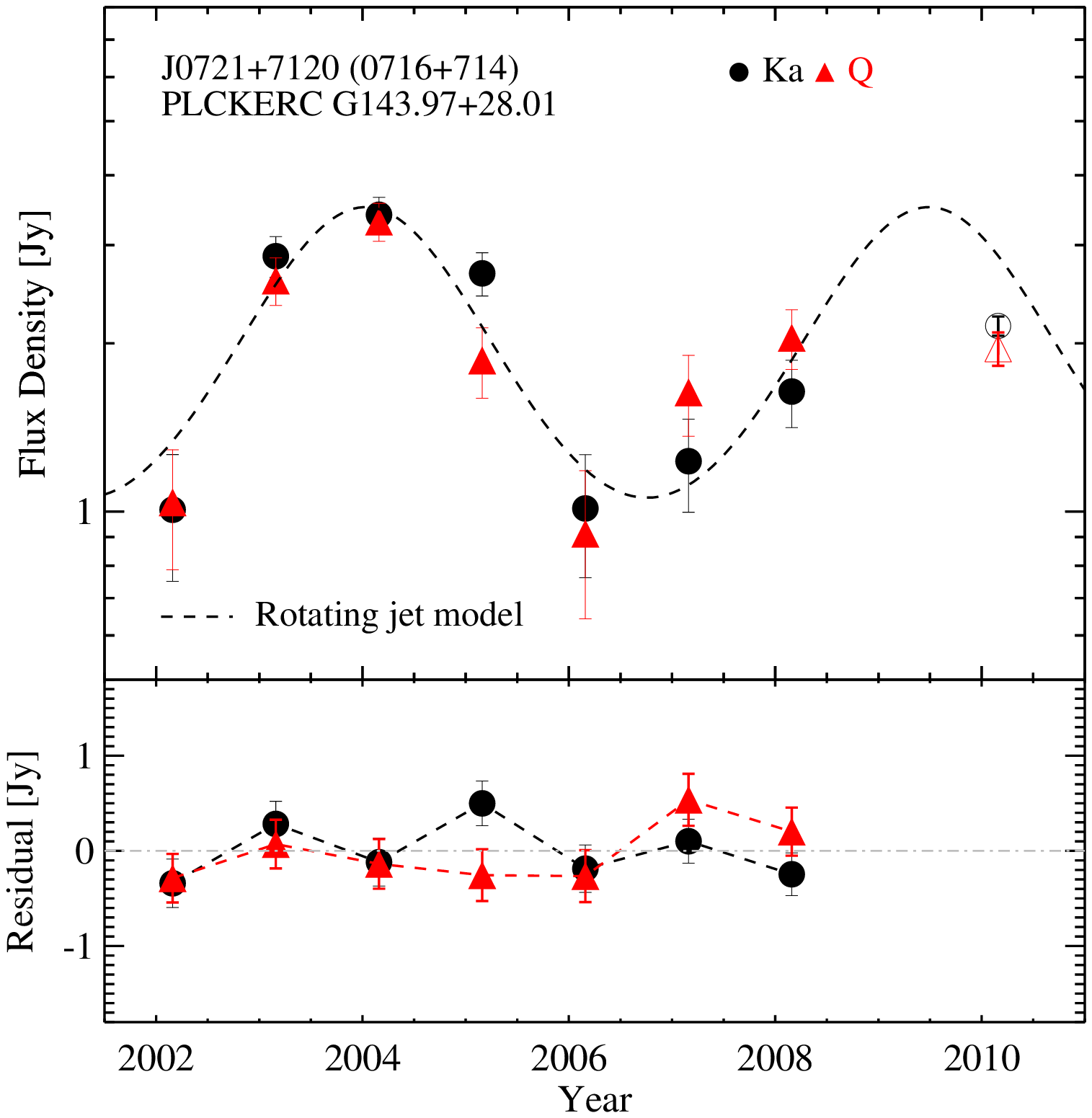} \\
\includegraphics[width=0.315\textwidth]{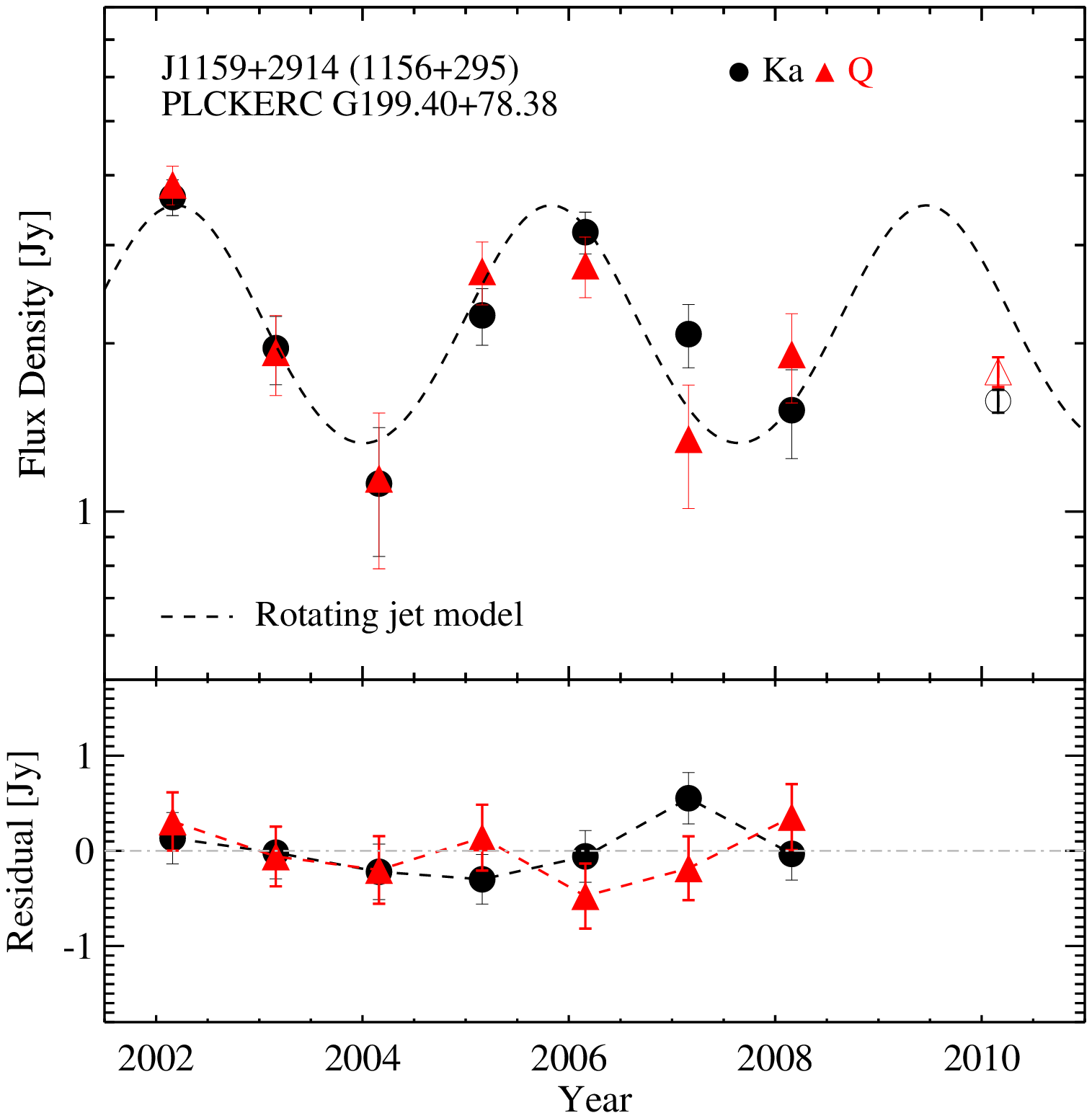} &
\includegraphics[width=0.315\textwidth]{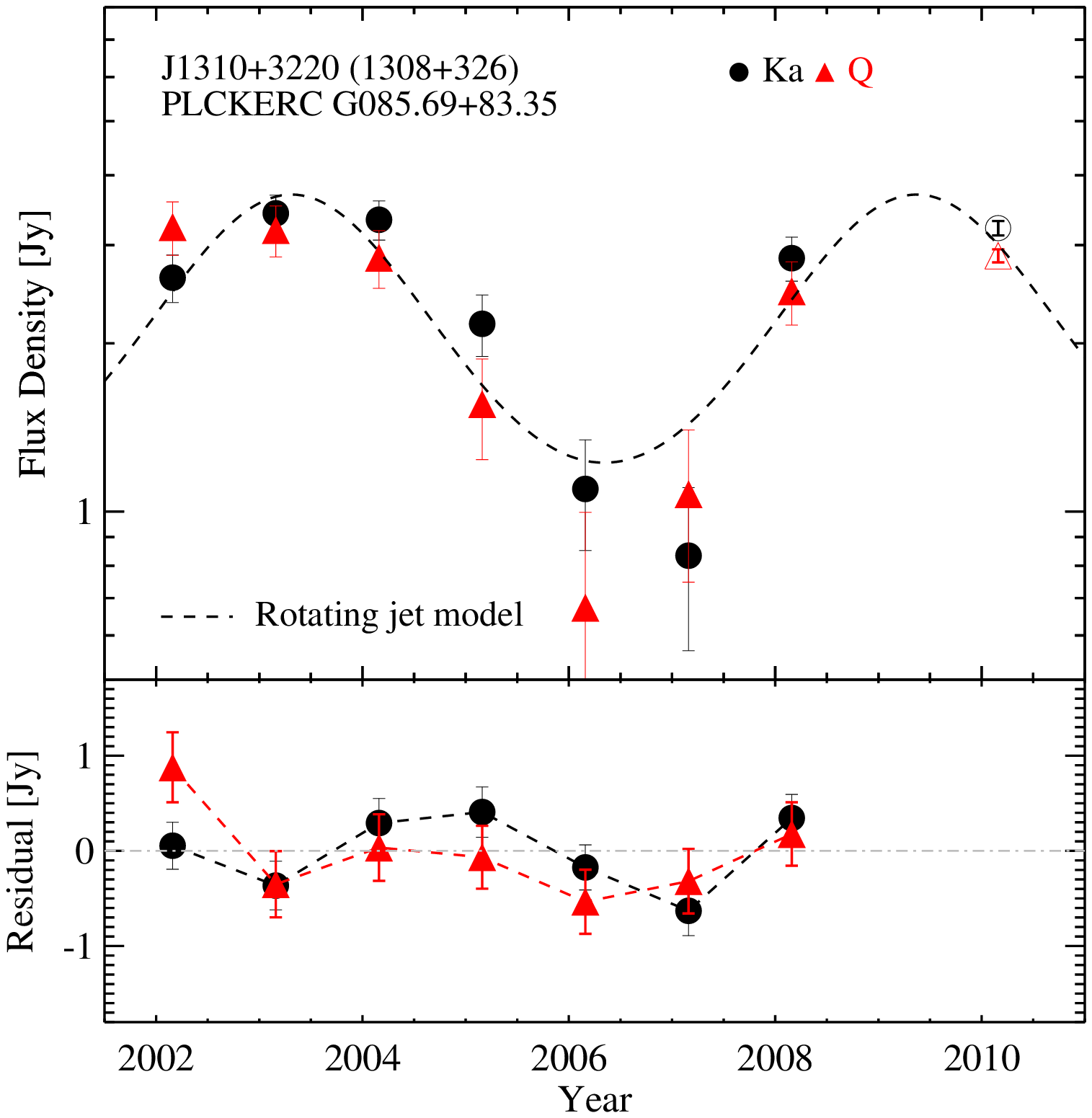} &
\includegraphics[width=0.315\textwidth]{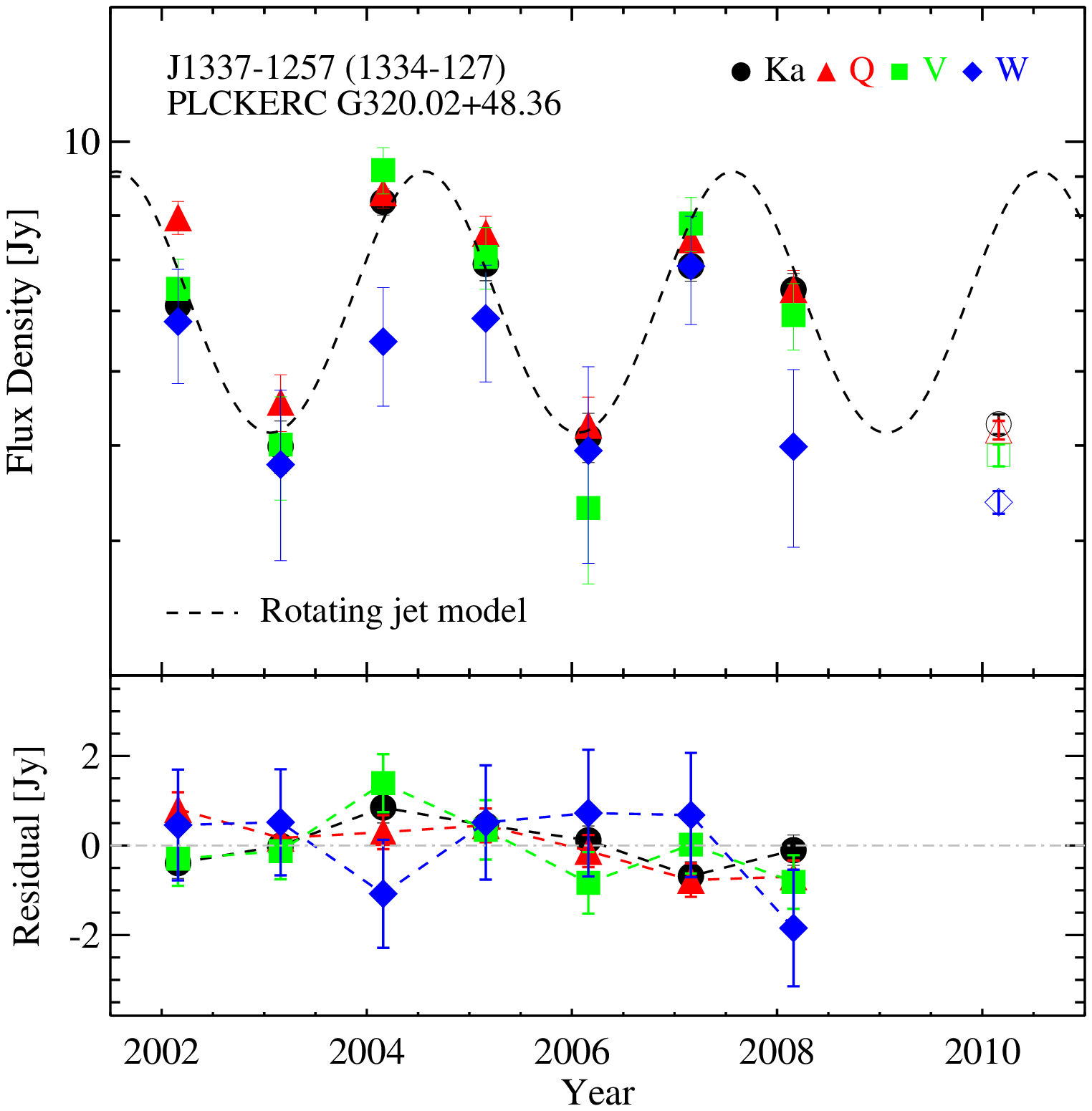} \\
\includegraphics[width=0.315\textwidth]{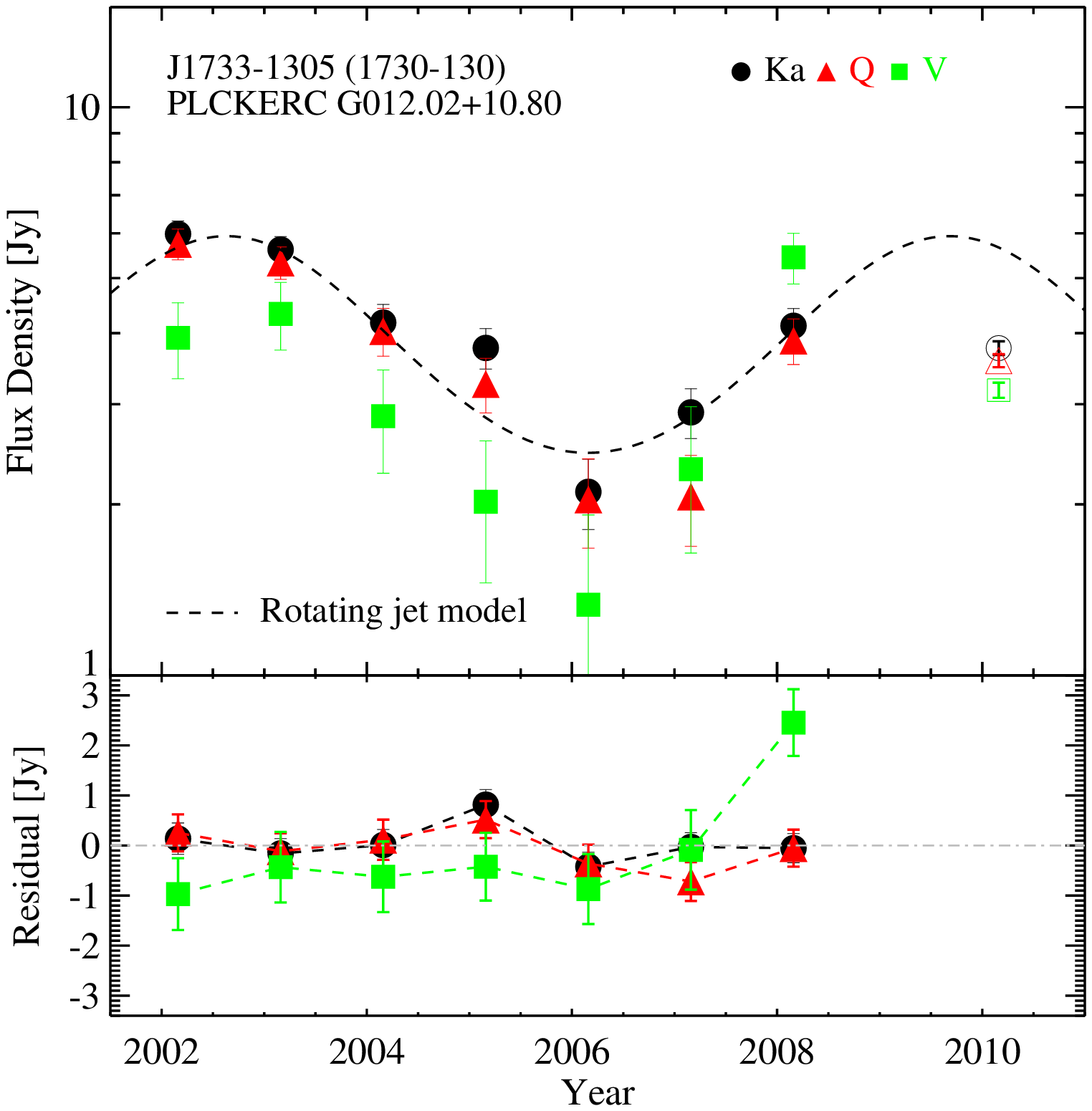} &
\includegraphics[width=0.315\textwidth]{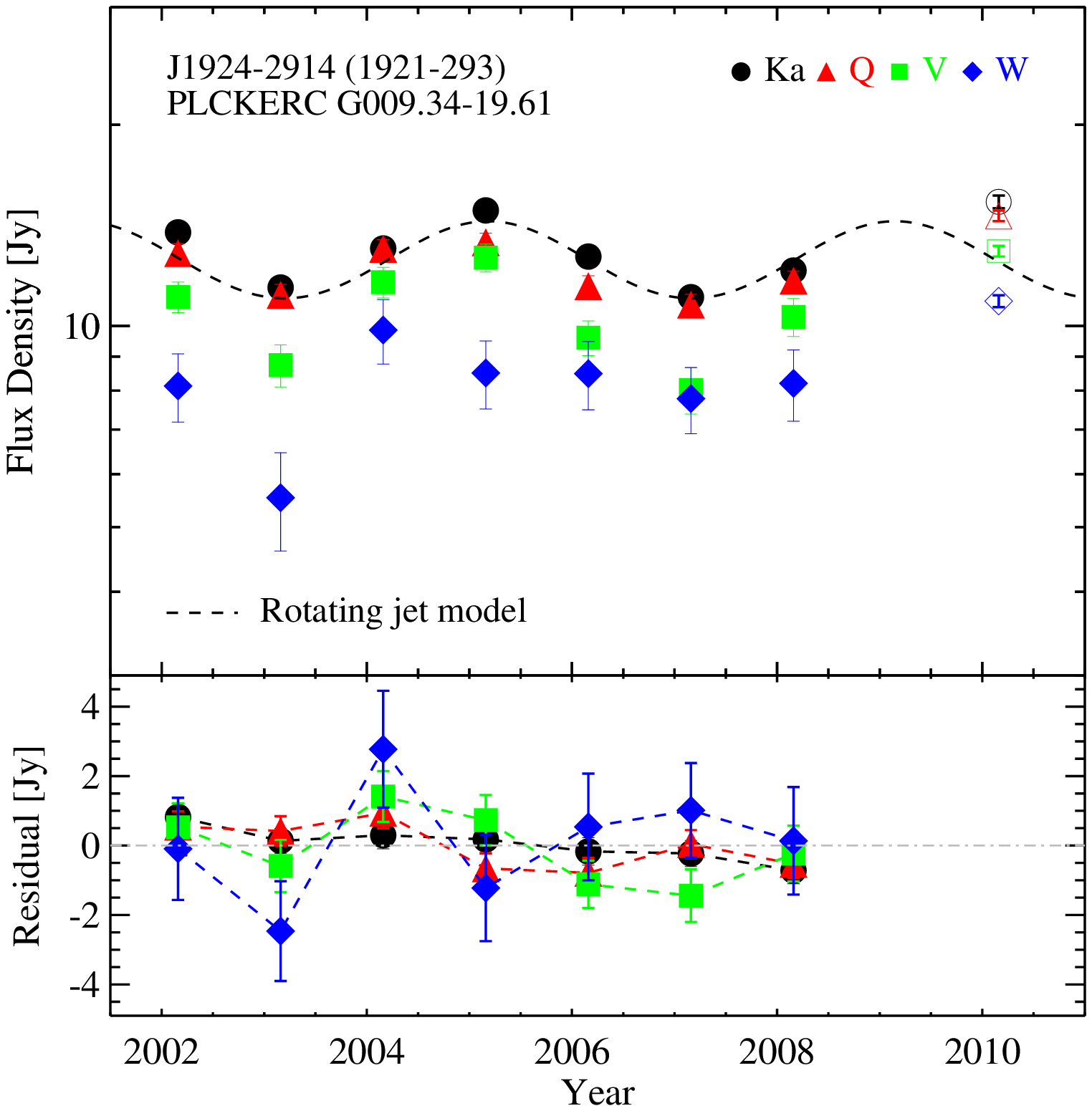} &
\includegraphics[width=0.315\textwidth]{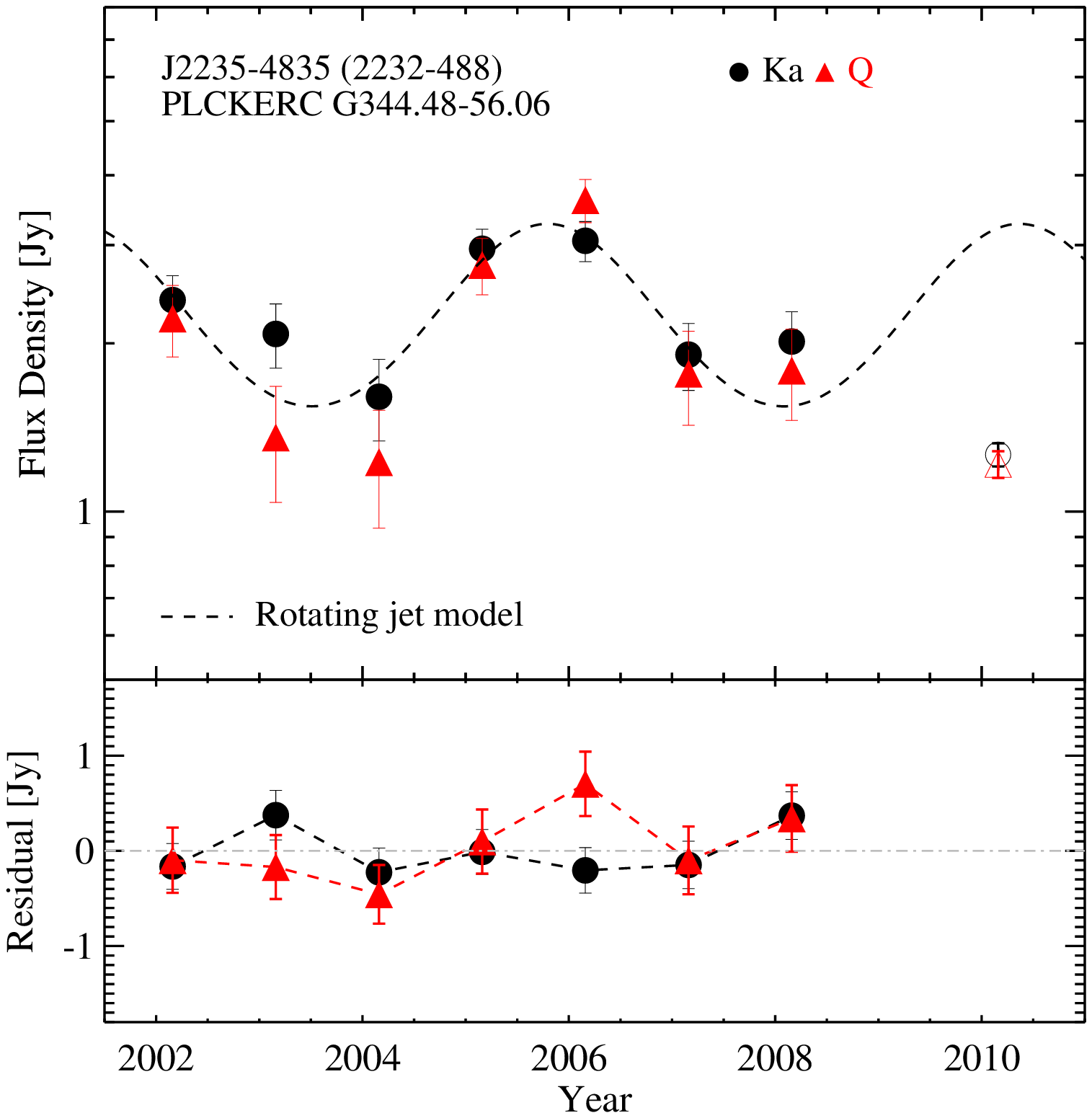}
\end{tabular}
\caption{{\it Upper panels}: \WMAP/\Planck\ light curves for the sources that are better fit by the rotating
jet model. The \WMAP\ data are shown by filled symbols and the later \Planck\ data are shown by open symbols. {\it Lower panels}: Residuals between the normalized \WMAP\ flux densities (Eq.~\ref{eq:scaling}) and the rotating jet model. Grey dash-dot lines indicate the zero level. \label{fig:rot}
}
\end{figure*}

We limit our analysis to sources that show strong correlated variability between
 Ka band and at least one other band (\S\,\ref{sec:cor}), which ensures a
better detection of common patterns between bands. In order to use information
from all the bands to determine the best fit parameters of a pattern and its fit
quality, we normalize all the data points to a common level at $\tilde S^{b} =
S^{b} f^b$, where \begin{equation} f^b = \frac{\langle S^{\rm Ka}\rangle +
\langle S^{\rm Q}\rangle}{2\langle S^b\rangle} \,. \label{eq:scaling}
\end{equation} This normalization is somewhat arbitrary, and was chosen
mainly for comparability with external data (see \S\,\ref{sec:longterm}). Using
an initial guess for both the VLTL and rotating patterns, we minimize the
residual $\chi^2$ with respect to the model using the Nelder-Meade simplex
method implemented in the GNU scientific
library\footnote{http://www.gnu.org/software/gsl/}. We then calculate the
residual $X_{\rm res}$, defined as \begin{equation}\label{eq:Xrms} X_{\rm res} =
\sqrt{\frac{\chi^2}{N-p} - 1} \quad, \end{equation} where $p = 4$ is the number
of fitted parameters. We accept a model fit if $X_{\rm res} < 1$, or if it
reduces the mean of $X_{\rm rms}$ (defined as in Eq.~\ref{eq:Vrms}) from all the
bands by a significant factor, i.e., $X_{\rm res} < \frac13 \bar X_{\rm rms}$.

The one-year averaged \WMAP\ flux densities have equal weighting in each year,
and therefore it is correct to space the points by exactly one year when fitting
models, even if the exact epoch of each (and hence the reference time of the
fit) is uncertain by a few months. We do not use the \Planck\ ERCSC flux
densities in the fitting as they have different weighting, and the time spacing
relative to the \WMAP\ observations is difficult to determine. 

\begin{figure*}
\centering
 \begin{tabular}{ccc}
\includegraphics[width=0.315\textwidth]{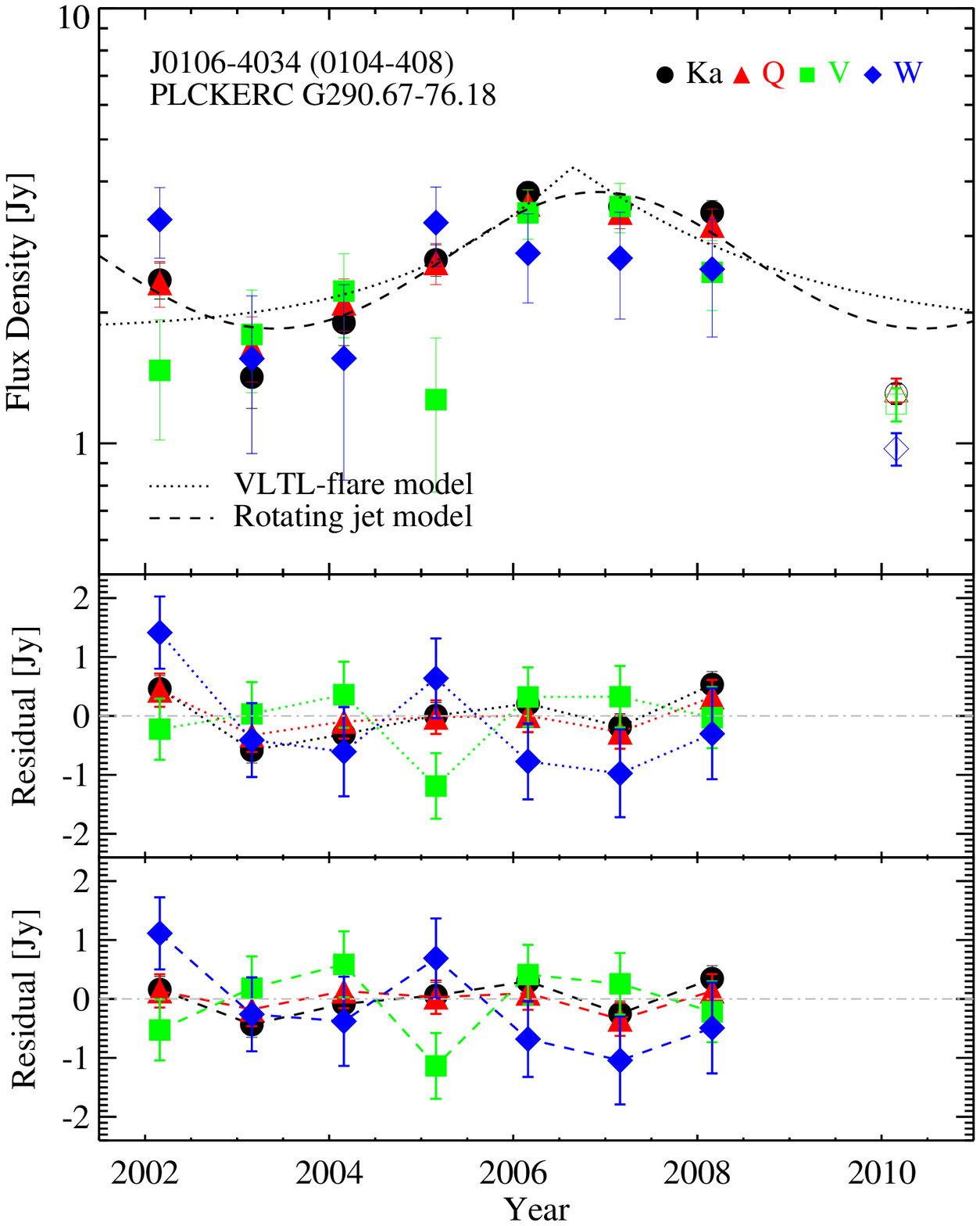} &
\includegraphics[width=0.315\textwidth]{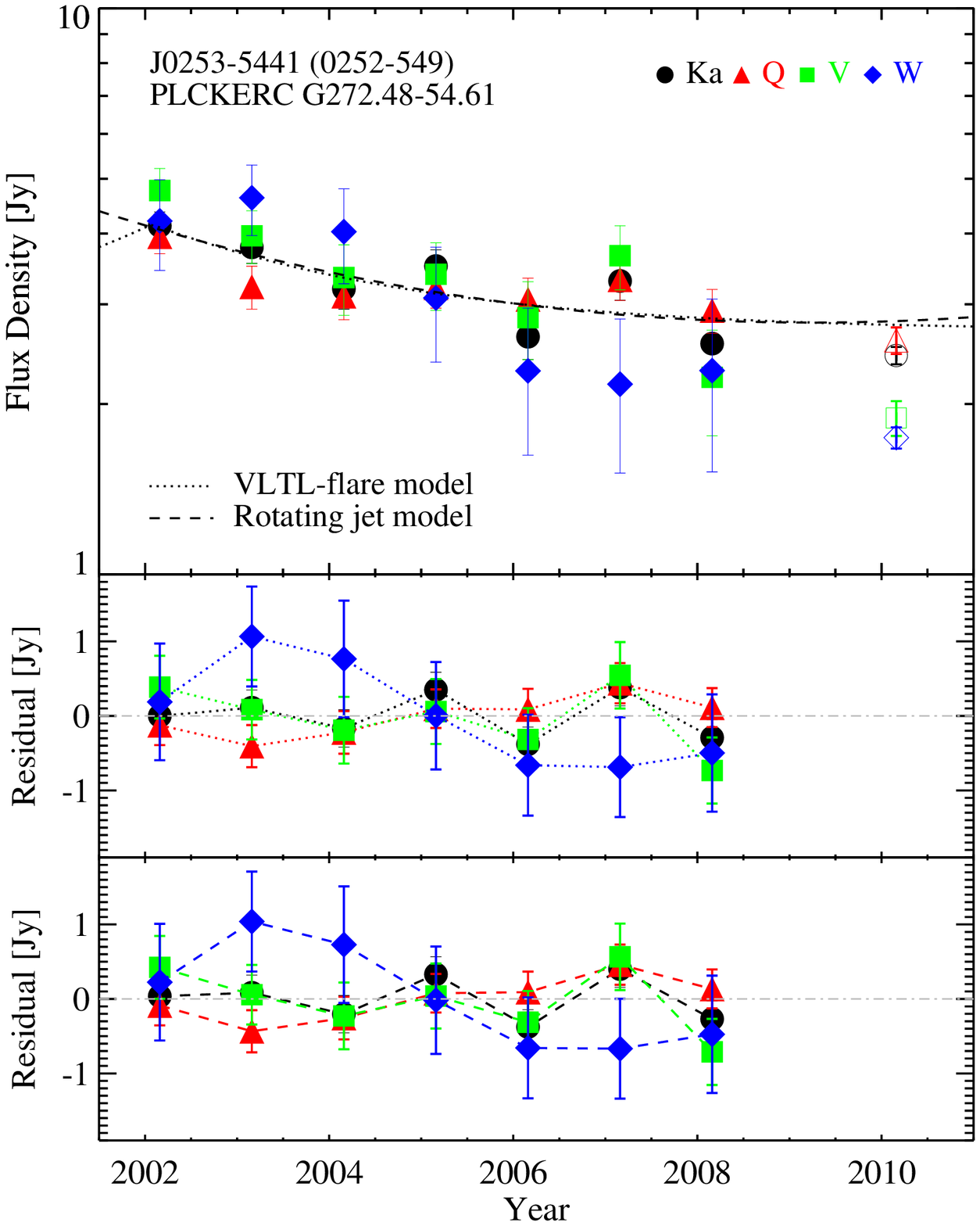} &
\includegraphics[width=0.315\textwidth]{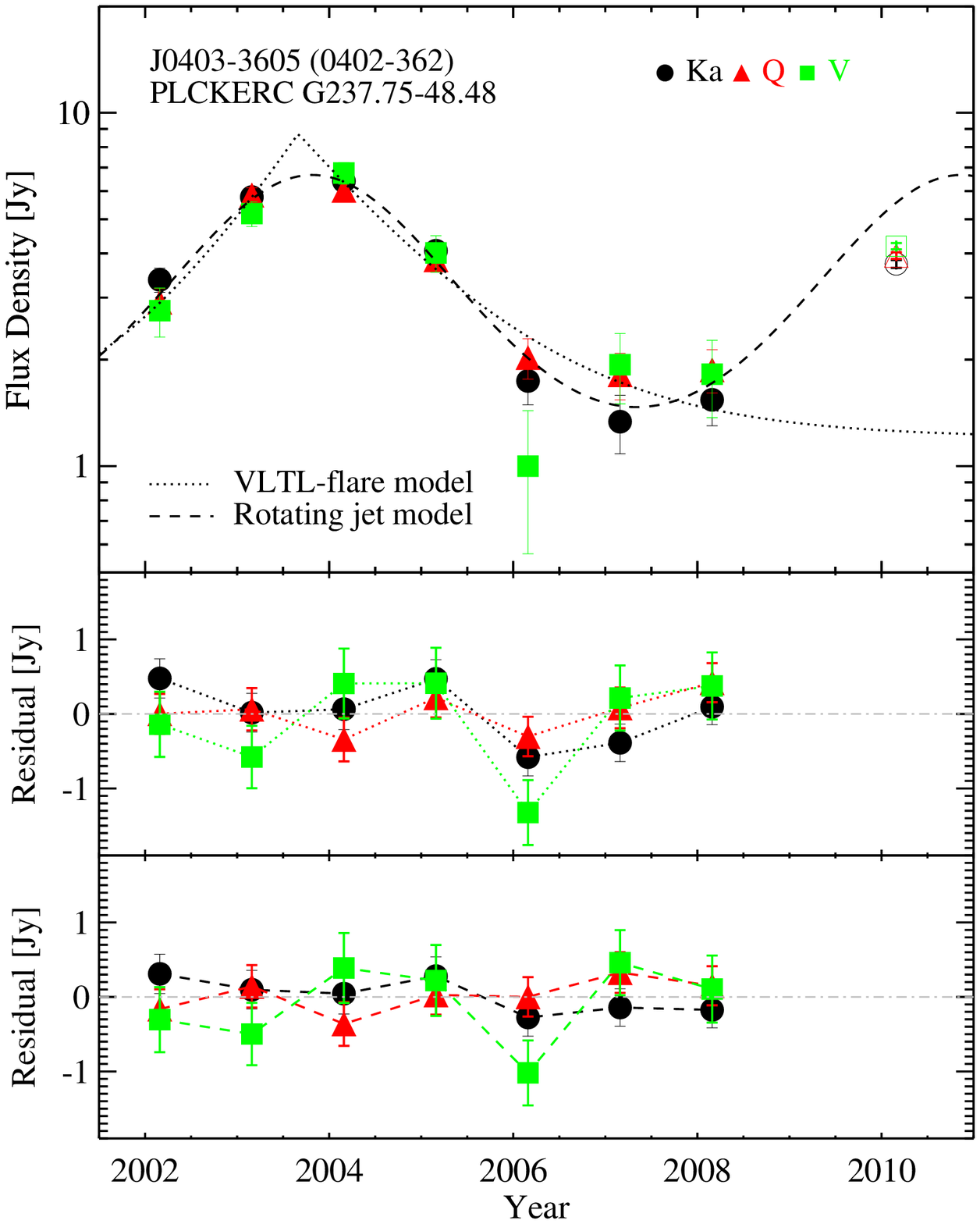} \\
\includegraphics[width=0.315\textwidth]{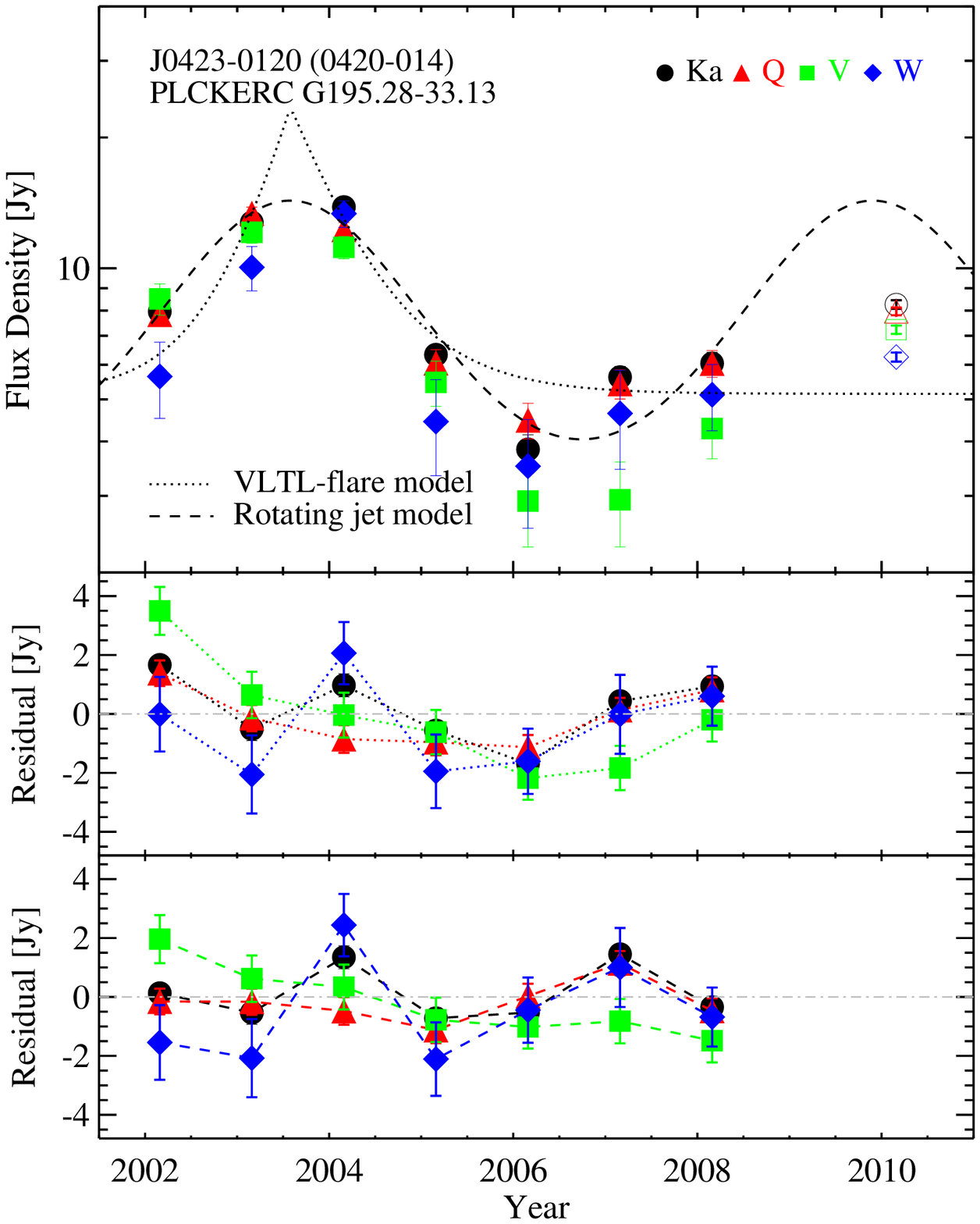} &
\includegraphics[width=0.315\textwidth]{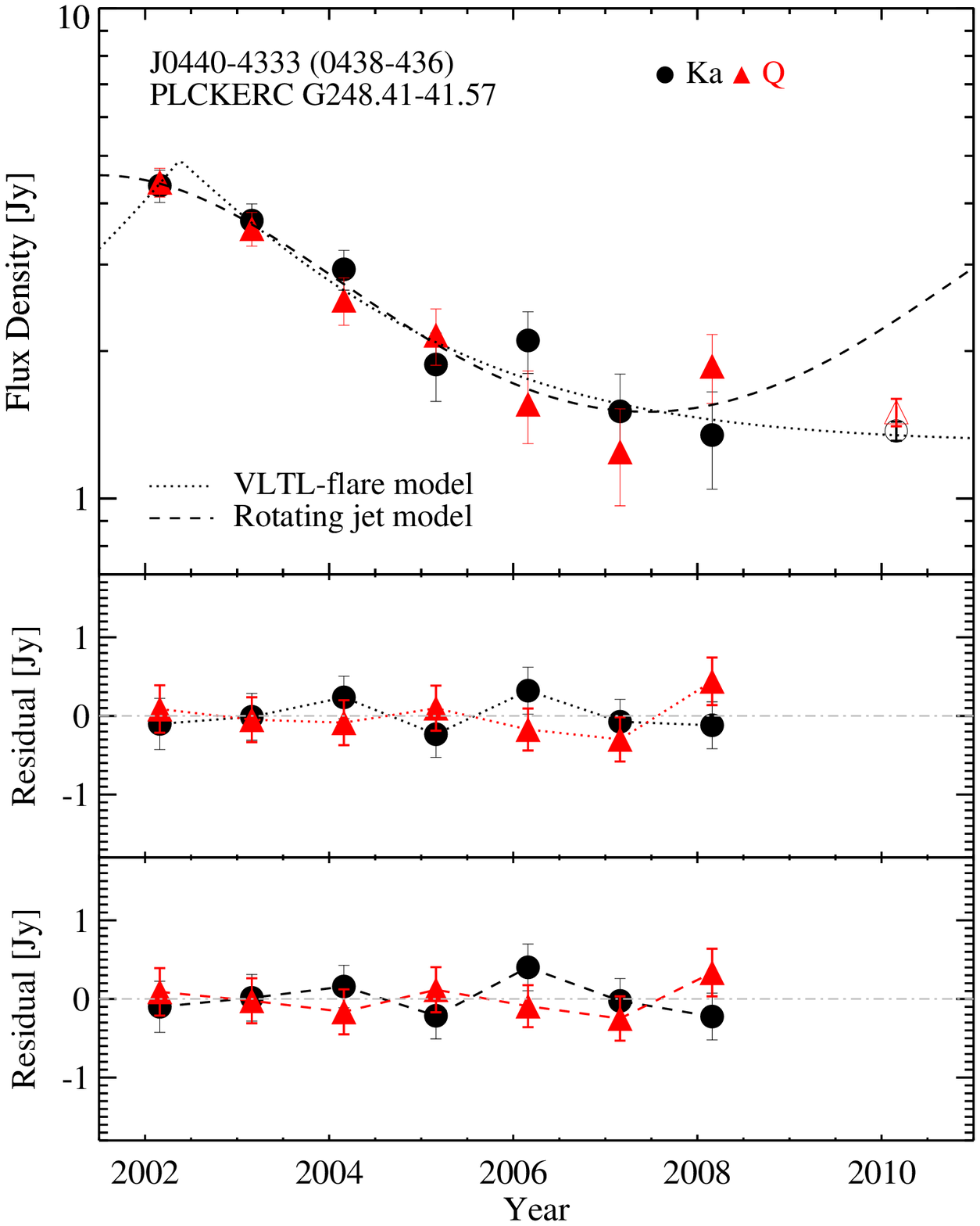} &
\includegraphics[width=0.315\textwidth]{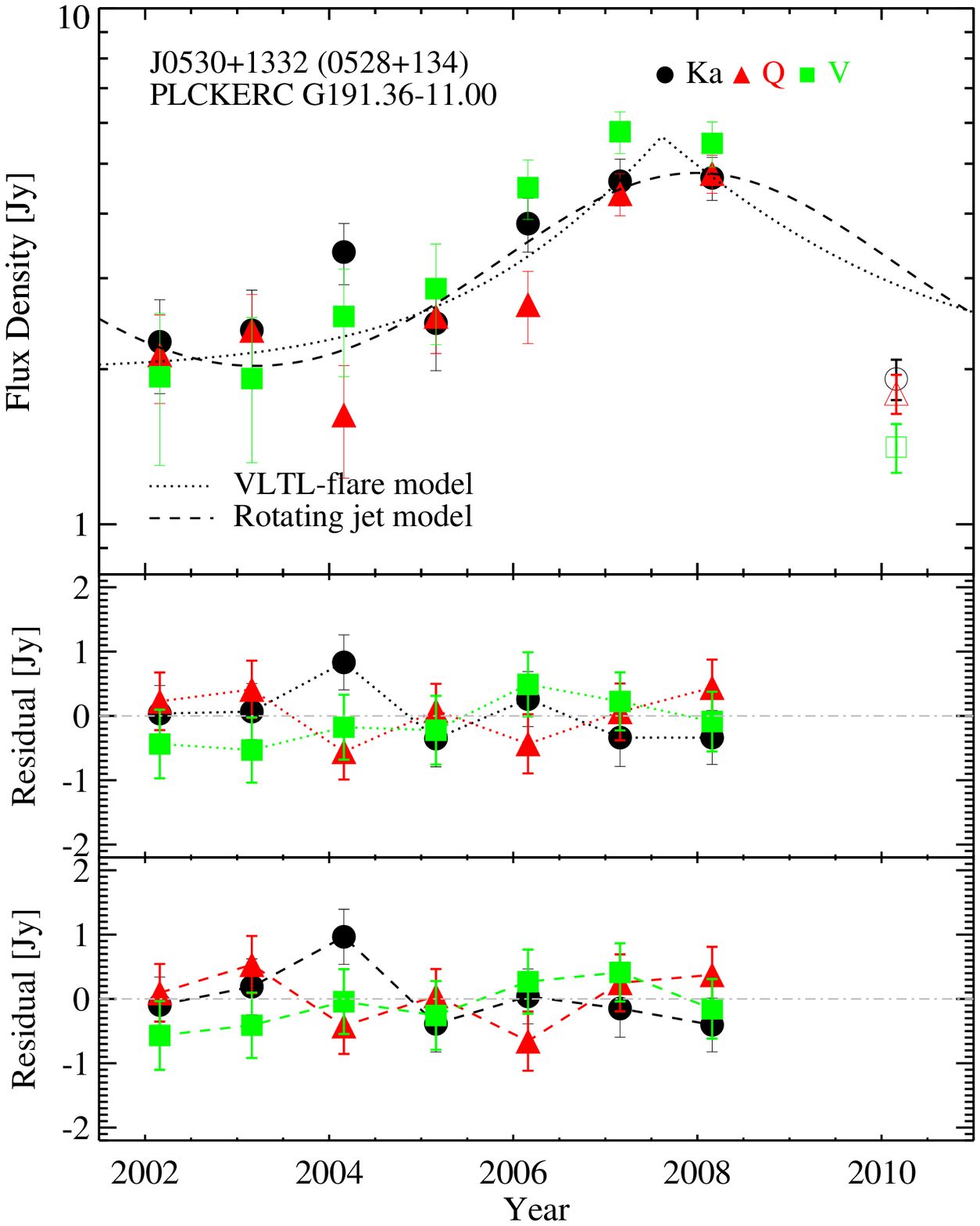} \\
\includegraphics[width=0.315\textwidth]{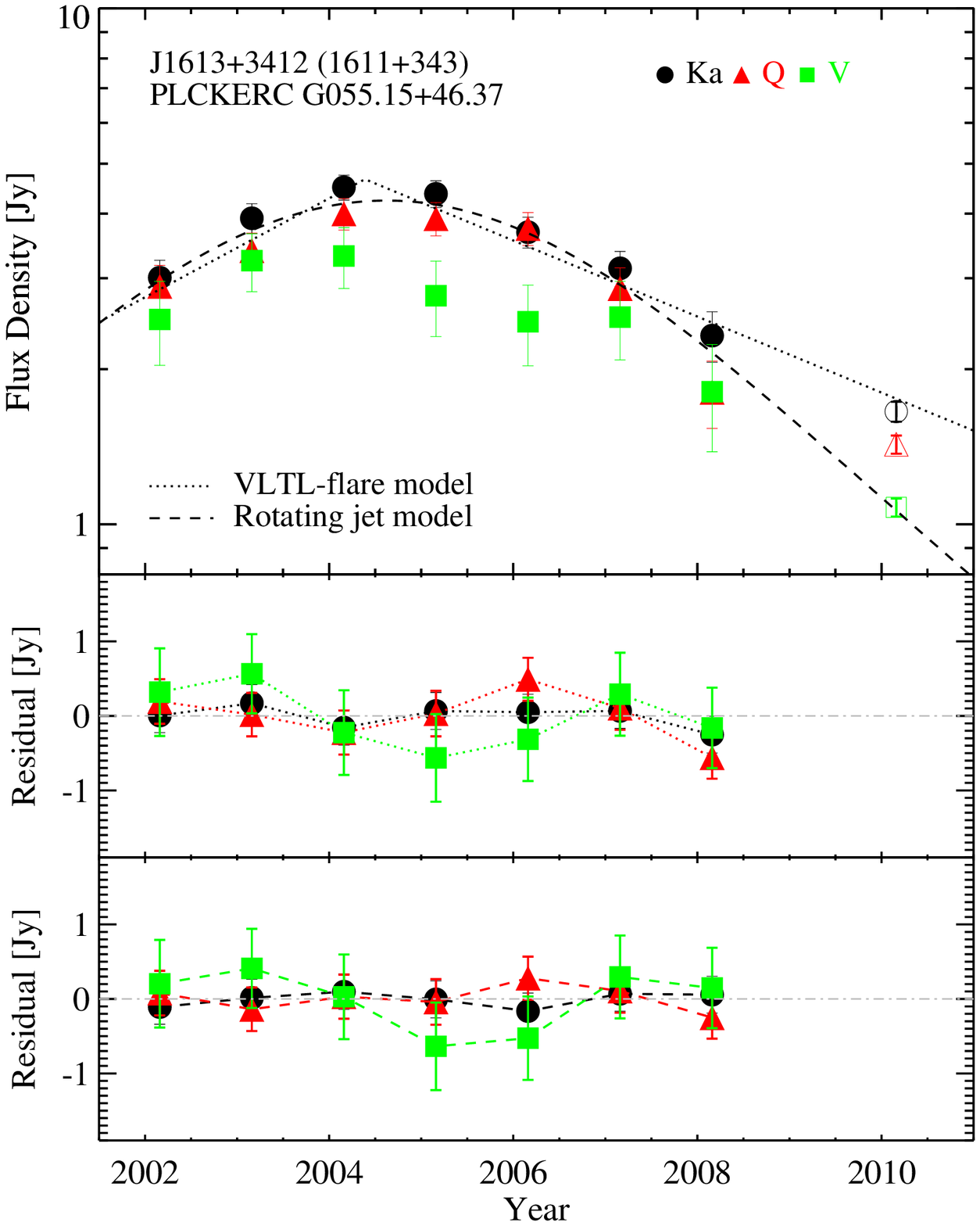} &
\includegraphics[width=0.315\textwidth]{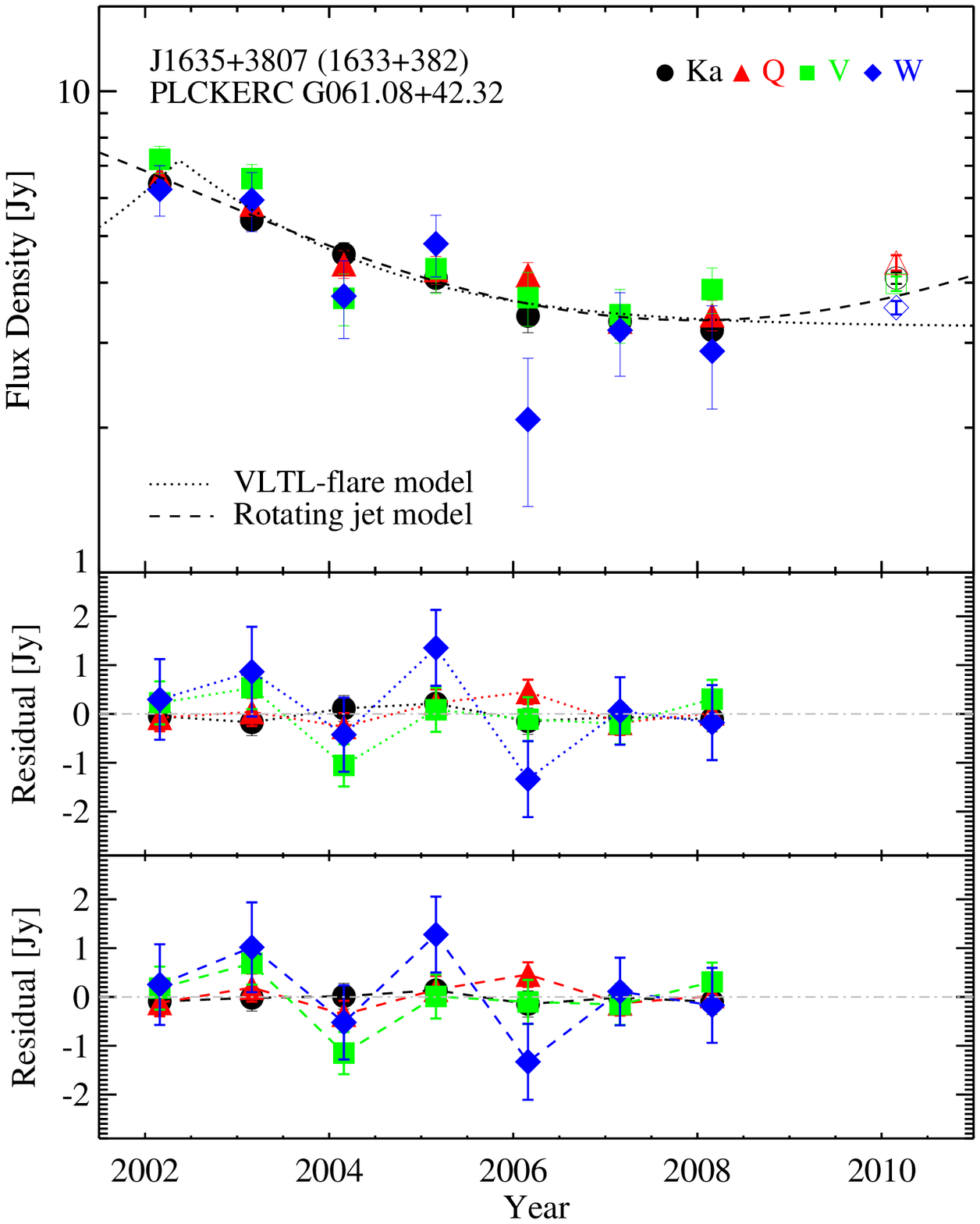} &
\includegraphics[width=0.315\textwidth]{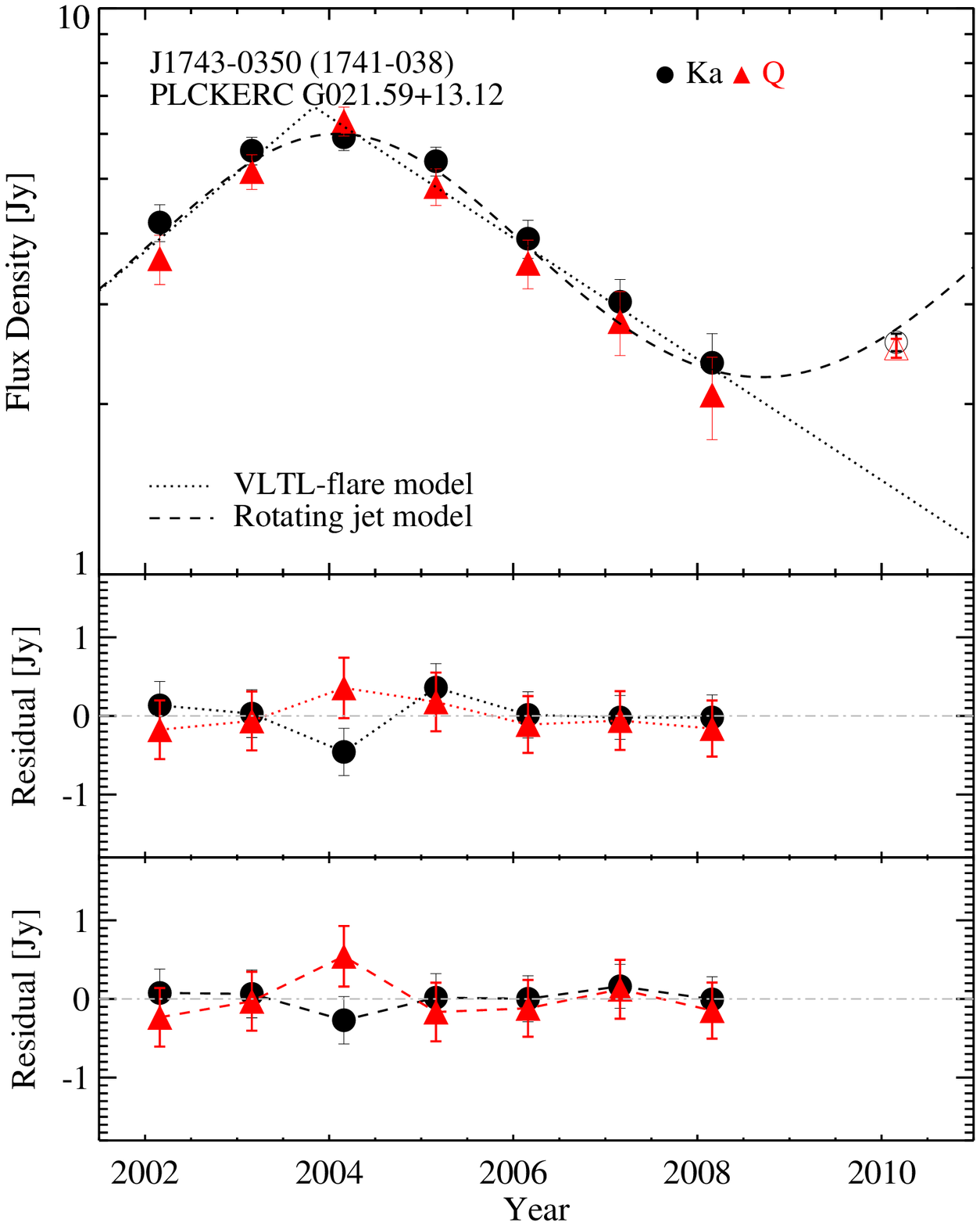} 
\end{tabular}
\caption{{\it Top panels}: \WMAP/\Planck\ light curves for the sources that
are well fit by both the VLTL-flare model and the rotating
jet model. The \WMAP\ data are shown by filled symbols and the later \Planck\ data are
shown by open symbols. {\it Middle panels}: Residuals between the normalized \WMAP\
flux densities (Eq.~\ref{eq:scaling}) and the VLTL-flare model. {\it Bottom panels}: Residuals between the normalized \WMAP\
flux densities and the rotating jet model. Grey dash dot lines indicate the zero level in the residual plots. \label{fig:ambiguous}}
\end{figure*}

\begin{figure}
\centering
\includegraphics[width=0.315\textwidth]{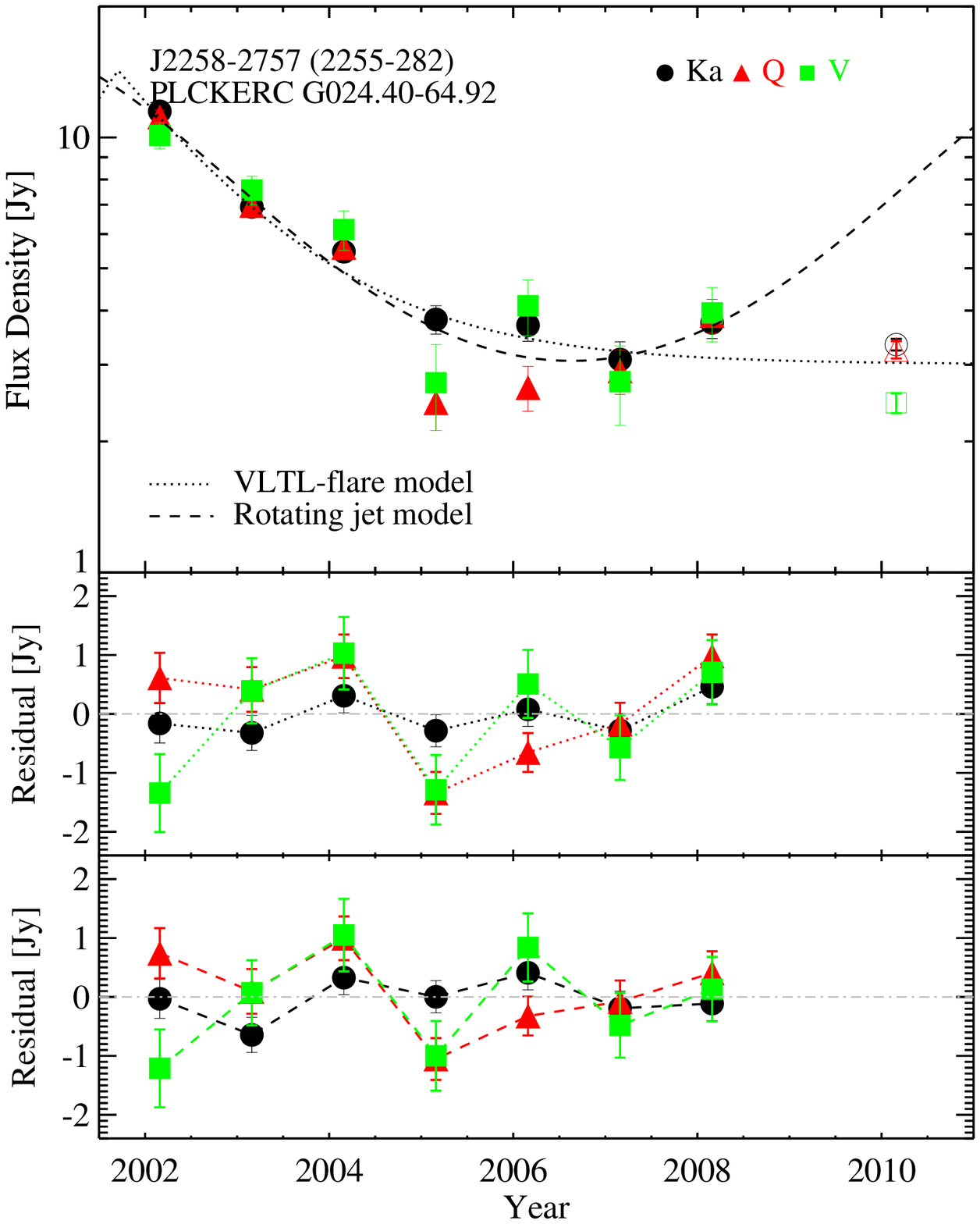} 
\addtocounter{figure}{-1} 
\caption{Continued.}
\end{figure}

\subsection{Results}

Among the $32$ sources in our sample that show strong correlated variability between Ka-band and at least another band, we
find $19$ ($60\%$) whose long-term averaged variability pattern is well
represented by a simple four-parameter model. The other $13$ sources show more
complex variability patterns.

\subsubsection{Simple Patterns}

A preference for a rotating jet pattern is found for sources J0428$-$3756,
J0457$-$2324, J0721+7120, J1159+2914, J1310+3220, J1337$-$1257, J1733$-$1305,
J1924$-$2914, and J2235$-$4835. Most of these sources show strongly correlated
variability in more than two bands. The rotation periods found are in the range
3--5\,yr. The shortest rotation period of $3$\,yr is found for
J1337$-$1257, which showed more than two complete oscillations within our observing
period. The \WMAP/\Planck\ light curves of these sources, as well as the fitted pattern, are shown in Fig.~\ref{fig:rot}. Although the fitting is done on the scaled \WMAP\ flux densities, we plot the original \WMAP\ flux densities to demonstrate that the model traces the variation trend of the original data very well. The residuals shown in
the bottom panels are, however, the differences between the normalized flux densities
and the fit.

For sources J0106$-$4034, J0253$-$5441, J0403$-$3605, J0423$-$0120, J0440$-$4333,
J0530+1332, J1613+3412, J1635+3807, J1743$-$0350, and J2258$-$2757, both the
rotating jet model and the VLTL-flare model yield a reasonable match to the
long-term flux variation pattern. Fig.~\ref{fig:ambiguous} gives their
\WMAP/\Planck\ light curves, together with the pattern fit. With the
exception of J0423$-$0120, the rotation
periods of these sources in the rotating jet model are found to be ${\ge}\,7$\,yr, and the
flare time scales in the VLTL-flare model are ${\sim}\,0.5{-}2$\,yr. For
five sources with rotation periods ${>}\,10\,$yr ($\tau>1.4\,$yr) we mostly see
only the increasing or decreasing part of the pattern in the time
window of our observations.

The \Planck\ ERCSC flux densities, though not used in the fitting, are found to be in good agreement with the model fits in some cases, e.g., J0457$-$2324, J1310+3220. When both models fit the \WMAP\ data, the \Planck\ flux densities sometimes prefer one model over the other, e.g., J0403$-$3605 for the rotating-jet pattern, and J2258$-$2757 for the VLTL-flare pattern. In about 50$\%$ of the cases, the \Planck\ data are in disagreement with the fitted pattern. This is likely due to the unmodelled short-term variability and the difference between the actual \Planck\ observation times and the referenced time (set to be exactly 2 yr later than the last \WMAP\ data point).

\subsubsection{Sources with complex variability}
\label{sec:complex}

There are 13 sources for which neither the VLTL-flare model nor the rotating jet
model fits the variation pattern well. These sources can be divided into two
classes: (1) those showing highly-correlated variability across all bands; and
(2) those showing significant differences between the variability patterns at
different bands. 

\begin{figure*}
\centering
 \begin{tabular}{ccc}
\includegraphics[width=0.315\textwidth]{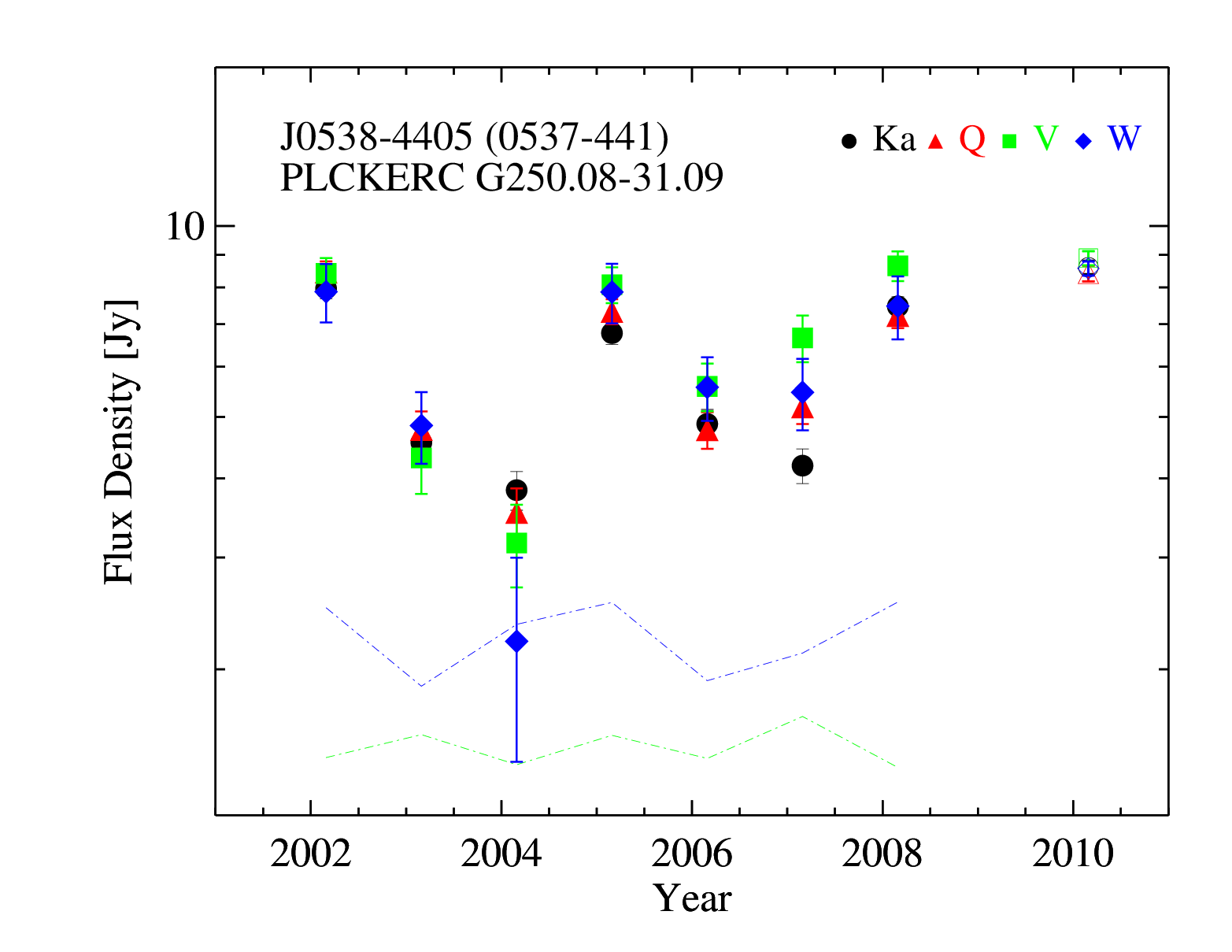} &
\includegraphics[width=0.315\textwidth]{figures/lc/J1229+0202.eps} &
\includegraphics[width=0.315\textwidth]{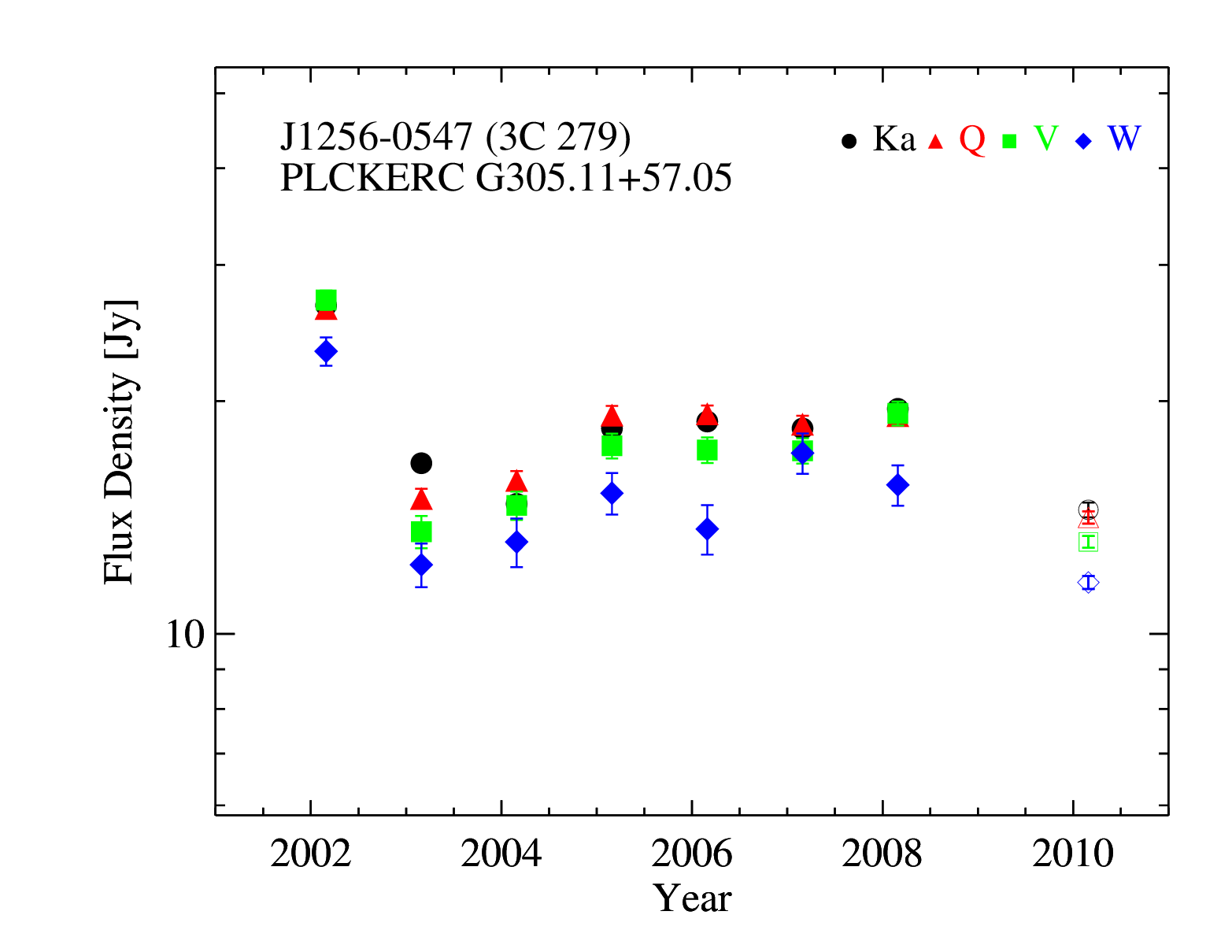} \\
\end{tabular}
 \caption{\WMAP/\Planck\ light curves for the sources J0538$-$4405, 3C\,273 and 3C\,279.
 The \WMAP\ flux densities are shown by filled symbols and the
\Planck\ flux densities are shown by open symbols of the same shape. The dashed
lines indicate the $3\sigma$ level at each band if in the plotted range.\label{fig:case1}}
\end{figure*}

\begin{figure*}
\centering
 \begin{tabular}{ccc}
\includegraphics[width=0.315\textwidth]{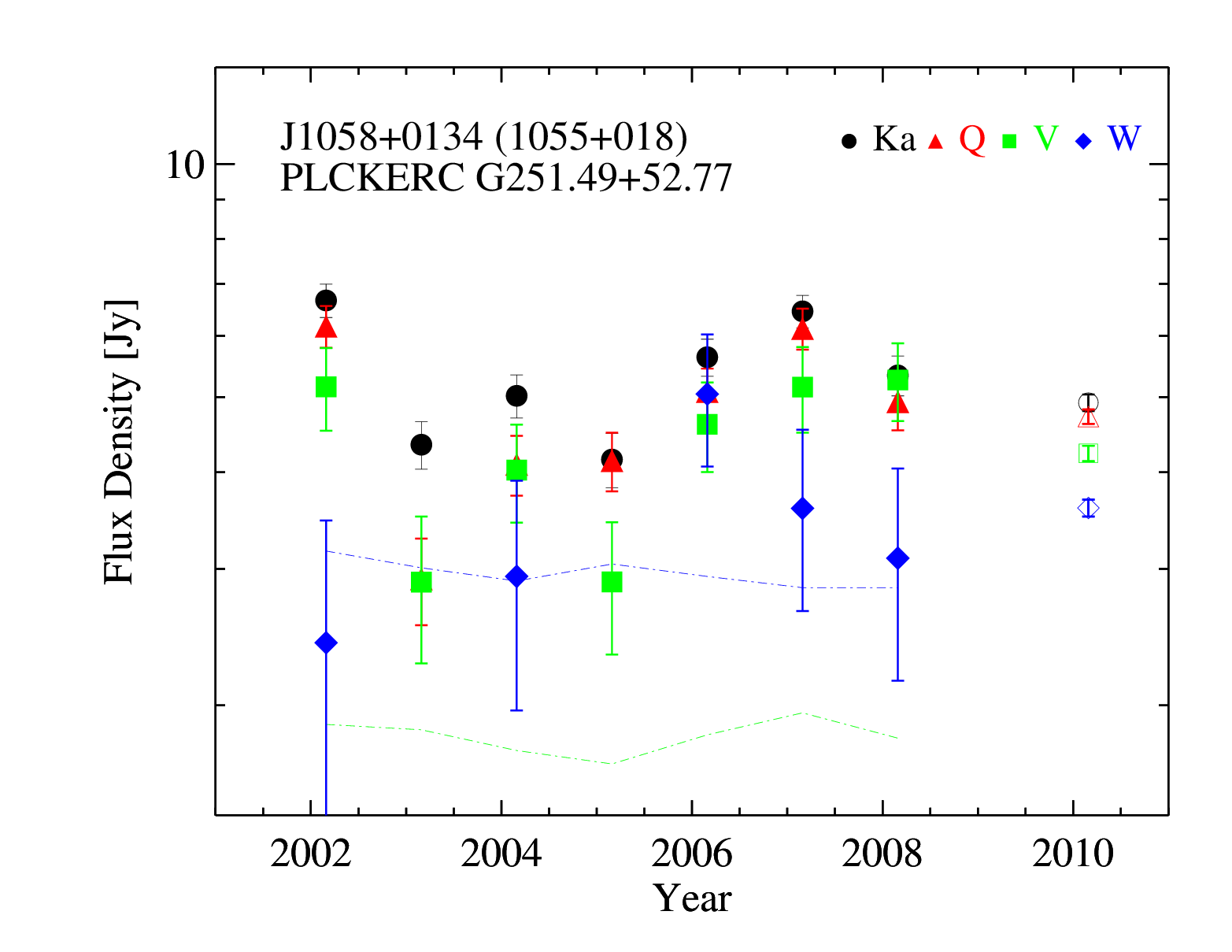} &
\includegraphics[width=0.315\textwidth]{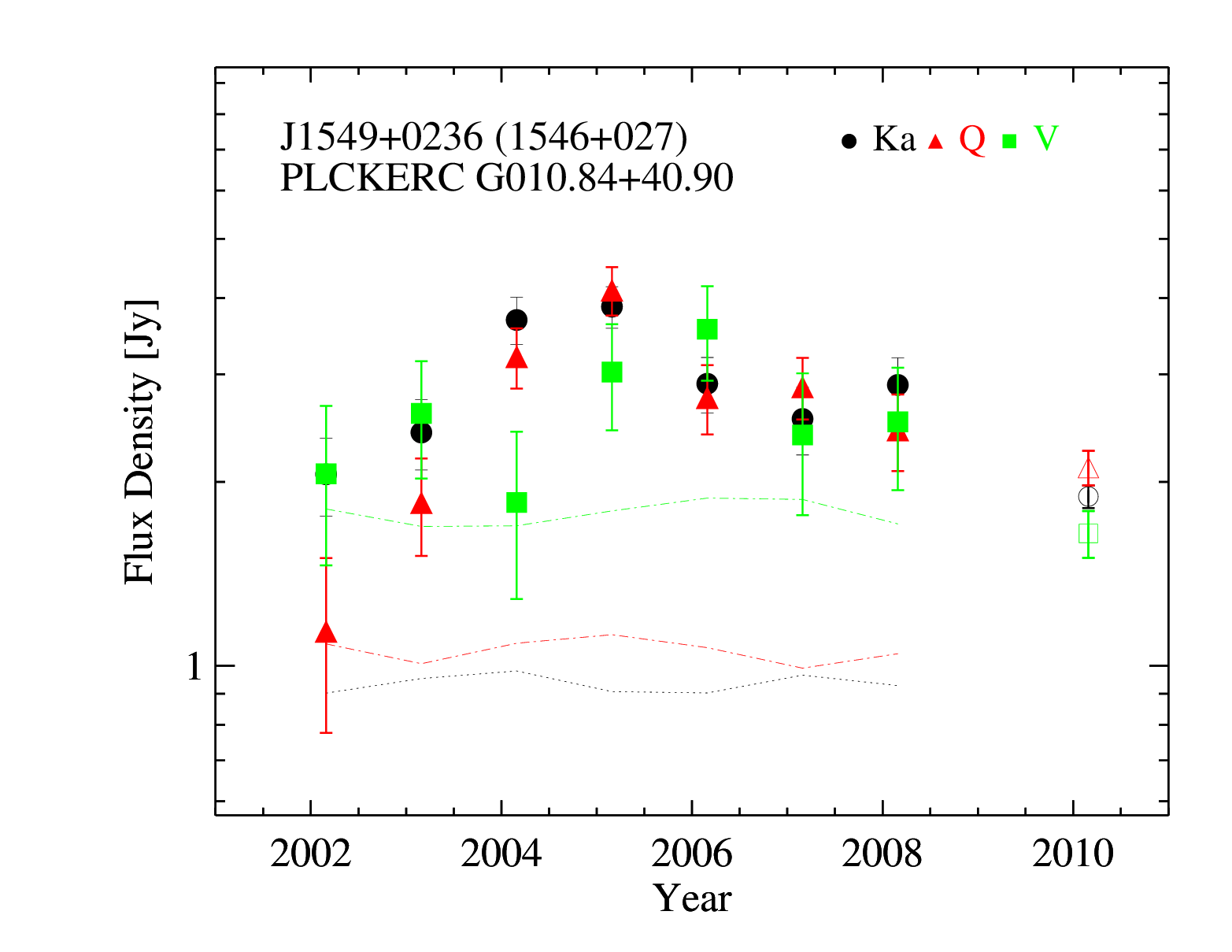} &
\includegraphics[width=0.315\textwidth]{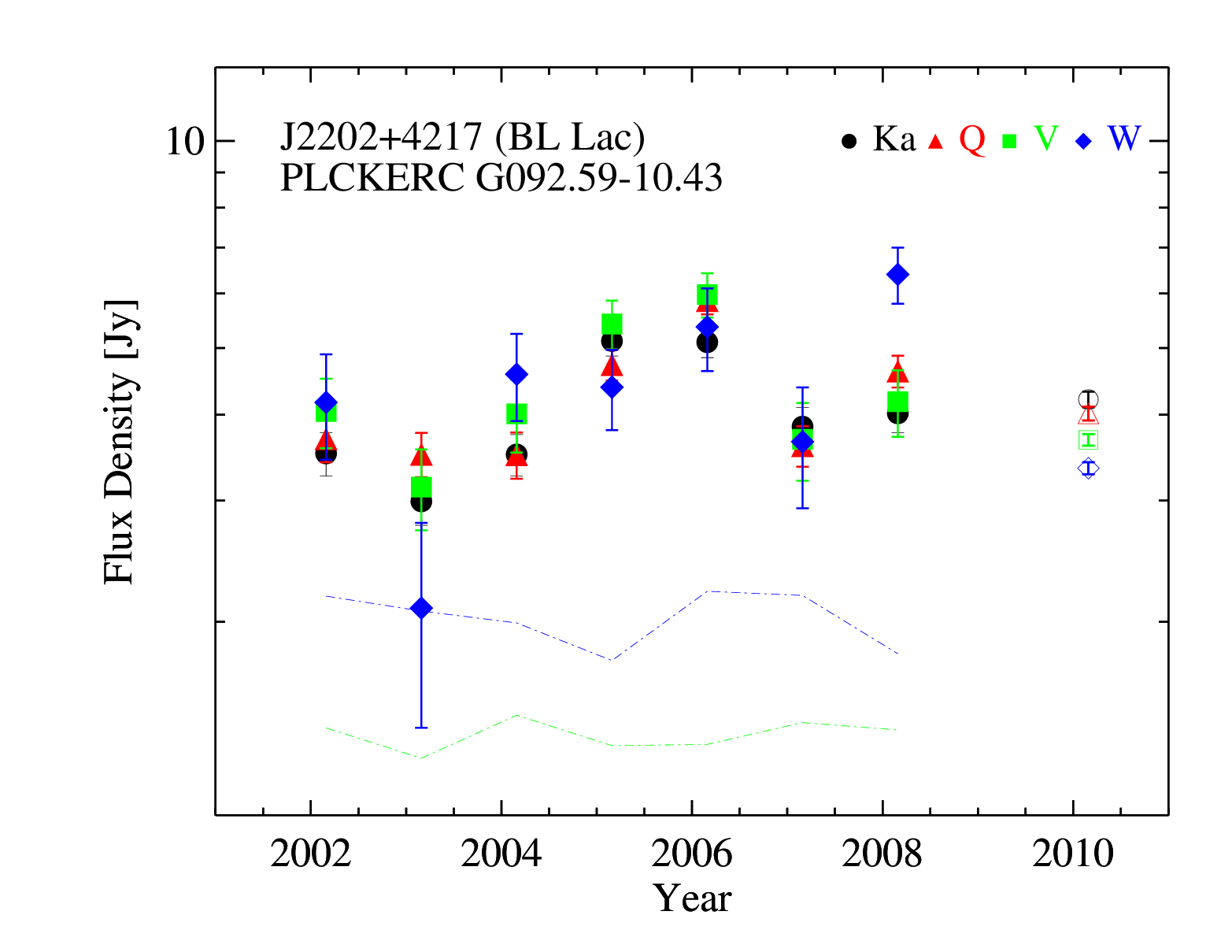} \\
\includegraphics[width=0.315\textwidth]{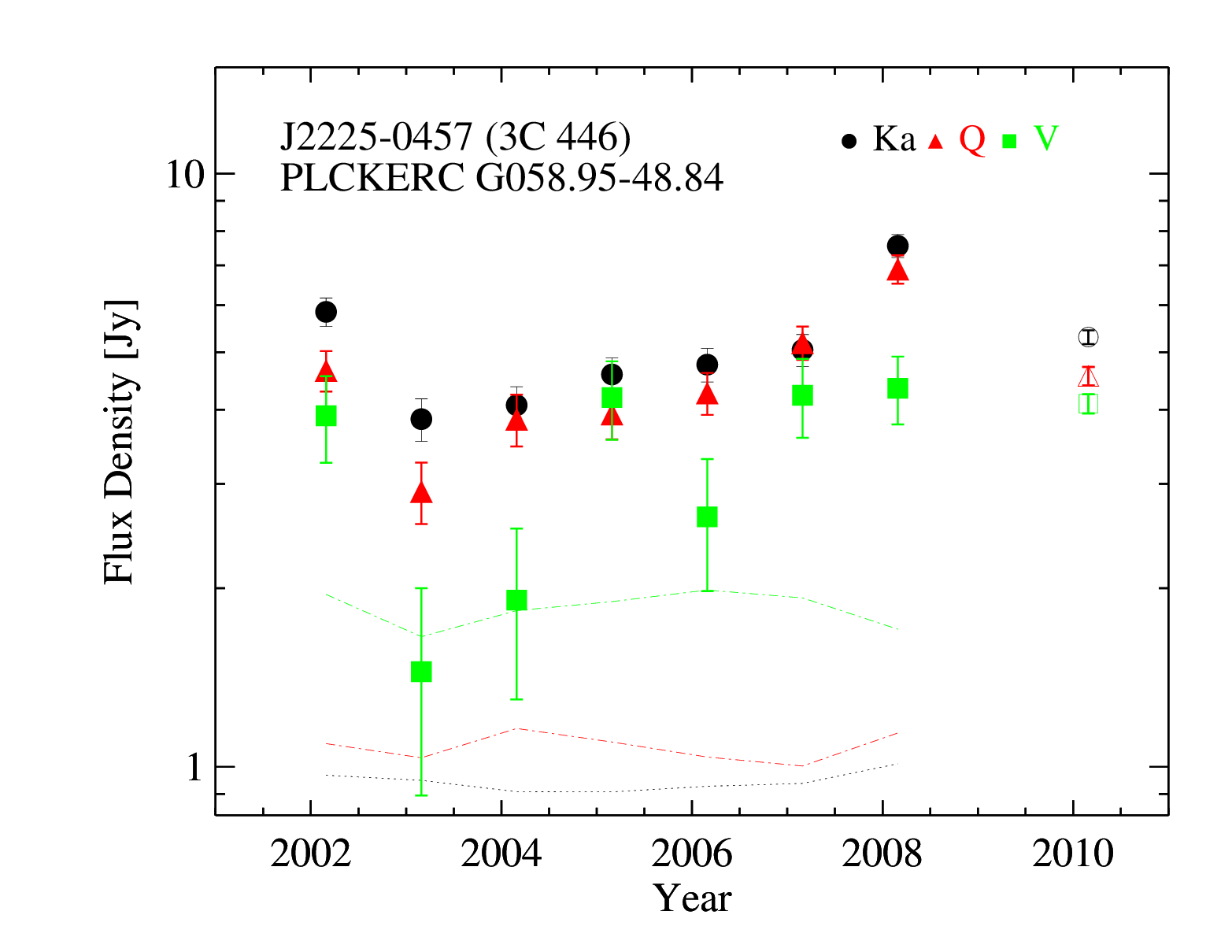} &
\includegraphics[width=0.315\textwidth]{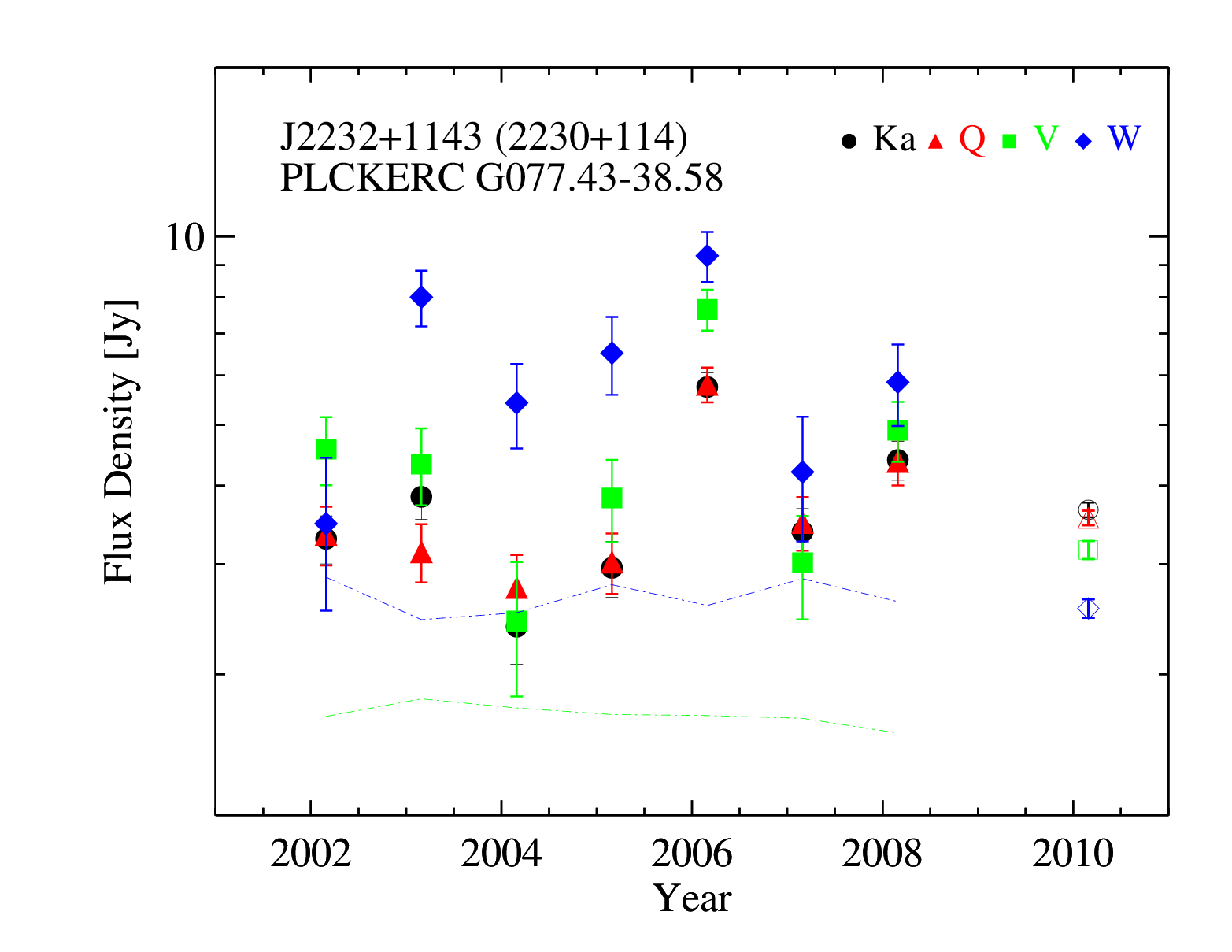} &
\includegraphics[width=0.315\textwidth]{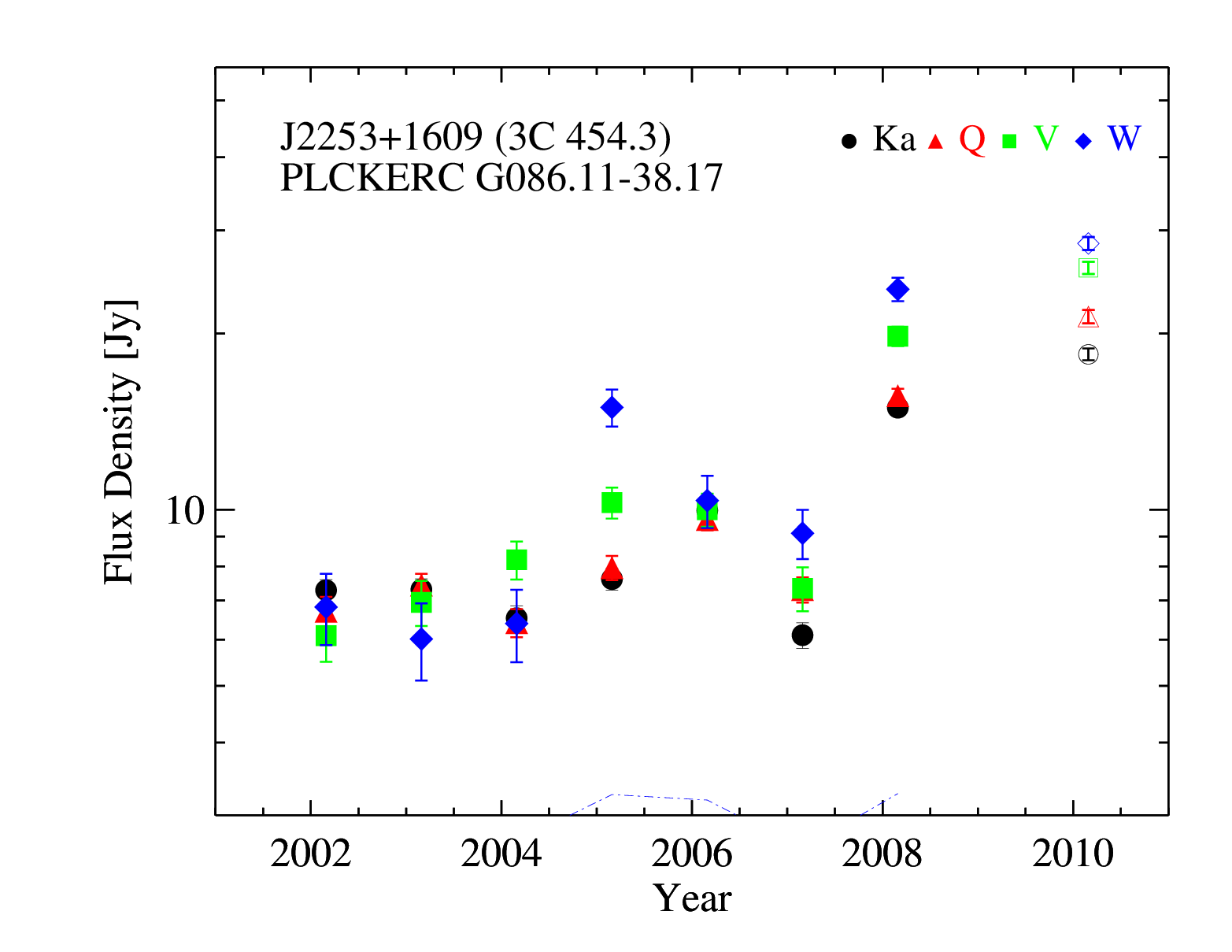} \\
\end{tabular}
 \caption{\WMAP/\Planck\ light curves for the sources J1058+0134, J1549+0236, BL\,Lac, 3C\,446, J2232+1143 and 3C\,454.3. 
 The \WMAP\ flux densities are
shown  by filled symbols and  the \Planck\ flux densities are shown by  open
symbols of the same shape. The dashed lines indicate the $3\sigma$ level at each
band if in the plotted range.\label{fig:case2}}
\end{figure*}

In the first category, we find bright FSRQs like 3C\,273, 3C\,279, and
J0538$-$4405 (Fig.~\ref{fig:case1}). These sources would probably be fitted by a
correlated pattern comprising several VLTL flares, or a rotating pattern plus
long-term flares. Sources in the second category all show moderately
correlated, partially inverted spectra at higher frequencies. A famous example
is 3C\,454.3, for which our data capture the well-studied millimetre flare from
2005 \citep{Krichbaum2008}. A very strong inversion in the V--W spectral index
is also found in source J2232+1143 between 2002 and 2004. In addition, the light
curves of BL Lac, 3C446, J1058+0134, and J1549+0236 all display similar spectral
behaviour (see Fig.~\ref{fig:case2}). Such signatures of spectral dependence of
variability are well known at shorter time scales, and are expected from physical models of both shock-induced flares and geometric
effects in rotating helical jets \citep{Villata1999, Ostorero2004}.

\subsubsection{Comparison with long-term monitoring data}
\label{sec:longterm}

A limitation of our analysis is the small dynamic range in time
scales within which our method can distinguish between different patterns; time
scales smaller than one year are not resolved, and for time
scales longer than a few years our fits become ambiguous, as discussed above. 

For the three sources J0423-0120, J0721+7120, and J1310+3220 that are
included in the 37\,GHz long-term monitoring program on the Mets\"{a}hovi radio
telescope (A.~L\"ahteenm\"aki, private communication, see also
\citealt{Terasranta1998}), we can test the validity of our
finding over a long time period ($>20$\,yr). In order to make the Mets\"{a}hovi
data comparable to our \WMAP\ data, we averaged them into one year bins (see
Appendix~\ref{app:Metsahovi}). Fig.\ref{fig:lightcurve_metsahovi} shows the the
\WMAP\ Ka and Q band light curves along with the binned Mets\"{a}hovi 37\,GHz
data. In all three cases, we find excellent agreement between the Mets\"{a}hovi
binned data and the \WMAP\ data. This confirms that long-term averaged data
provide a good and consistent representation of the long-term variability
trends, regardless of the details of the averaging method.

The three examples both highlight the potential strengths and reveal the various
problems of our approach. For source J0423$-$0120, neither the Mets\"{a}hovi data nor
the \Planck\ data support the rotating-jet pattern that is compatible with
the \WMAP\ data; rather, the binned Mets\"{a}hovi data
suggest that the source underwent a single, strong activity phase between 2001
and 2005, well fitted by a VLTL flare profile with $\tau\approx 0.5\,$yr,
superimposed on a flux density which remained roughly constant
between 1980 and 2010.
For source J0721+7120, both the Mets\"{a}hovi and the \Planck\ data are
consistent with the fitted rotational pattern between 2002 and 2010, but before
1998 the long-term average flux density of the source was continuously
in a low state. This could be interpreted as the onset of a new rotating pattern
as expected, e.g., in lighthouse scenarios \citep{Steffen1995}. An alternative
explanation would be to assume two consecutive long-term flares (or flaring
phases) of $\tau \sim 1\,$yr in the years after 2002.
Finally, for source J1310+3220, the Mets\"{a}hovi and \Planck\ data are
consistent with a continuation of the rotating pattern fitted to the \WMAP\ data
alone at least for the period 1998--2010, covering almost two full oscillations.
Additionally, the long-term averaged 37\,GHz light curve observed between 1983
and 1990 fits the same pattern, with a consistent phase, as expected of
geometric variability from stable rotating jet; the departure from the pattern
between 1990 and 1998 could be explained by a single
superimposed flare ($\tau\sim 1\,$yr). 

\begin{figure*}
\centering
\includegraphics[width=0.8\textwidth,height=0.16\textheight]%
{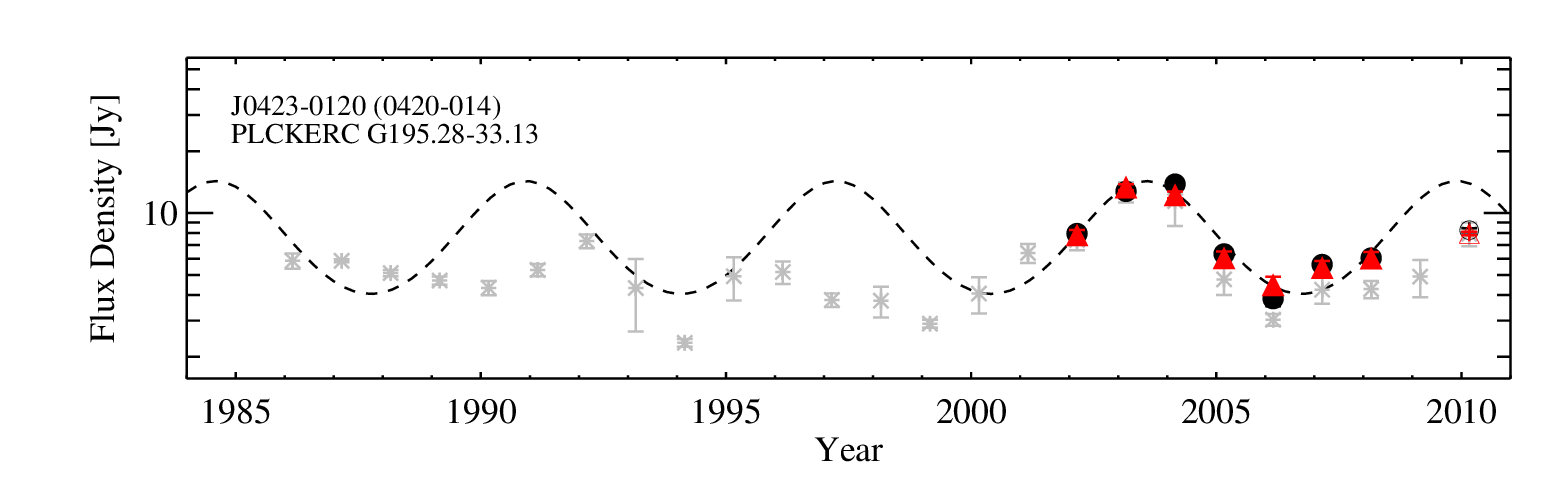} \\
\includegraphics[width=0.8\textwidth,height=0.16\textheight]%
{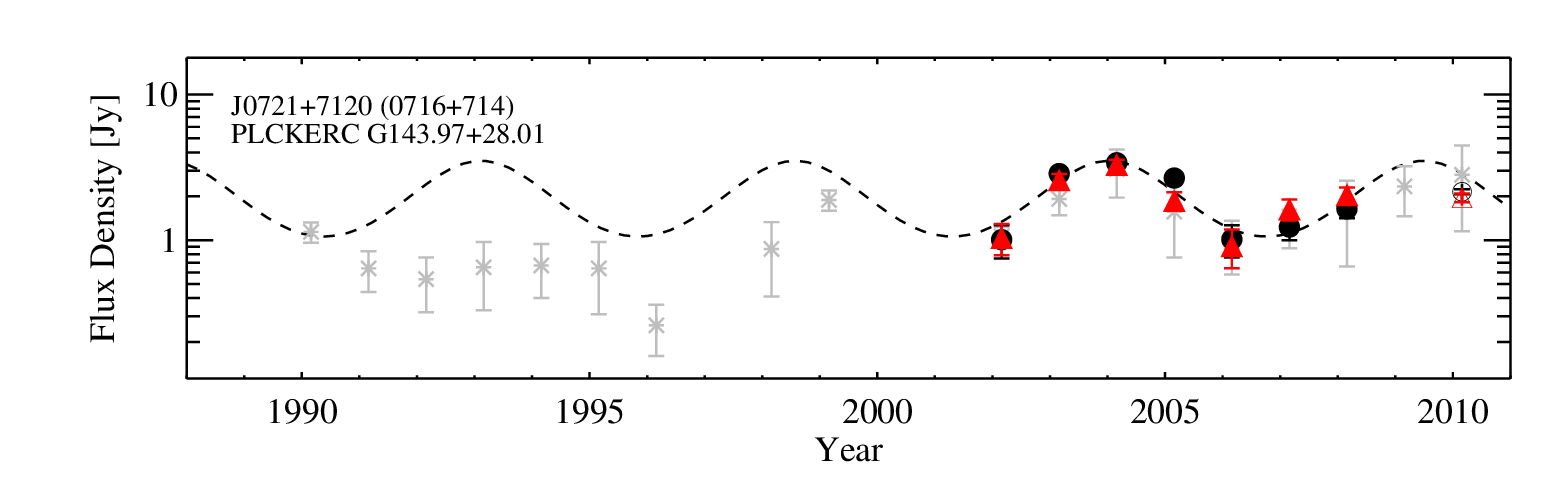} \\
\includegraphics[width=0.745\textwidth,height=0.22\textheight]%
{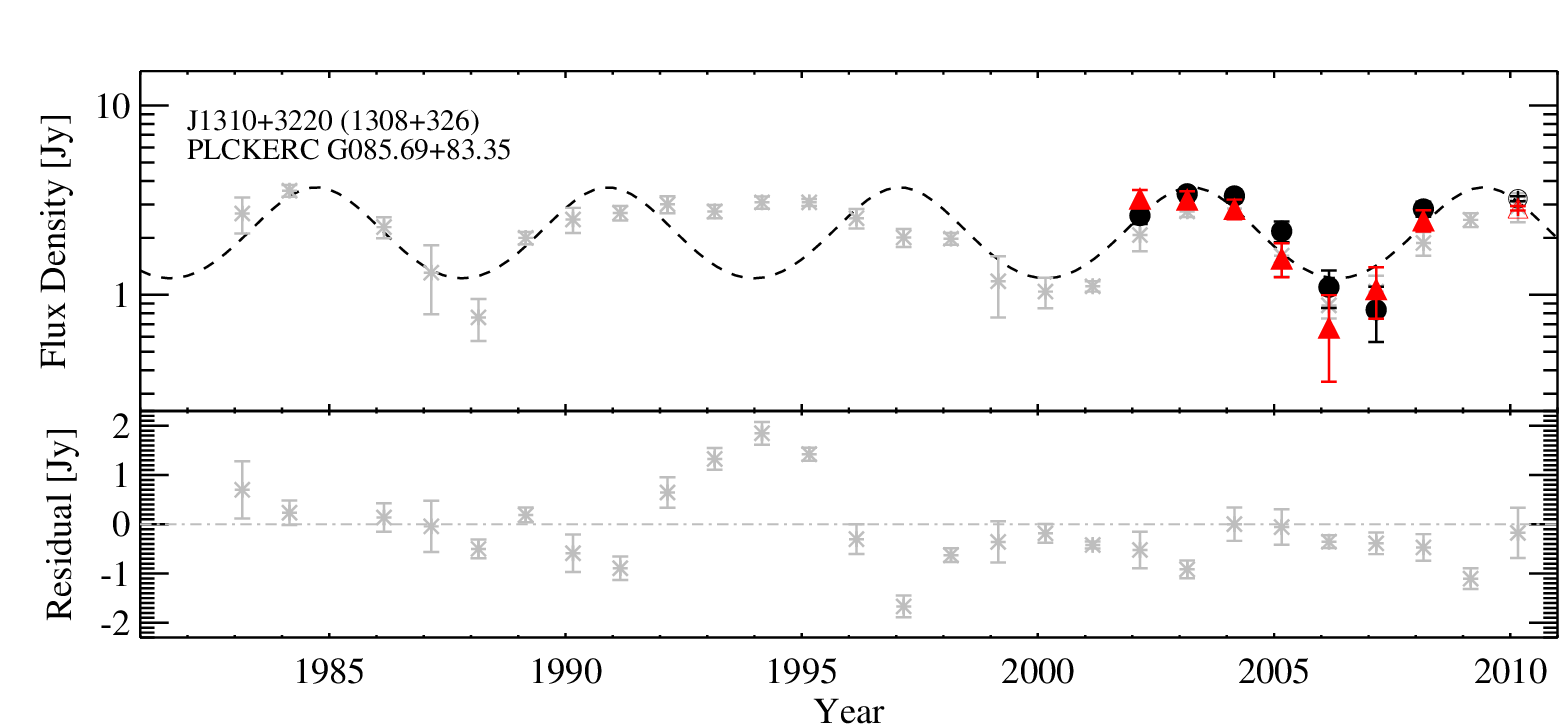} 
 \caption{\WMAP/\Planck\ light curves for the sources J0423$-$0120, J0721+7120,
and J1310+3220 at 33\,GHz ({\it black filled circles}) and 41\,GHz {\it (red
filled triangles}), with  the Mets\"{a}hovi 37\,GHz long-term monitoring data added
({\it grey stars}). For J1310+3220, the residuals between the Mets\"{a}hovi
data and the best fit model are shown. \label{fig:lightcurve_metsahovi}}
\end{figure*}

\section{Variability in unbeamed sources}
\label{sec:unbeamed}

Our sample includes some bright radio galaxies, which in the
standard unification model \citep[e.g.,][]{Urry1995} are the unbeamed
counterparts of
blazars. 
We note that for a source with inclination angle $\theta \gg \Gamma^{-1}$,
the contribution of the
compact, highly relativistic jets is not only unbeamed, but \textit{deboosted}
with a Doppler factor $D\sim 0.1$ if the source is exactly viewed from the side.
This suppresses its contribution to the total emission of the source by a factor
${\gsim}\, 10^{6}$ relative to a beamed blazar. Therefore, in such sources all
the
emission
must  come from the extended jets and lobes, which
are subdominant in highly beamed blazars.

\begin{figure}
\centering
\includegraphics[width=0.45\textwidth]{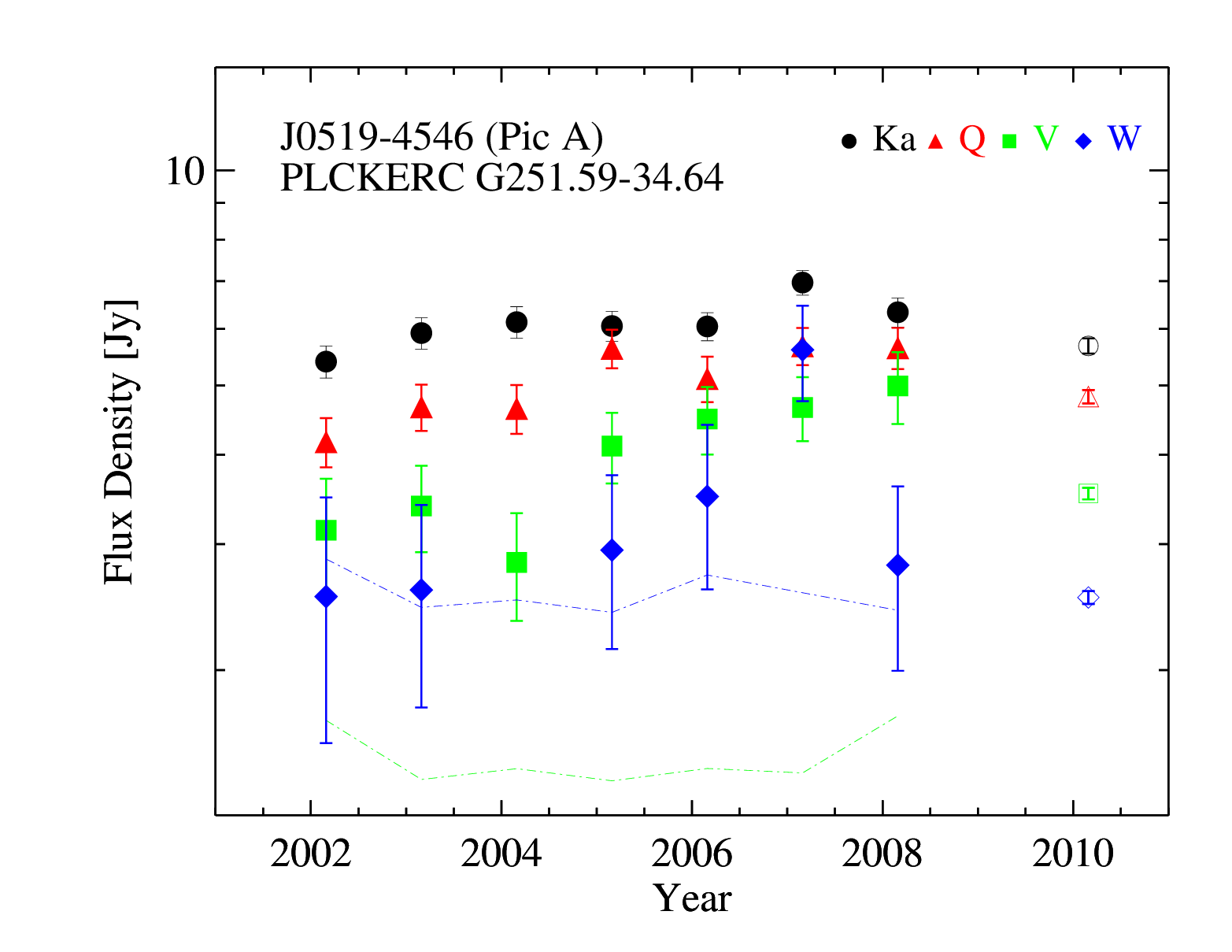}
\caption{\WMAP/\Planck\ light curves of the FR\,II radio galaxy Pic\,A.
\label{fig:picA}}
\end{figure}

The extended radio galaxies in our sample, most notably
Cen\,A, Cyg\,A, and M\,87, are all classified as non-variable at all
frequencies, which is expected because the emission regions in
these sources extend over several kiloparsecs, too large to show significant variability on
the time scales we probe. This is a consistency check on
the reliability of our photometry.

\subsection{A millimetre-wavelength flare in the giant radio galaxy Pictor A}

The giant radio galaxy
Pic\,A, however, is an exception: it shows moderate to strong variability in the Ka, Q, and V bands,
with $\chi^2$ values of $17.0$, $18.3$, and $15.6$, respectively.
The flux densities in the \WMAP\ Ka, Q, and V band were generally increasing
until 2008. \Planck\ ERCSC data suggest that in 2010 the
flux density returned to its 2002 value for all bands (Fig.~\ref{fig:picA}). 
The $\chi^2$ values suggest a \textit{combined} probability of ${\sim}\, 10^{-6}$ for constant flux
density in all bands, so the variation is highly
significant. From the shape of the variation, which is
essentially coherent in all bands, we can estimate an observed flare time
scale of ${\sim}\, 3$\,yr, assuming an exponential rise, and a somewhat
shorter decay time scale.

Pic\,A is a Fanaroff-Riley class II (FR\,II) radio galaxy at a redshift
$z=0.035$. It shows two prominent radio lobes, with bright hot spots
terminating the jets at distances of about 140 kpc from the central AGN.  They
are connected to the AGN by extended jets, which appear
to propagate with mildly relativistic speeds \citep{Meisenheimer1989}. Within
these central jets, at a distance of ${\sim}\,$30 kpc from the AGN, Chandra
observations provided evidence for a transient X-ray knot, which
apparently decayed within the time scale of ${\sim}\, 1$\,yr
\citep{Marshall2010}. 
Assuming that the X-ray radiation was produced by synchrotron radiating
electrons, \citet{Marshall2010} concluded that the magnetic field in the
emission region must have been ${\sim}\, 2\,$mG, about 100 times larger than the
canonical estimates for the jets of radio galaxies.  This magnetic field, however, is still too weak to explain the time scales of the flare observed by \WMAP\ at millimetre wavelengths (also assuming synchrotron origin), therefore we consider it unlikely that the two flares originate in the same region of the jet.

An alternative explanation is that the increase in millimetre
flux density is not from the extended jet, but from the central
flat-spectrum core, which is almost as bright as the dominant western
hot spot at $20\,$GHz \citep{Perley1997,Burke2009, Sajina2011}.
  Interpreting the variable component as unboosted ($D\sim 1$)
  emission from a central blazar, and assuming standard jet
  parameters, the emission region would have an inclination angle to
  the line of sight of $\theta \approx 25^\circ$, and the observed factor-of-two
  increase in flux density  within three years
  could be explained by a change of inclination angle by
  ${\approx}\,3^\circ$ towards the observer.  However, this would
  require a rotational speed significantly larger than that inferred
  from VLBI observations of the central jet of Pic\,A, if these are
  interpreted in a precessing jet model \citep{Tingay2000}. An
  interpretation of the flare in Pic\,A as due to a shock in the jet
  would be more consistent with the msialigned-blazar picture: with a
  Doppler factor $D\sim 10$ the Pic\,A flare would be comparable in
  {\it intrinsic} time scale and peak frequency with the exceptional outburst in
  3C\,454.3 observed in 2005.
 
\section{Discussion}
\label{sec:discussion}

\subsection{Geometric variability and helical jets}
\label{sec:helical}

Rotating jet patterns have been discussed in the literature in connection with
VLBI observations of morphologically ``curved'' or ``bent'' jets, and of
apparent helical motion of parsec-scale  jet components \citep{Zensus1997,
Kellermann2004}. Among the sources for which we find a variability pattern
matched with the rotating jet model, curved and potentially helical VLBI jets
have been observed in J0423$-$0120 \citep{Wagner1995, Britzen2001}, J0530+1332
\citep{Pohl1996, Cai2006}, J0721+7120 \citep{Bach2005}, J1159+2914
\citep{Hong2004, Zhao2011}, J1337$-$1257
\citep{Lister2009}, J1613+3412 \citep{Piner1997},  and J1924$-$2914 \citep{Shen1999, Lister2009}. Helical
structures have also been found in the VLBI jet of FSRQ  J1310+3220
\citep{Cassaro2002}, for which a rotating pattern with essentially constant
period is largely consistent with the flux variation pattern shown in the
Mets\"{a}hovi 37\,GHz data (\S\,\ref{sec:longterm}). 

Although our data are consistent with geometric variability,
all the above-mentioned sources also show flaring behaviour on short
time scales. An extreme example is the BL\,Lac object J0721+7120, which
showed rapid variability on time scales of a day or less \citep{Ostorero2006}.
In some sources this flaring activity may have obscured an underlying rotating
jet pattern, or in some cases our geometric model may be too simple.
For example,  BL\,Lac shows complex long-term variability
in our data, but its variations have been sucessfully modelled over much
longer time scales as a purely geometric effect, assuming a rotating
inhomogeneous helical jet \citep{Bach2006, Villata2009}. 

Long-term rotation cycles have been identified in
VLBI observations of 3C\,454.3\,\citep{Qian2007} and
3C\,279\,\citep{Carrara1993};
both sources show strong variability correlation between bands, but a complex
variability pattern in our data (\S\,\ref{sec:complex}). Source J0423$-$0120
shows a pattern compatible with the rotating jet model in the \WMAP\ data, but
the 25-year Mets\"{a}hovi 37 GHz light curve does not support a rotational
pattern.

There are a few helical jet candidates that are not included in our variability
pattern test due to weak correlation of variability between \WMAP\ bands. These
include 3C\,84 \citep[NGC\,1275,][]{Krichbaum1992}, 3C\,345 \citep{Steffen1995,
Qian2009}, J1800+7828 \citep{Cassaro2002}, and the famous binary black hole
candidate OJ\,287 \citep{Silanpaa1988, Villata1998, Valtonen2009,
Valtonen2011}. The FSRQ J1833$-$2103 is a gravitationally lensed rotating jet
source \citep{Guirado1999, Nair2005} that is also excluded from our pattern
test, owing to the lack of correlation between flux variations in the
W-band and the other \WMAP\ bands, although visual inspection suggests a
rotating-jet pattern with a period of $<10$\,yr. 

In summary, rotating jets are common features in AGN,
but only some of them display a characteristic sinusoidal pattern
in the long-term averaged \WMAP\ light curves.

\subsection{Relation to other frequency bands}
\label{sec:Fermi}

The long-term variability
and possible periodic behaviour of AGN have been studied using many methods
(e.g., structure function, wavelet, periodogram, DCCF; see
\S\,\ref{sec:blazars}). Evidence for periodic behaviour has mostly come from
optical data, in particular for bright sources where archival optical data
provide time series extending over many decades \citep{Villata1998,
Fan2000}. These periodicities have generally been interpreted as resulting from
geometric effects, potentially in connection with shocks \citep{Lainela1999}.
Whilst large samples of radio time series spanning more than 20\,yr have
not provided strong support for periodicities, WEBT
campaigns on BL\,Lac and AO\,0235+16 (J0238+1637) have shown that time lags
between radio and optical variability can be explained as a geometric
effect, where rotation successively moves regions emitting at
different frequencies into positions of maximal Doppler
boosting \citep{Ostorero2004, Bach2006, Villata2009}.

Gamma-ray emission in blazars is likely to a have a different origin
from the radio emission \citep[e.g.,][]{Ghisellini1998, Rachen2000,
  Boettcher2012}, so comparison of the long-term variability in radio
and gamma-ray data could provide an important cross-check for the
geometric interpretation, which predicts that these data should show
long-term correlations like those seen in the radio and optical
data. Among our sources, 133 are present in the second source
catalogue from the {\it Fermi} Large Area Telescope (2FGL,
\citealt{Fermi_2LAC}), 105 of which show significant variability in
gamma-rays. The uniform sampling of {\it Fermi}-GST makes it an ideal
instrument for such study.

\subsection{Caveats}
\label{sec:caveats}

We find simple, sinusoidal patterns on time scales ${<}\, 10$\,yr in
only nine out of 198 sources, i.e., $4.5\%$ of the full sample. This is
consistent with the results of \citet{Ciaramella2004}, who found
periodic behaviour in about $4\%$ of their irregularly
sampled blazar light curves spanning several decades.  For most
sources, the time scales are too long or the variability is too
complex to fit our simple models.  More complex, physically motivated
models are possible, and have been used to model selected blazar light
curves. However, the added flexibility of the more complex models allows
both geometric models \citep{Villata1999, Ostorero2004, Bach2006} and
shock-based models \citep{Hovatta2008, Nieppola2009} to fit the
data. For example, the exceptional outburst observed during 2005 in
3C\,454.3, has been modelled successfully in both geometric \citep{Villata2007} and
shock-based scenarios \citep{Rachen2010}.  One approach to
discriminate between models of different complexity is to adopt
Bayesian classification methods \citep{Rachen2012}. It is possible
that such methods, in conjunction with more data and better models, may
eventually elucidate the true cause of long-term variability in
blazars.

The present data merely demonstrate that a surprisingly simple sinusoidal
pattern, indicative of geometric variability, is a good representation
of the long-term averaged variability behaviour for more than half of
the sources in our selected sample of 32 sources. It is not yet
possible to exclude any of the more complex explanations for blazar
variability that have been proposed.

\section{Conclusions}
\label{sec:conclusions}

We have used data obtained from the \WMAP\ and \Planck\ satellites to analyse the 
long-term flux-density variations of a sample of extragalactic radio sources selected from 
the \Planck\ ERCSC. We have used flux densities averaged over one-year periods, which suppresses 
short-term variability and makes the data set particularly
suitable for studying underlying, long-term variability patterns. 

At \WMAP\ Ka, Q, V, and W band, we found 82, 67, 32, and 15
sources to be variable at ${>}\,99\%$ confidence level. The number of
variable sources decreases at higher frequencies owing to the higher flux density uncertainties
at these frequencies. The amplitudes of
variation are found to be comparable in the different bands, and are not correlated with
either the flux densities or the spectral indices of the sources. At Ka
band, we modelled the source number counts in each year, and found
that although the fitted parameters for the model vary from year to year as a
consequence of individual source variability, the variation is within the
uncertainty of the fitted parameters, and therefore the number counts derived
from the source sample remain stable over the years. Almost all of our
sources show correlated variability between bands.
Essentially all of the sources in our sample are radio loud AGNs,
and most of them are blazars. We do not find any significant difference between
the two subclasses of blazars, FSRQs and BLLs.

As expected from theory and VLBI observations, both shock induced flares and
geometric variability due to helical or precession jet motion may contribute to
the total variability at long time scales. In more than half 
of the sources that show strong correlated variability between Ka and another band, the long-term averaged variability shows the sinusoidal
patterns expected from simple jet rotation, with periods greater than ${\approx}\,3\,$yr.
 A simple exponential flare model is less successful in
representing our data, but definite conclusions
on the nature of variability cannot be drawn because more complex models cannot be ruled out.

Most extended radio galaxies in our sample
show no sign of variability. An exception is the extended
radio galaxy Pic\,A, for which we found significant changes in the flux density
between 2002 and 2010, which could arise from the  central ``misaligned blazar''  nucleus if the inner jet has a small angle to the line of sight.

The unexpected observation of a millimetre-flare in an extended radio galaxy underlines the potential of cosmology probes and other multi-survey instruments for research on variability and transient phenomena, as they allow comparisons between many, identically sampled full sky surveys.

\begin{acknowledgements}
We thank Dave Clements, Joaquin Gonz\'{a}lez-Nuevo,
Anne L\"ahteenm\"aki, Anthony Lasenby, Luigi Toffolatti, and the
anonymous referee for useful comments and discussion. We gratefully acknowledge
the use of partially unpublished observation data from the Mets\"{a}hovi
telescope 37\,GHz long-term blazar monitoring program. We also make use of data
obtained with the 100-m telescope of the Max-Planck-Institut f\"ur
Radioastronomie (MPIfR) at Effelsberg, and the 30\,m telescope of the Institut
de Radioastronomie Millim\'{e}trique (IRAM) at Pico Veleta. This research has
made use of the NASA/IPAC Extragalactic Database (NED) which is operated by the
Jet Propulsion Laboratory, California Institute of Technology, under contract
with the National Aeronautics and Space Administration, and  the SIMBAD
database, operated at CDS, Strasbourg, France. We acknowledge the use of the
Legacy Archive for Microwave Background Data Analysis (LAMBDA). Support for
LAMBDA is provided by the NASA Office of Space Science. Some of the results in
this paper have been derived using the HEALPix package. 

The development of \Planck\ has been supported by: ESA; CNES and CNRS/INSU-IN2P3-INP (France); ASI, CNR, and INAF (Italy); NASA and DoE (USA); STFC and UKSA (UK); CSIC, MICINN and JA (Spain); Tekes, AoF and CSC (Finland); DLR and MPG (Germany); CSA (Canada); DTU Space (Denmark); SER/SSO (Switzerland); RCN (Norway); SFI (Ireland); FCT/MCTES (Portugal); and DEISA (EU). A full description of
the Planck Collaboration and a list of its members, indicating which technical
or scientific activities they have been involved in, can be found at \url{http://
www.rssd.esa.int/Planck}.

MLC thanks the Spanish Ministerio de Econom\'ia y Competitividad for a Juan de la Cierva fellowship and acknowledges partial financial support from projects AYA2010-21766-C03-01 and CSD2010-00064. CD acknowledges an STFC Advanced Fellowship and an EU IRG grant under the FP7.

\end{acknowledgements}

\bibliographystyle{aa}

\bibliography{Planck_bib,xi_refs}

\begin{appendix}

\section{Quantifying variability in astronomical data sets}

\subsection{Statistical significance and variability index}
\label{app:var}

To quantify significant changes in the measured flux density of a source, recent
astronomical literature has adopted a so-called \textit{de-biased variability
index},
\begin{equation}\label{eq:Vrms_Ak}
V_{\rm rms} = \frac{1}{\langle S\rangle} \sqrt{ \frac{ \sum{(S_i - \langle
S\rangle)^2 - \sum{\sigma_i^2}}}{N} }\quad.
\end{equation}
The expression is usually attributed to \cite{Akritas1996}, but the form of Eq.~\ref{eq:Vrms_Ak} was first applied to 
time series by \cite{Sadler2006}.
For the special case that all measurement errors are equal, $\sigma_i = \sigma$,
$V_{\rm rms}$ can be written as 
\begin{equation}
\label{eq:Vrms_chi}
 V_{\rm rms} = \frac{\sigma}{\langle S\rangle} \sqrt{\frac{\chi^2}{N} - 1}.
\end{equation}
Assuming $N$ is large and the bias through diminishing one degree of freedom
is small, this equation relates to the statistical significance
of variability in a plausible way: it shows that for $\chi^2/N > 1$, i.e., an
average scatter in excess of the measurements error, $V_{\rm rms}$ is positive, and is 
given as a fraction of the average total flux density by scaling with $\sigma/\langle S\rangle$.
For $\chi^2/N < 1$, $V_{\rm rms}$ becomes
imaginary, indicating that the data scatter is less than
expected. Data sets showing a large fraction of sources
with imaginary $V_{\rm rms}$ are likely to have overestimated the measurement errors
and thus underestimated the fraction of truly
variable sources.

From the derivation of this expression in \cite{Akritas1996}, we note
that, in spite of the notation of individual errors, Eq.~\ref{eq:Vrms_Ak} is
based on a valid statistical estimator only for the case of equal errors.
Moreover, it has been derived under an assumption of large sample sizes (see also
\citealt{Franzen2009}). Comparing with Eq.~\ref{eq:Vrms_chi}, we
immediately see where this constraint comes from: (a) the failure to reduce the
degrees of freedom in the normalized $\chi^2$; and (b) the use of the arithmetic
mean $\sigma/\langle S\rangle$, which does not minimize $\chi^2$ except in case
of equal errors. Correcting for this, i.e., doing the transformations
\begin{equation}
\langle S\rangle \to  \wmean{S} \quad,\qquad 
\frac{\chi^2}{N} \to \frac{\chi^2}{N-1}\quad,
\end{equation}
and introducing $\wmean{\sigma}$ as an estimate for the average error
in the scaling factor,
we obtain Eq.~\ref{eq:Vrms}.

\subsection{Dealing with incomplete data}
\label{app:limit}

In this paper, we are limited not only by the small sample size, but also by the small fraction of above 3$\sigma$ detections in a time line in many cases. 
However, given that enough ``good'' (${>}\,3\sigma$) 
data are available to achieve a reasonable estimate for the mean
source flux density, and upper limits in the years in between can provide valuable
information on the variability of the source, we chose to treat all upper limits
as data points with $S_i = 3\sigma_{\rm noise}$ and $\sigma_i = \sigma_{\rm
noise}$, and use Eq.~\ref{eq:chi2} to calculate $\chi^2$. 
Since a reasonable estimate on the average source flux density $S$ \textit{above} the noise level is vital for
a meaningful analysis, we require more good data than upper limits in order to
include a time line in our analysis, i.e., $N_{3\sigma} \ge 4$. 
We also note that the value obtained in
this way for variability strength and significance is a lower limit, as the distribution is cut on
one end by our noise level, which reduces the scatter.

\subsection{Pearson product correlation coefficient}
\label{app:cor}

For our correlation analysis, we calculated the Pearson product correlation
coefficient, defined as
\begin{equation}
 r ({\rm Ka},X) = \frac{1}{N-1} \sum_{i=1}^N 
\left(\frac{S_i^{\rm Ka}-\langle S^{\rm Ka}\rangle}{\sqrt{{\rm var}(S^{\rm Ka})}}\right) 
\left(\frac{S_i^{X}-\langle S^{X}\rangle}{\sqrt{{\rm var}(S^{X})}}\right)
\end{equation}
between the Ka band and other bands $X=$ Q, V, W showing significant variability. 
As all errors in the same band for the same source are
approximately equal, we took $\langle S\rangle$ and var$(S)$ as the
straight, unweighted arithmetic mean and variance of the data in each time line. For
approximate Gaussian distributions of $S$ in each band, and large enough
samples, the variable 
\[
t = r\,\sqrt{\frac{N-2}{1-r^2}}
\] 
follows Student's $t$-distribution for $N-2$ degrees of freedom. 
We note that none of the above assumptions really holds for our data sets, but more exact derivations of correlation confidence do not help as they also depend on the assumption of Gaussianity.

\section{Binning and de-biasing Mets\"ahovi monitoring data}
\label{app:Metsahovi}

To assure compatibility with the \WMAP\ results, we need to bin the Mets\"ahovi
data to match \WMAP\ operation years, starting from August of a given year through July
of the next year. We have to consider that, unlike \WMAP\ with its fixed scanning
pattern, monitoring programs like Mets\"{a}hovi do not sample data in an
unbiased way. Telescope time is usually focused on sources that show rapid
flaring; sources that are not in an active phase are observed less
frequently. Therefore, we have more data points in high states than in low
states of a source, leading to strong bias in simple averages. To
reduce the bias, we bin the data using a weighted average
\begin{equation}
\langle S\rangle_j = \frac{\sum_i S_{ji} \Delta t_{ji}}
{\sum_i \Delta t_{ji}}\;, 
\end{equation}
where $j$ denotes the number of the bin, and the sum runs over all data points
included in the bin with individual flux densities $S_{ji}$. Here, $\Delta t_i =
t_{i+1}-t_i$ represents the time difference between adjacent flux density
measurements in the monitoring, and is used as the weight factor in the average to compensate the bias introduced by the irregular sampling rate. In a similar way, we determine a weighted average of squared flux density
$\langle S^2\rangle_j$, and obtain the standard deviation 
\begin{equation}
 \Sigma_j = \sqrt{\langle S^2\rangle_j - \langle S\rangle_j^2},  
\end{equation}
which describes the amount of variation of the source flux density within the one-year
bin. We note that this should not be confused with a measurement error; rather it reflects the standard deviation of the distribution of data points within the
bin. It may therefore be regarded as a measure of the strength of short-term
variability during the respective one-year period.

\section{Brightness variation in relativistically beamed rotating emitters}
\label{app:pattern}

We consider a beamed emitter rotating with angular velocity $\omega$ around a cone of opening angle $\Phi$
about a central axis of a jet. Let $\Theta$ be
the angle of the line of sight to the central axis of the cone. The inclination
angle of the jet at time $t$ to the line of sight is then given by
\begin{equation}
 \cos\theta = \cos\Phi\,\cos\Theta + \sin\Phi\,\sin\Theta\,\cos\omega t \,,
\end{equation}
which can be simply written as $a + b\cos\omega t$. For a beamed emitter, $a\approx 1$ and $b\ll 1$.
Inserting this into the definition of the Doppler
factor and setting $a = 1$, we obtain
\begin{equation}
\label{eq:Dsine}
 D \approx \frac{2\Gamma}{1 + A\,\cos\omega t}
\end{equation}
with $A = b\beta/(1-\beta) < 1$, where we
used $1-\beta^2 = 1/\Gamma^2$, and eventually set $\beta=1$.  As $S_\nu \propto
D^{3+\alpha}$, this leads to Eq.~\ref{eq:rotpattern}.

If we consider a combined variability process which comprises both flare 
and geometric variability (see \S\,\ref{sec:longtermvar}), Doppler boosting acts
as a multiplicative factor on the light curve describing the flare-induced
process. For a rotating jet, this means that Eq.~\ref{eq:rotpattern} not
only describes the oscillation of the average flux density of the
jet, but also a modulation of the amplitude $\Delta S$ of all short-term flares
in it. For the special case of $(1-A)^{3+\alpha} <
\langle\sigma\rangle / \langle\Delta S\rangle$, this can lead to a pattern in
which short term variability, clearly detectable in the high states of
boosting, becomes insignificant in the low states. For a more detailed
treatment of rotating patterns with superimposed inhomogenities we refer to
\citet{Ostorero2004}, and references therein.

\end{appendix}

\Online

\begin{appendix} 
\begin{figure*}
\centering
\begin{tabular}{cccc}
\includegraphics[width=0.23\textwidth]{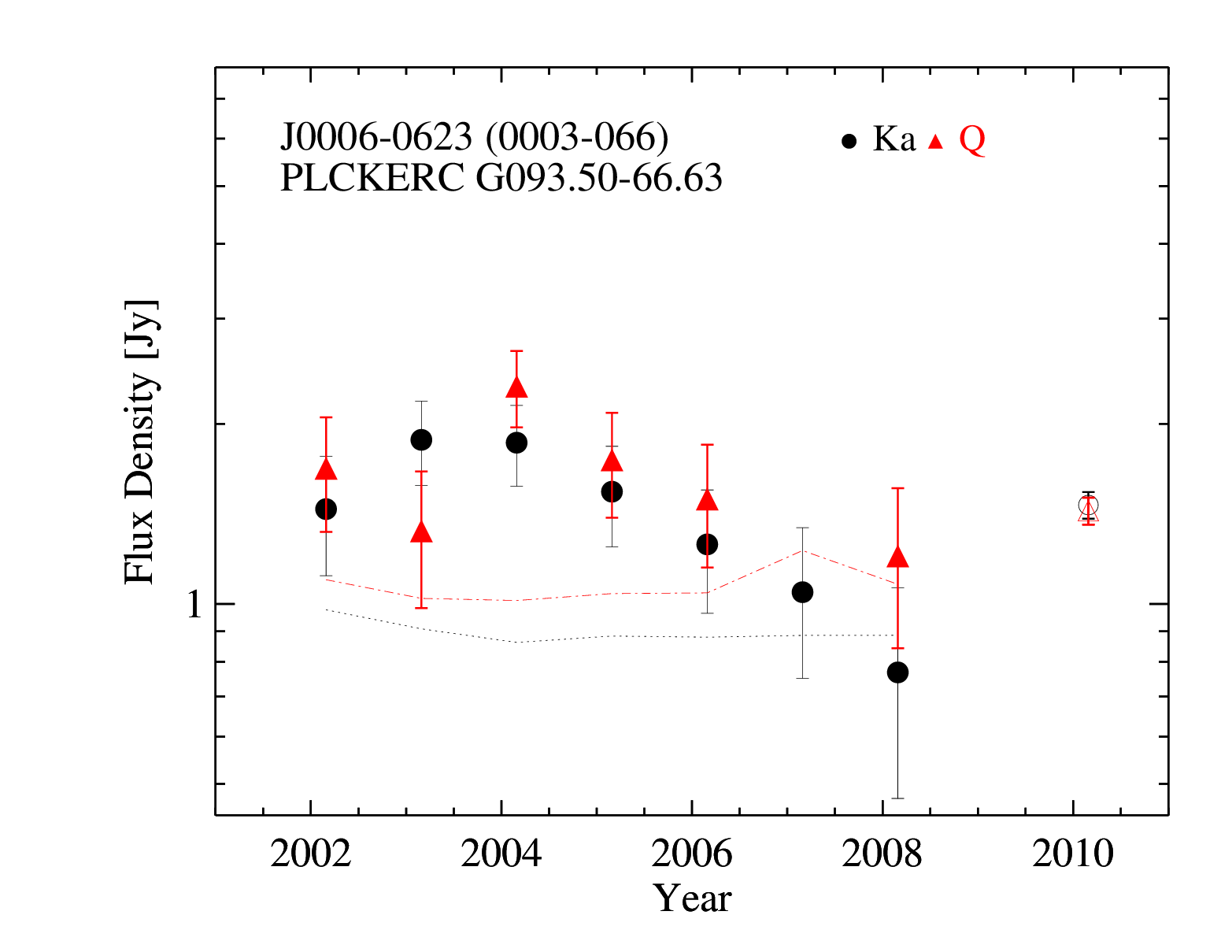} & \includegraphics[width=0.23\textwidth]{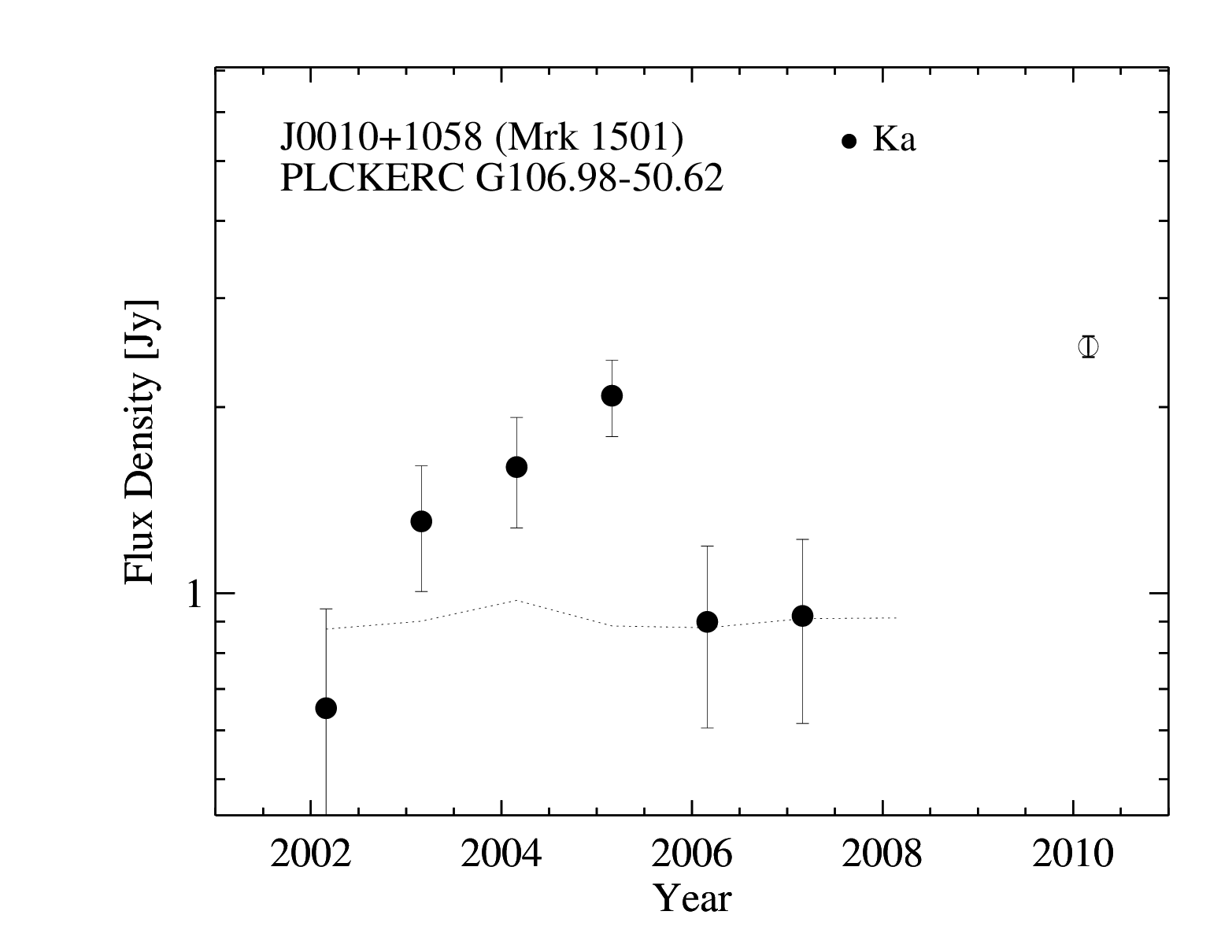}  & \includegraphics[width=0.23\textwidth]{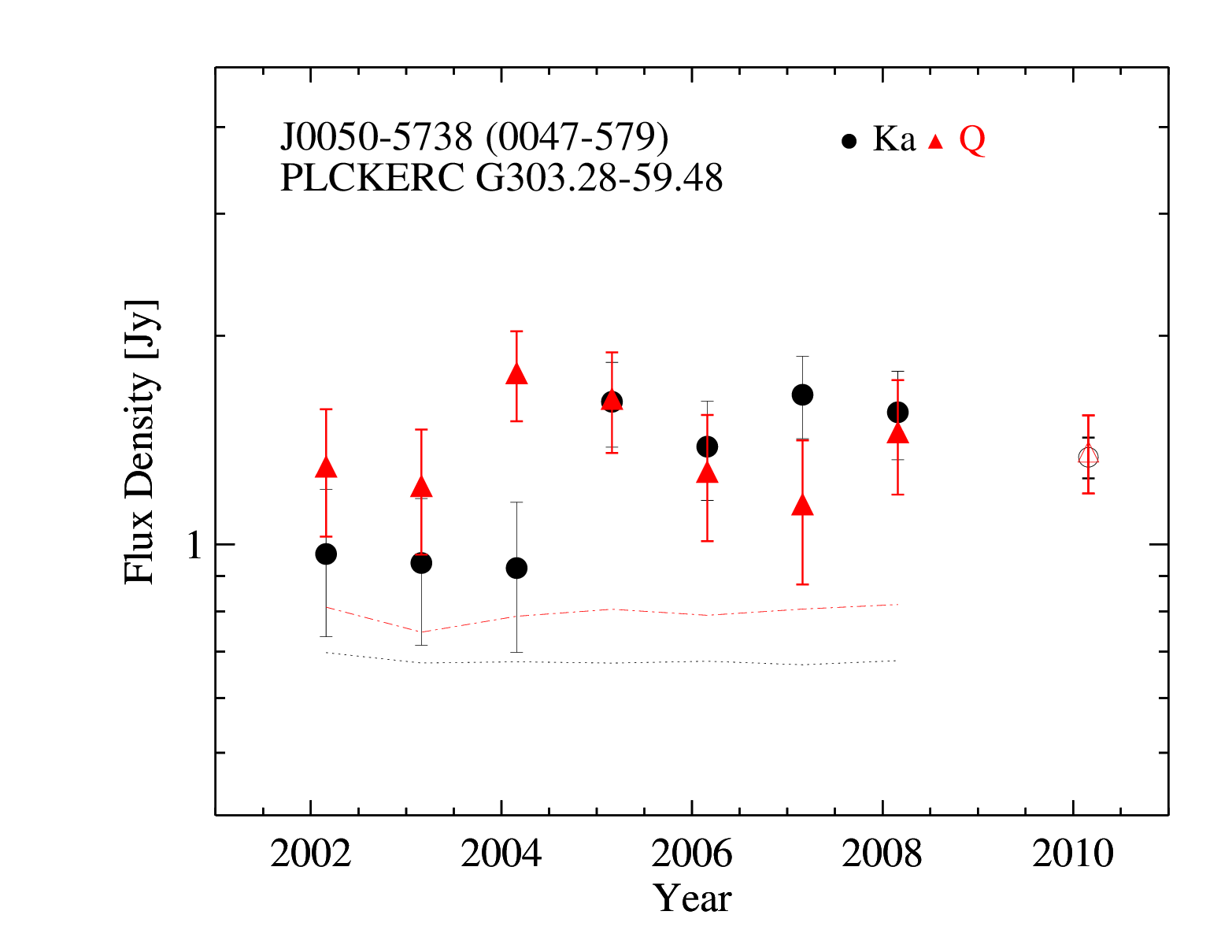} & \includegraphics[width=0.23\textwidth]{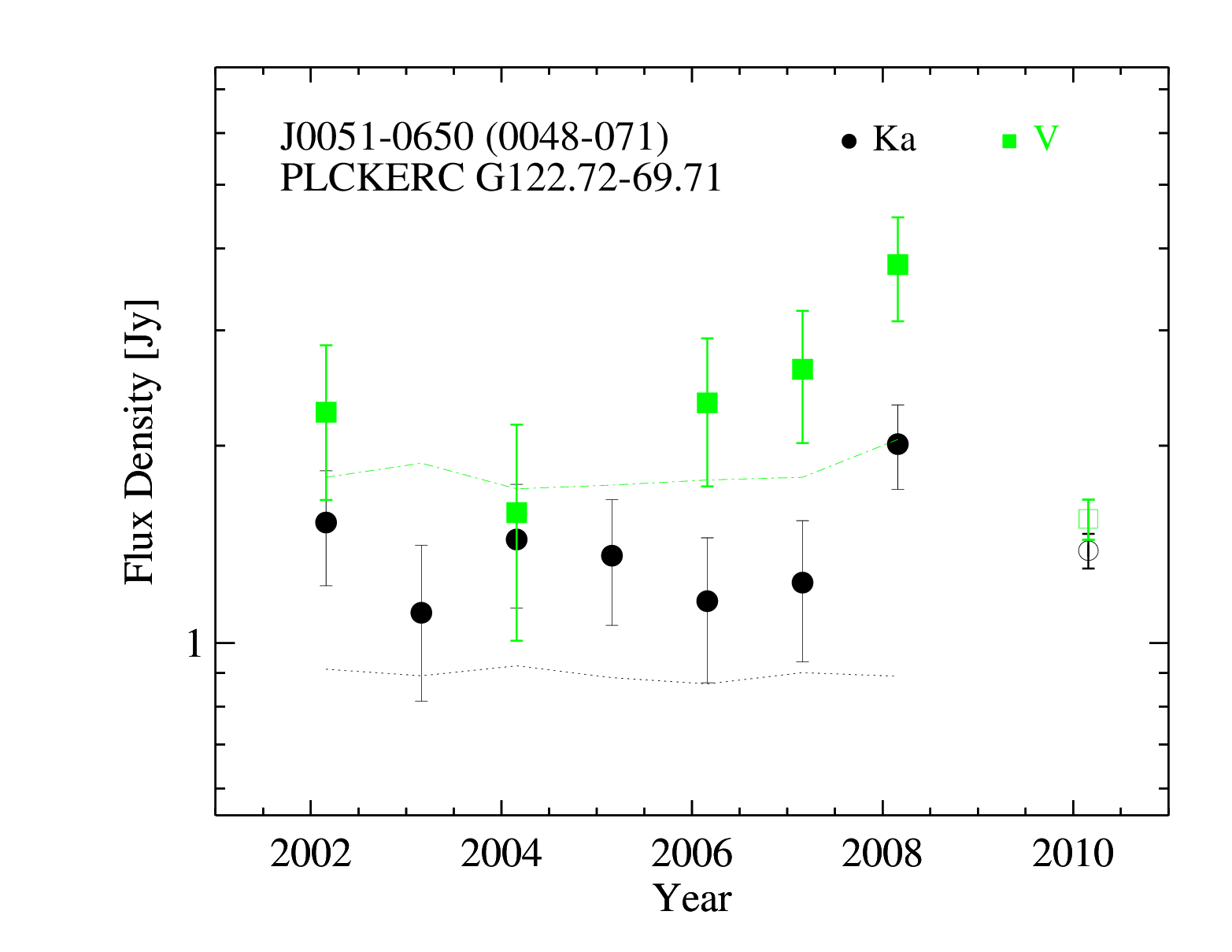}  \\
\includegraphics[width=0.23\textwidth]{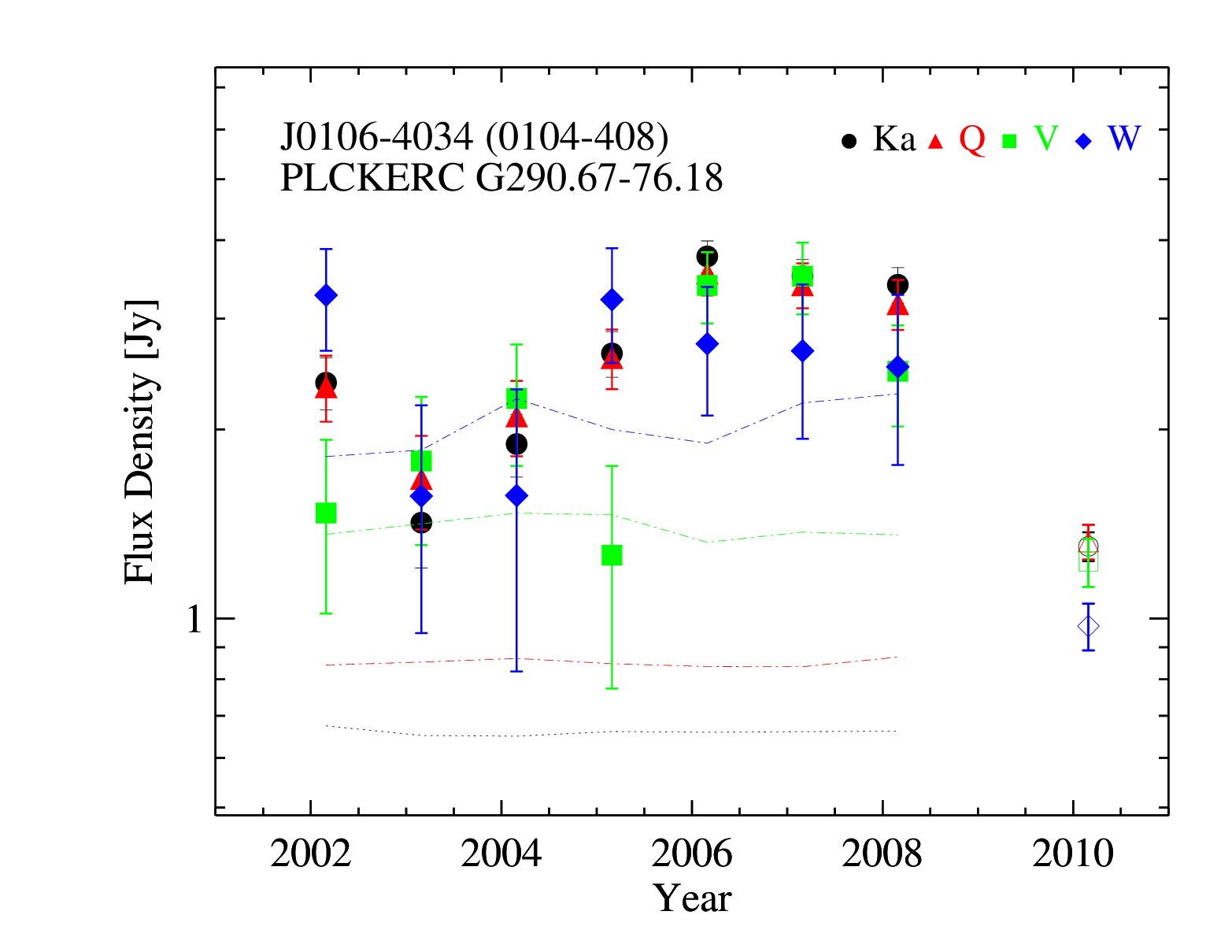} & \includegraphics[width=0.23\textwidth]{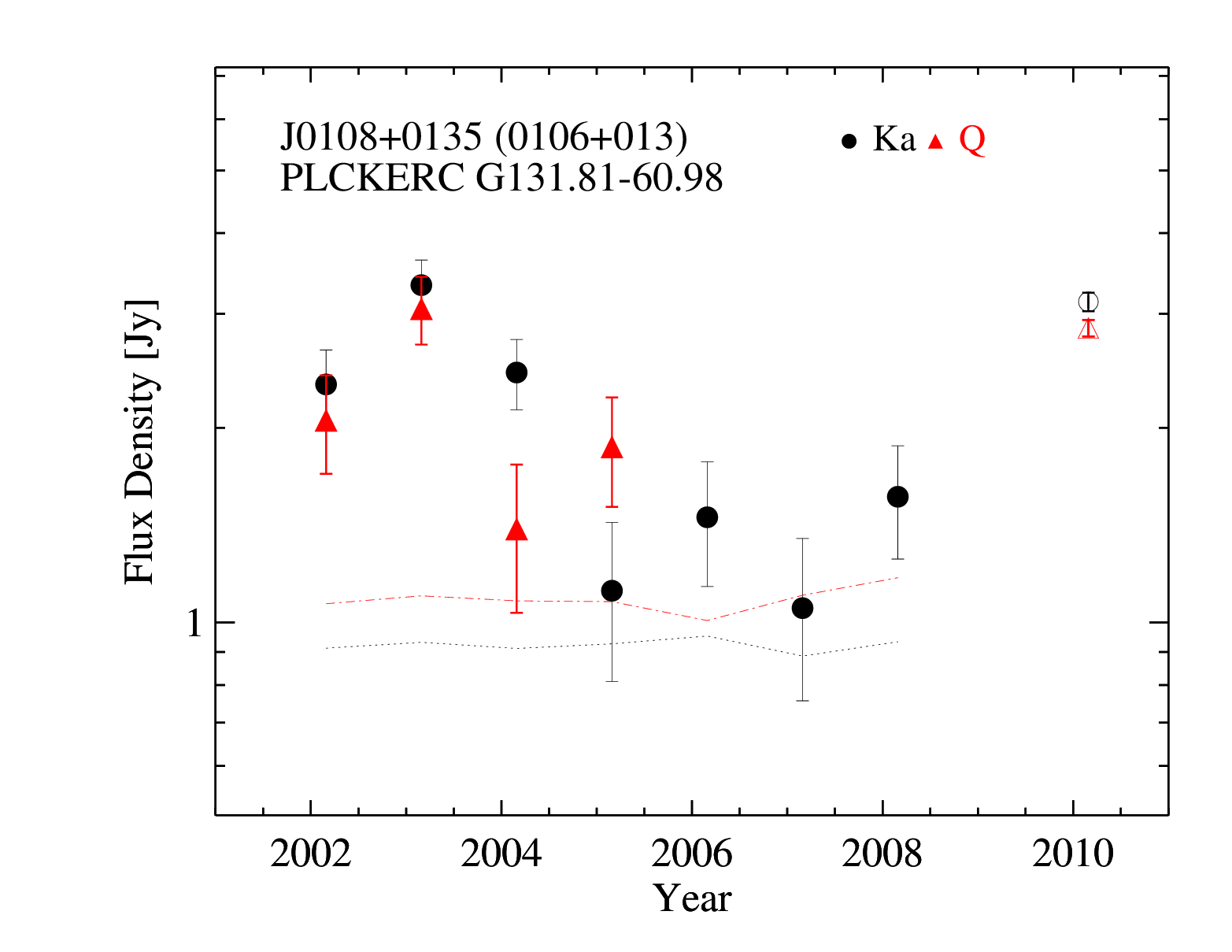}  & \includegraphics[width=0.23\textwidth]{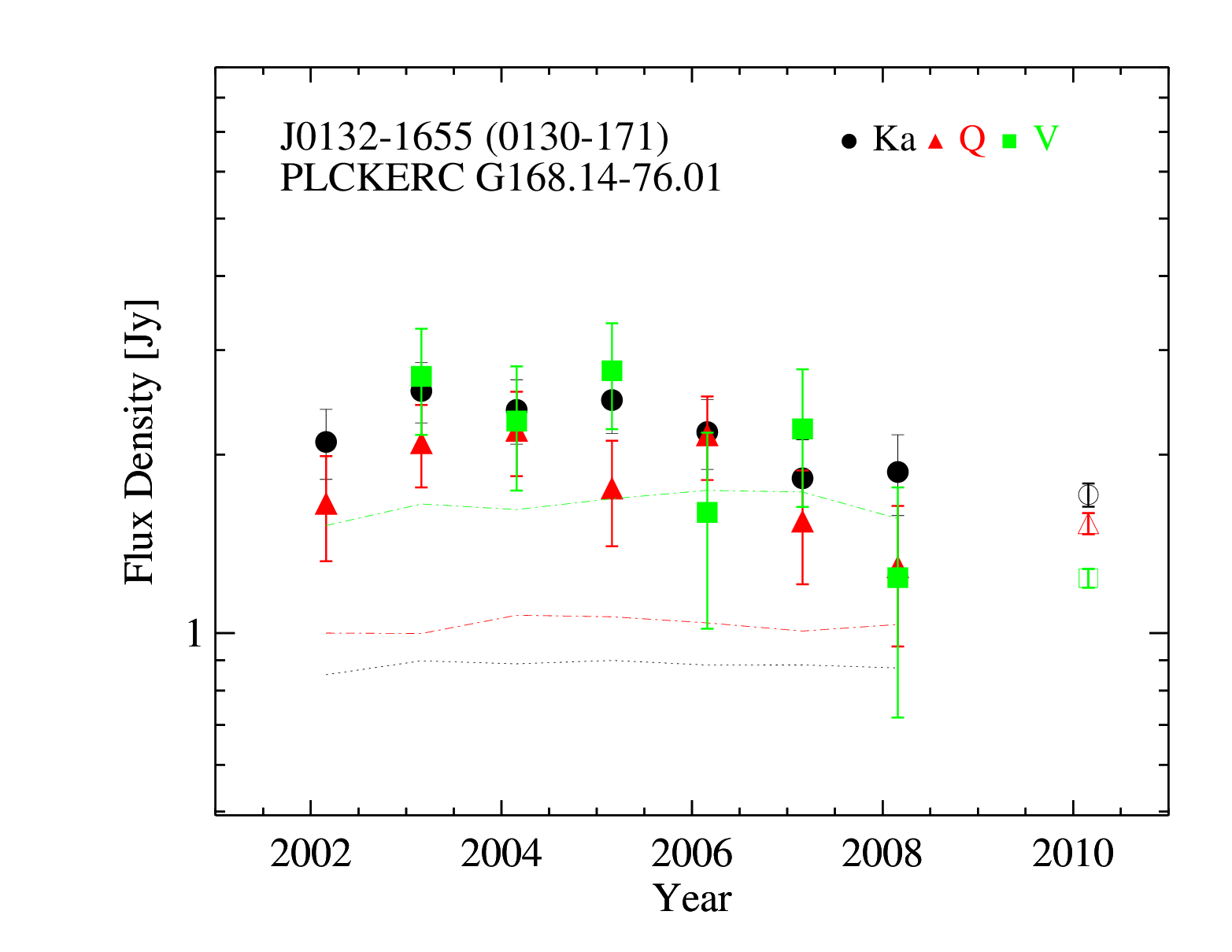} & \includegraphics[width=0.23\textwidth]{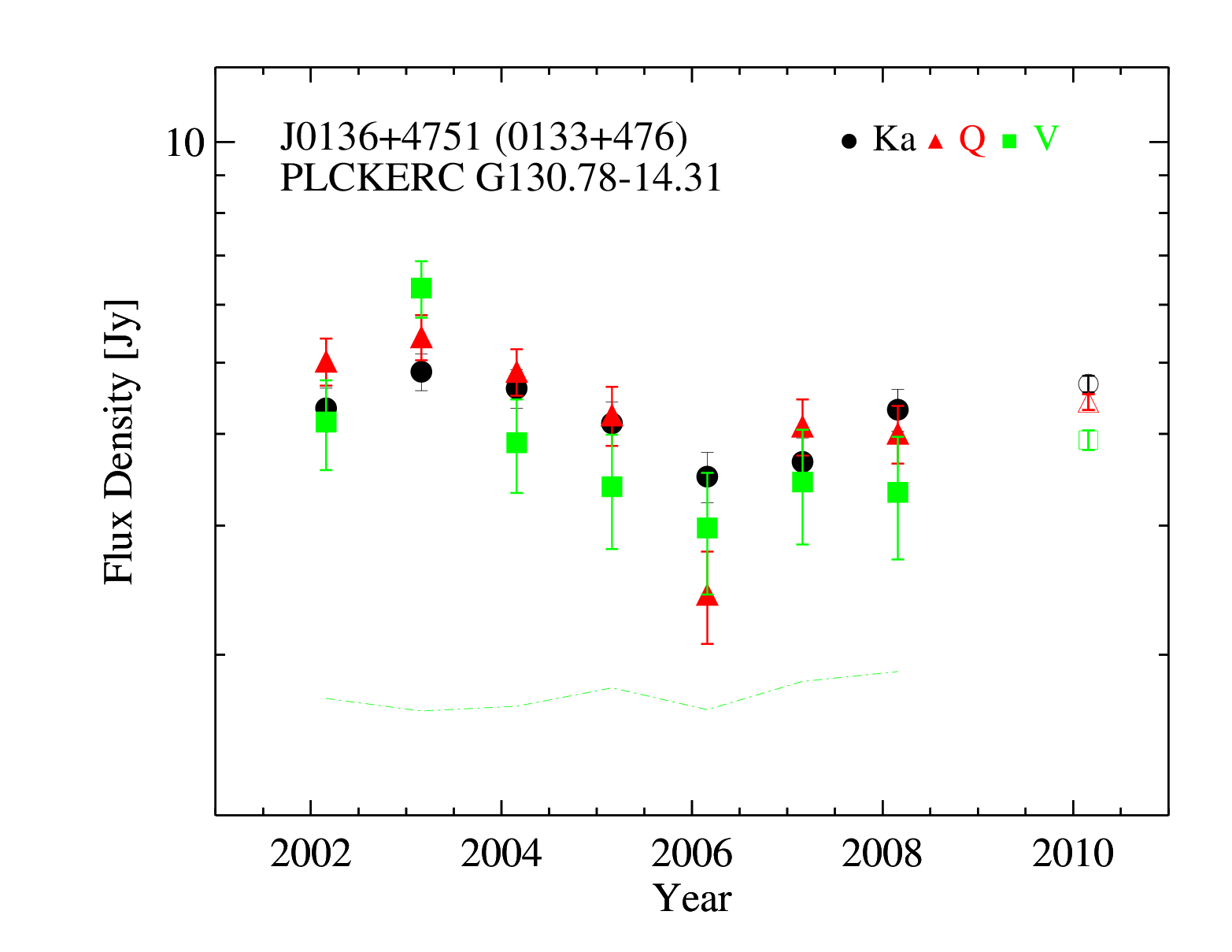}  \\
\includegraphics[width=0.23\textwidth]{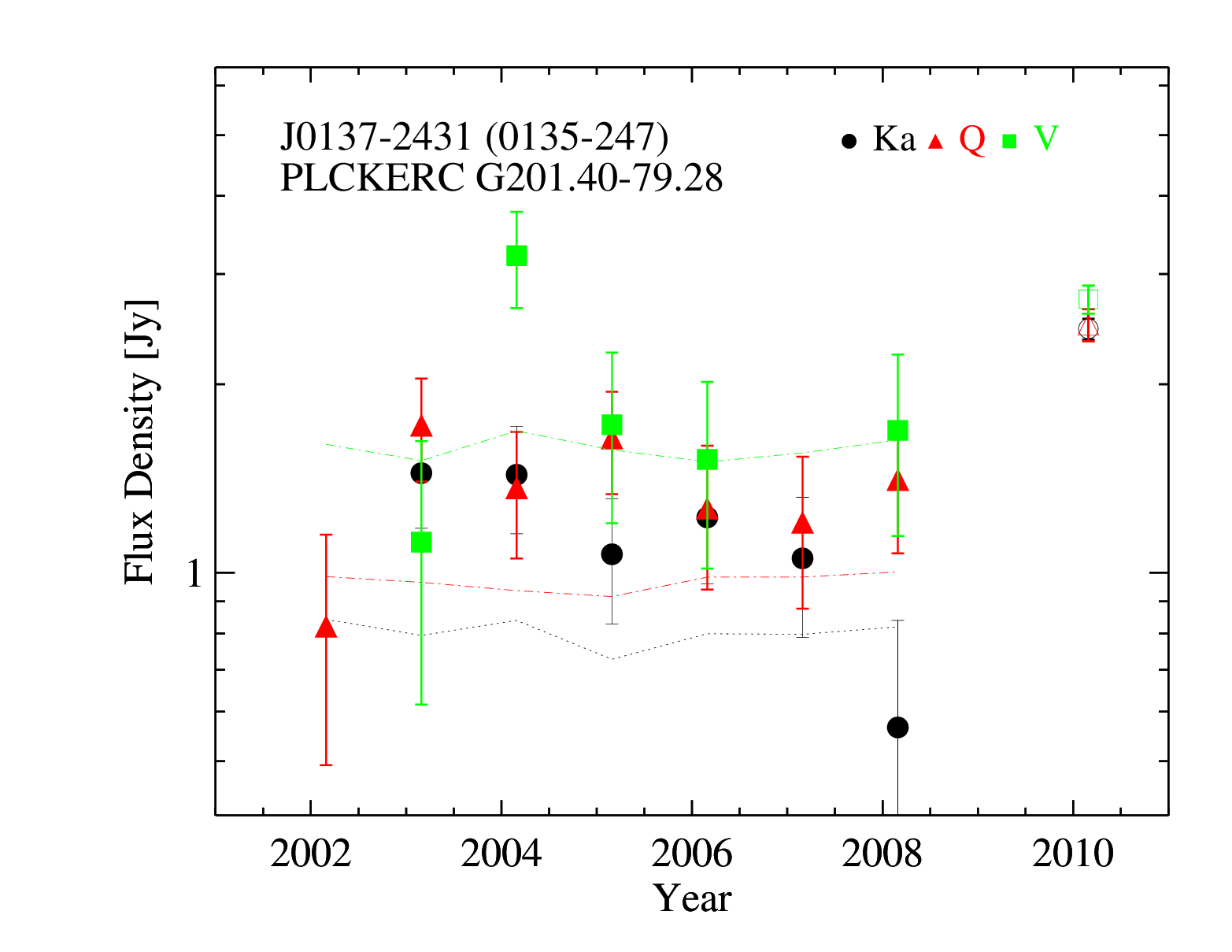} & \includegraphics[width=0.23\textwidth]{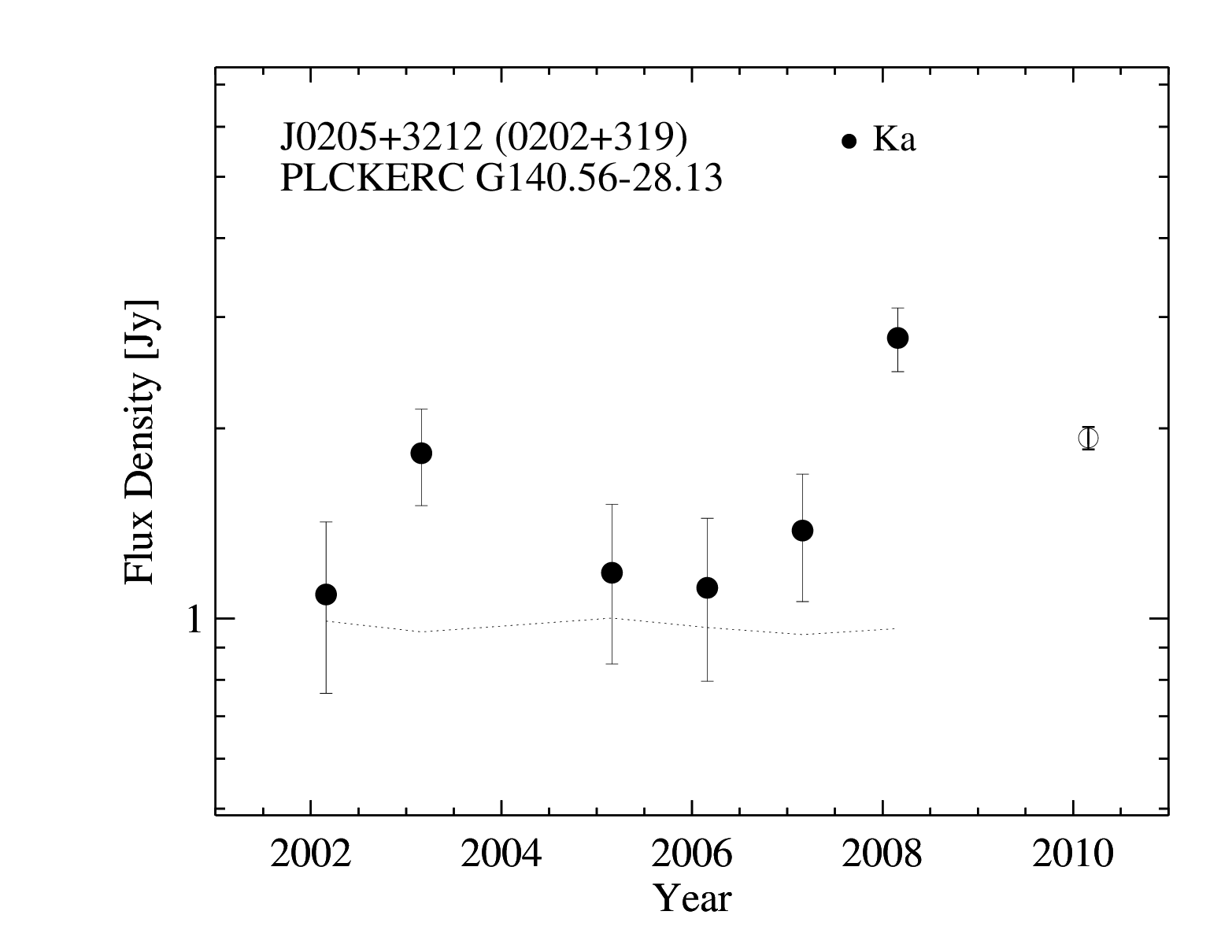}  & \includegraphics[width=0.23\textwidth]{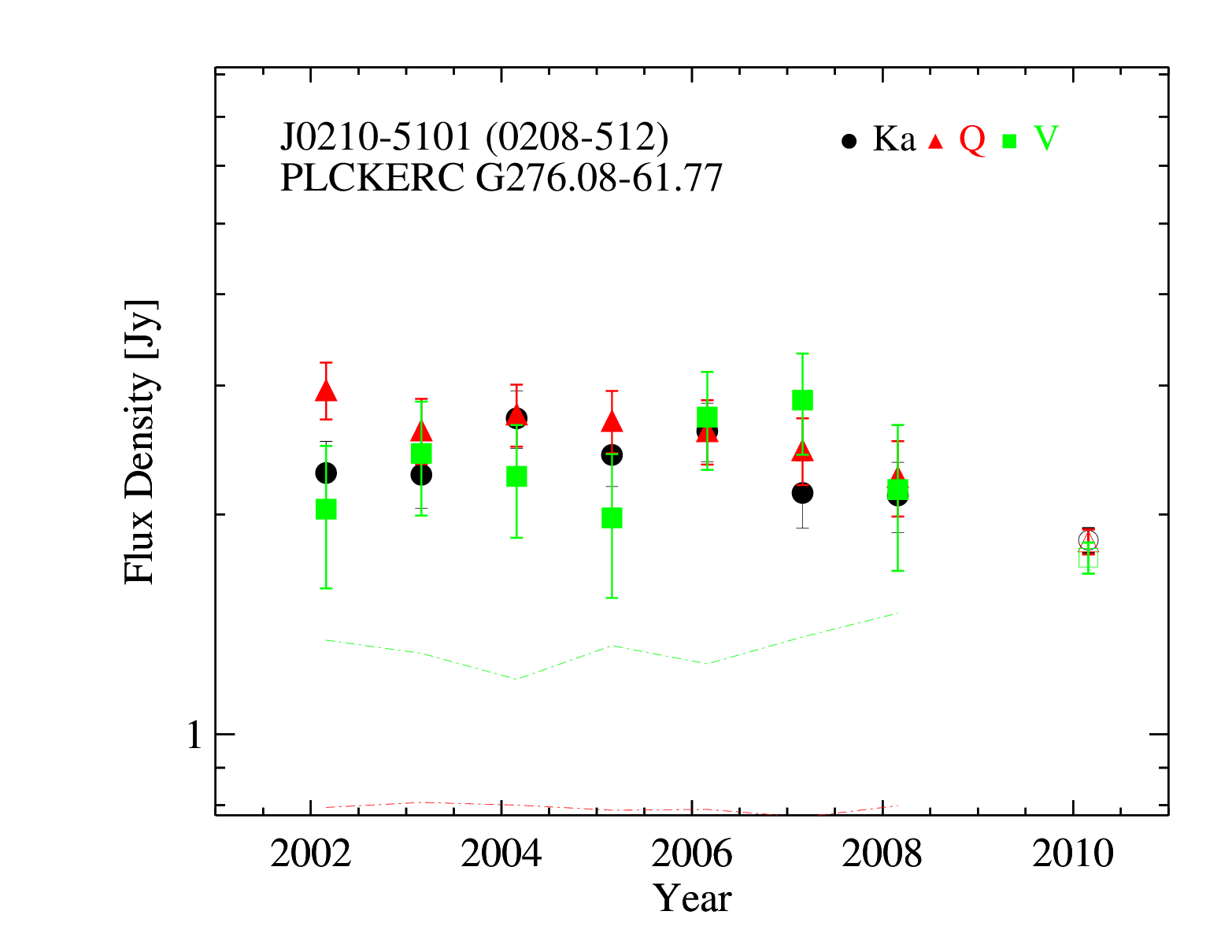} & \includegraphics[width=0.23\textwidth]{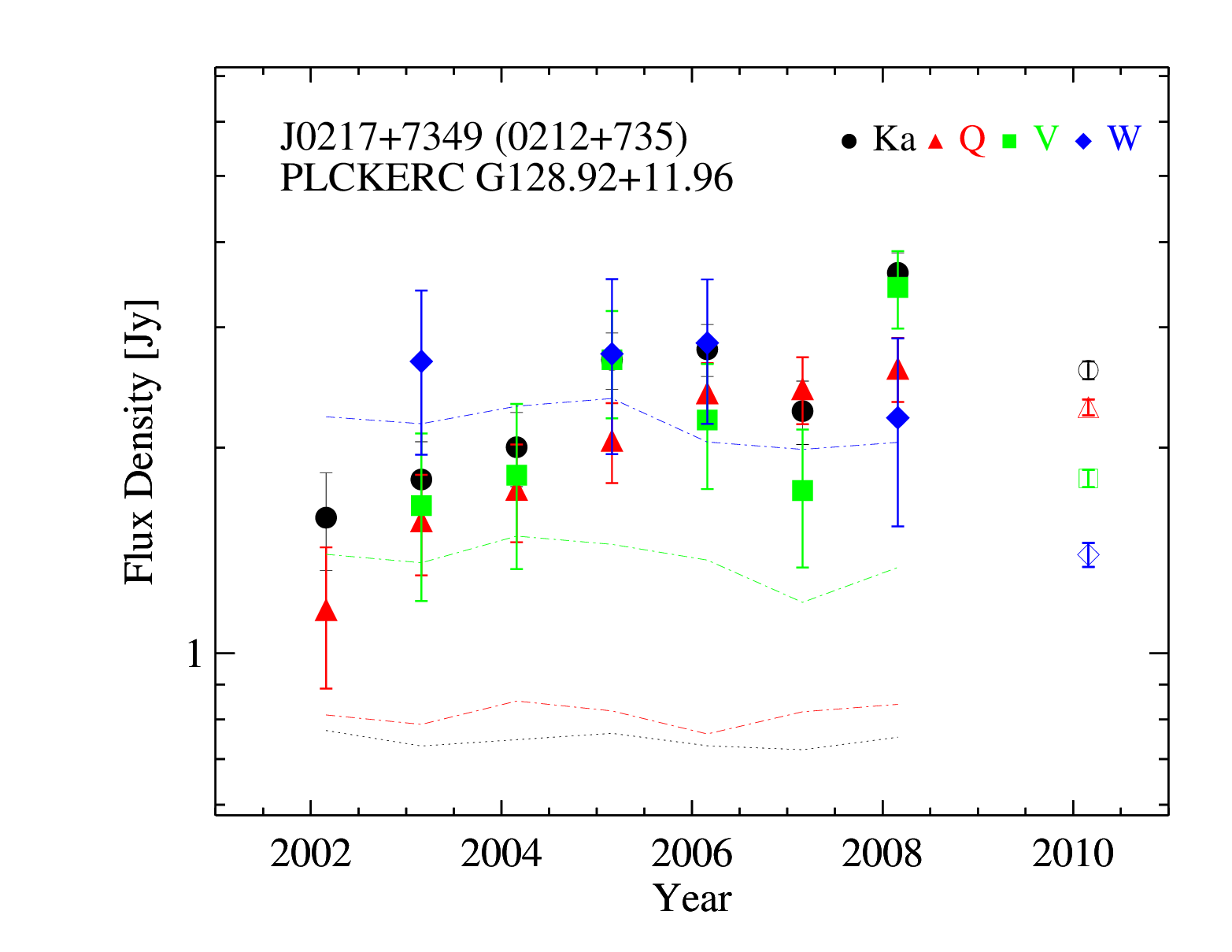}  \\
\includegraphics[width=0.23\textwidth]{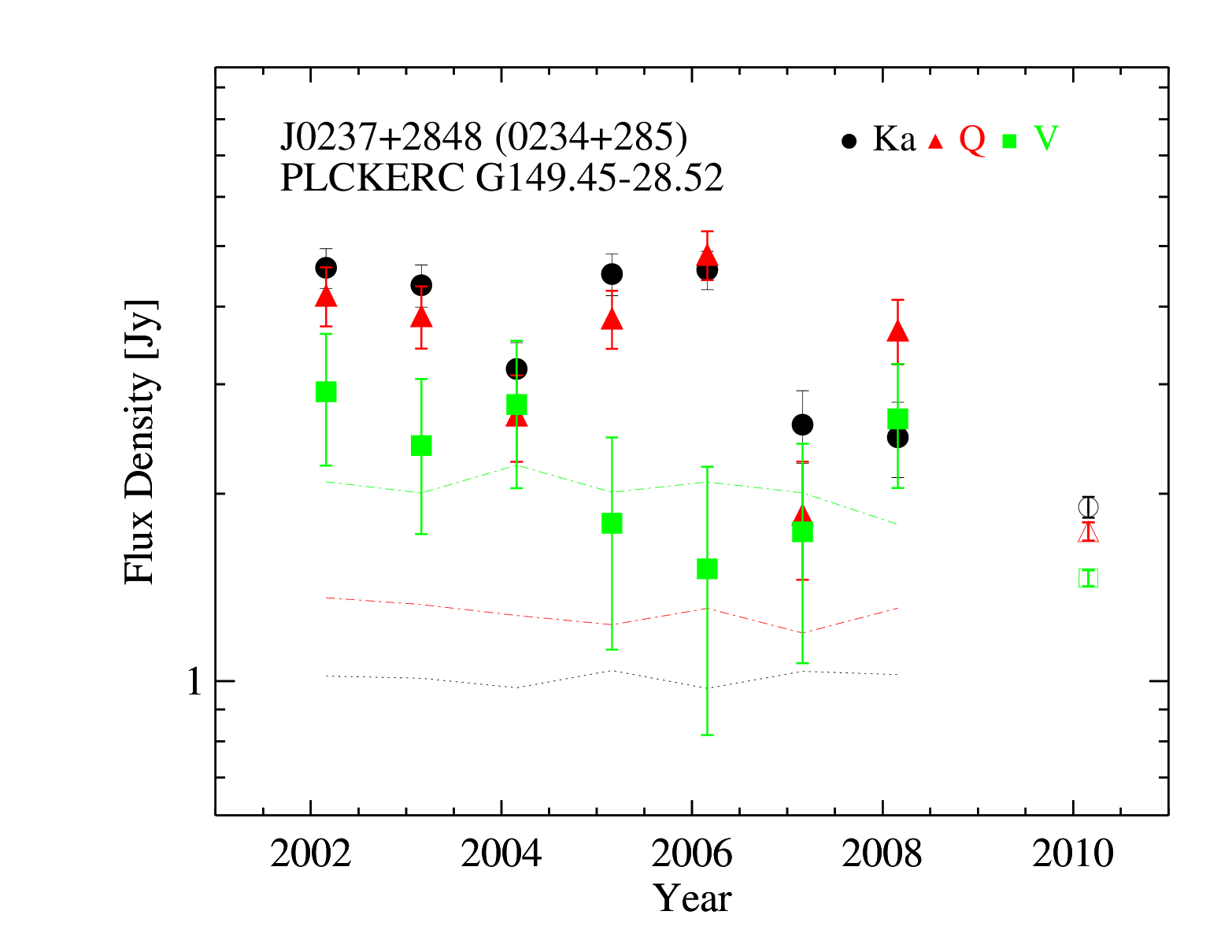} & \includegraphics[width=0.23\textwidth]{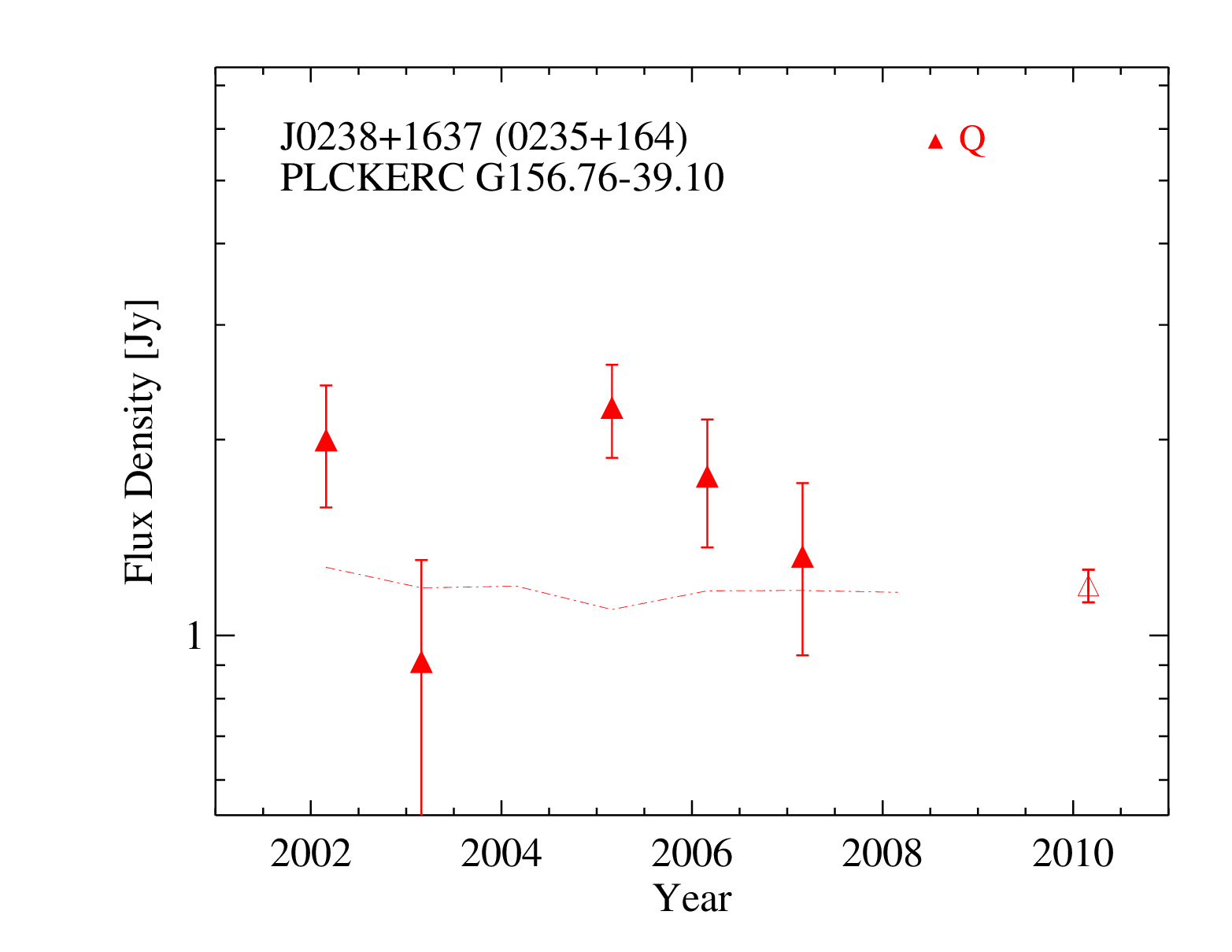}  & \includegraphics[width=0.23\textwidth]{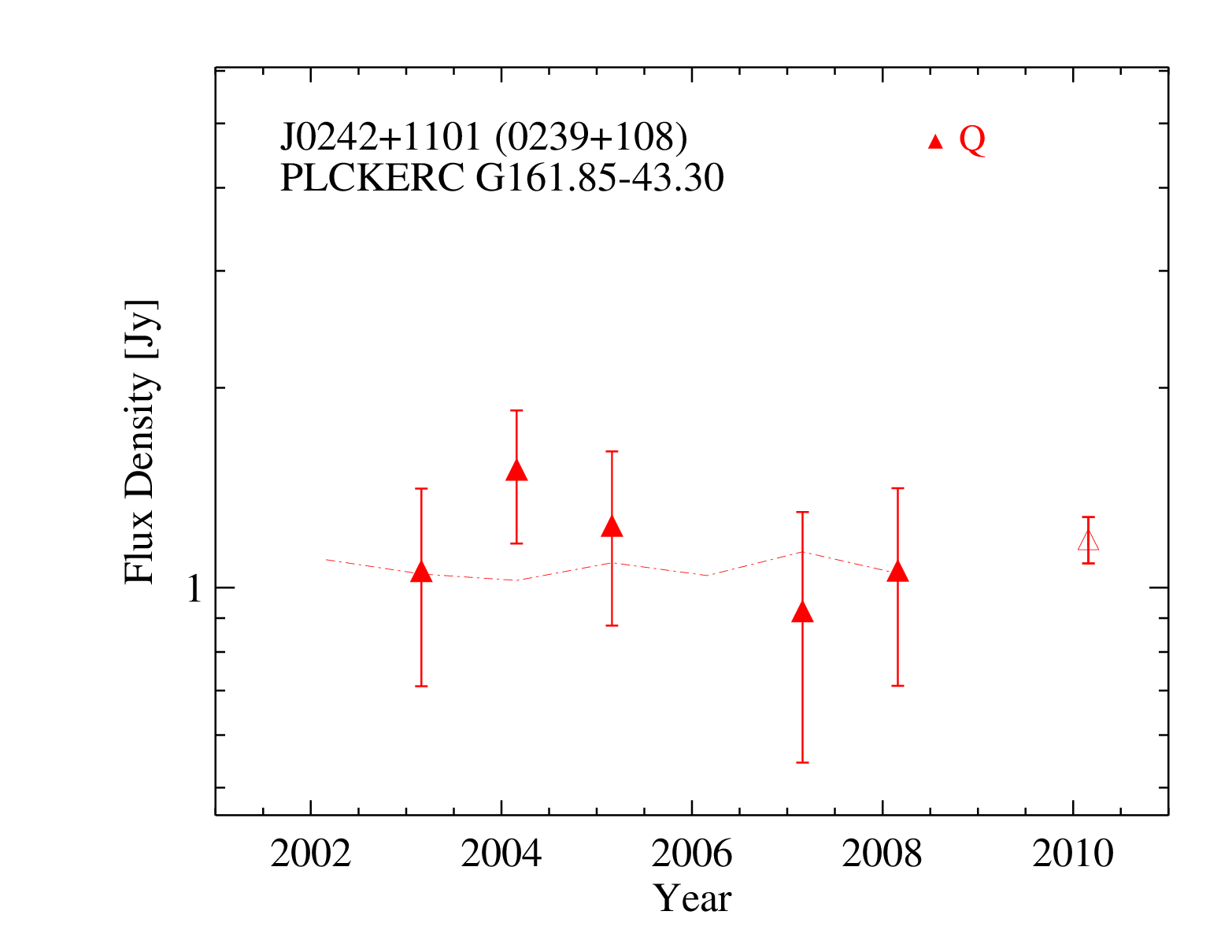} & \includegraphics[width=0.23\textwidth]{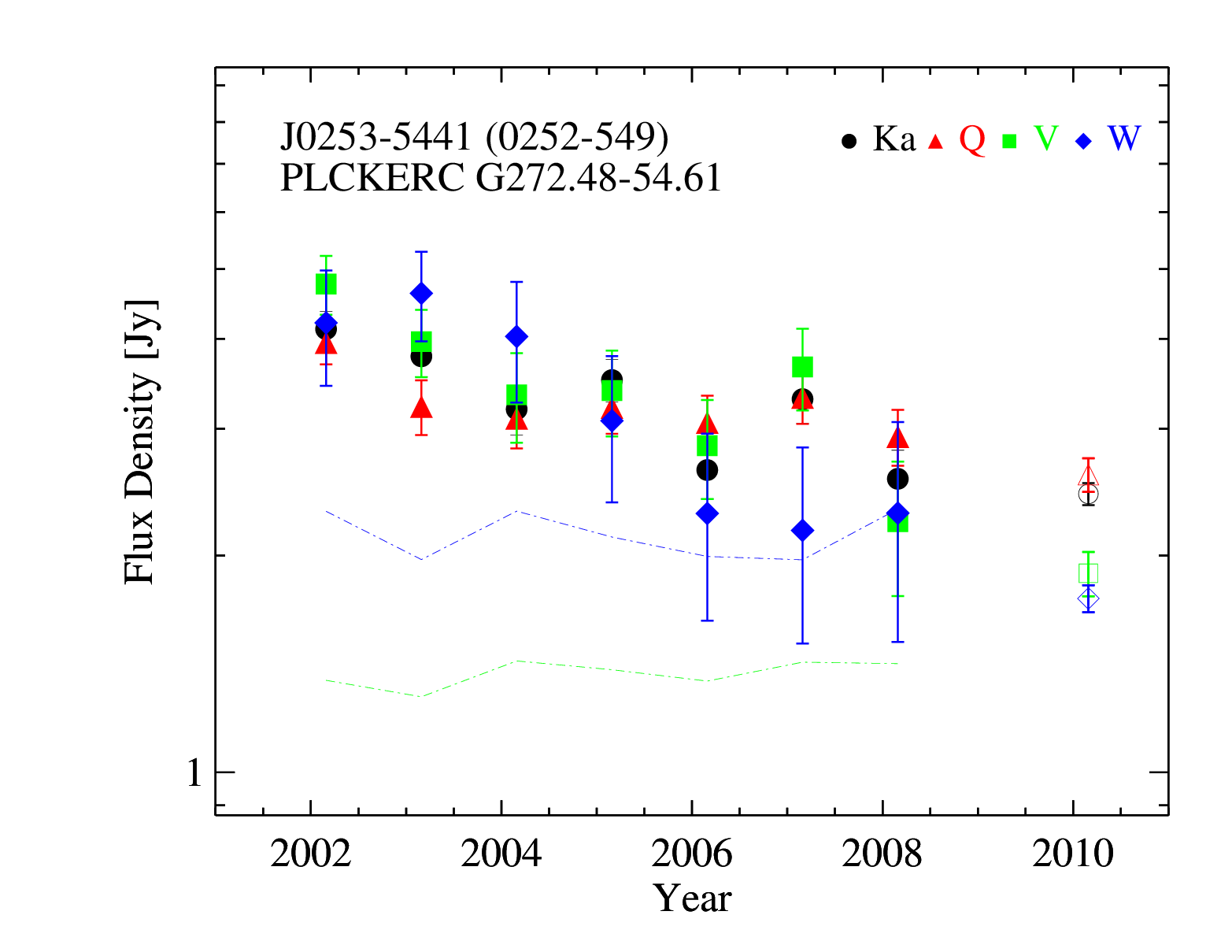}  \\
\includegraphics[width=0.23\textwidth]{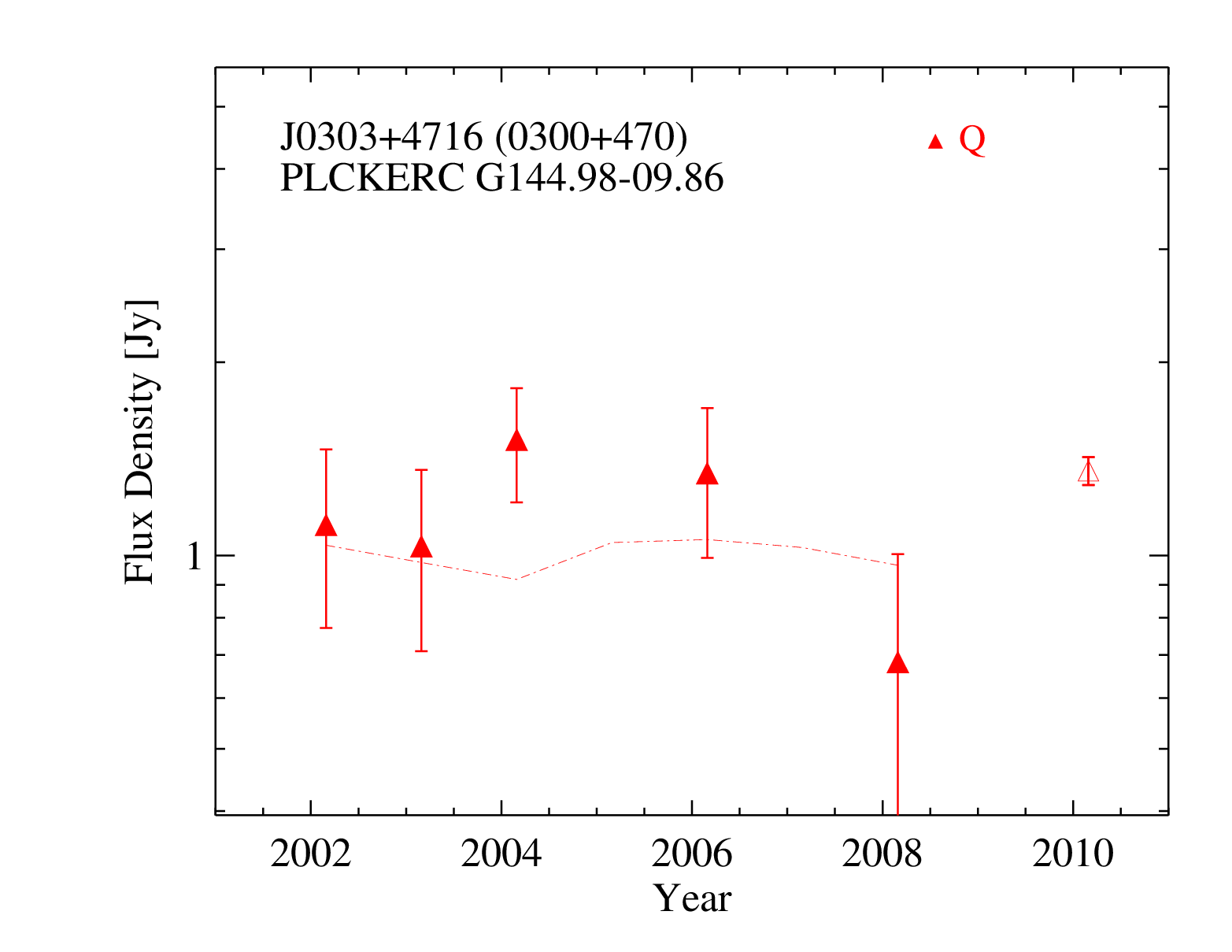} & \includegraphics[width=0.23\textwidth]{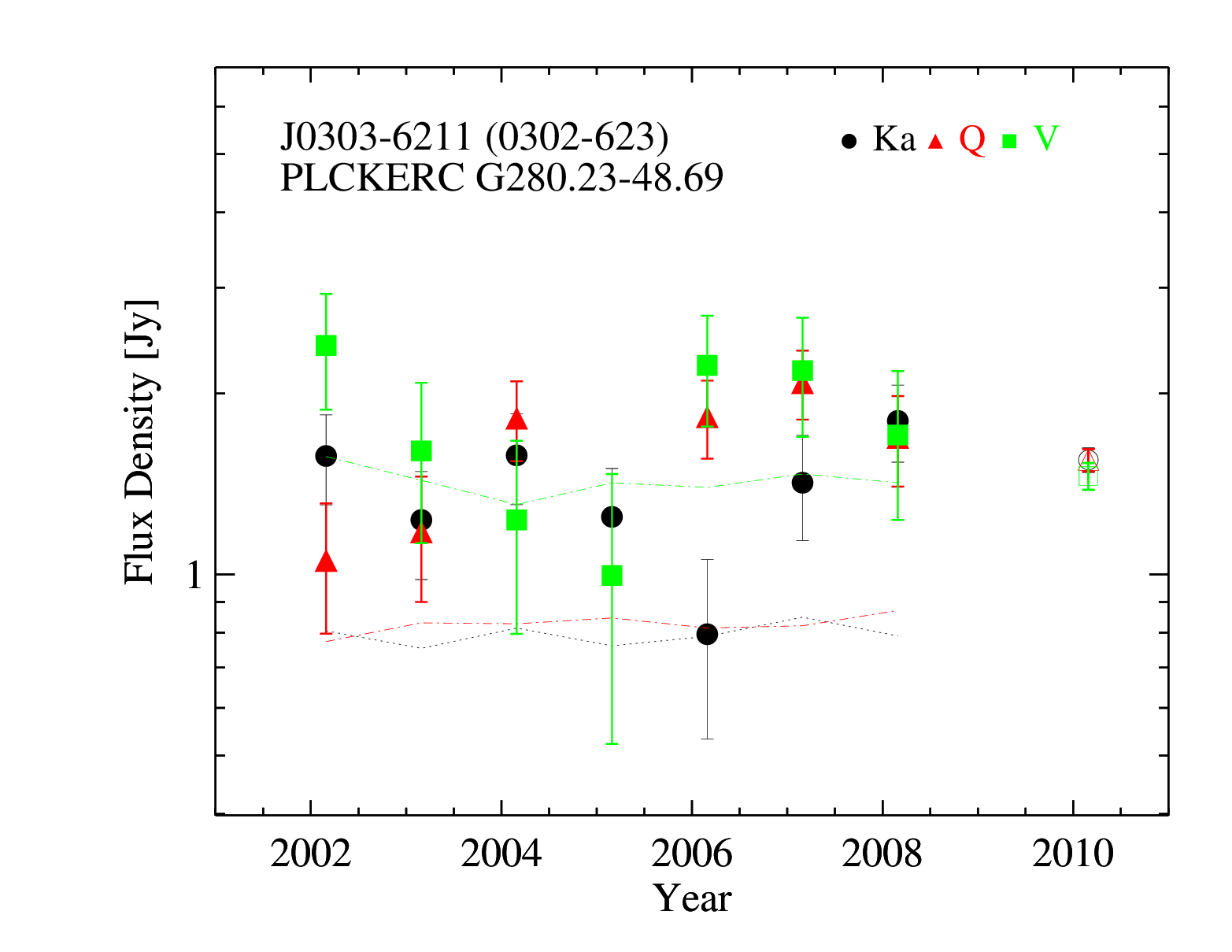}  & \includegraphics[width=0.23\textwidth]{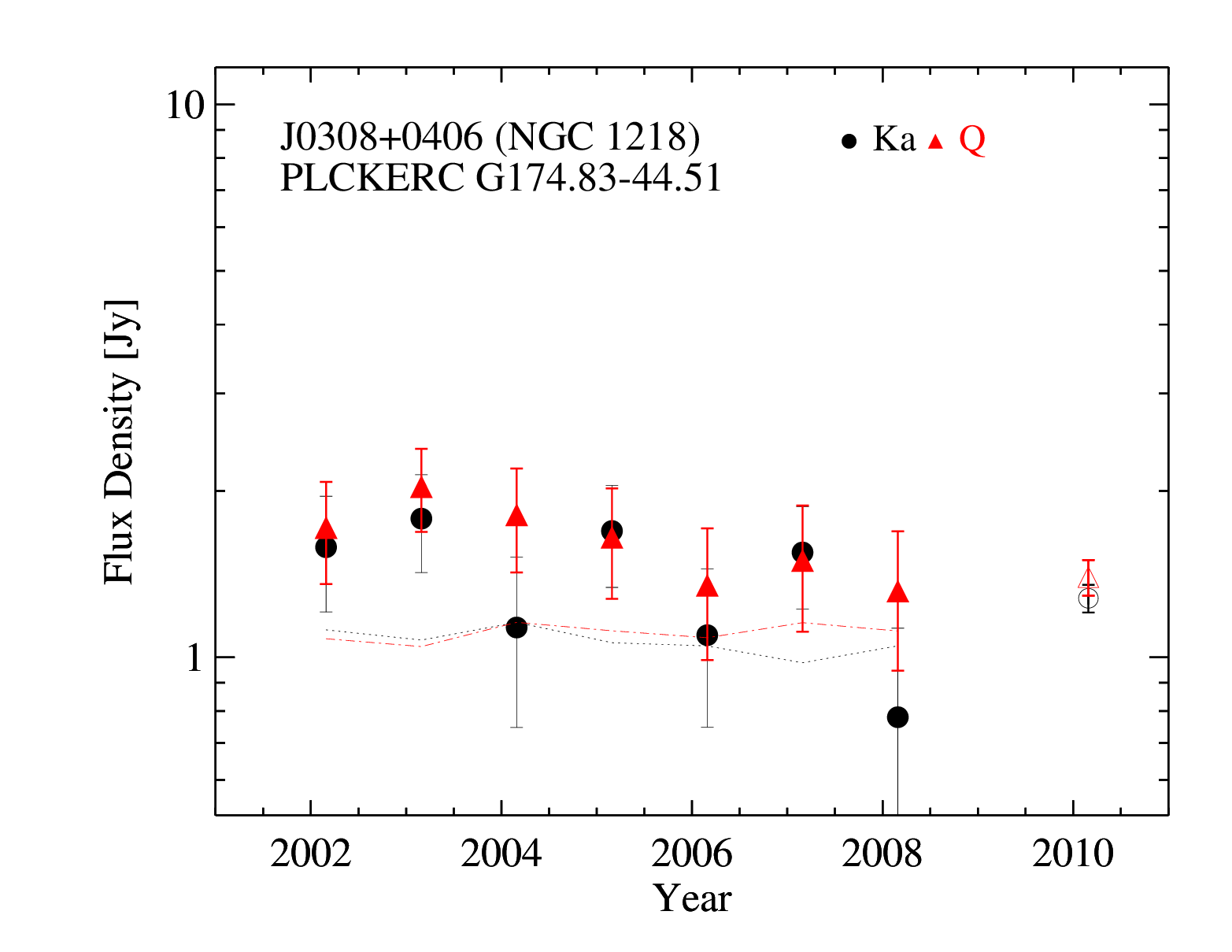} & \includegraphics[width=0.23\textwidth]{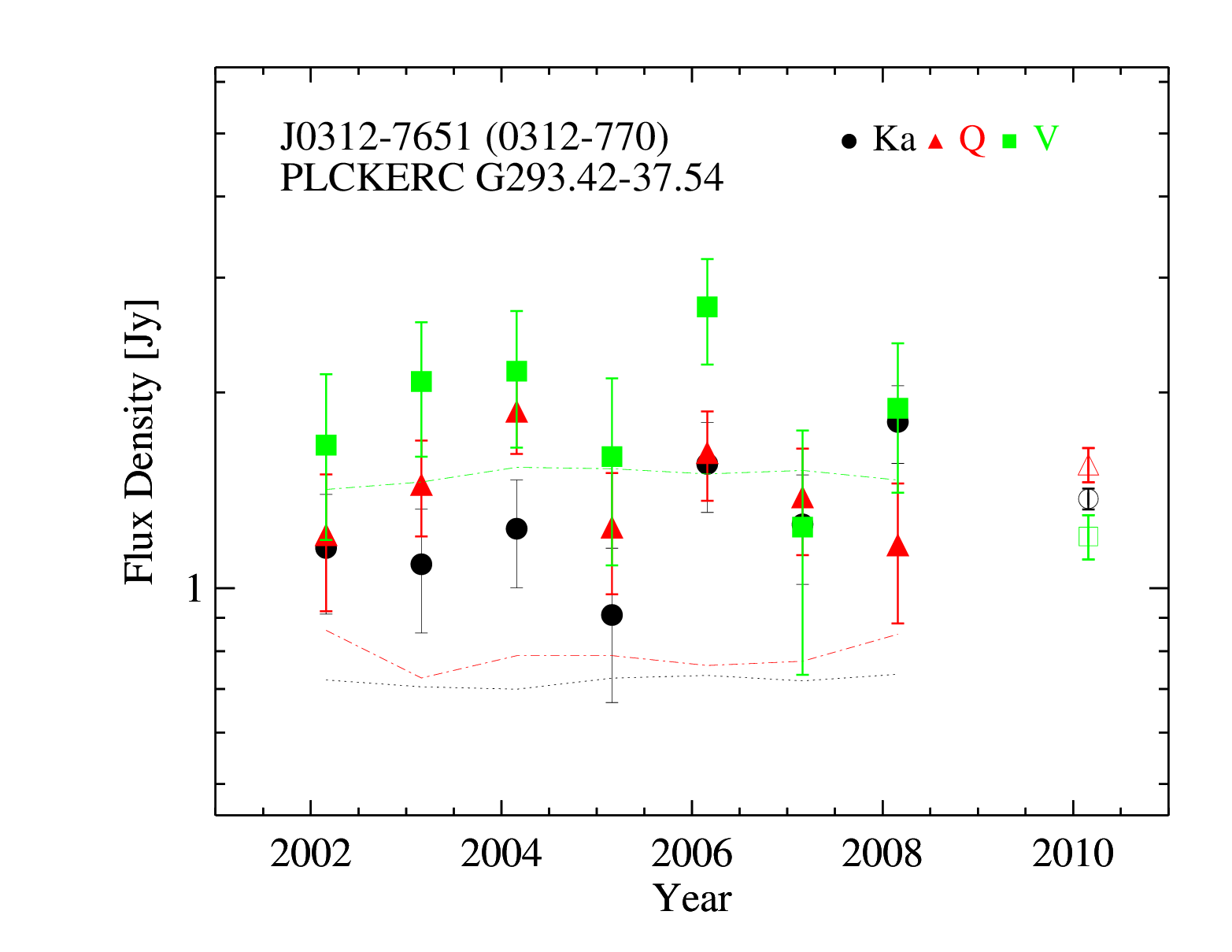}  \\
\includegraphics[width=0.23\textwidth]{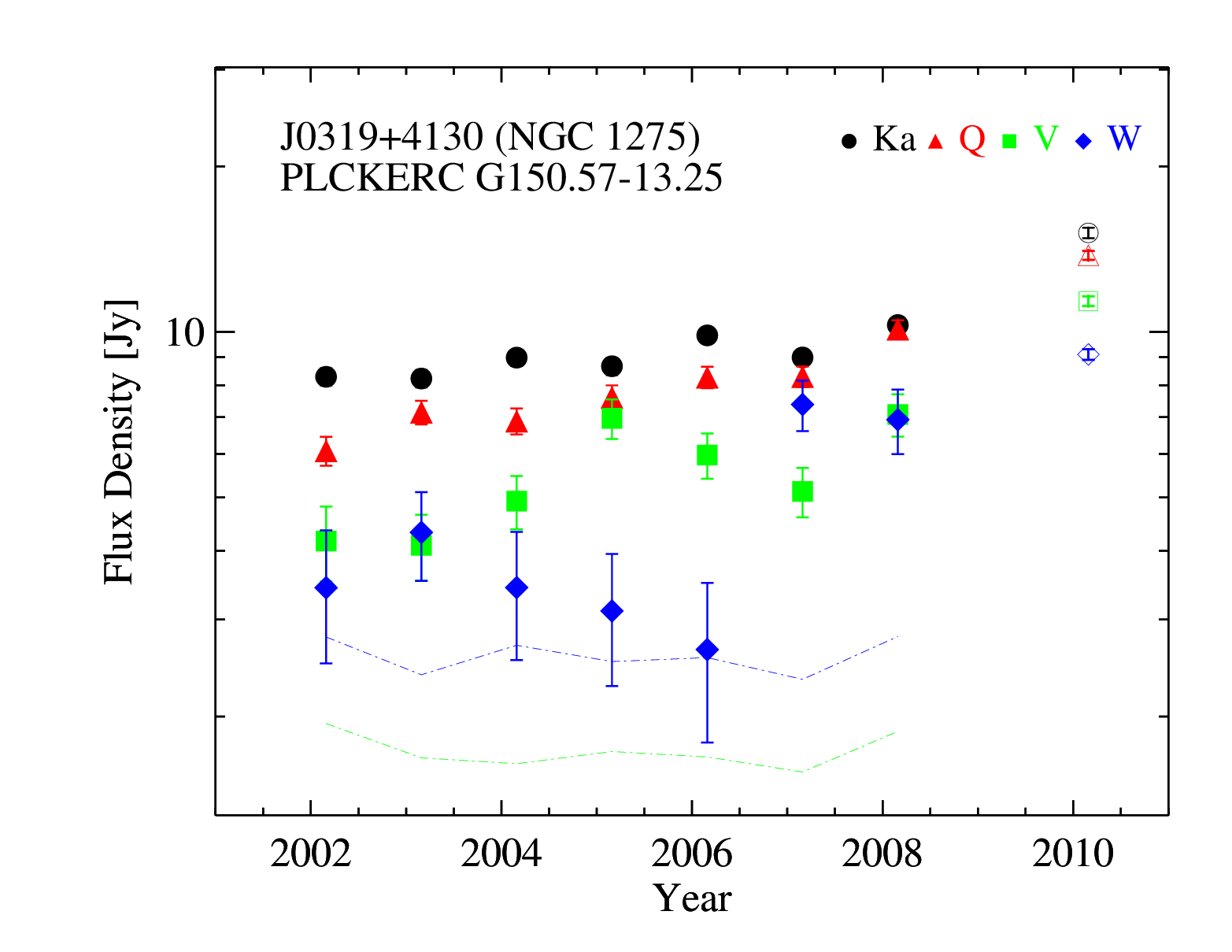} & \includegraphics[width=0.23\textwidth]{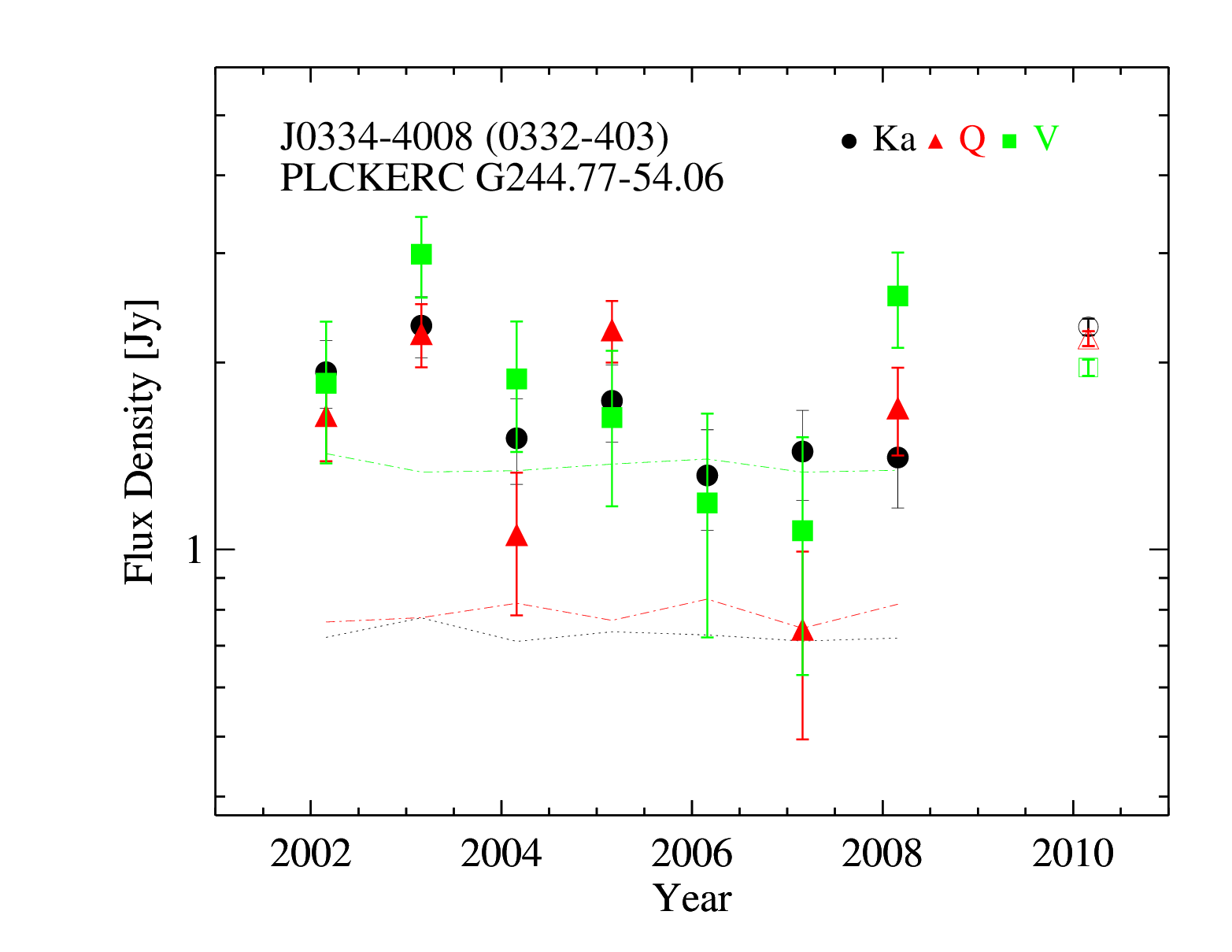}  & \includegraphics[width=0.23\textwidth]{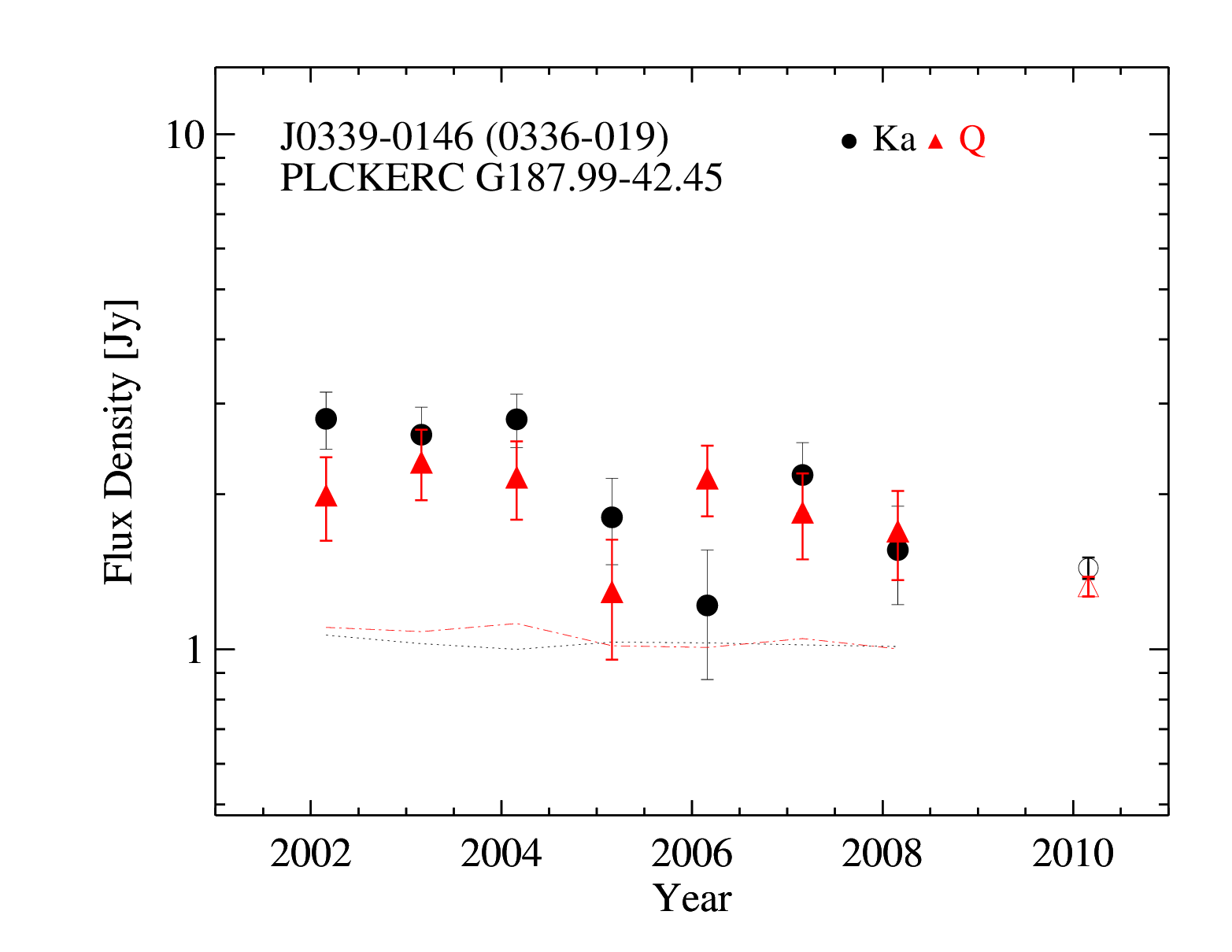} & \includegraphics[width=0.23\textwidth]{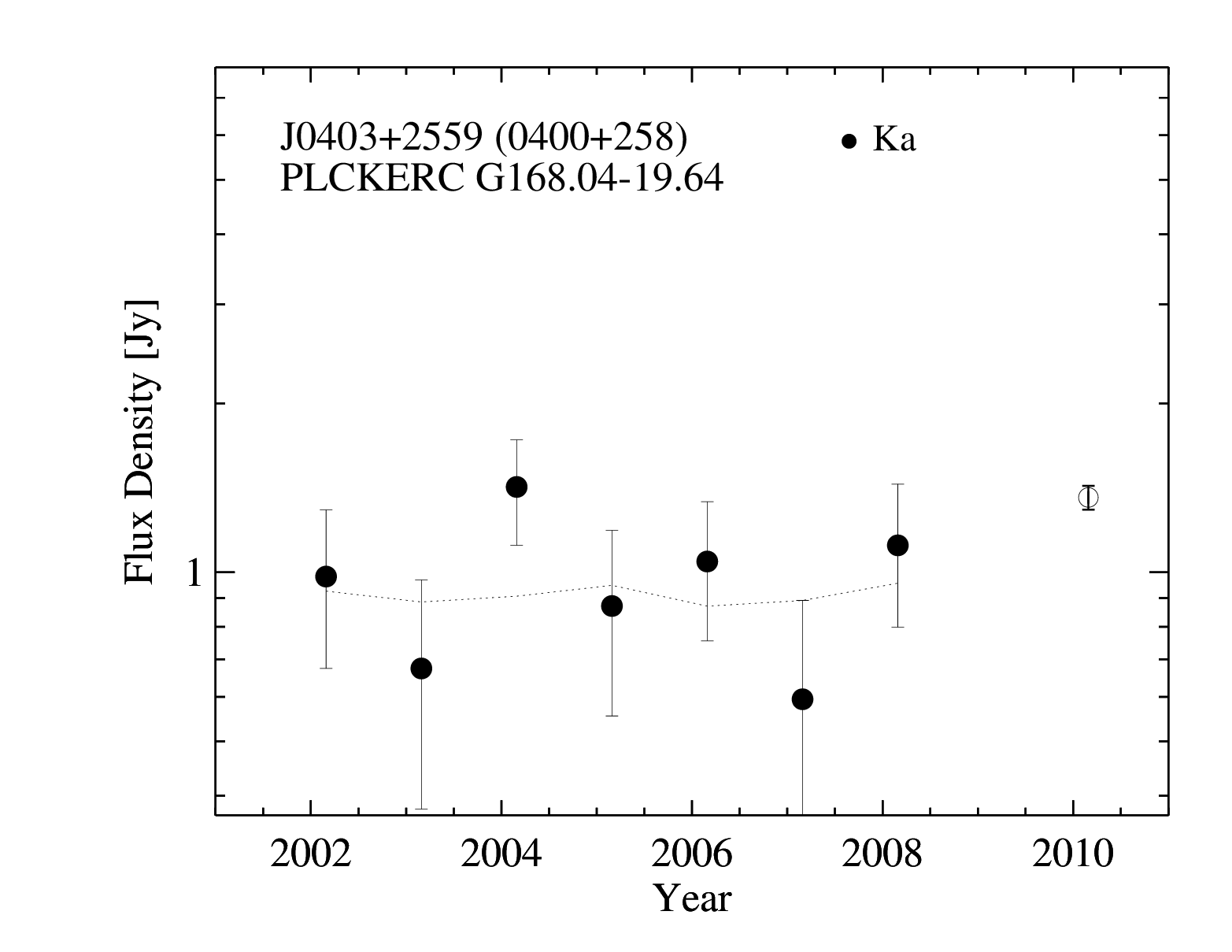}  \\
\includegraphics[width=0.23\textwidth]{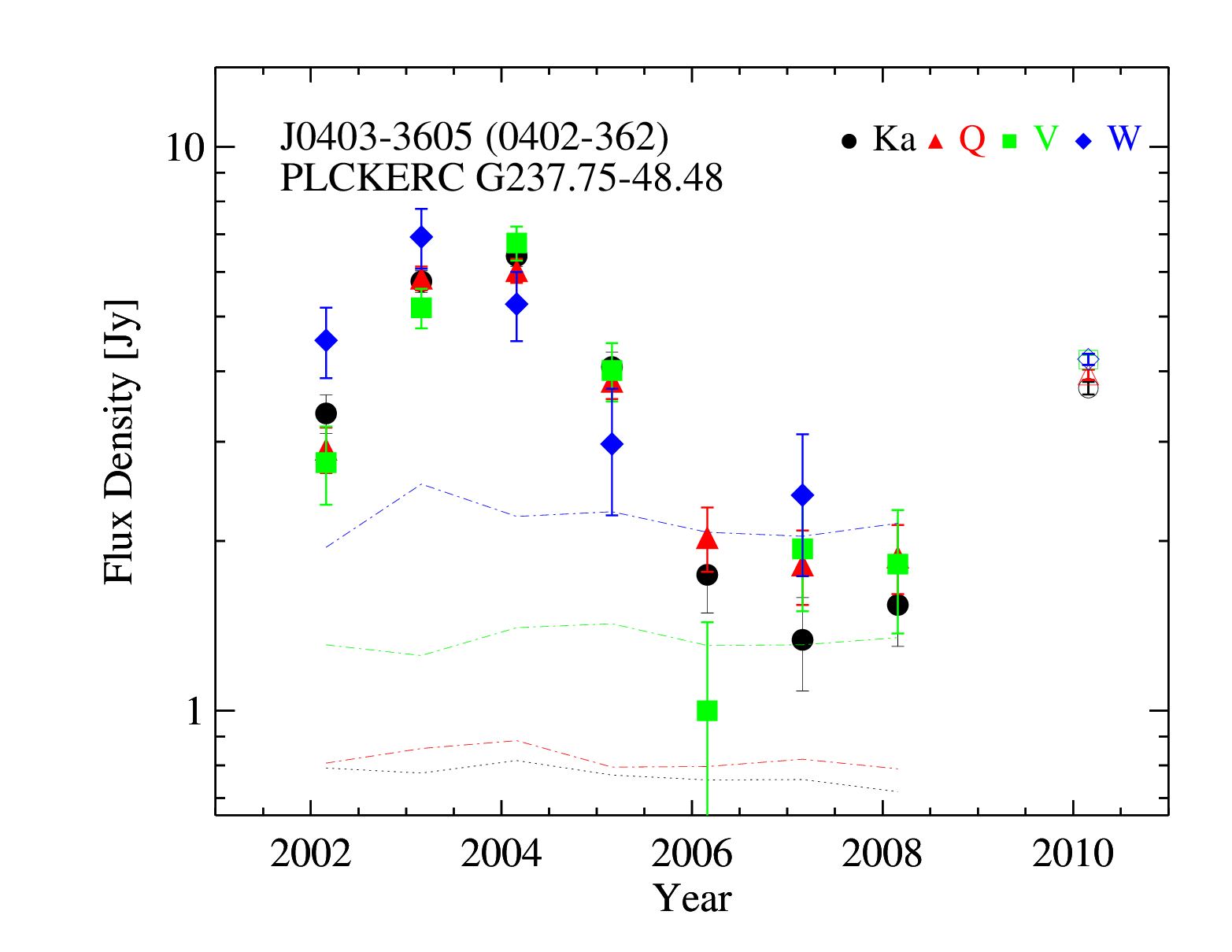} & \includegraphics[width=0.23\textwidth]{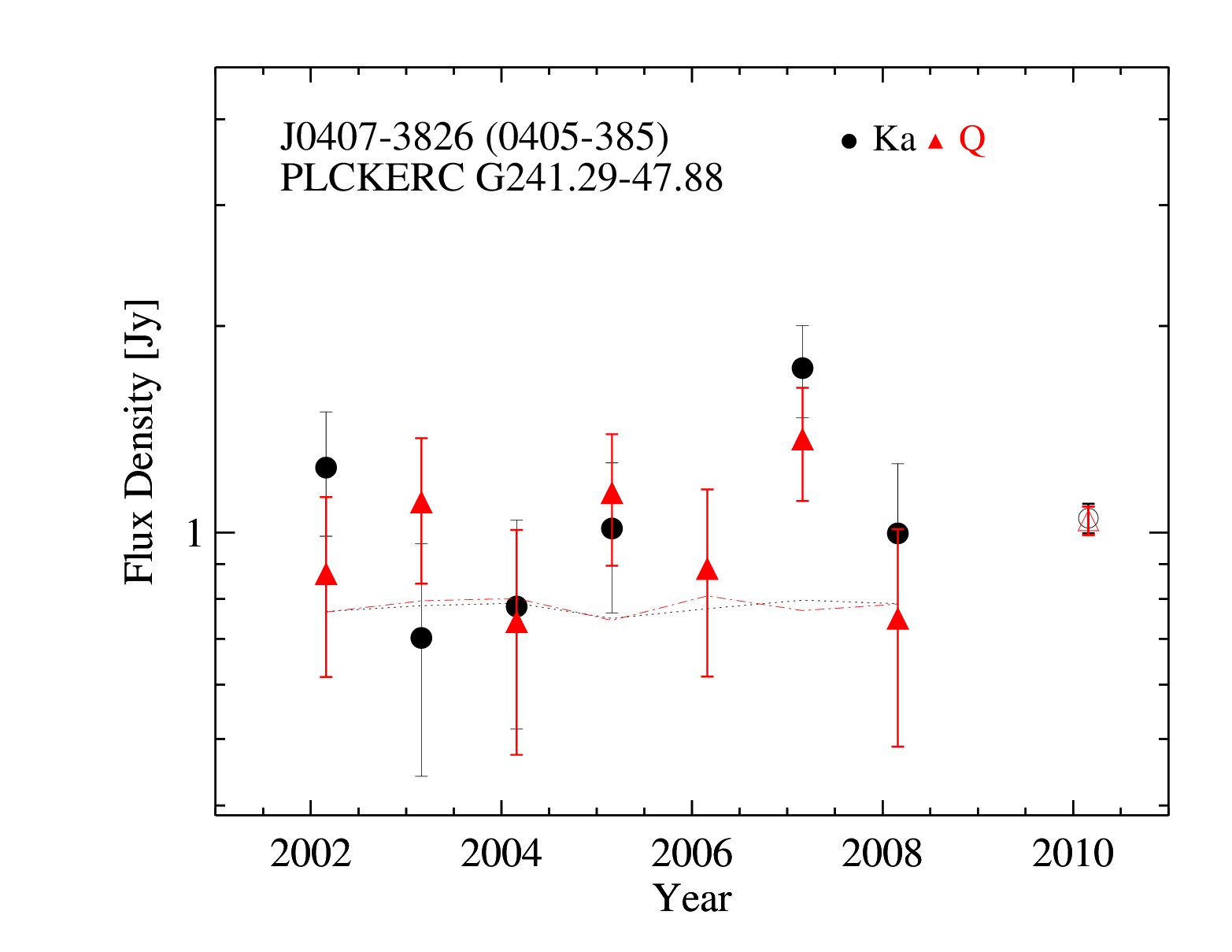}  & \includegraphics[width=0.23\textwidth]{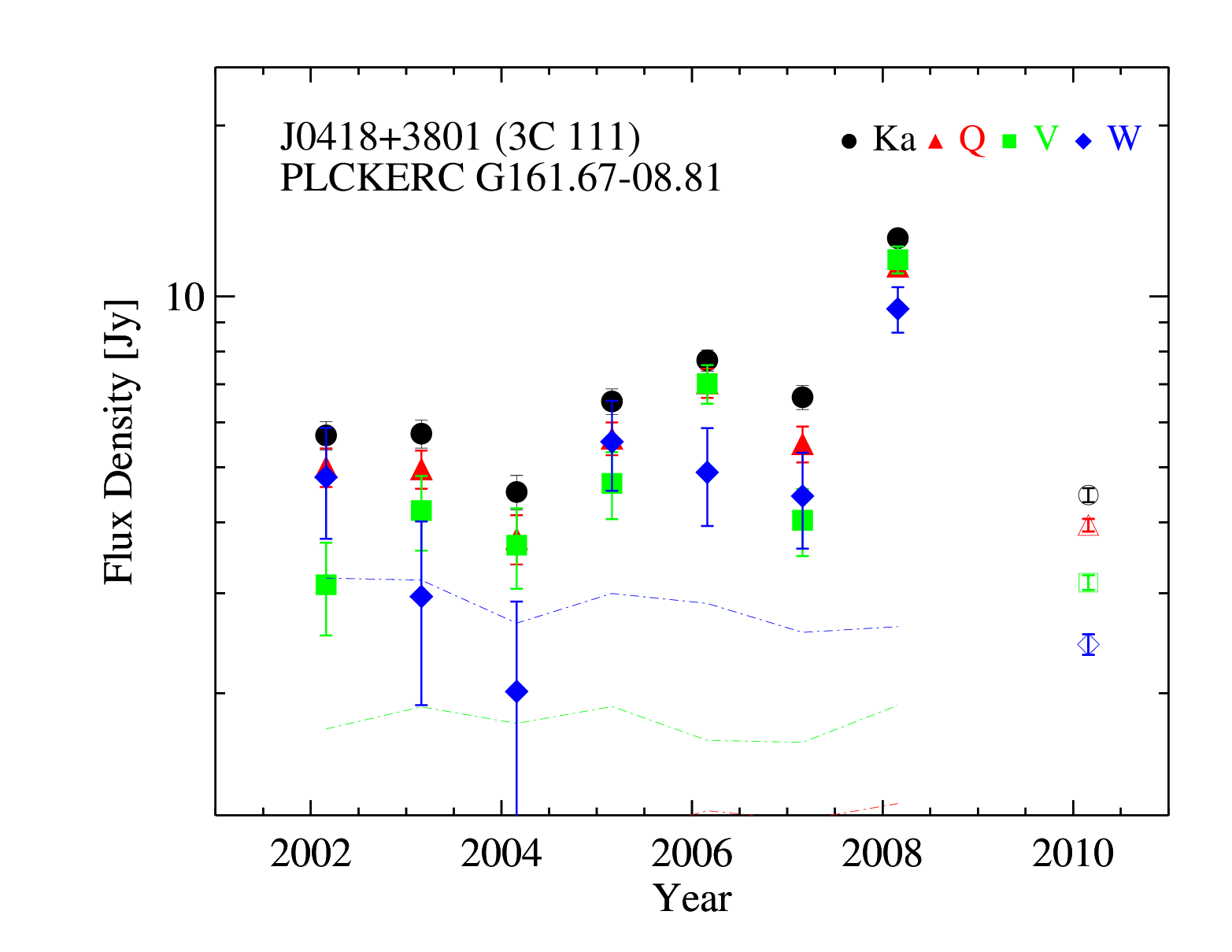} & \includegraphics[width=0.23\textwidth]{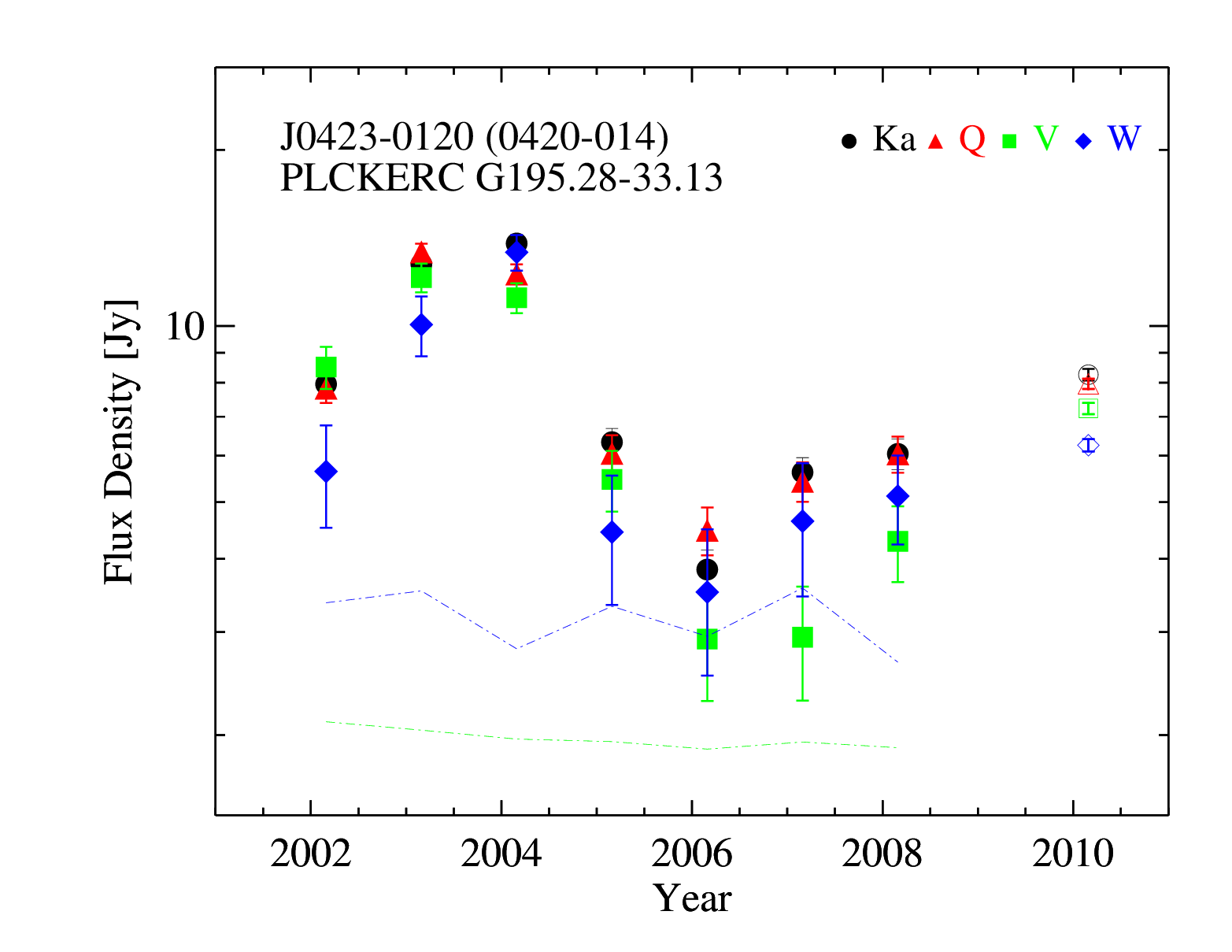}  \\
\end{tabular}
\end{figure*}

\begin{figure*}
\centering
\begin{tabular}{cccc}
\includegraphics[width=0.23\textwidth]{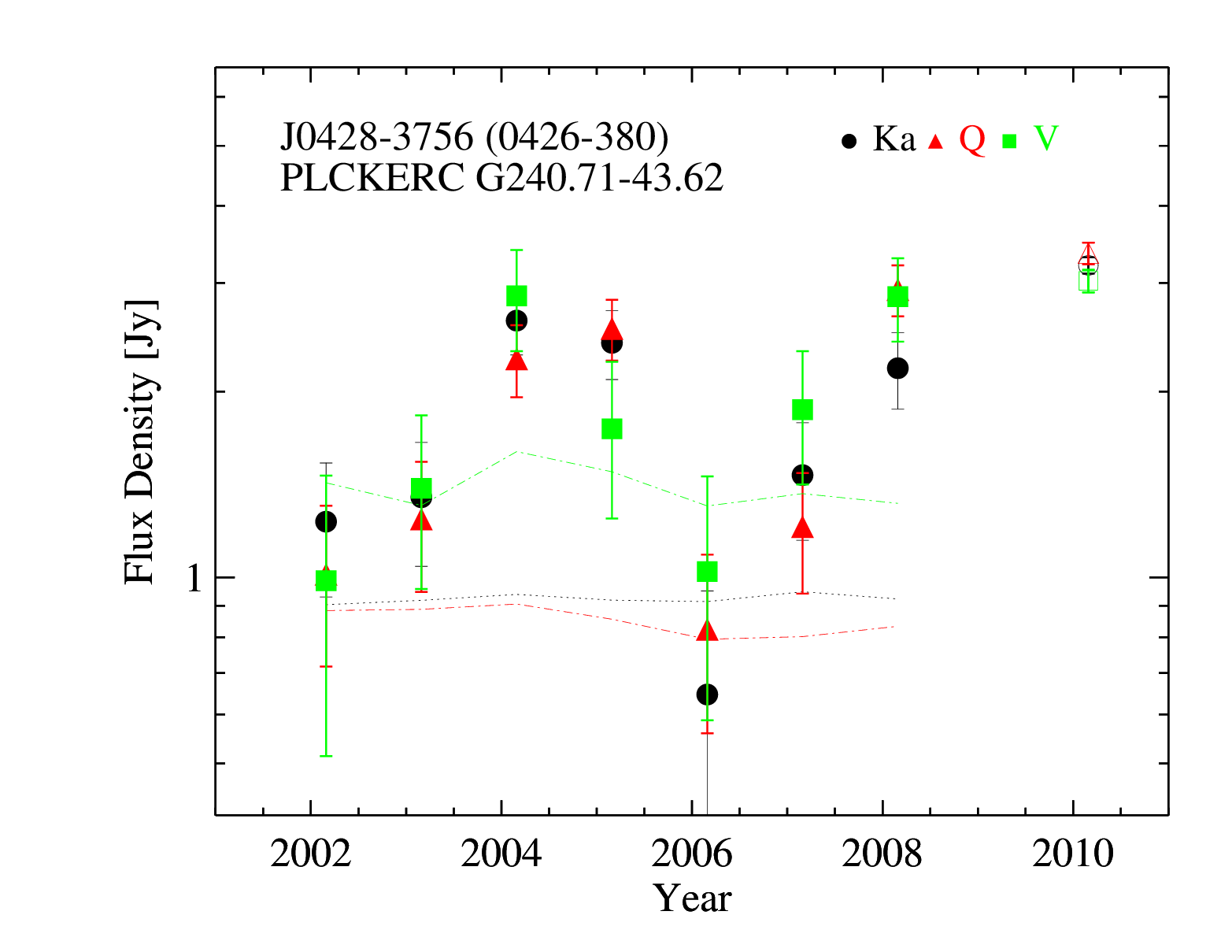} & \includegraphics[width=0.23\textwidth]{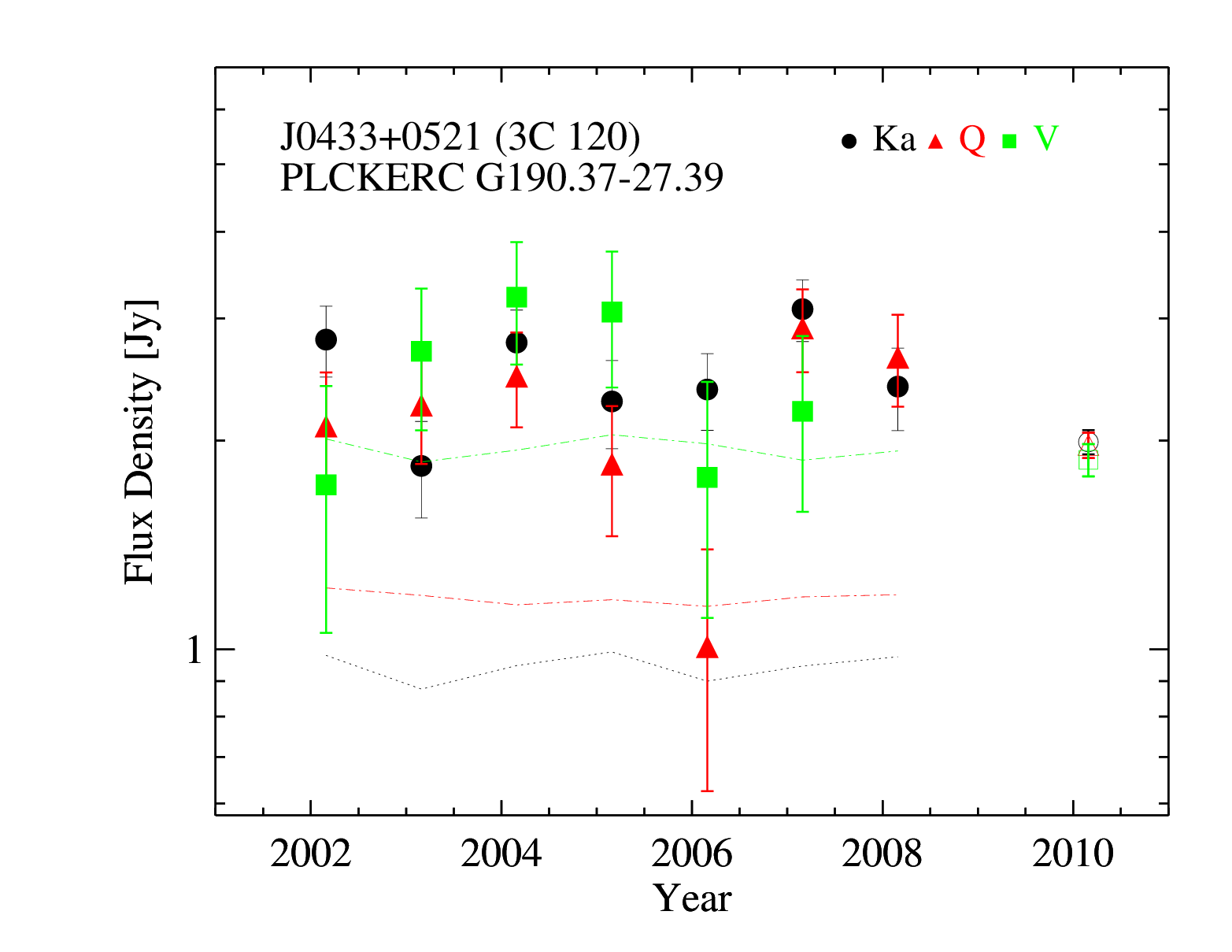}  & \includegraphics[width=0.23\textwidth]{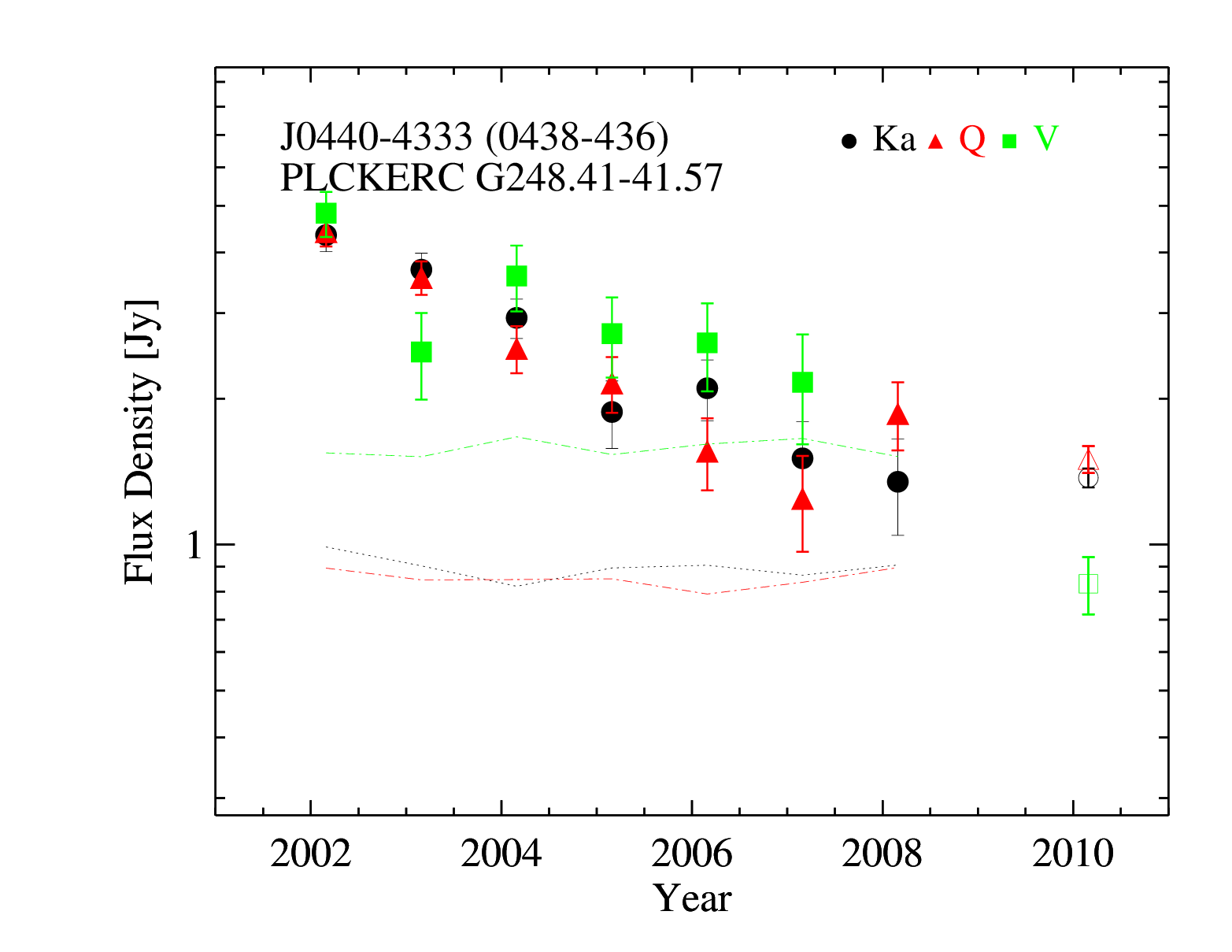} & \includegraphics[width=0.23\textwidth]{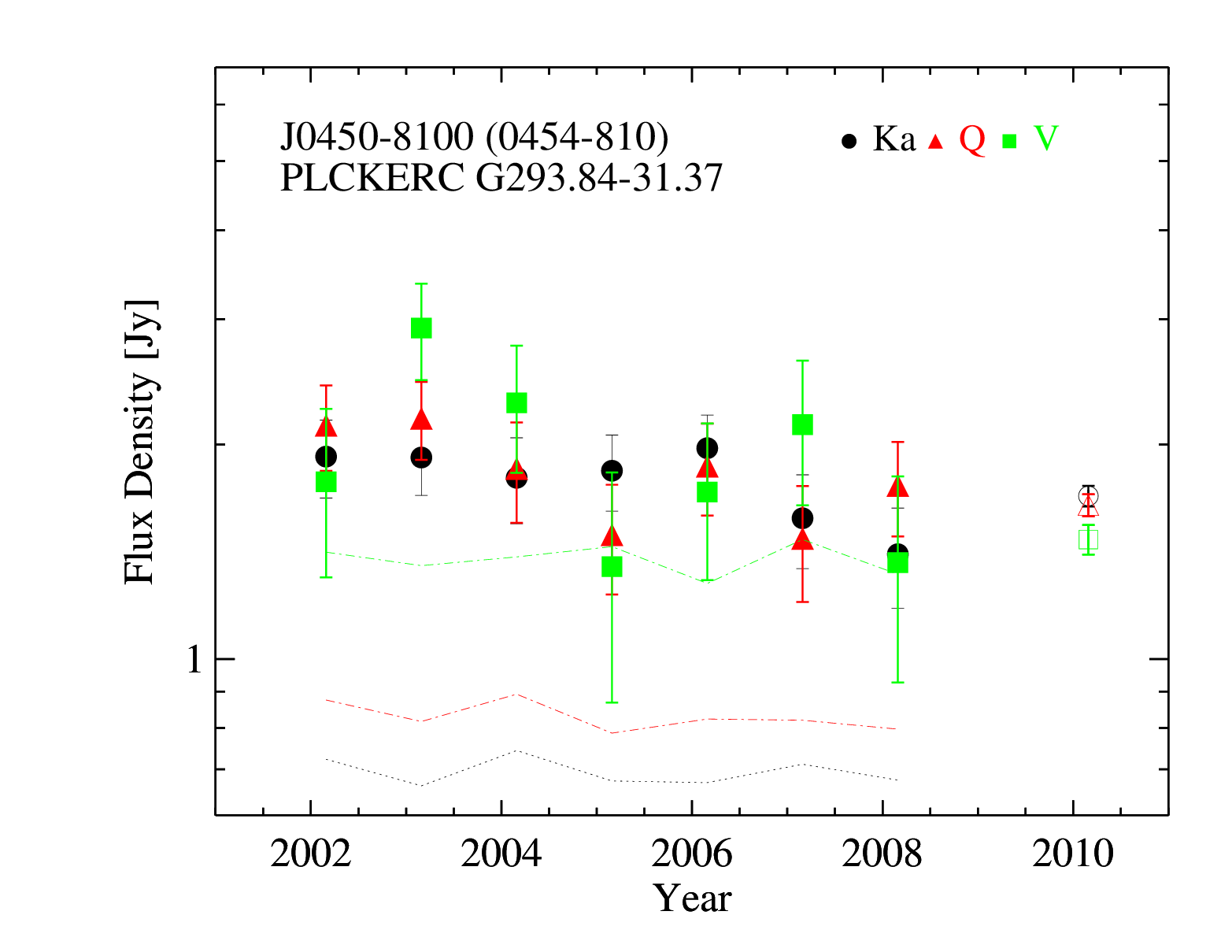}  \\
\includegraphics[width=0.23\textwidth]{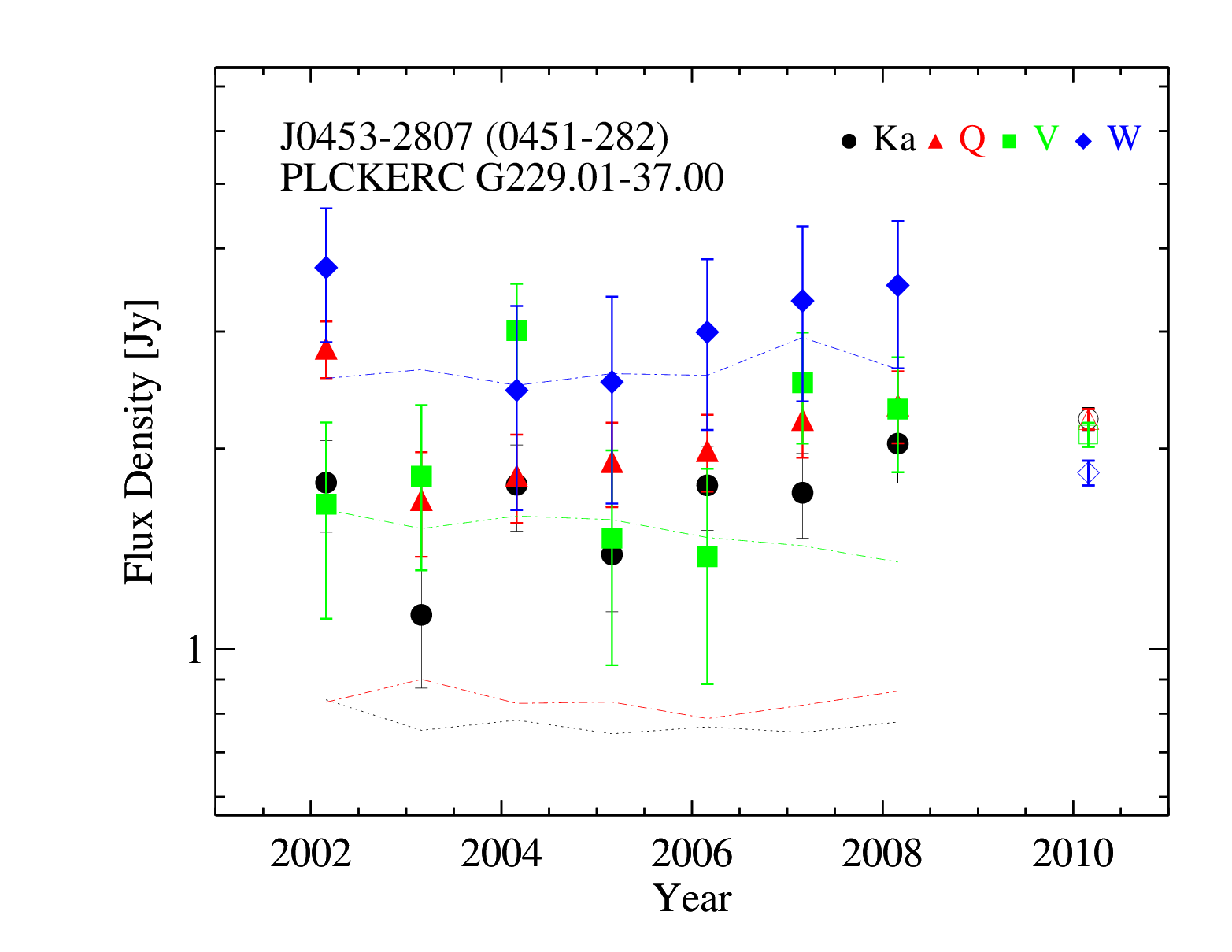} & \includegraphics[width=0.23\textwidth]{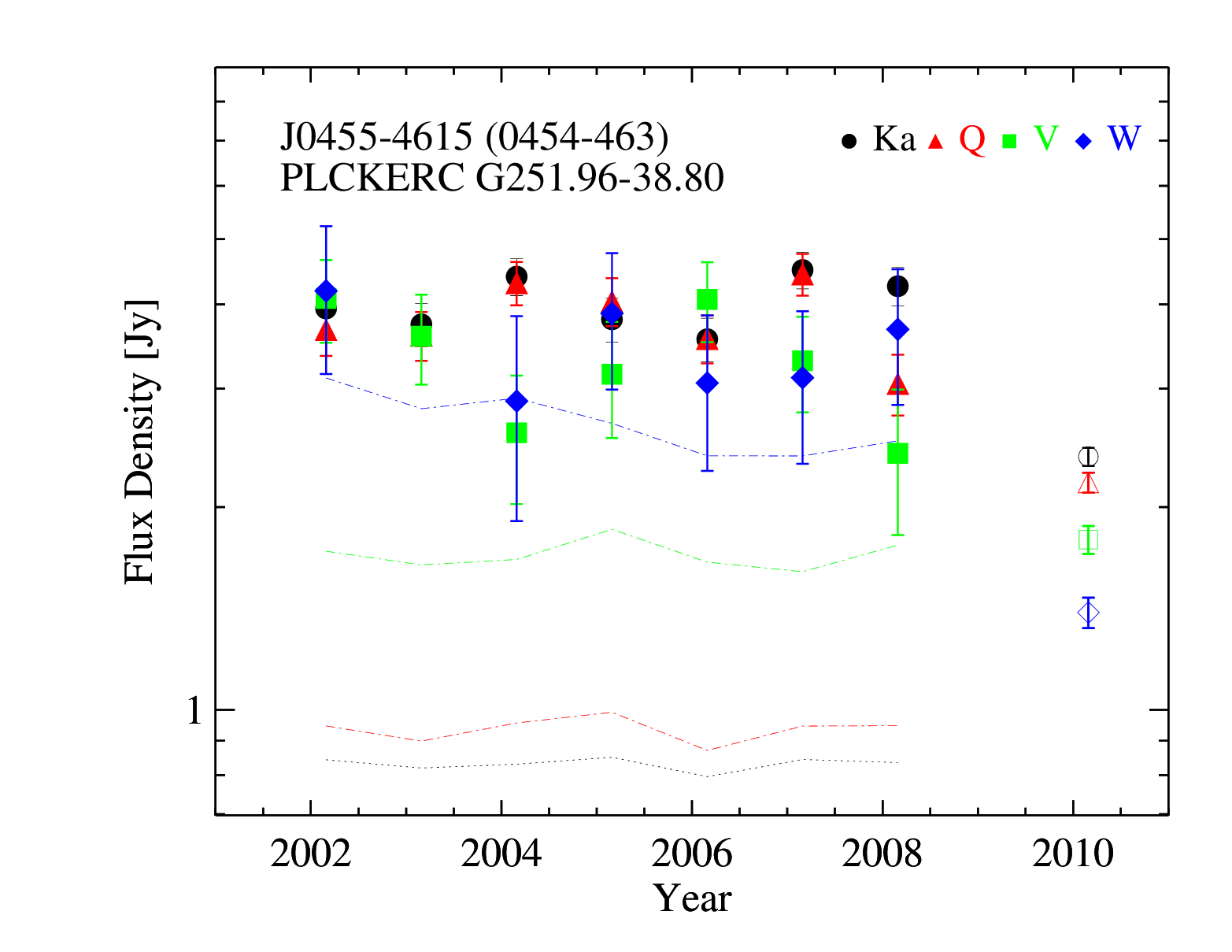}  & \includegraphics[width=0.23\textwidth]{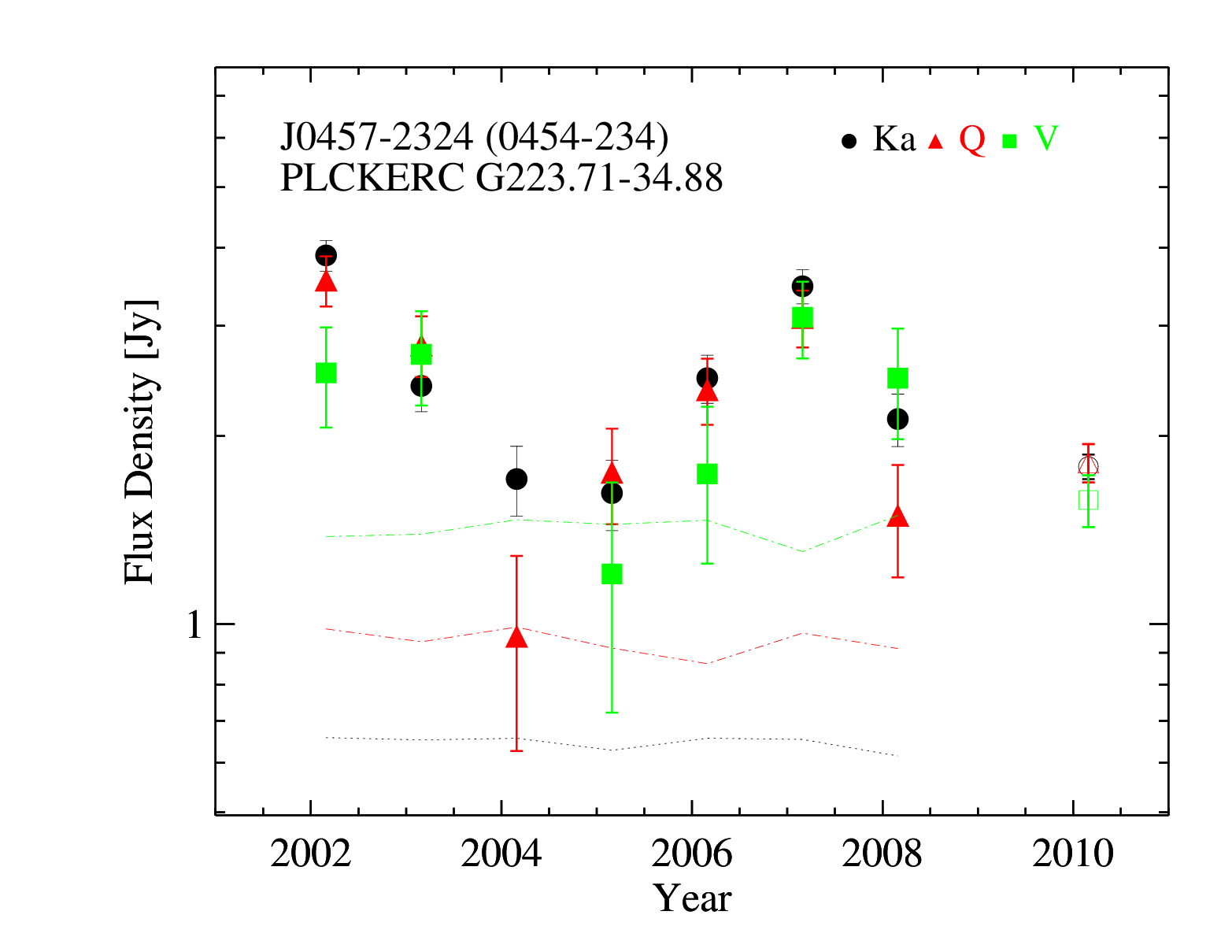} & \includegraphics[width=0.23\textwidth]{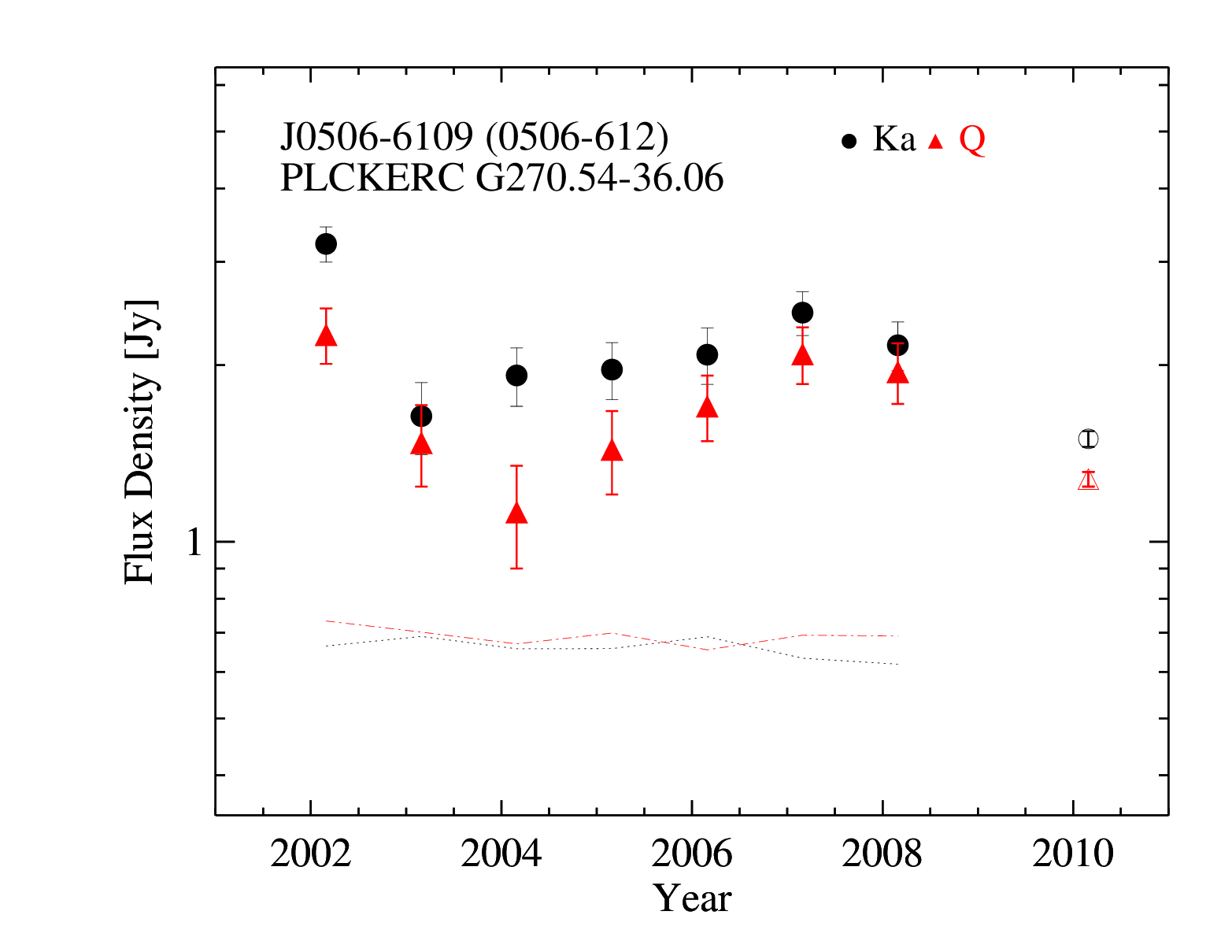}  \\
\includegraphics[width=0.23\textwidth]{figures/lc/J0519-4546.eps} & \includegraphics[width=0.23\textwidth]{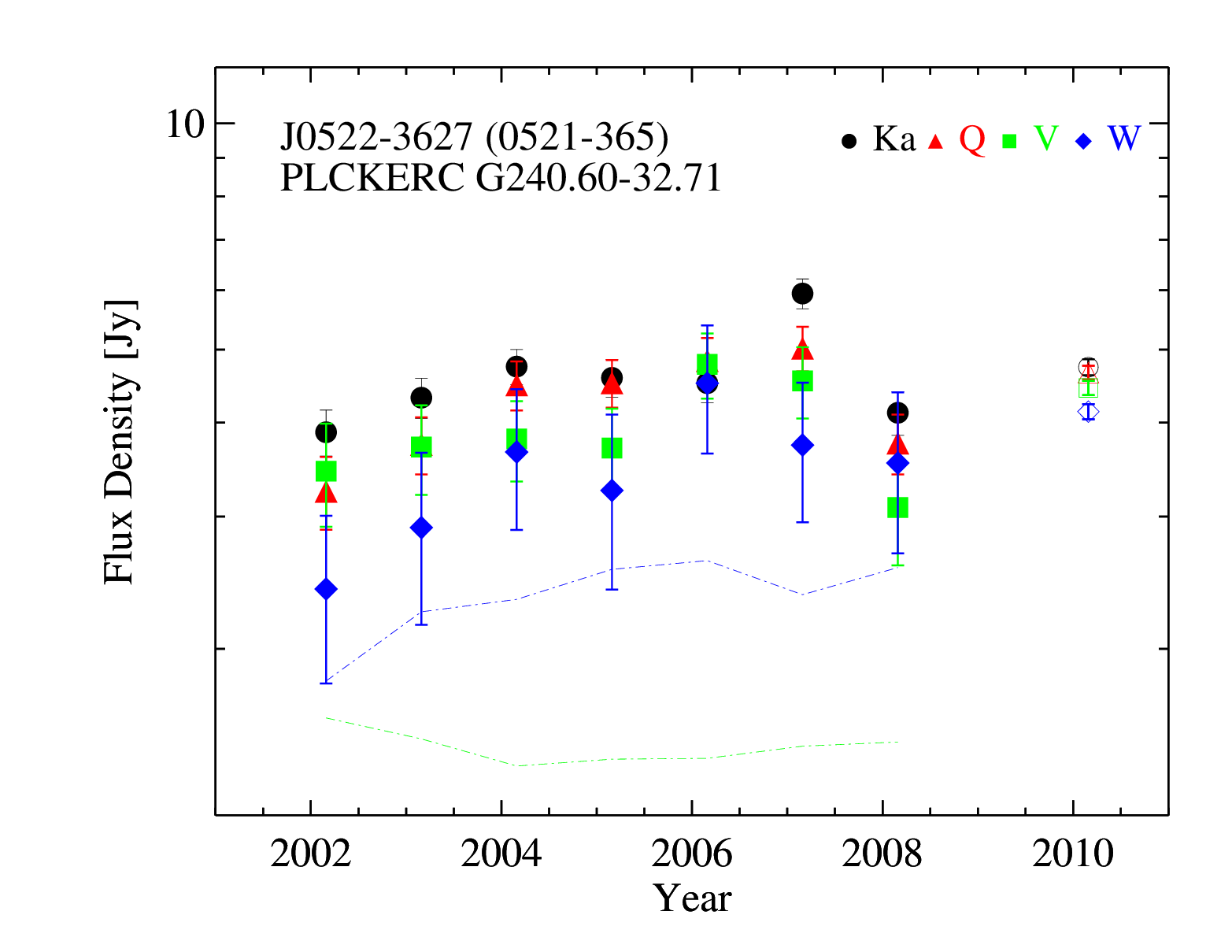}  & \includegraphics[width=0.23\textwidth]{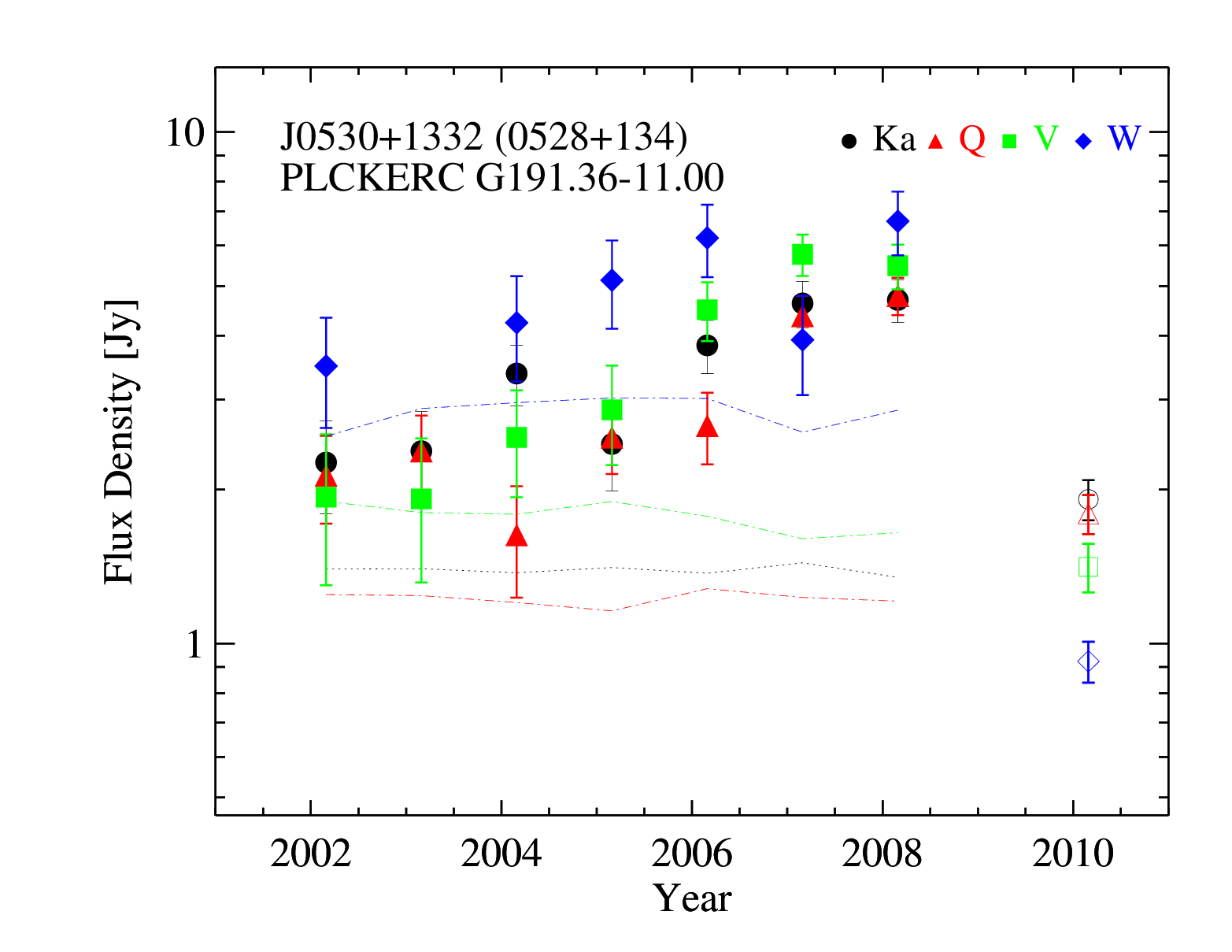} & \includegraphics[width=0.23\textwidth]{figures/lc/J0538-4405.eps}  \\
\includegraphics[width=0.23\textwidth]{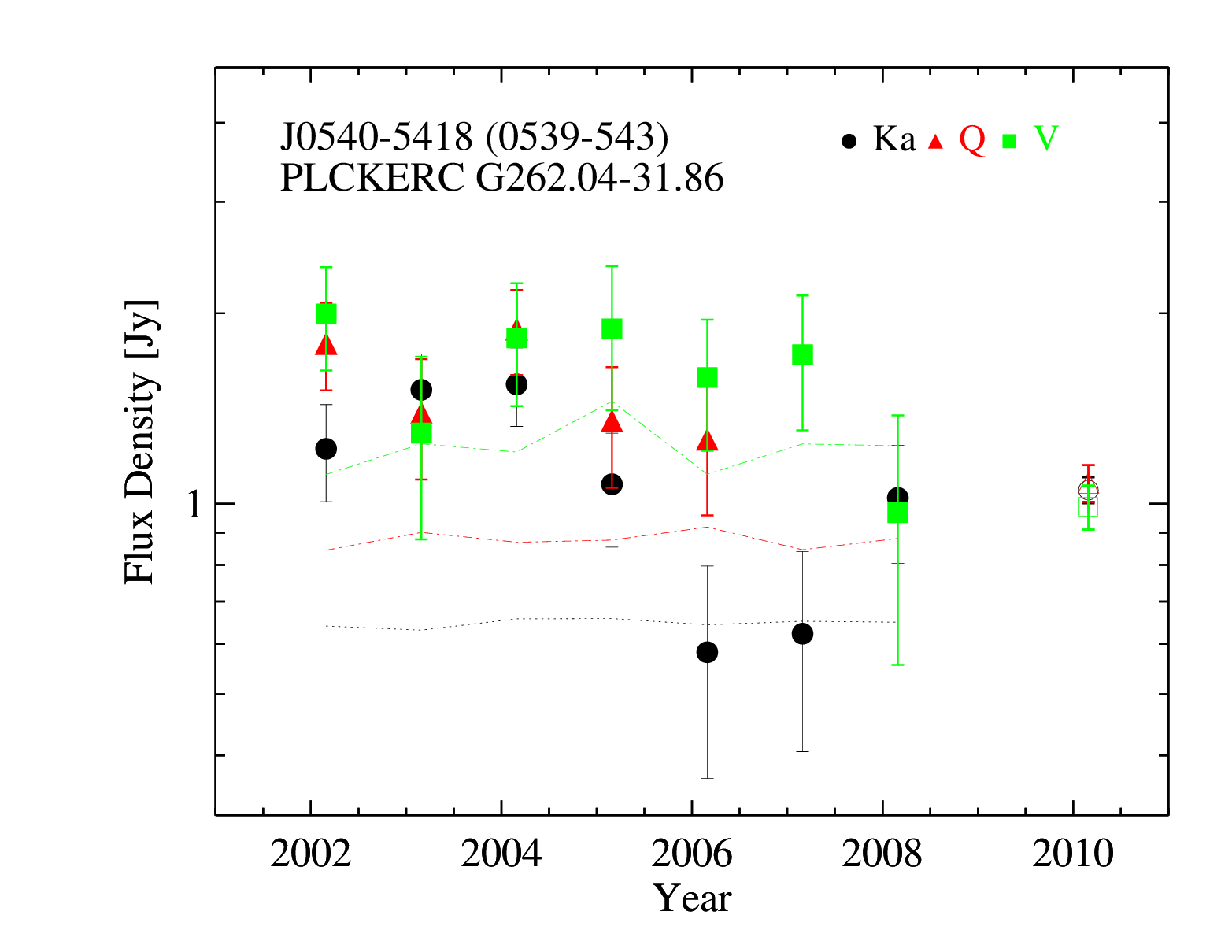} & \includegraphics[width=0.23\textwidth]{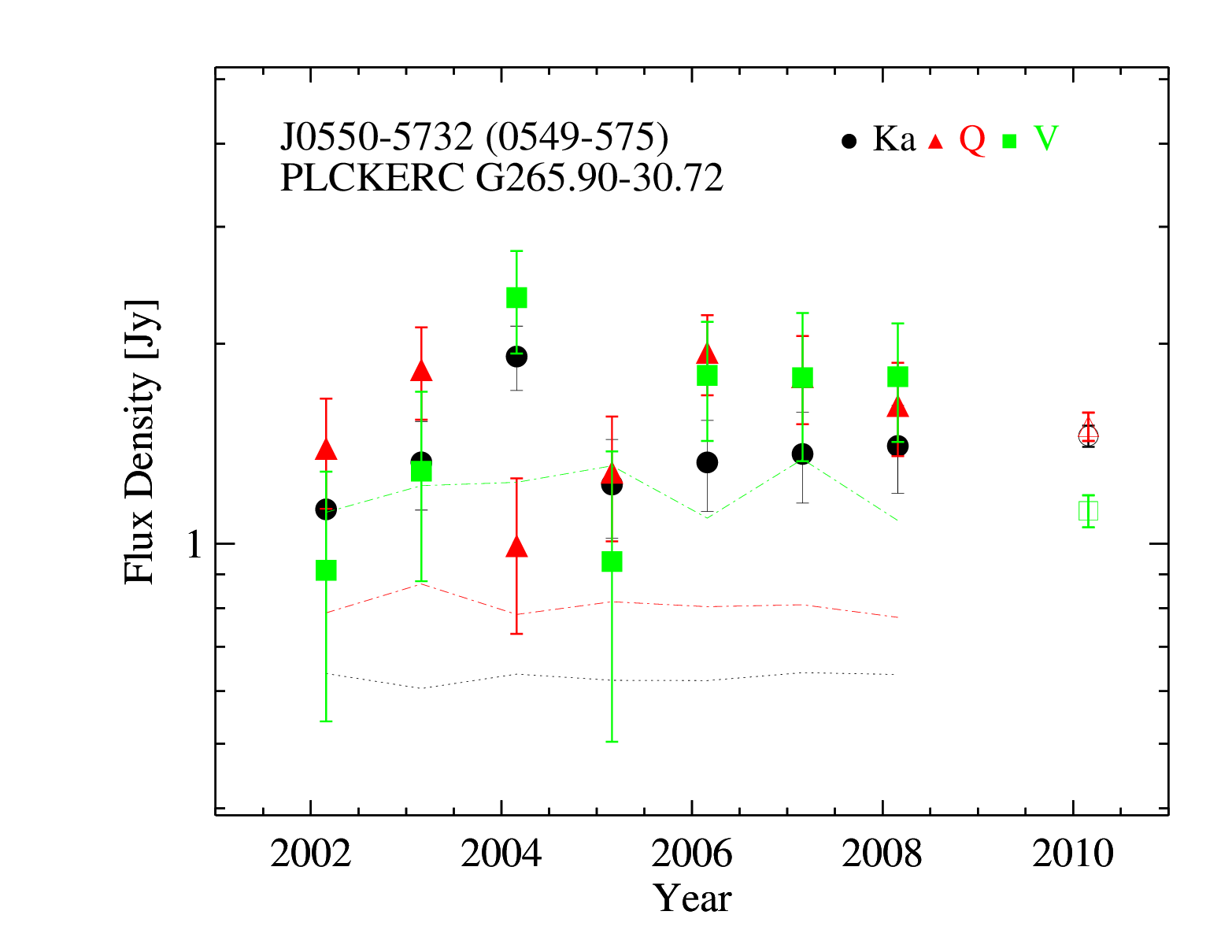}  & \includegraphics[width=0.23\textwidth]{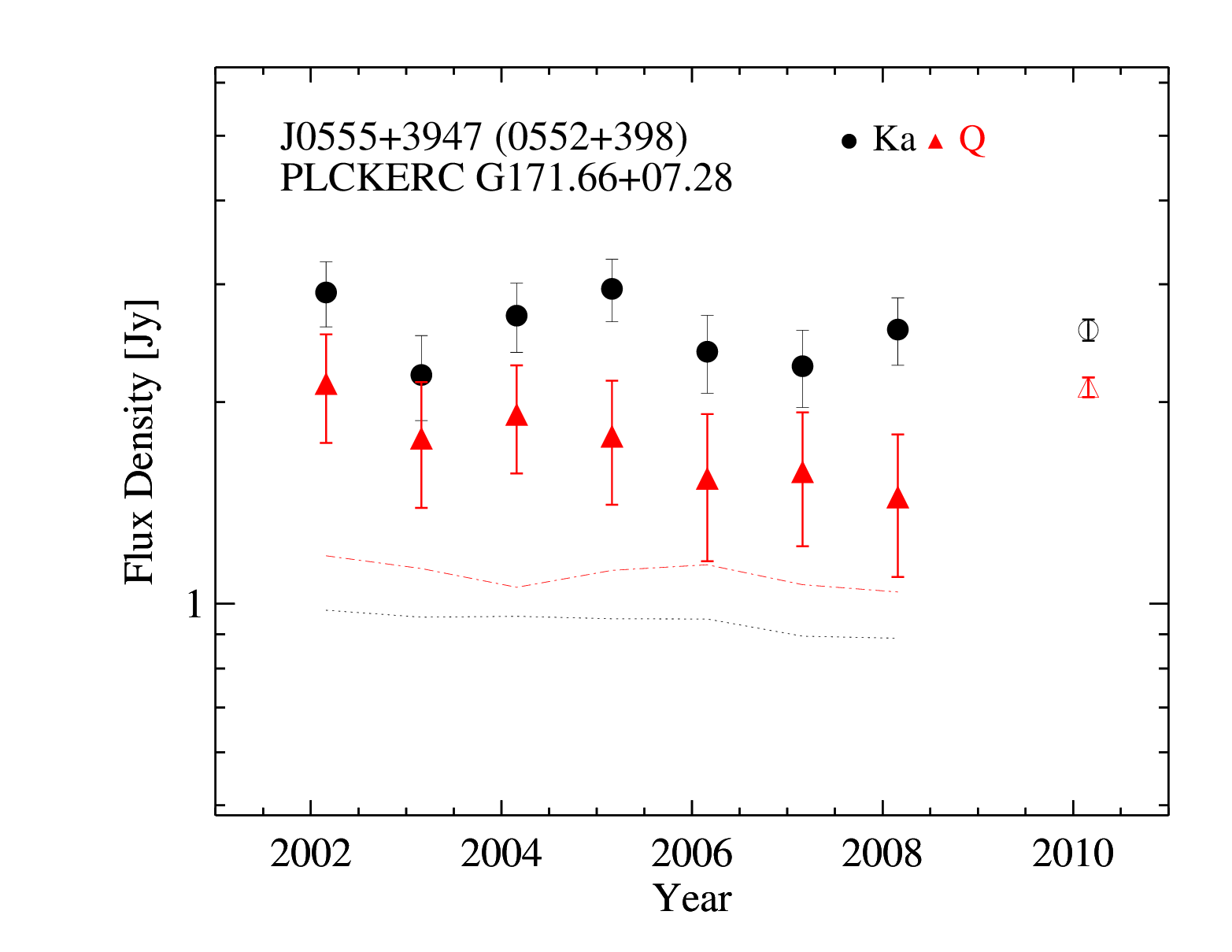} & \includegraphics[width=0.23\textwidth]{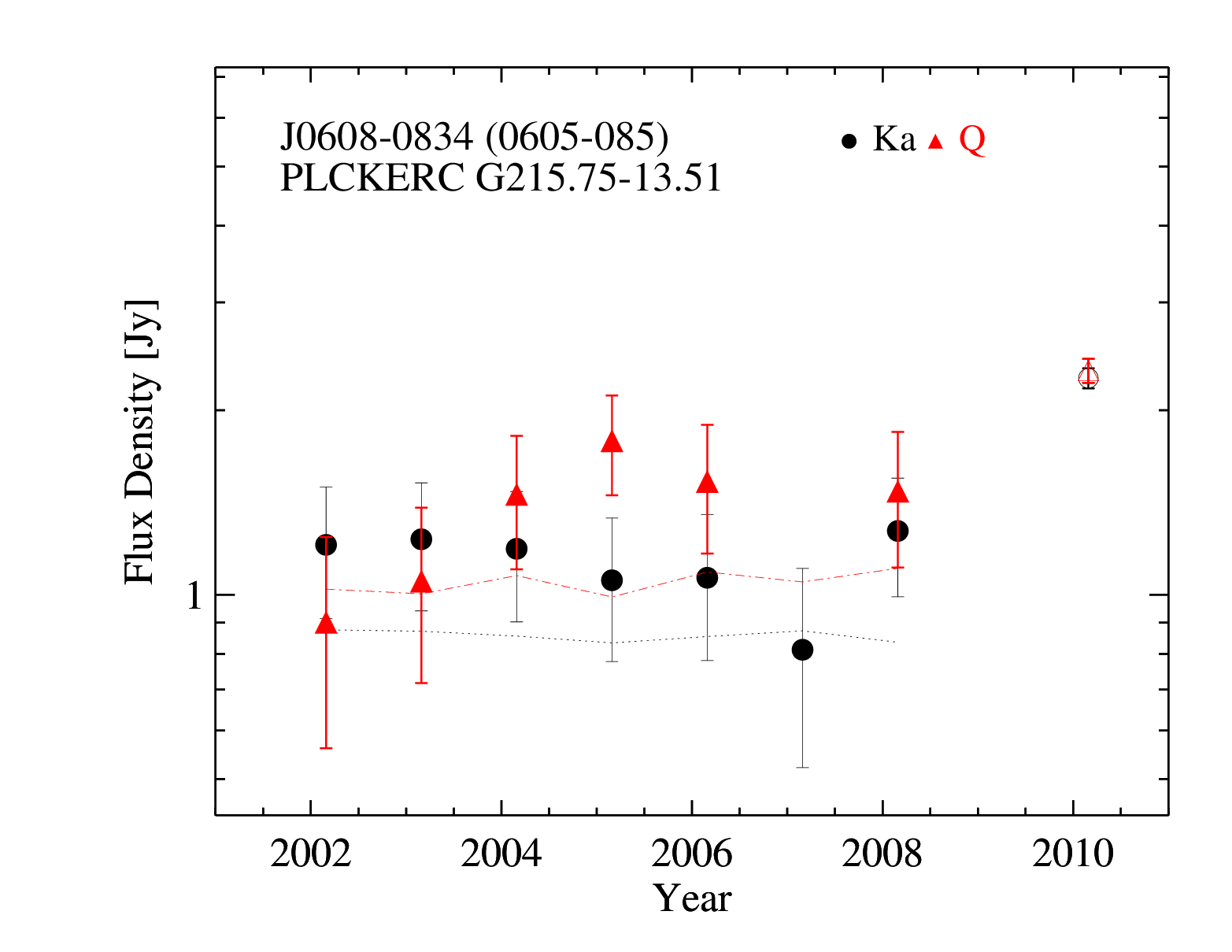}  \\
\includegraphics[width=0.23\textwidth]{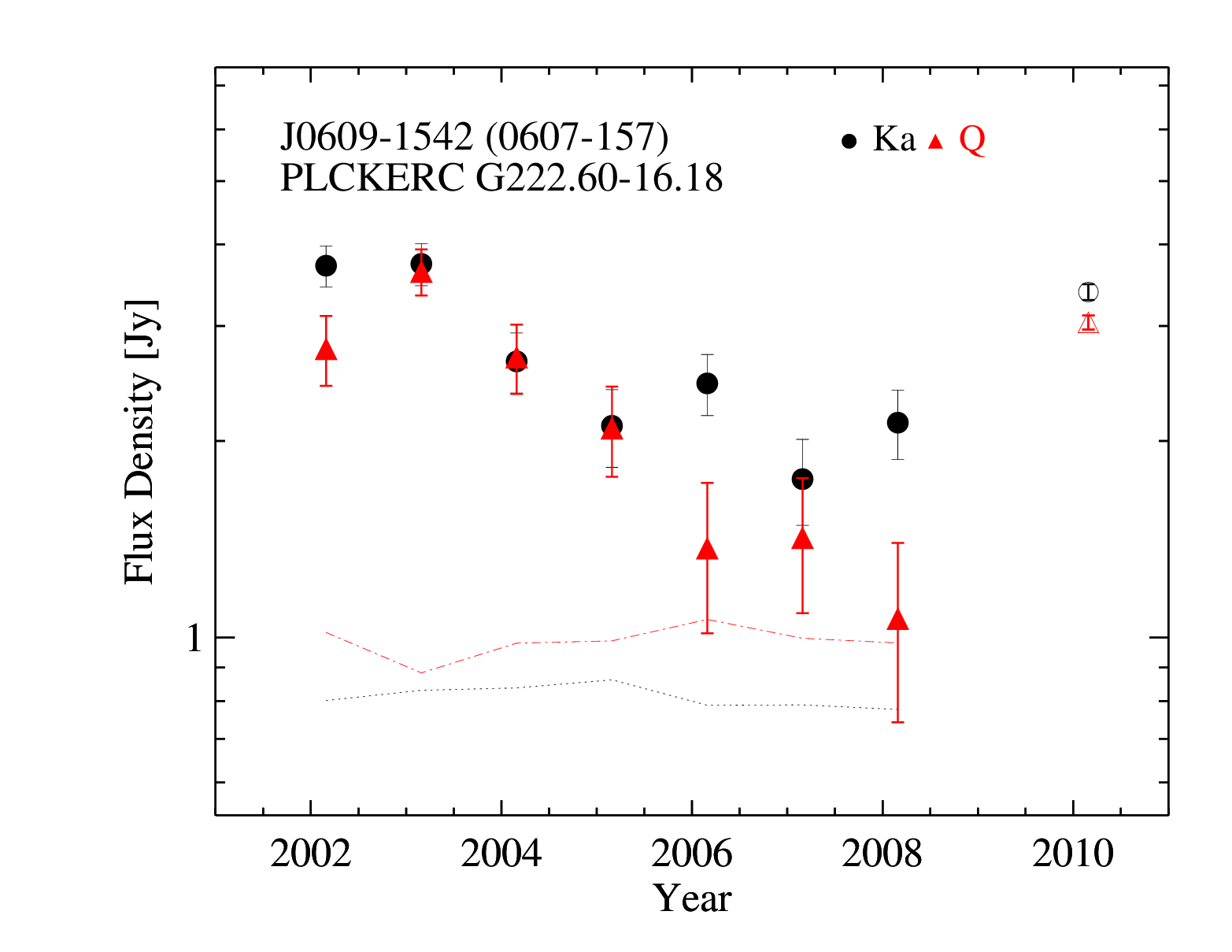} & \includegraphics[width=0.23\textwidth]{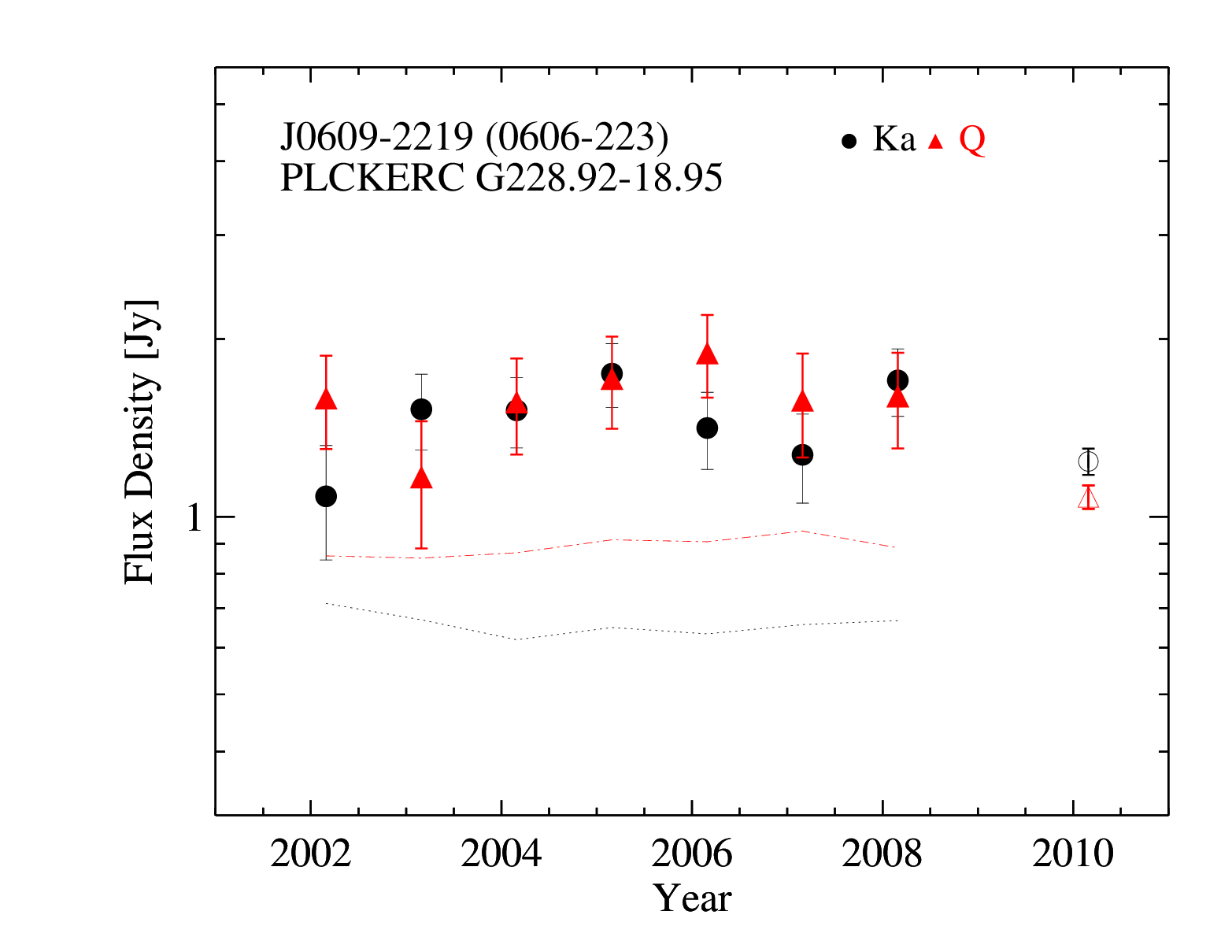}  & \includegraphics[width=0.23\textwidth]{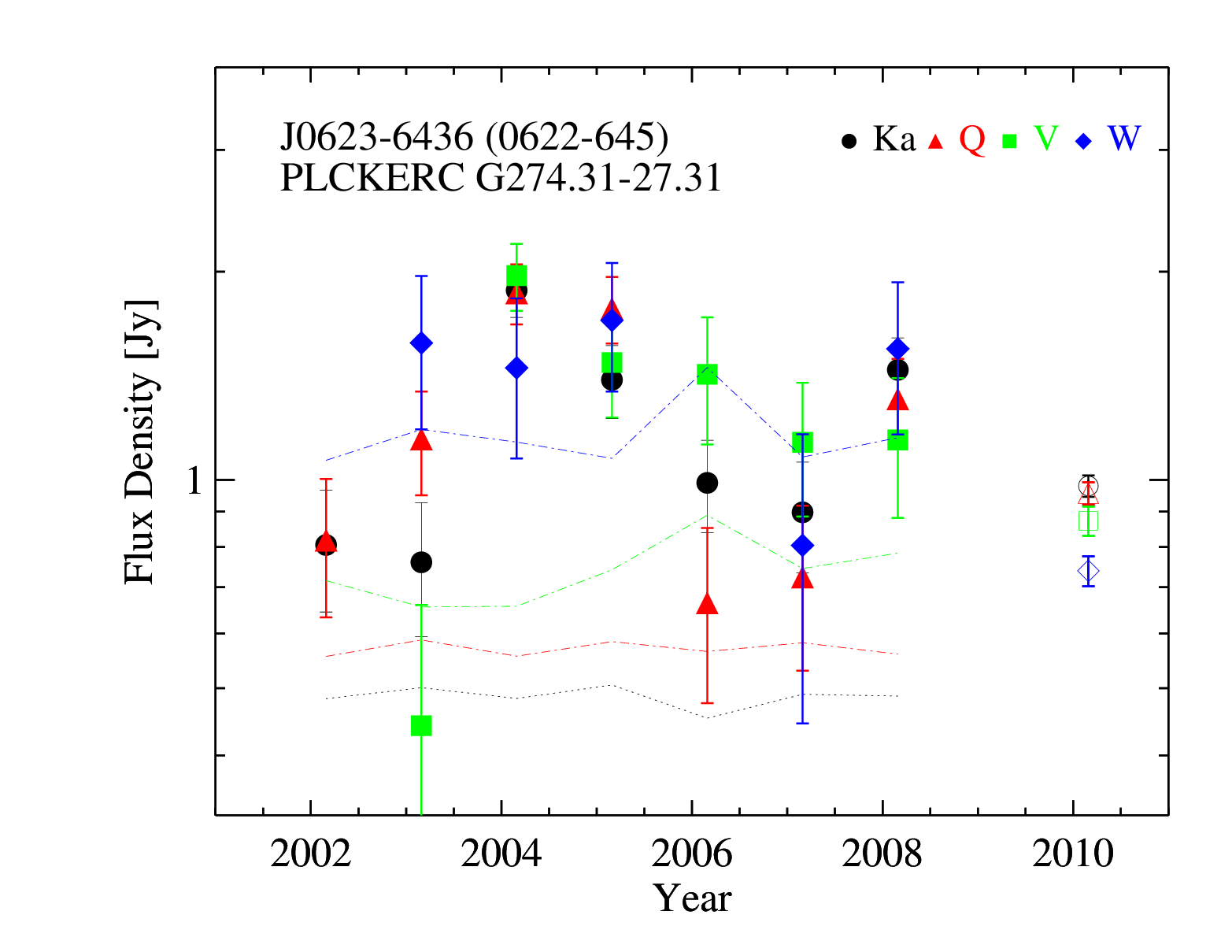} & \includegraphics[width=0.23\textwidth]{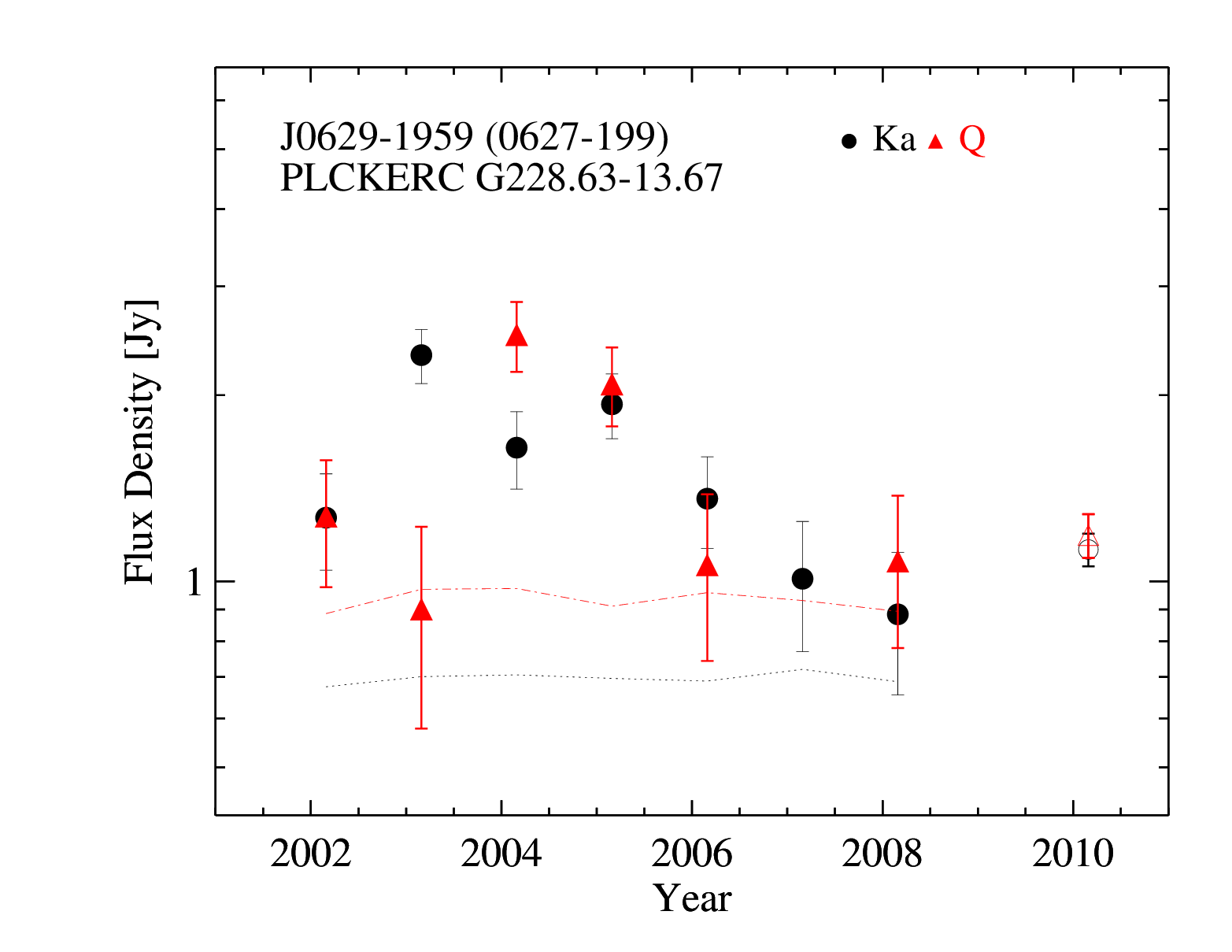}  \\
\includegraphics[width=0.23\textwidth]{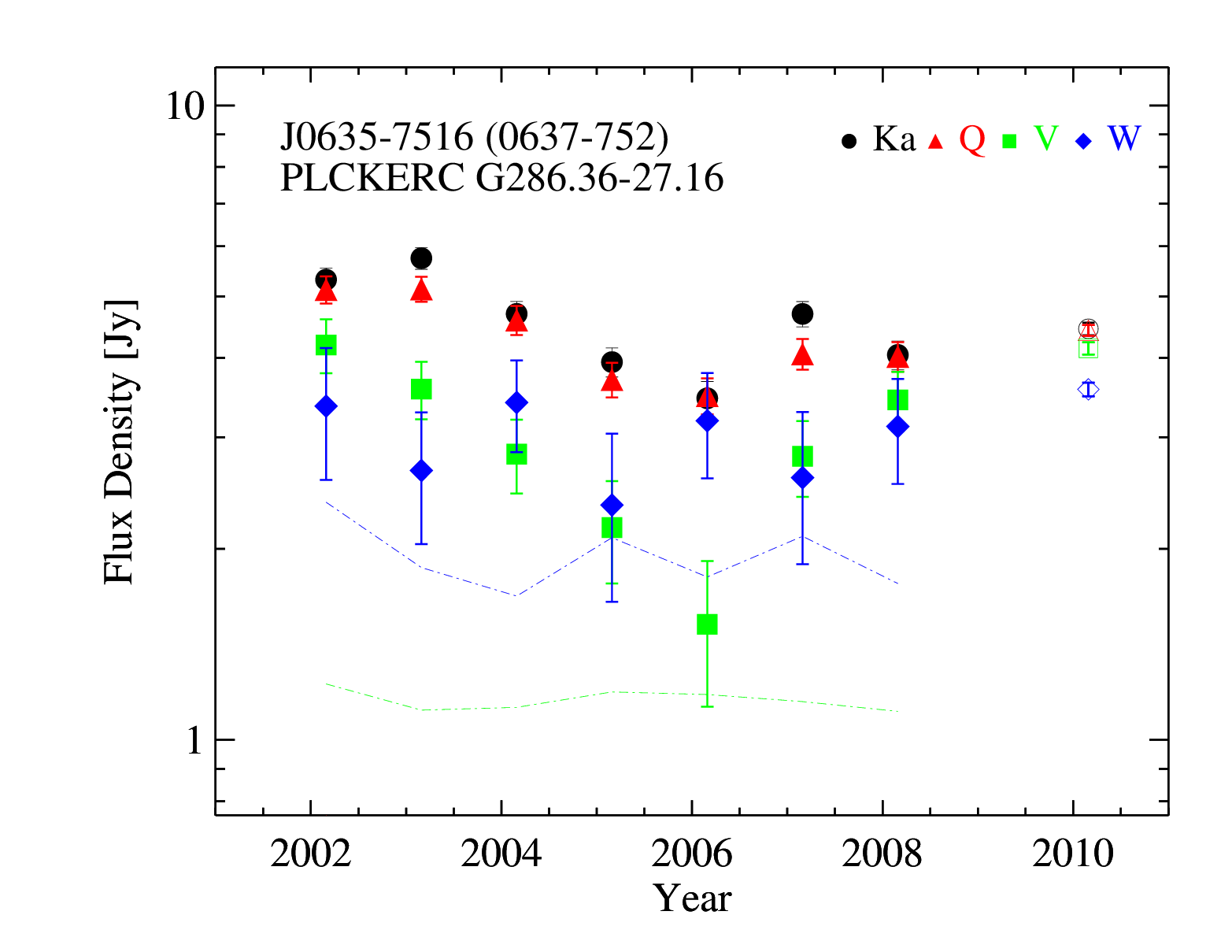} & \includegraphics[width=0.23\textwidth]{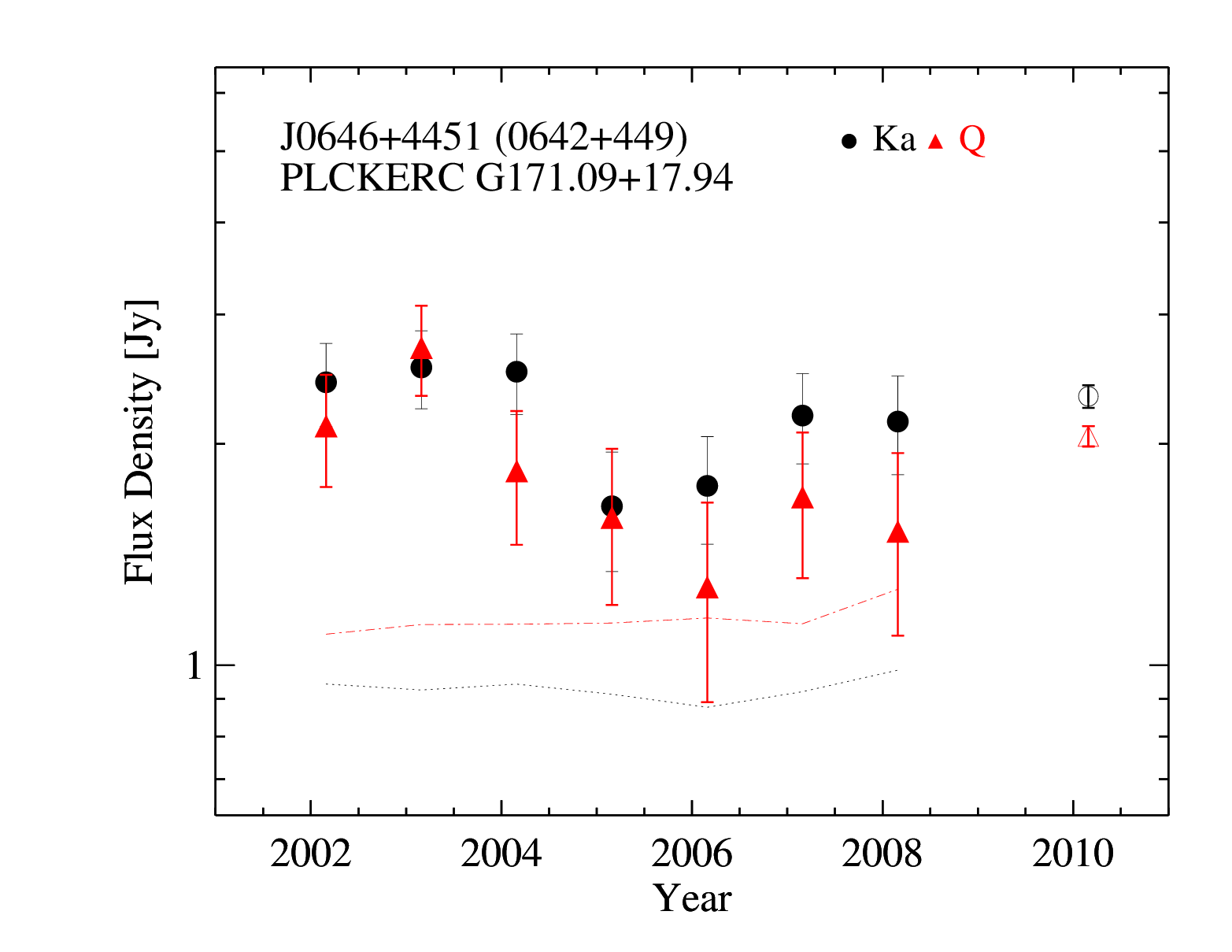}  & \includegraphics[width=0.23\textwidth]{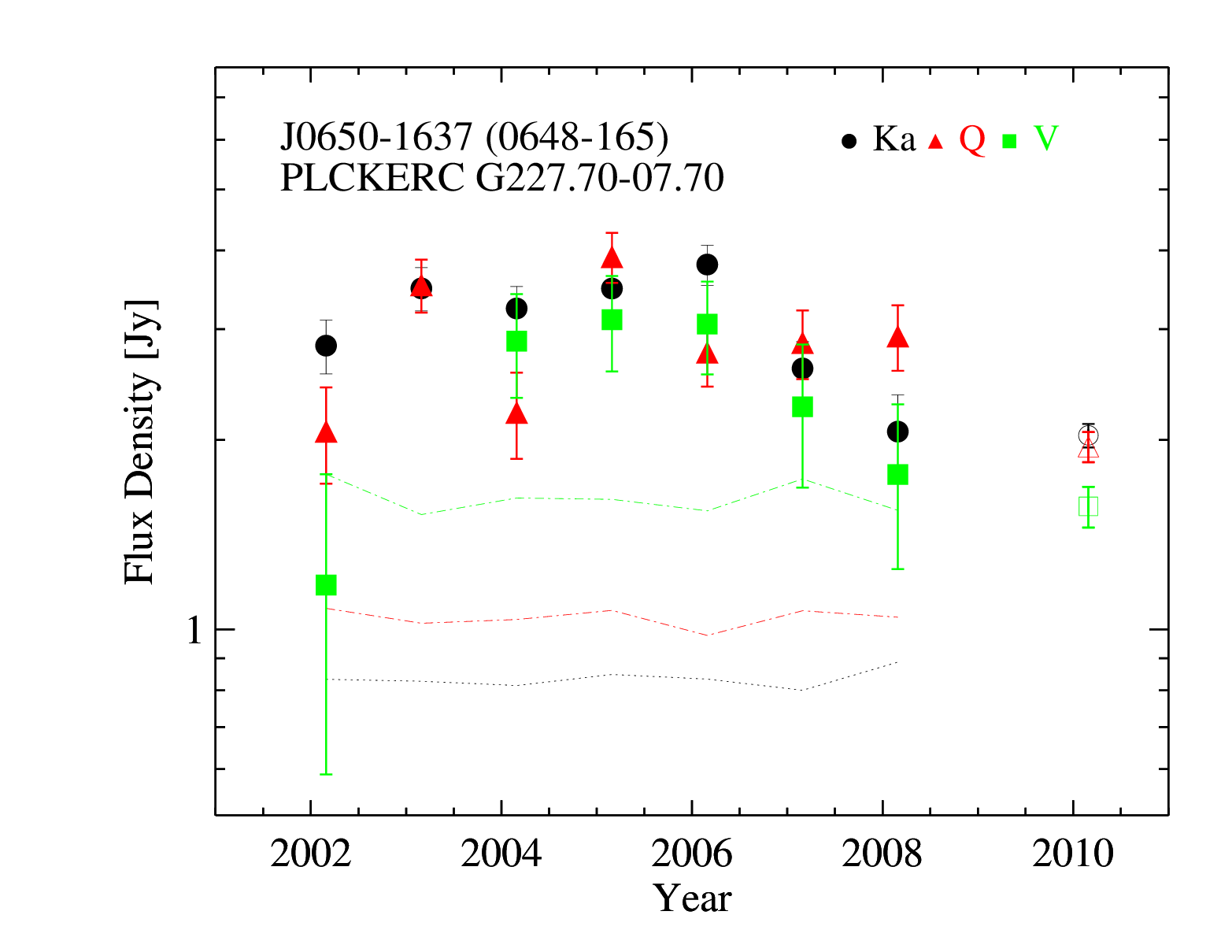} & \includegraphics[width=0.23\textwidth]{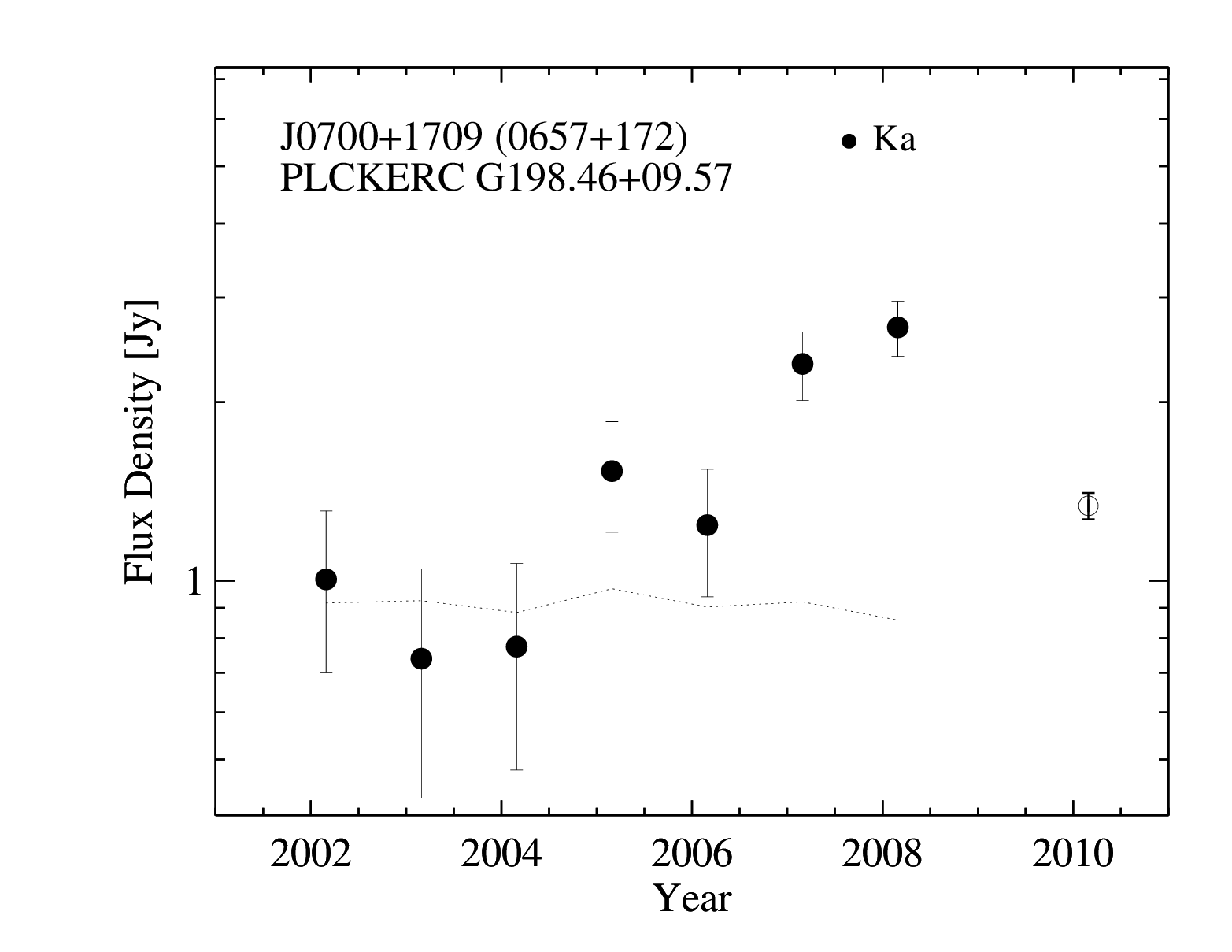}  \\
\includegraphics[width=0.23\textwidth]{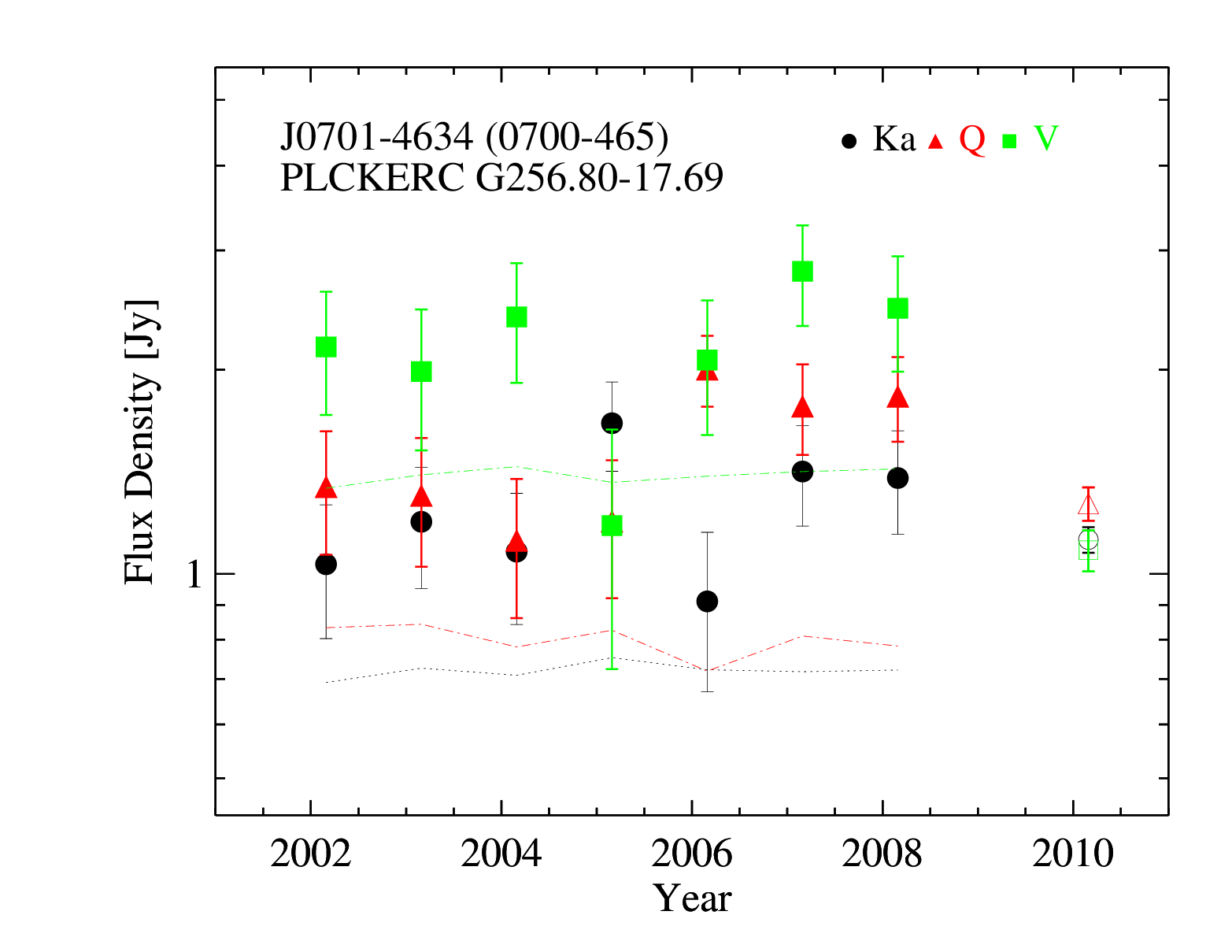} & \includegraphics[width=0.23\textwidth]{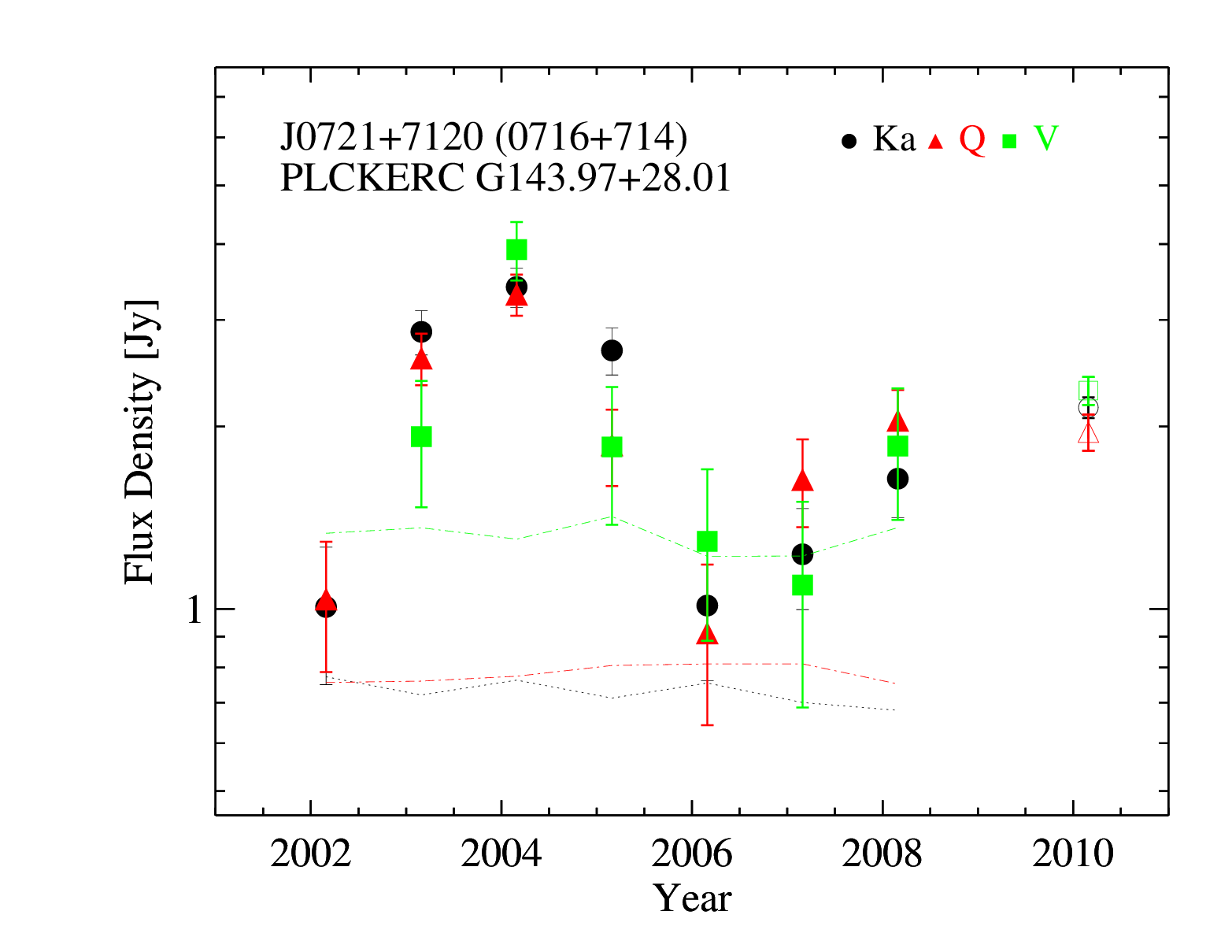}  & \includegraphics[width=0.23\textwidth]{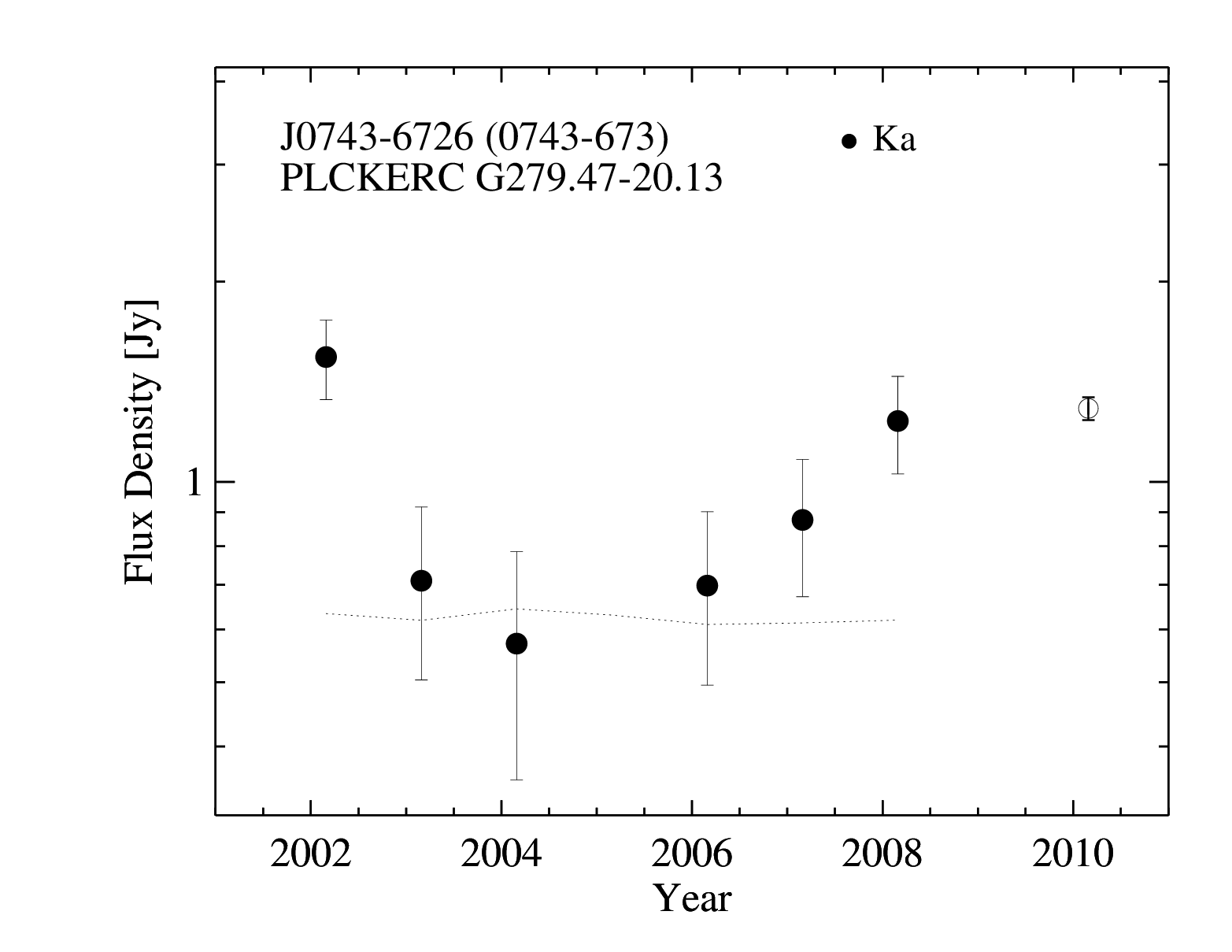} & \includegraphics[width=0.23\textwidth]{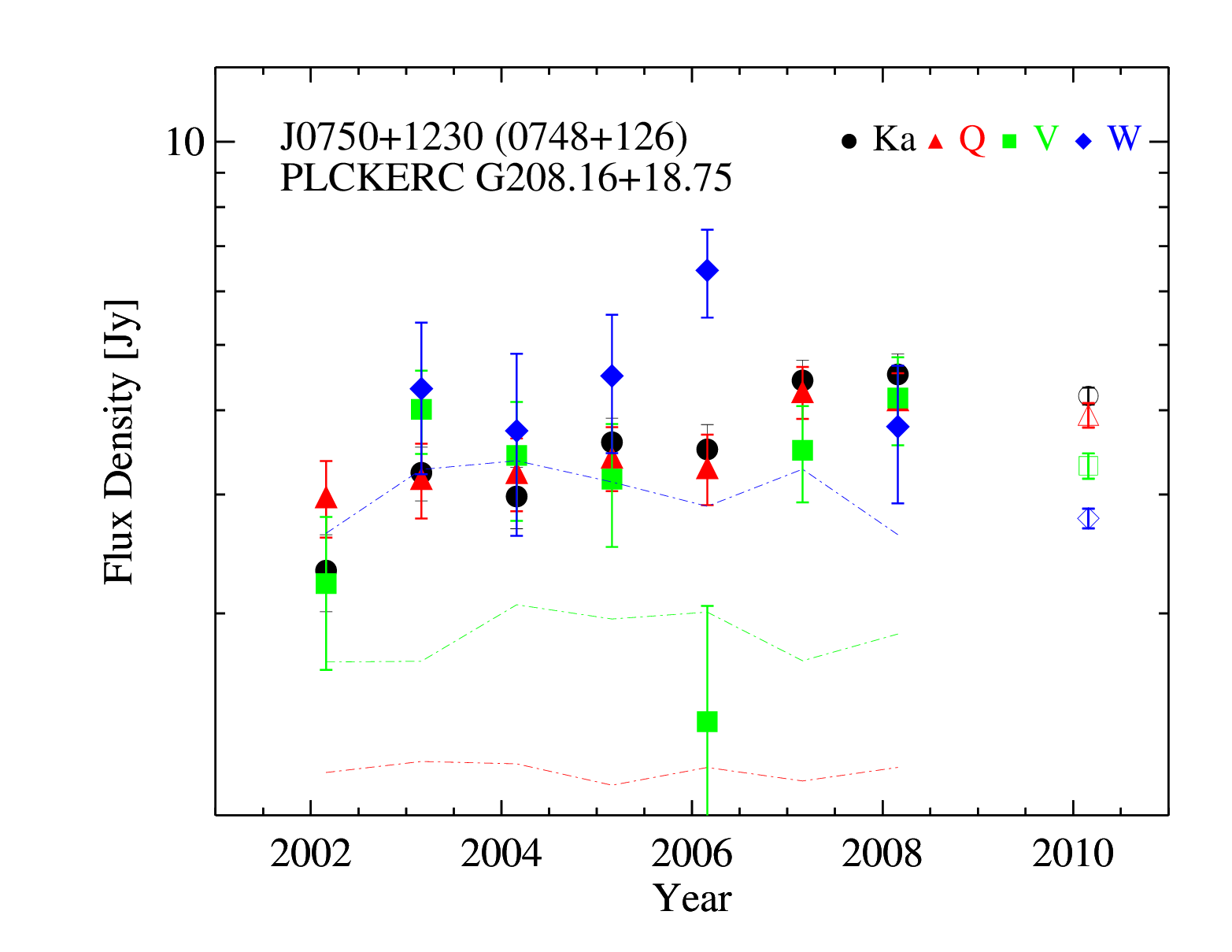}  \\
\end{tabular}
\end{figure*}

\begin{figure*}
\centering
\begin{tabular}{cccc}
\includegraphics[width=0.23\textwidth]{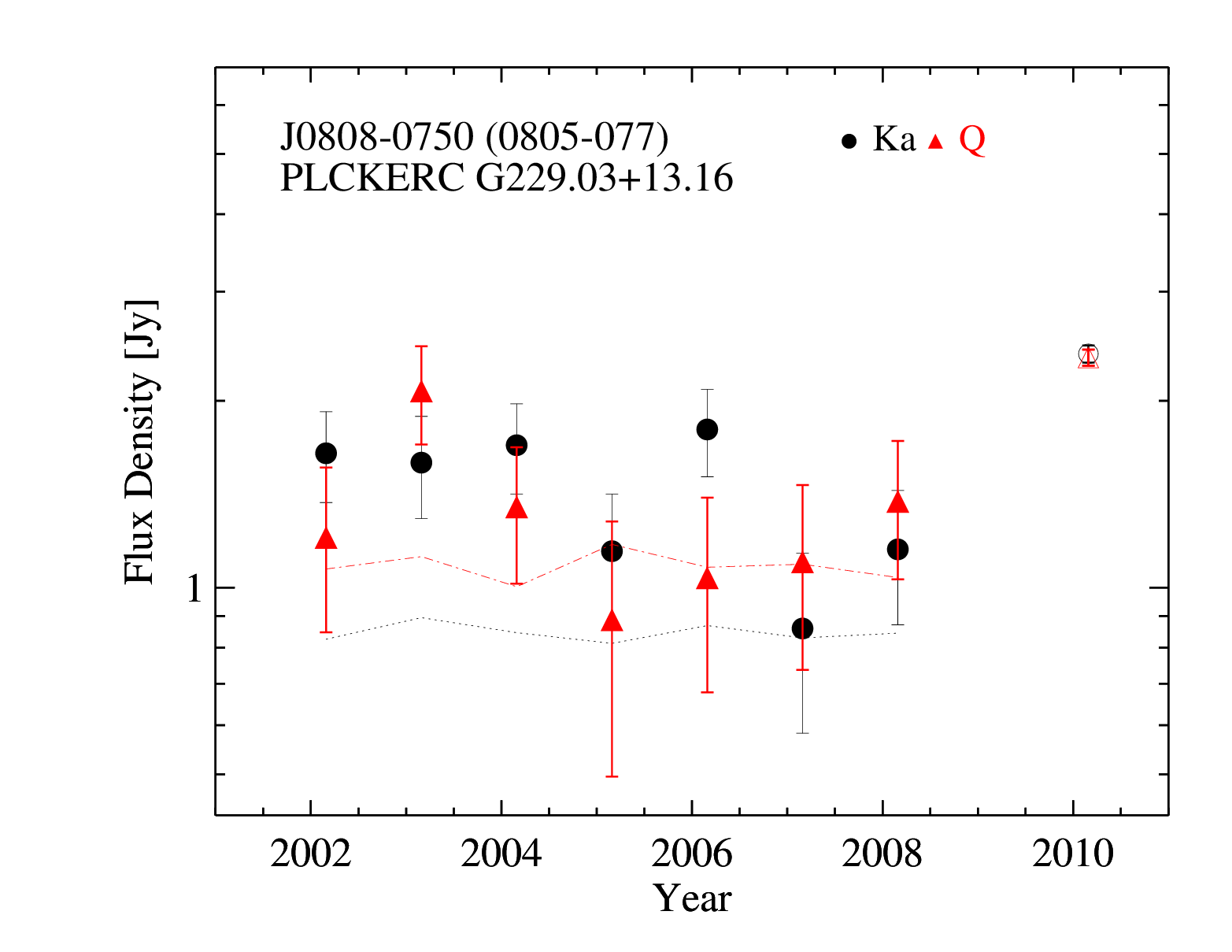} & \includegraphics[width=0.23\textwidth]{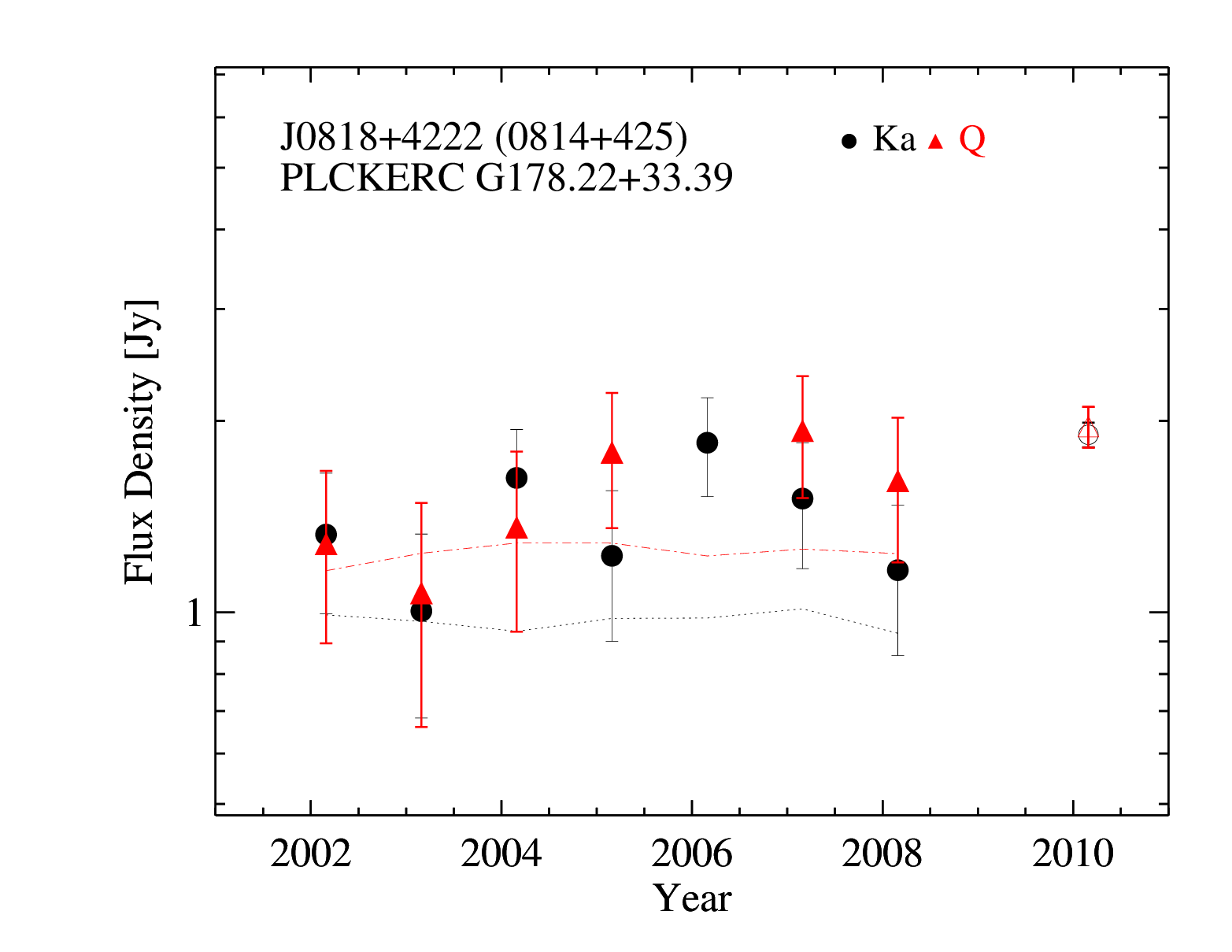}  & \includegraphics[width=0.23\textwidth]{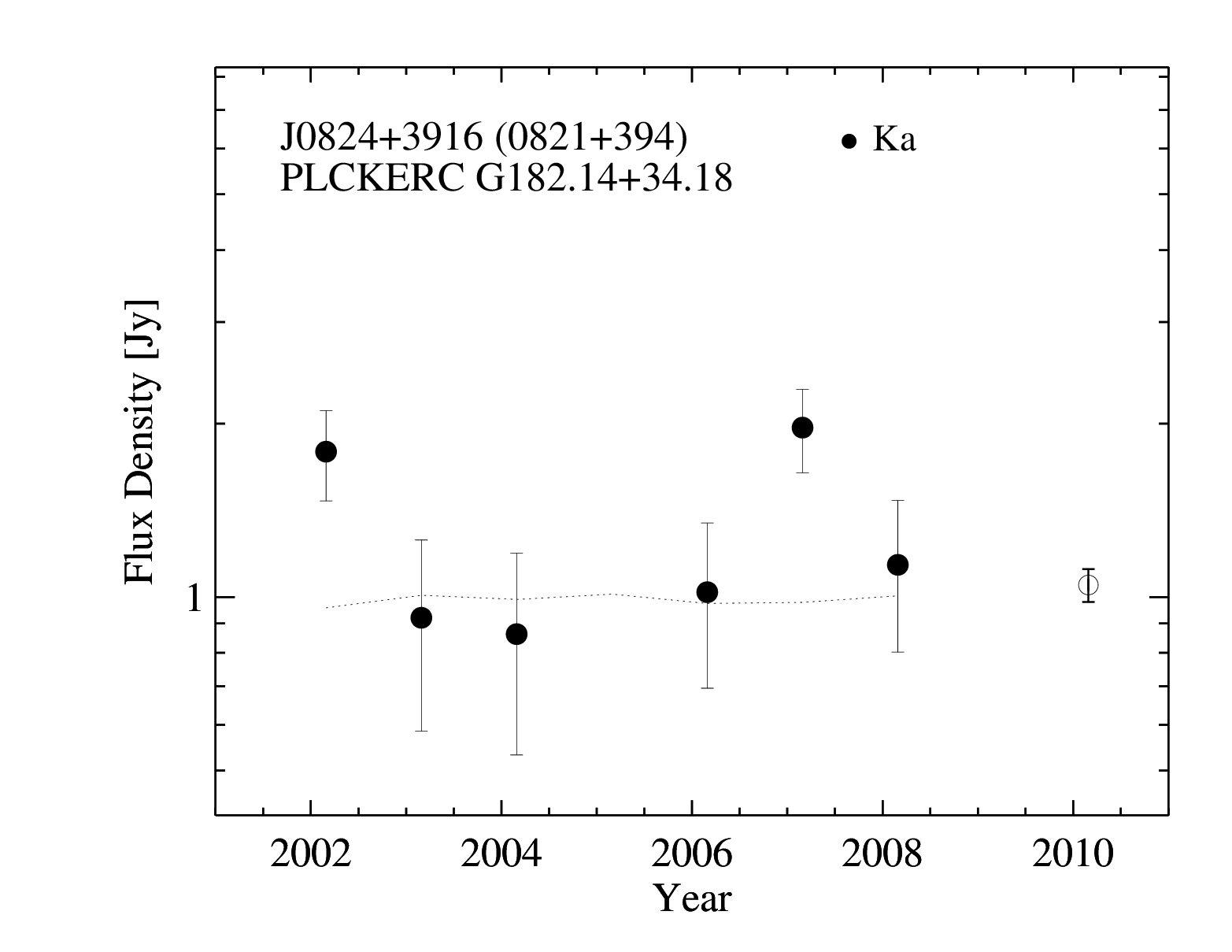} & \includegraphics[width=0.23\textwidth]{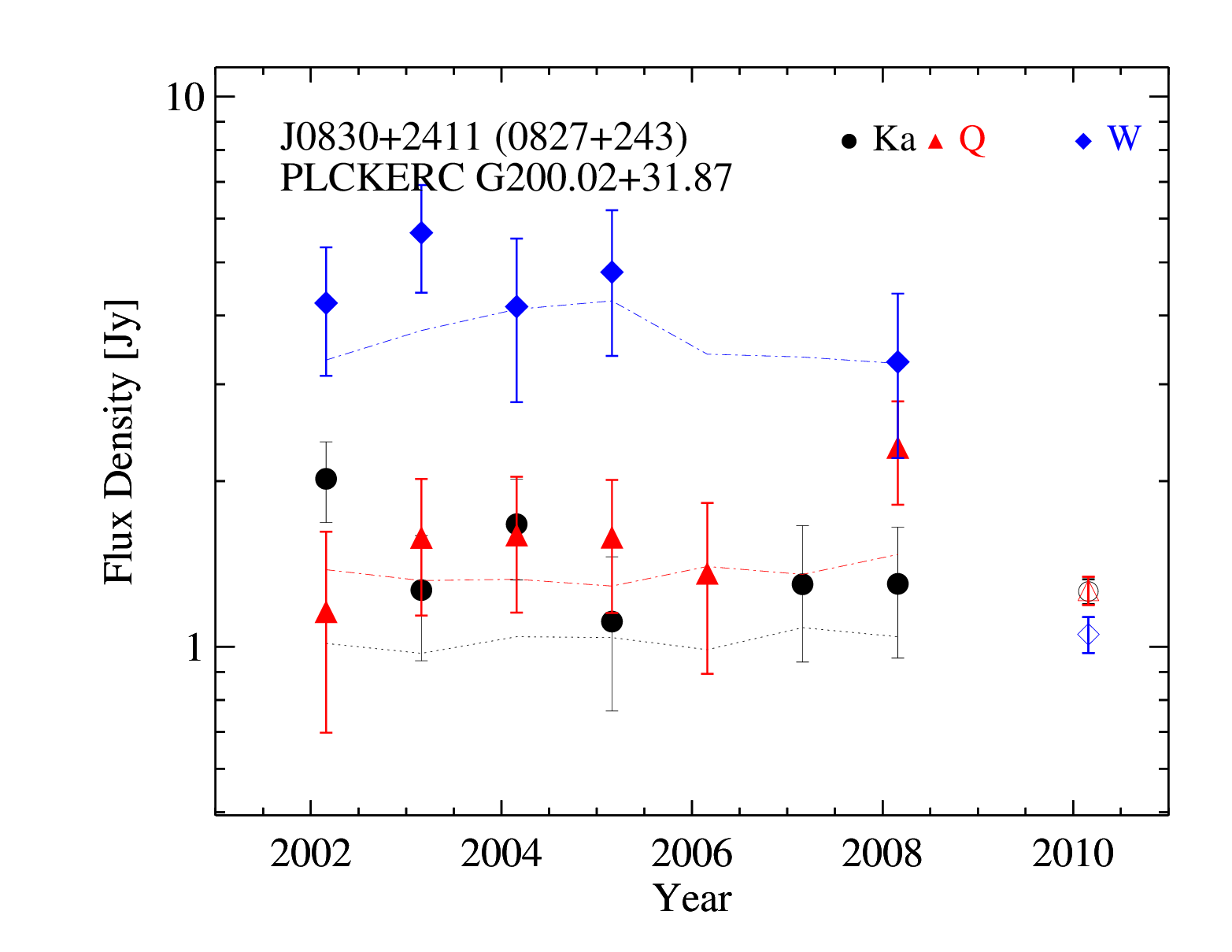}  \\
\includegraphics[width=0.23\textwidth]{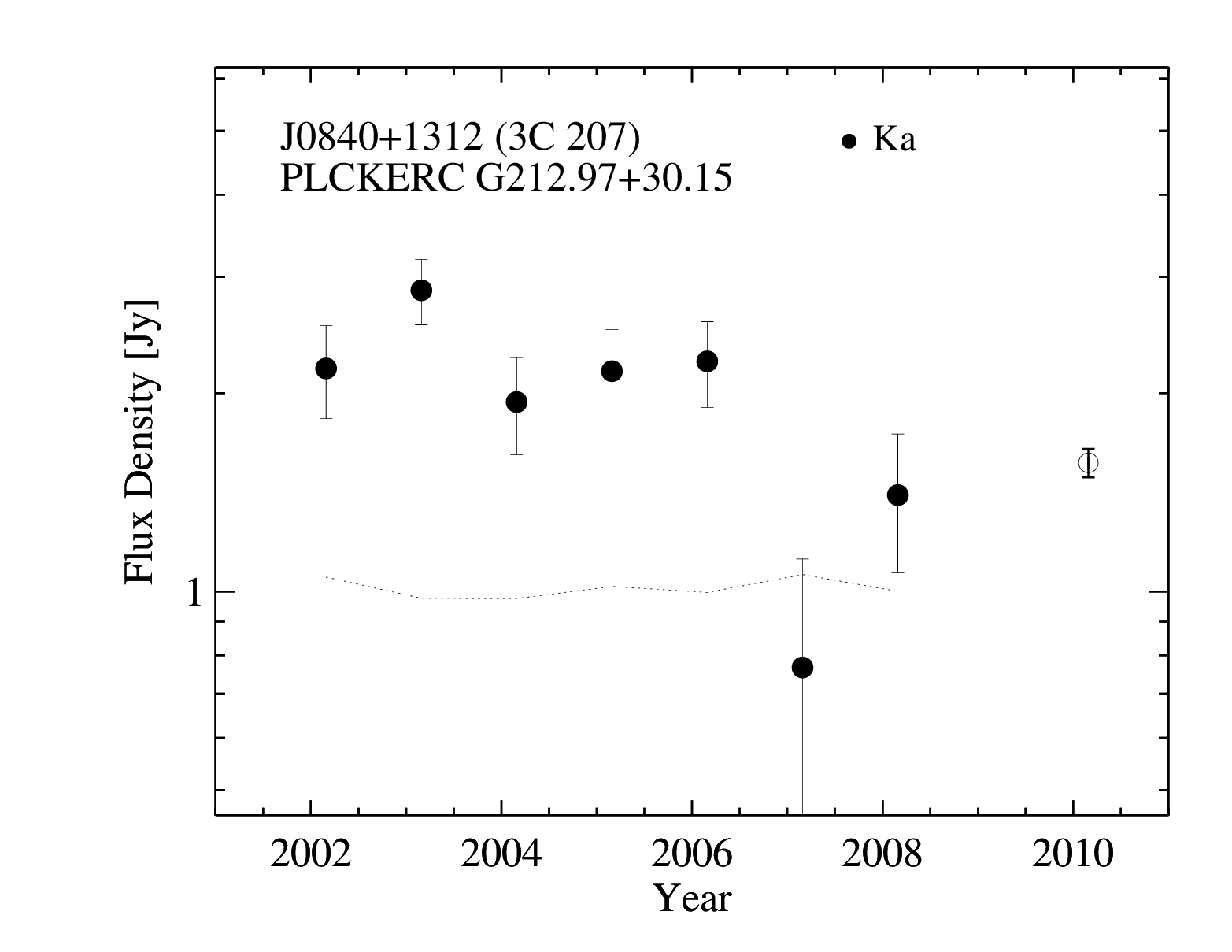} & \includegraphics[width=0.23\textwidth]{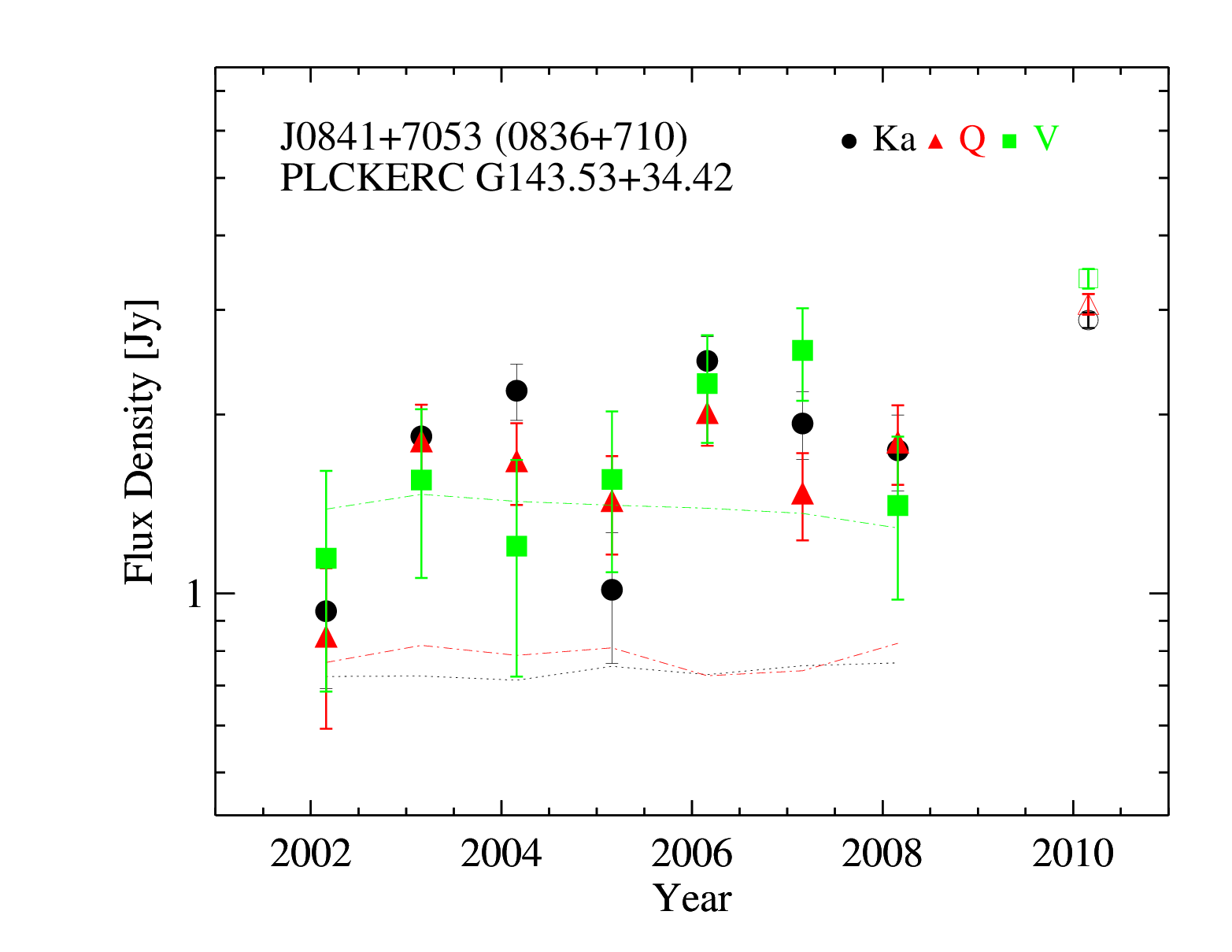}  & \includegraphics[width=0.23\textwidth]{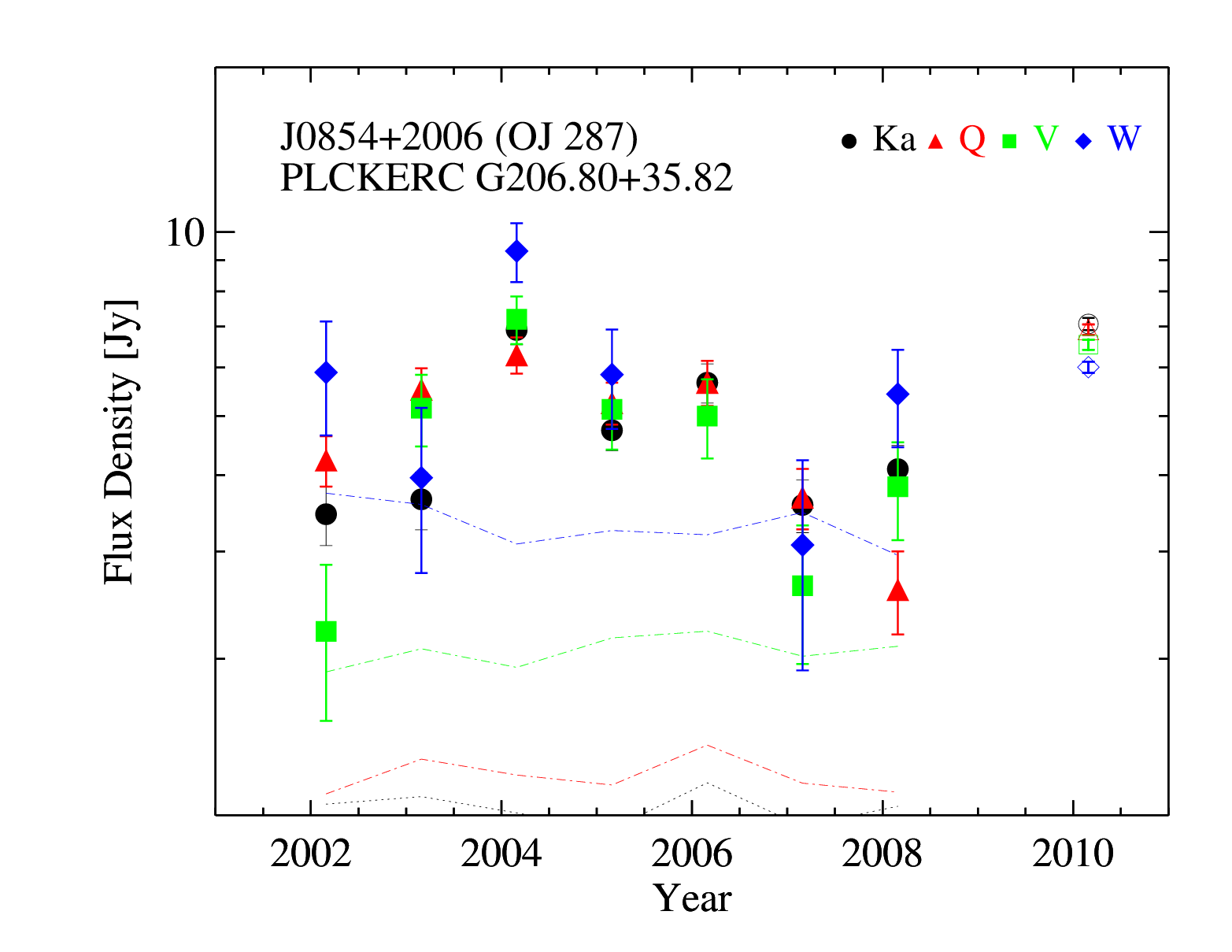} & \includegraphics[width=0.23\textwidth]{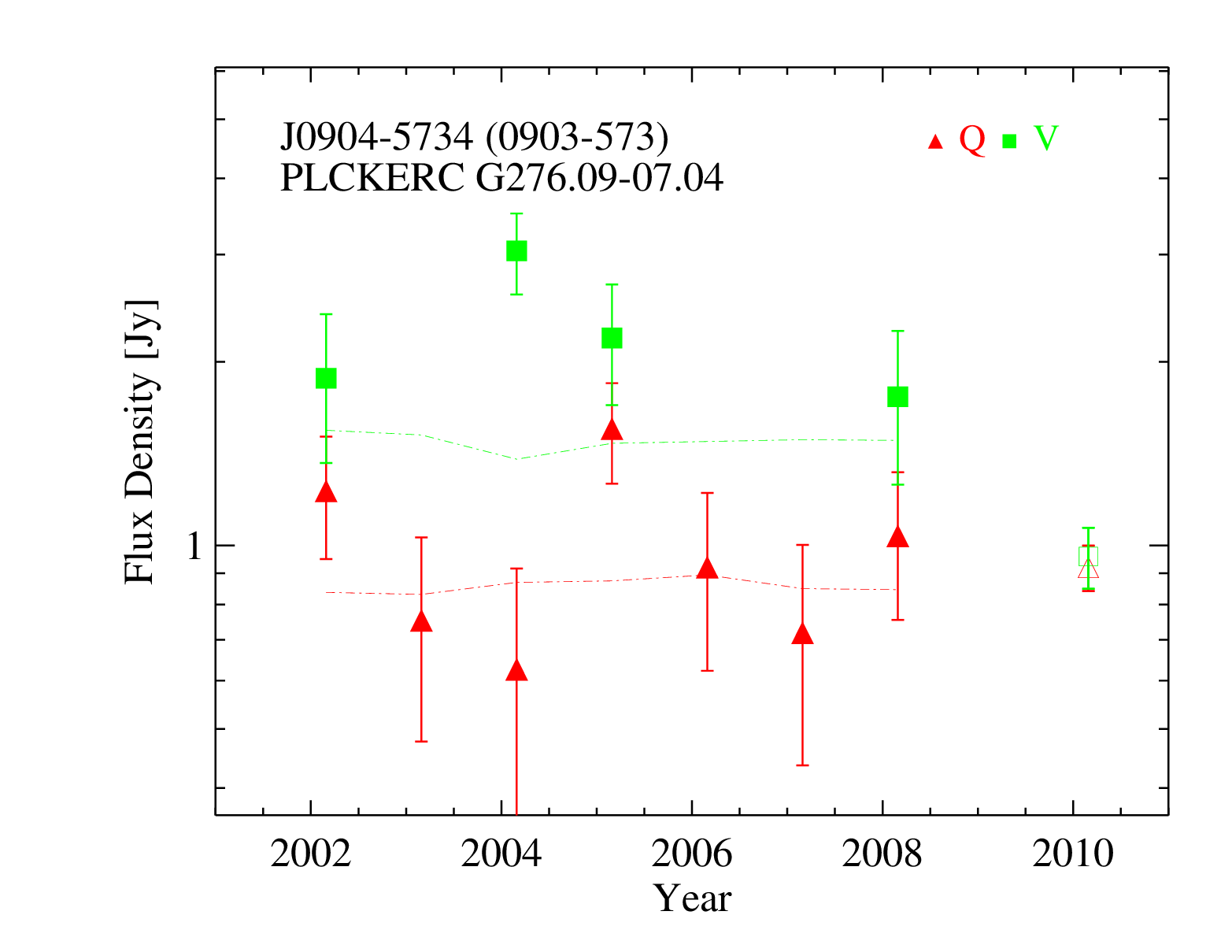}  \\
\includegraphics[width=0.23\textwidth]{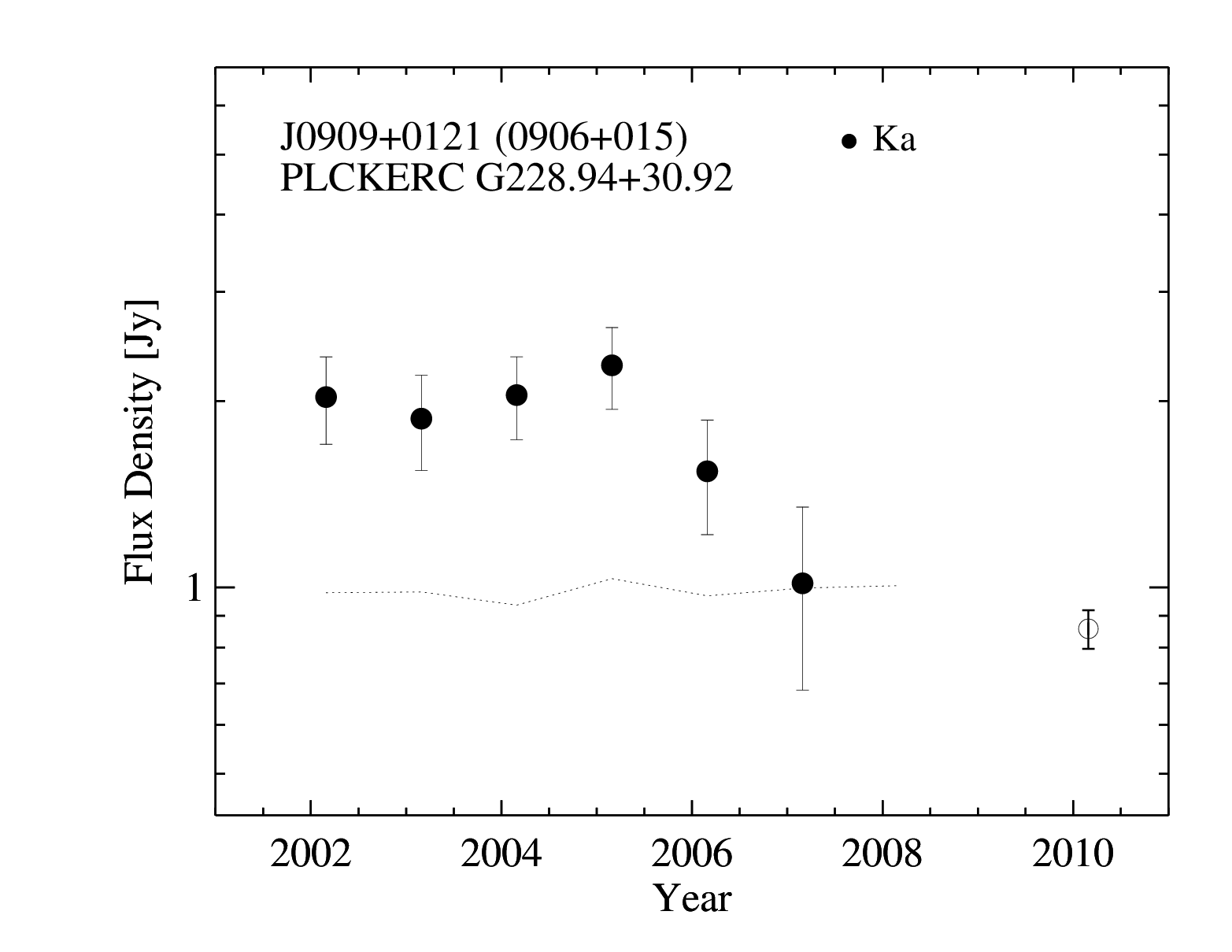} & \includegraphics[width=0.23\textwidth]{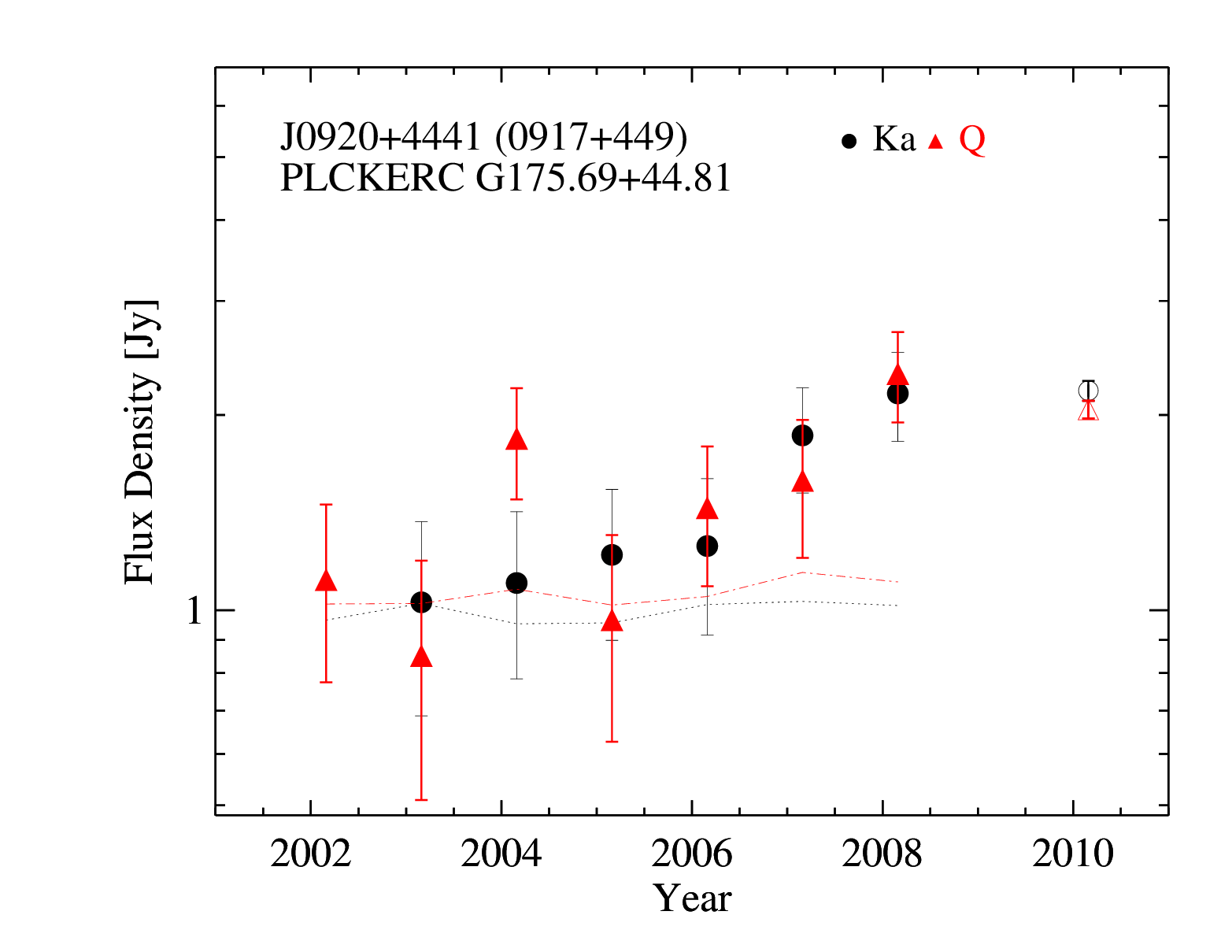}  & \includegraphics[width=0.23\textwidth]{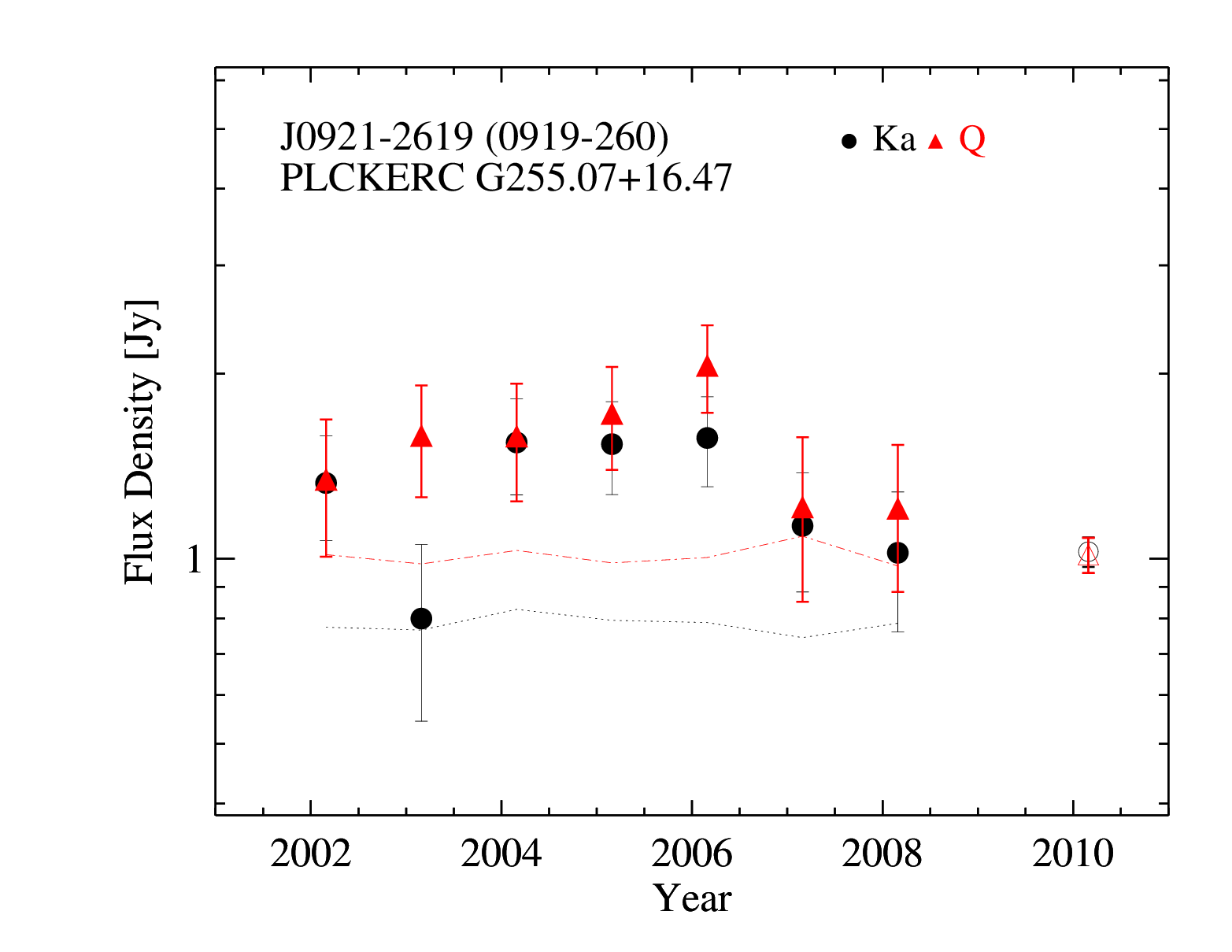} & \includegraphics[width=0.23\textwidth]{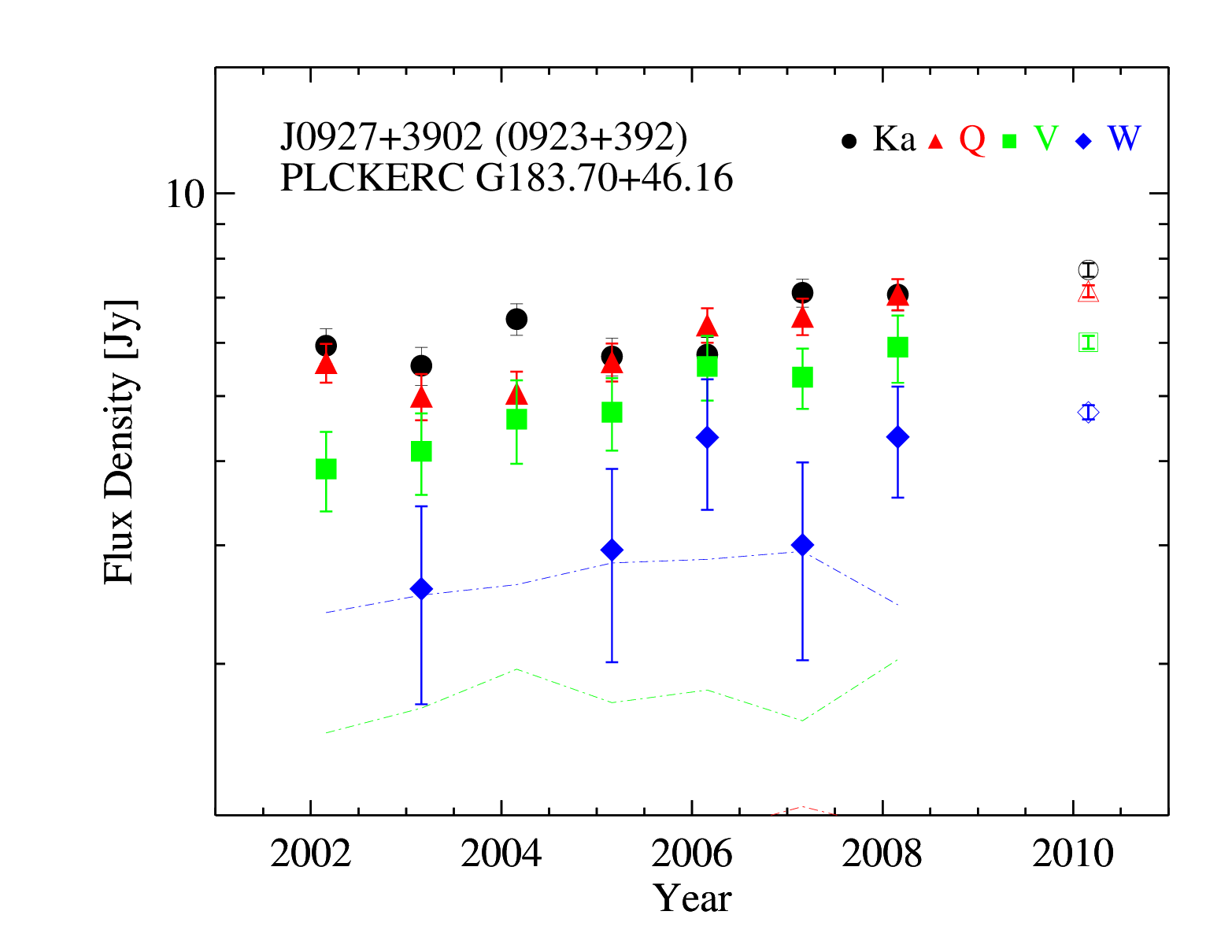}  \\
\includegraphics[width=0.23\textwidth]{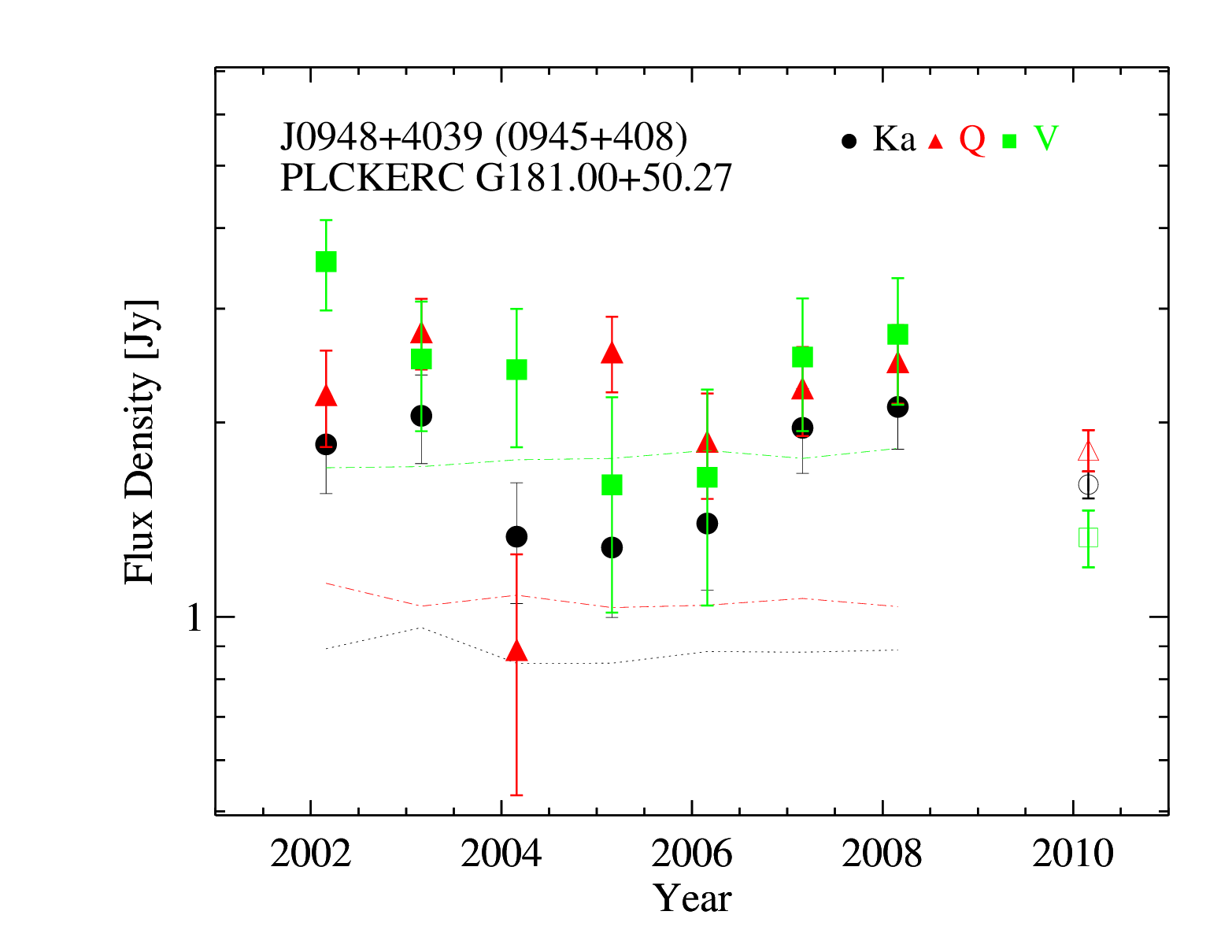} & \includegraphics[width=0.23\textwidth]{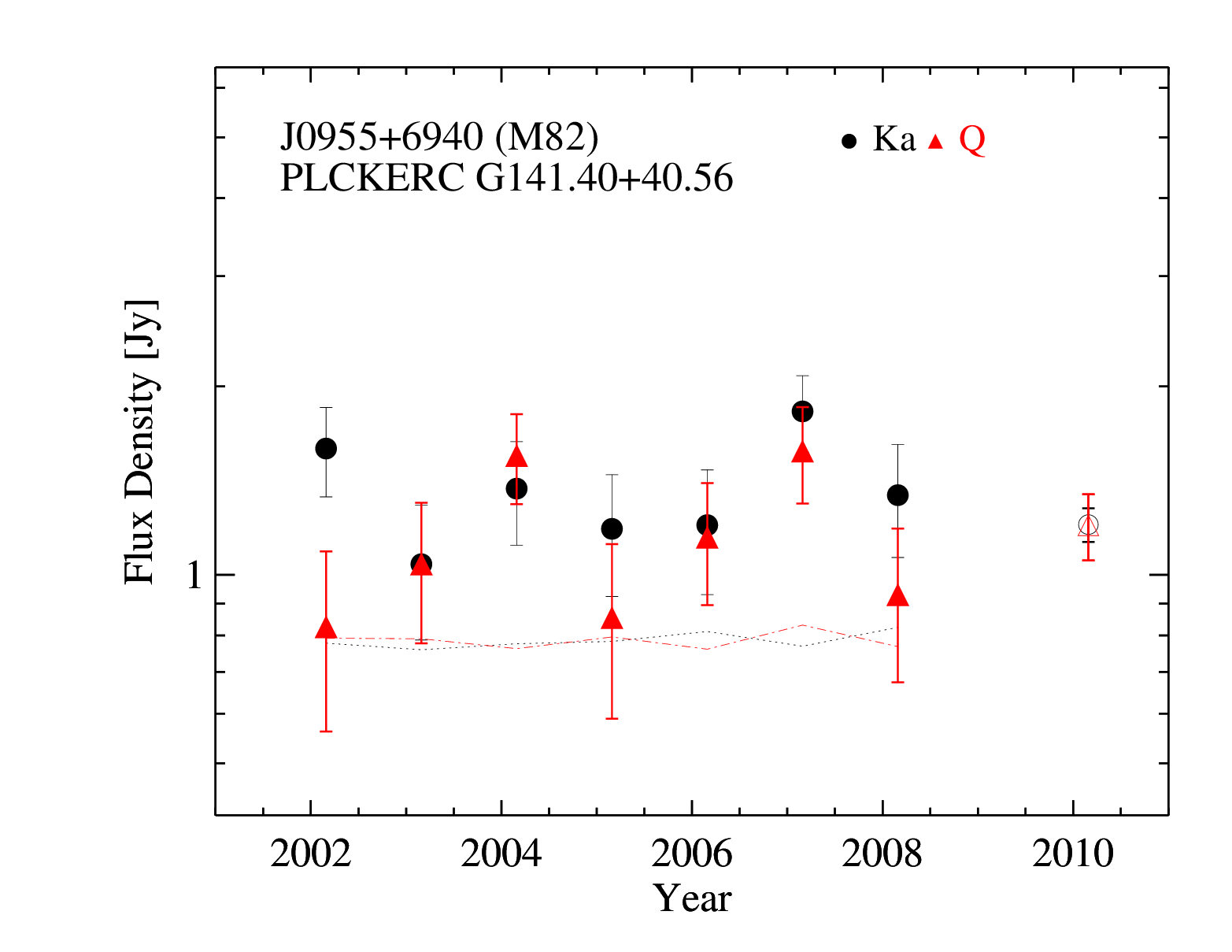}  & \includegraphics[width=0.23\textwidth]{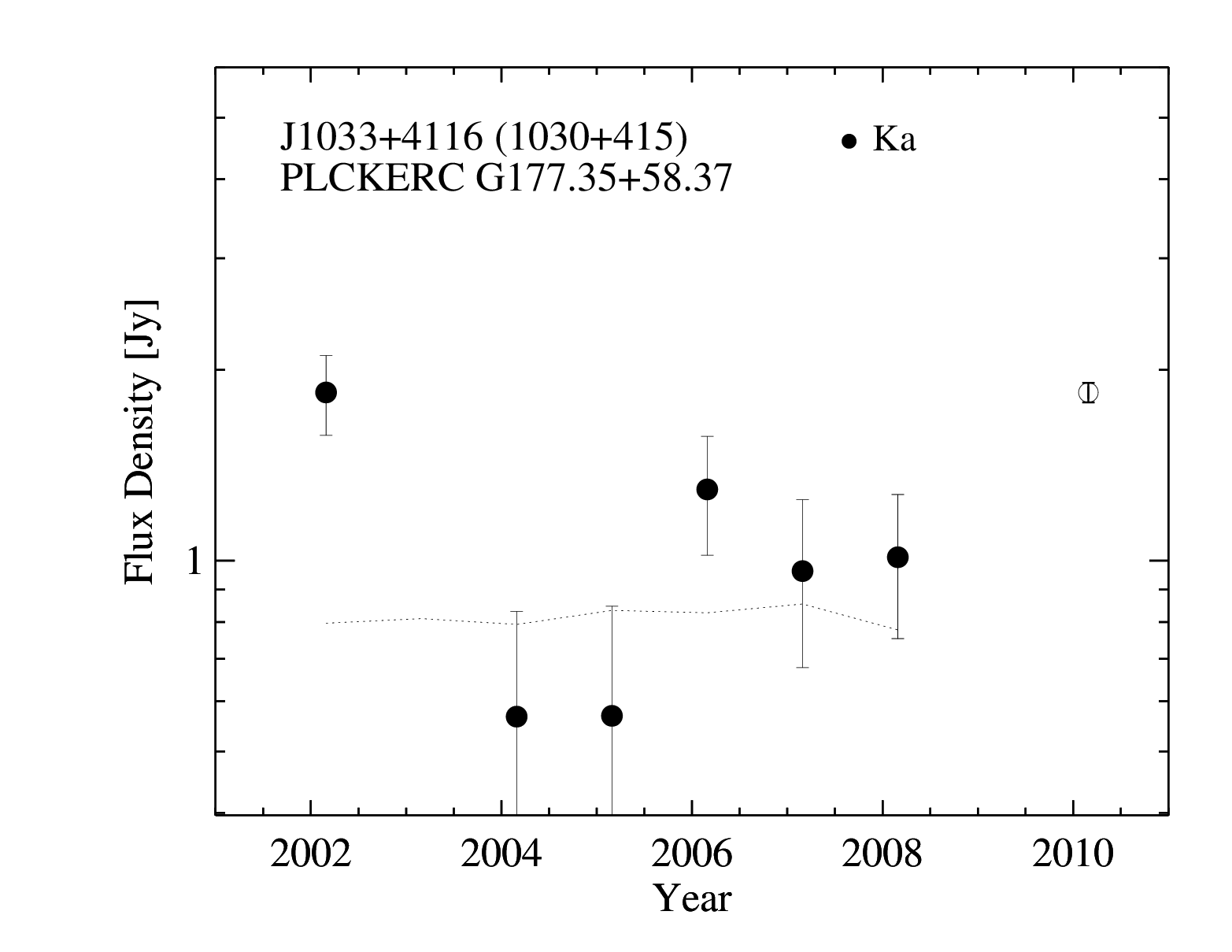} & \includegraphics[width=0.23\textwidth]{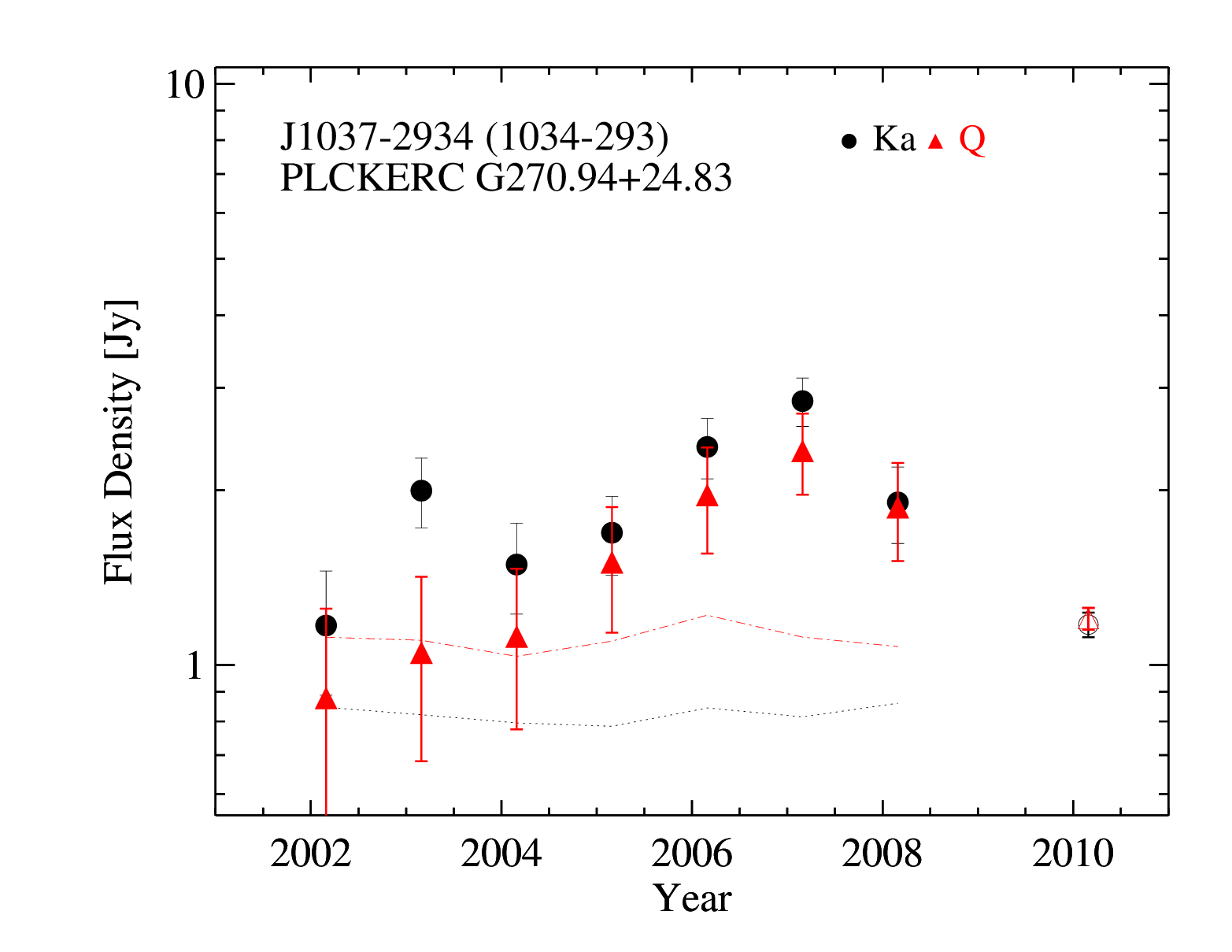}  \\
\includegraphics[width=0.23\textwidth]{figures/lc/J1058+0134.eps} & \includegraphics[width=0.23\textwidth]{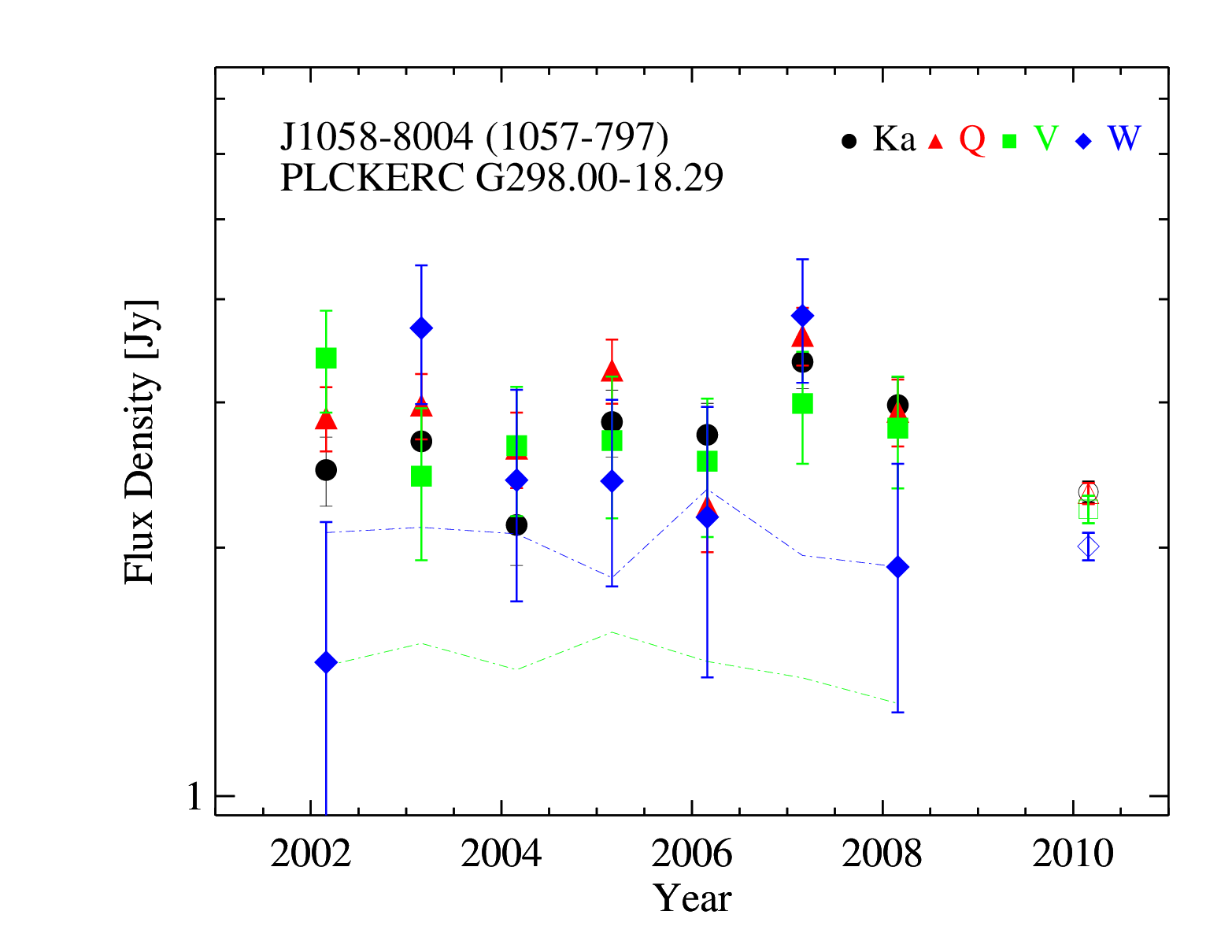}  & \includegraphics[width=0.23\textwidth]{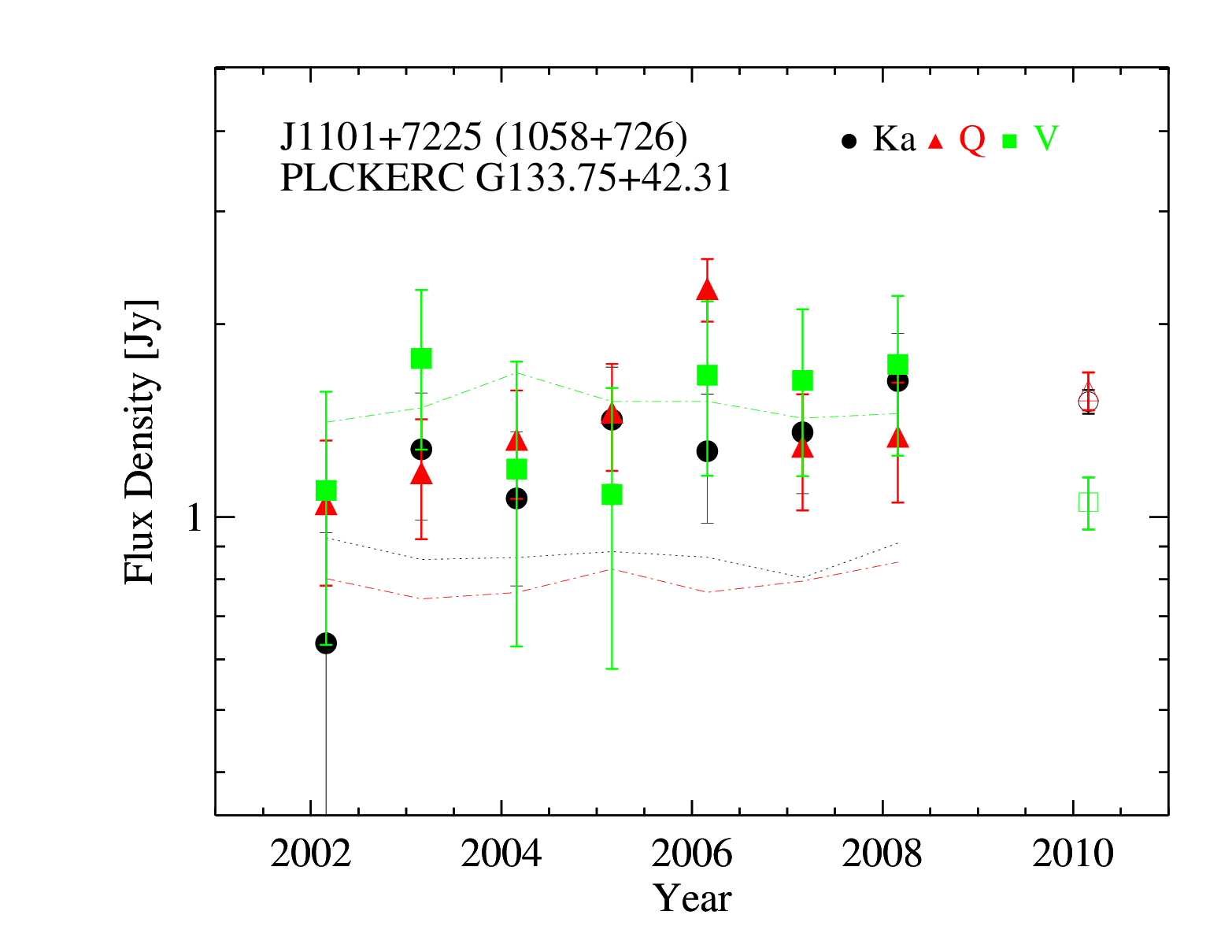} & \includegraphics[width=0.23\textwidth]{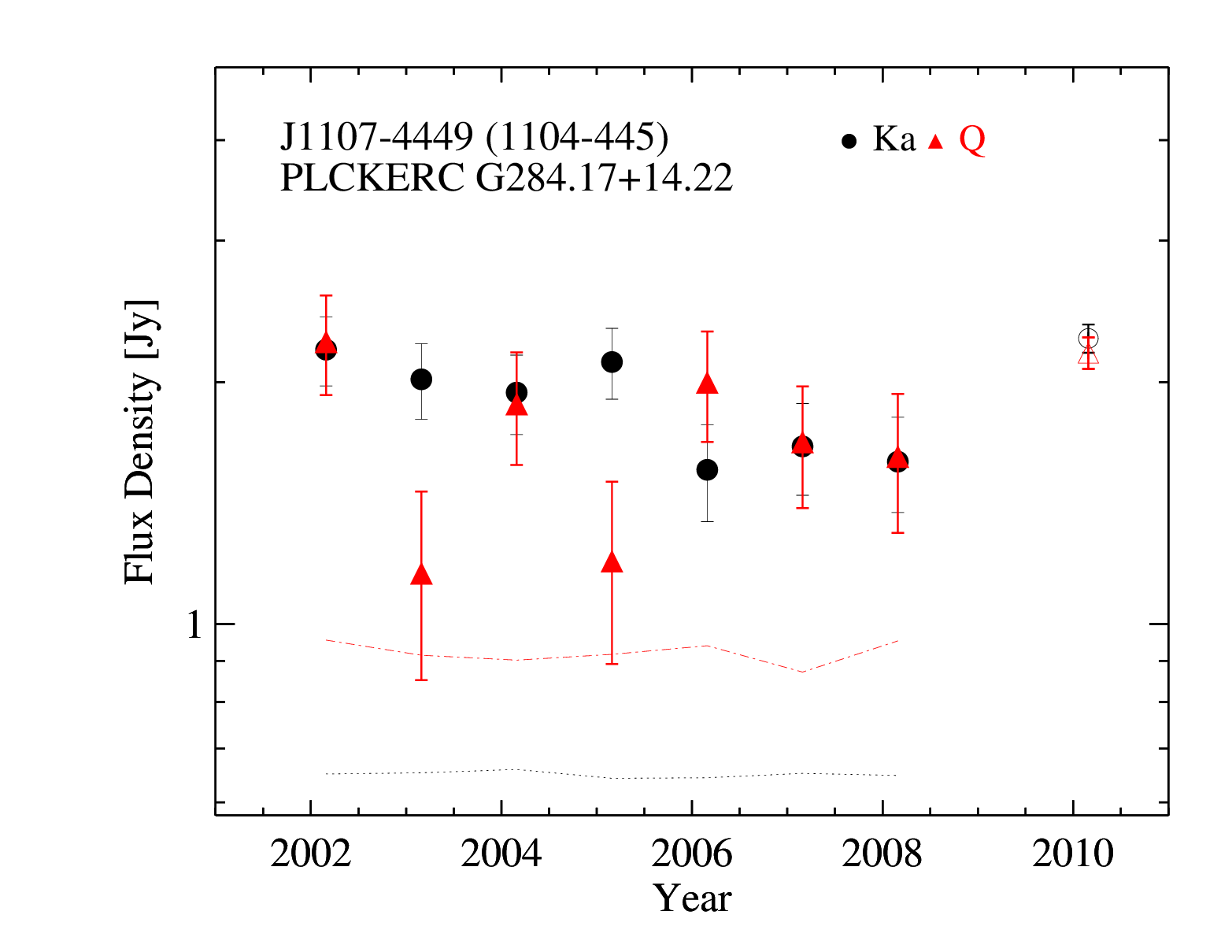}  \\
\includegraphics[width=0.23\textwidth]{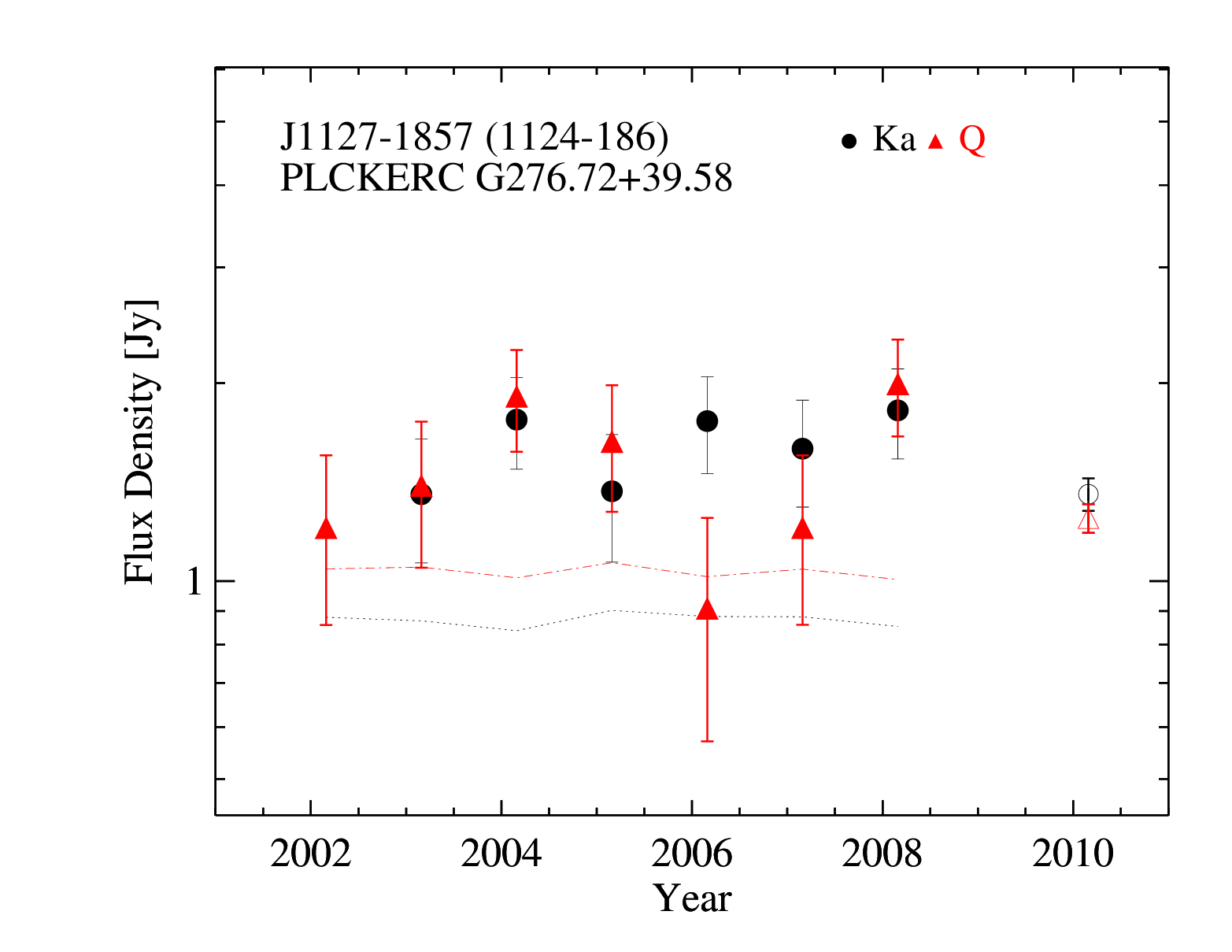} & \includegraphics[width=0.23\textwidth]{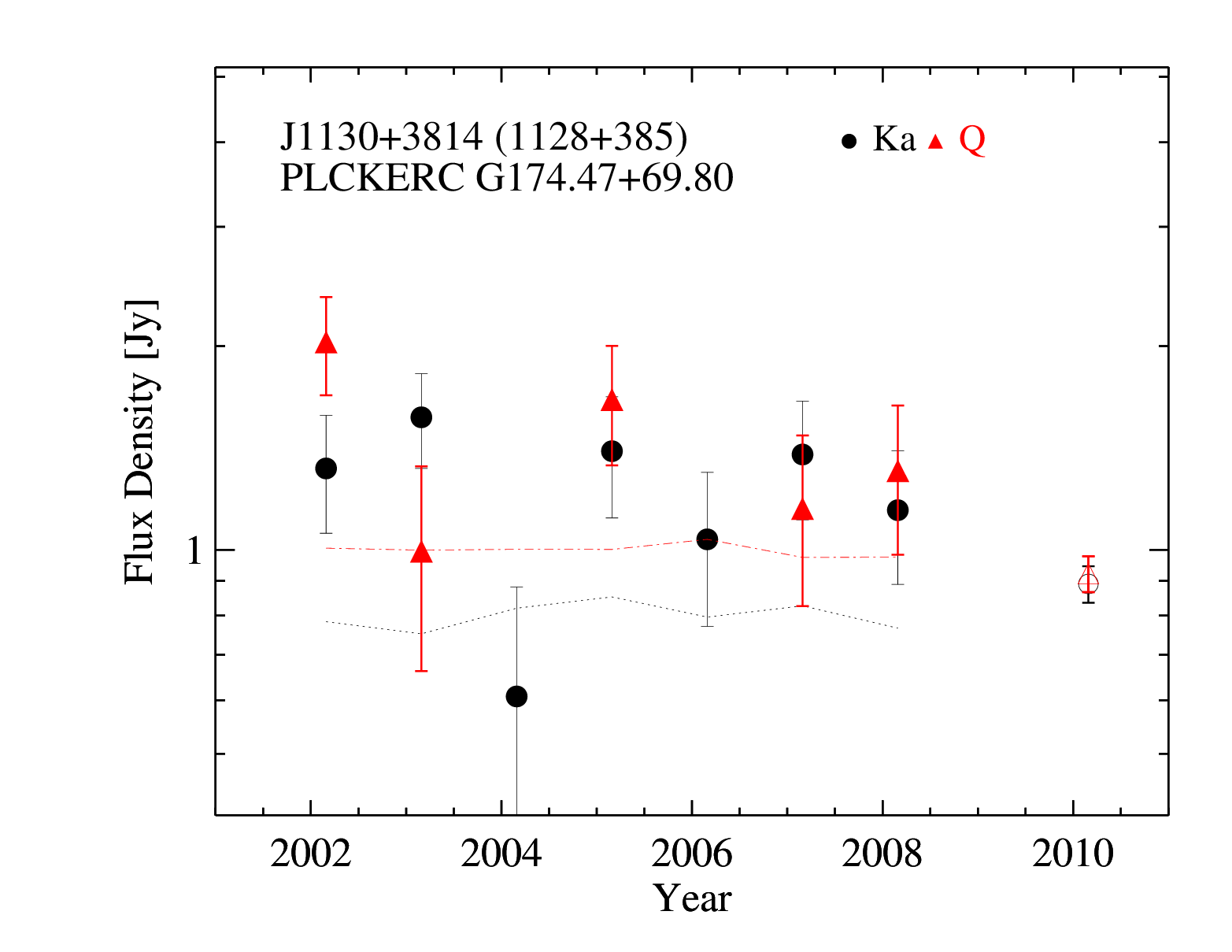}  & \includegraphics[width=0.23\textwidth]{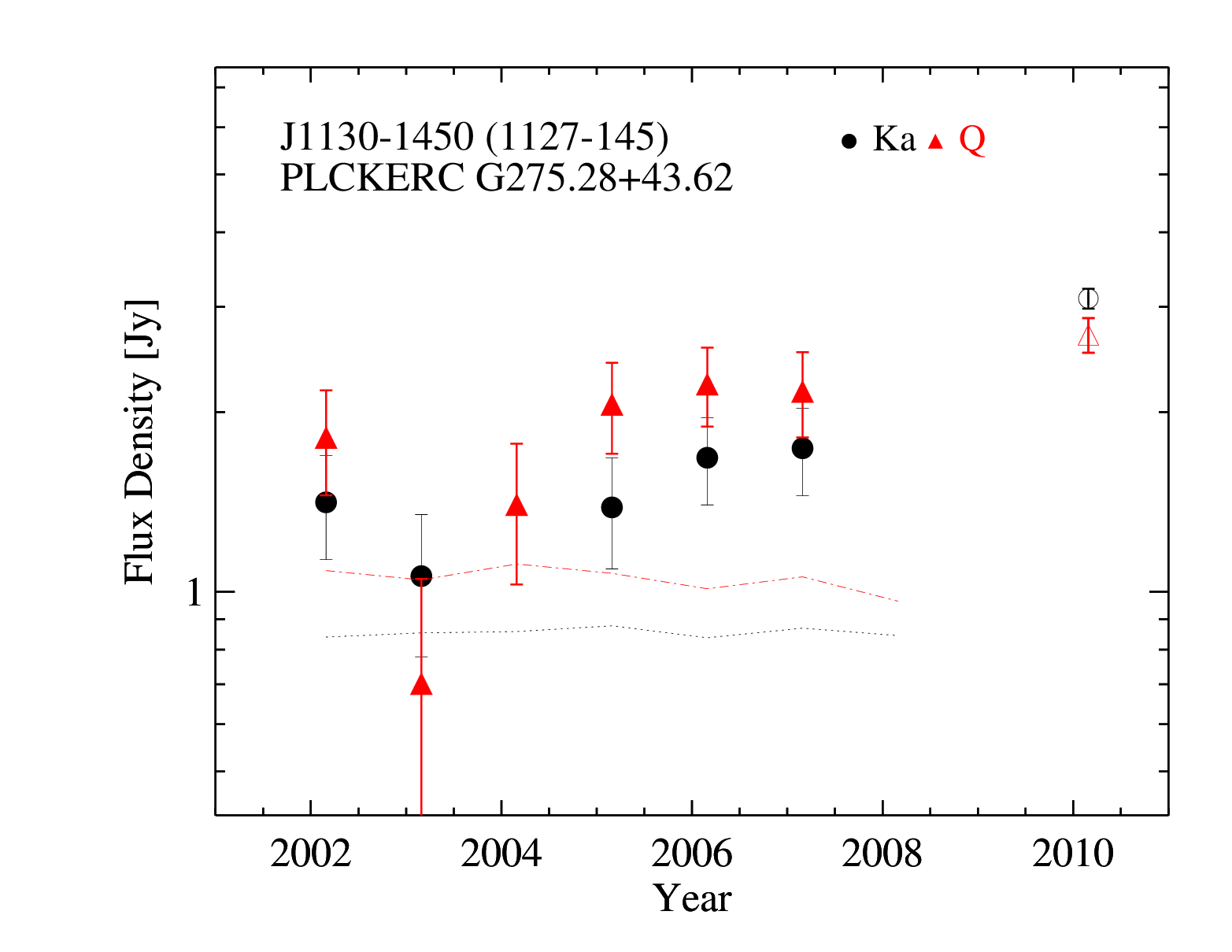} & \includegraphics[width=0.23\textwidth]{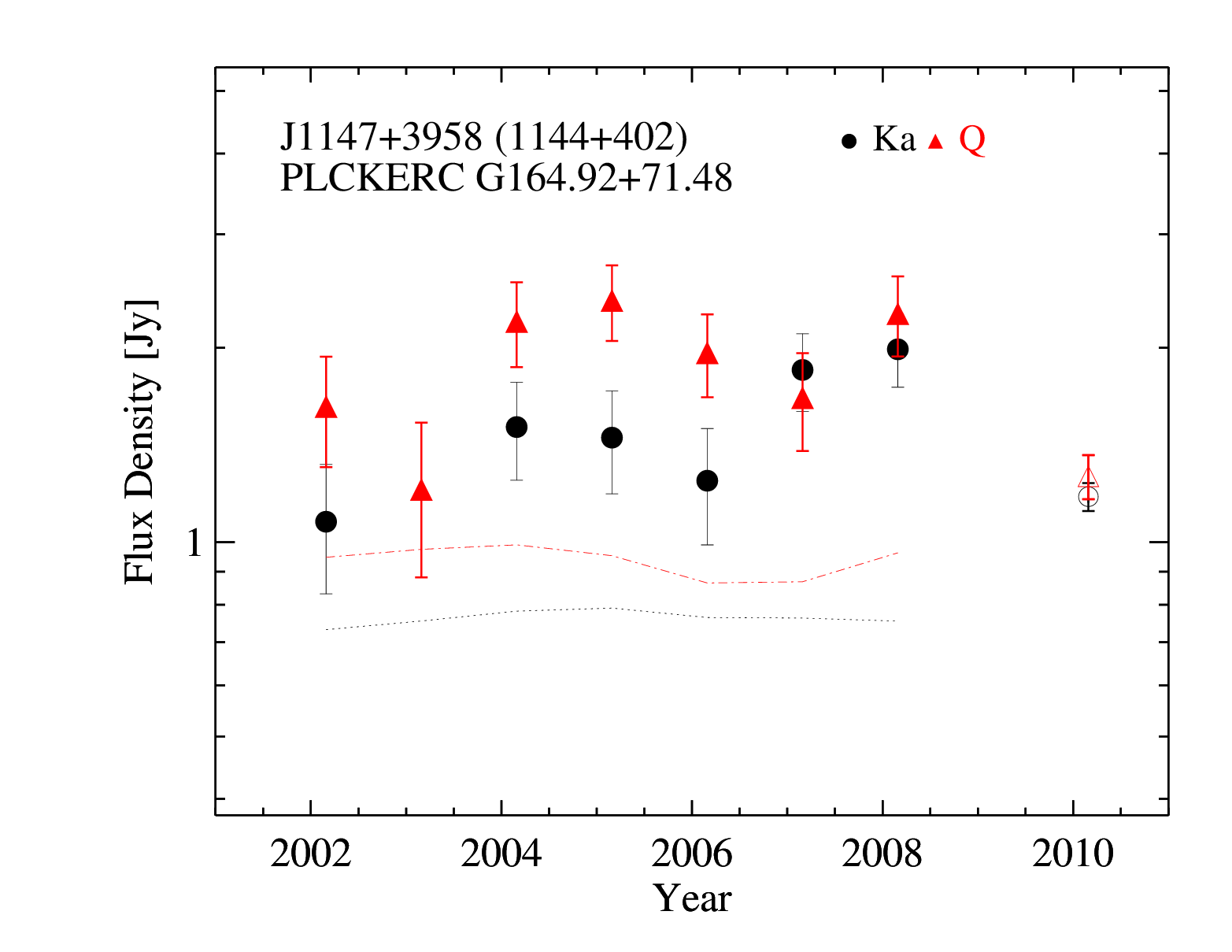}  \\
\includegraphics[width=0.23\textwidth]{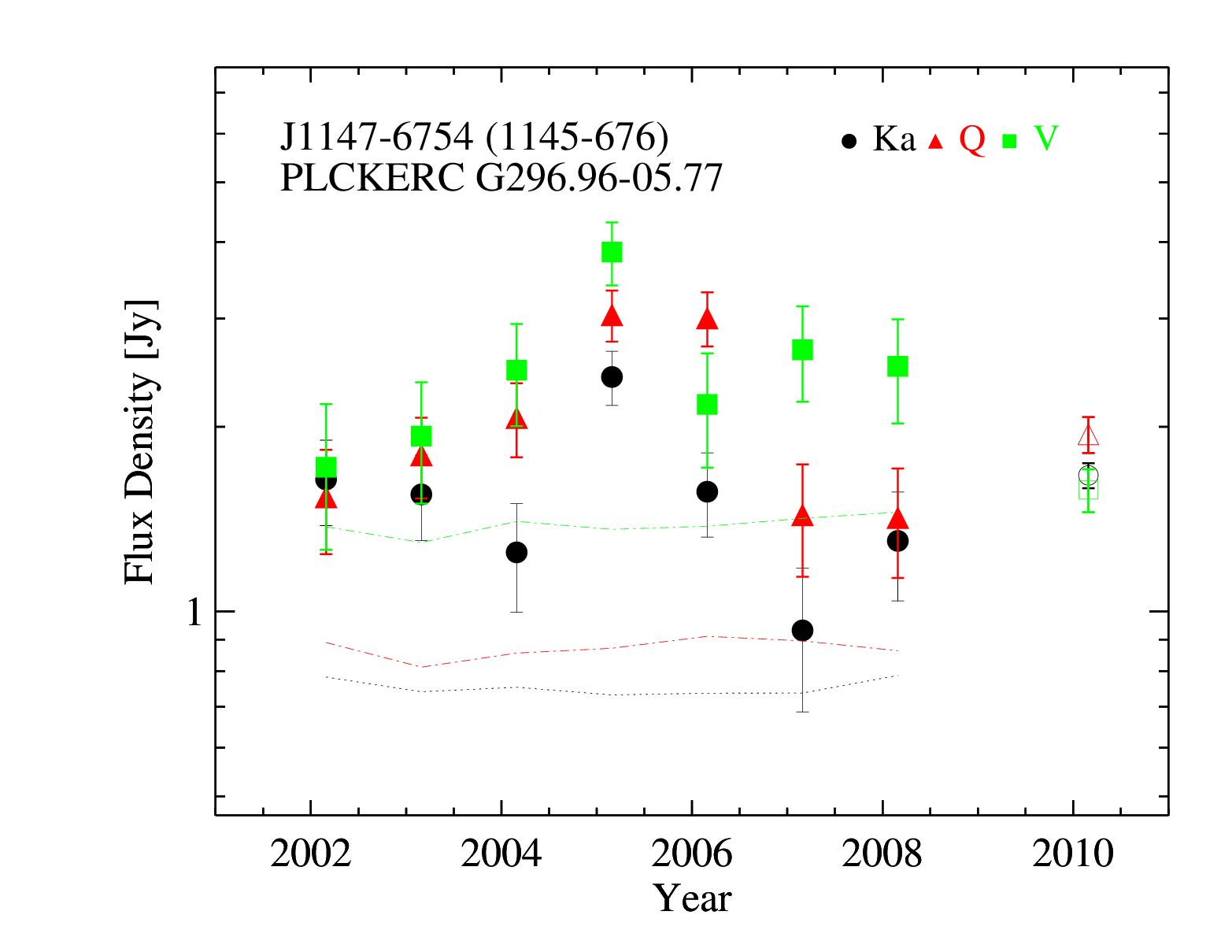} & \includegraphics[width=0.23\textwidth]{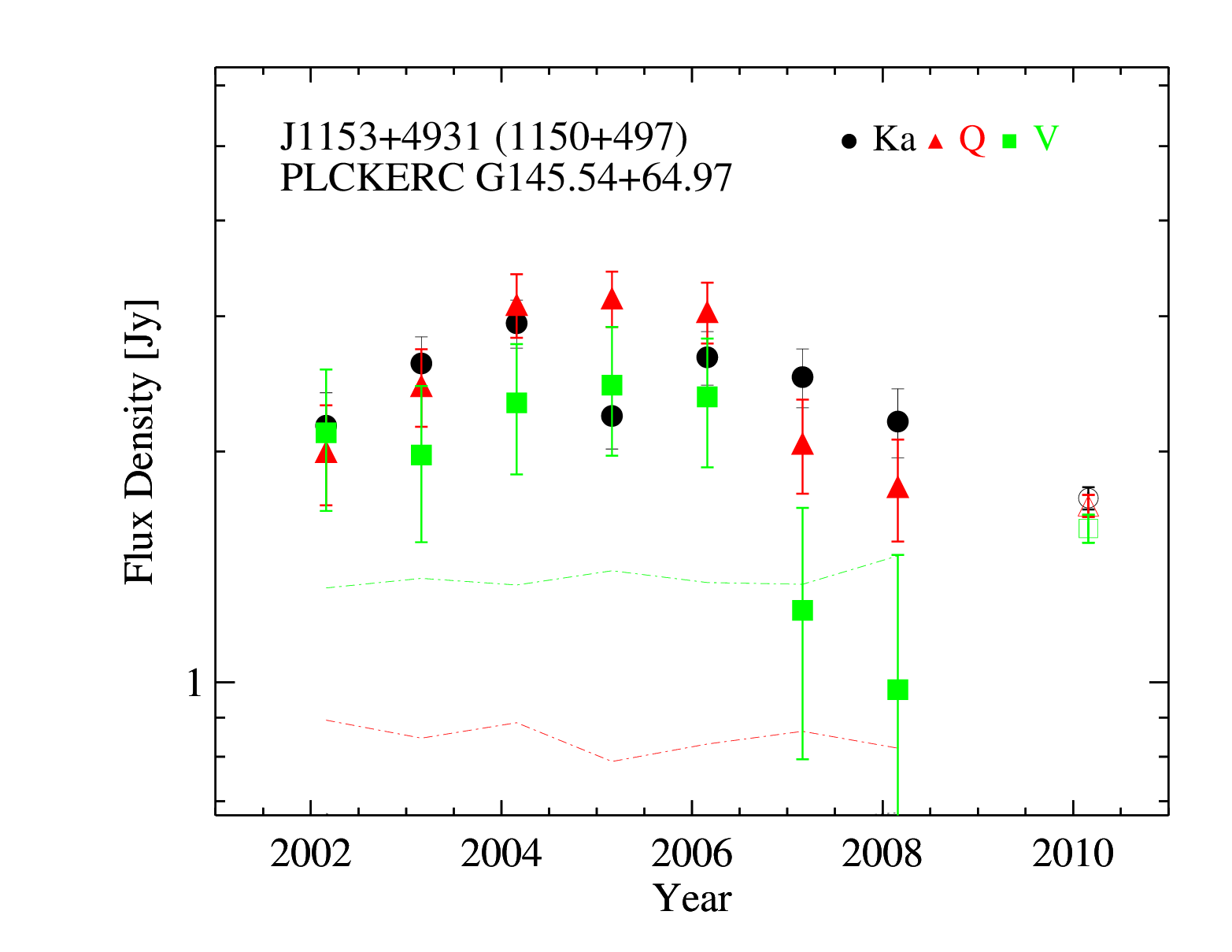}  & \includegraphics[width=0.23\textwidth]{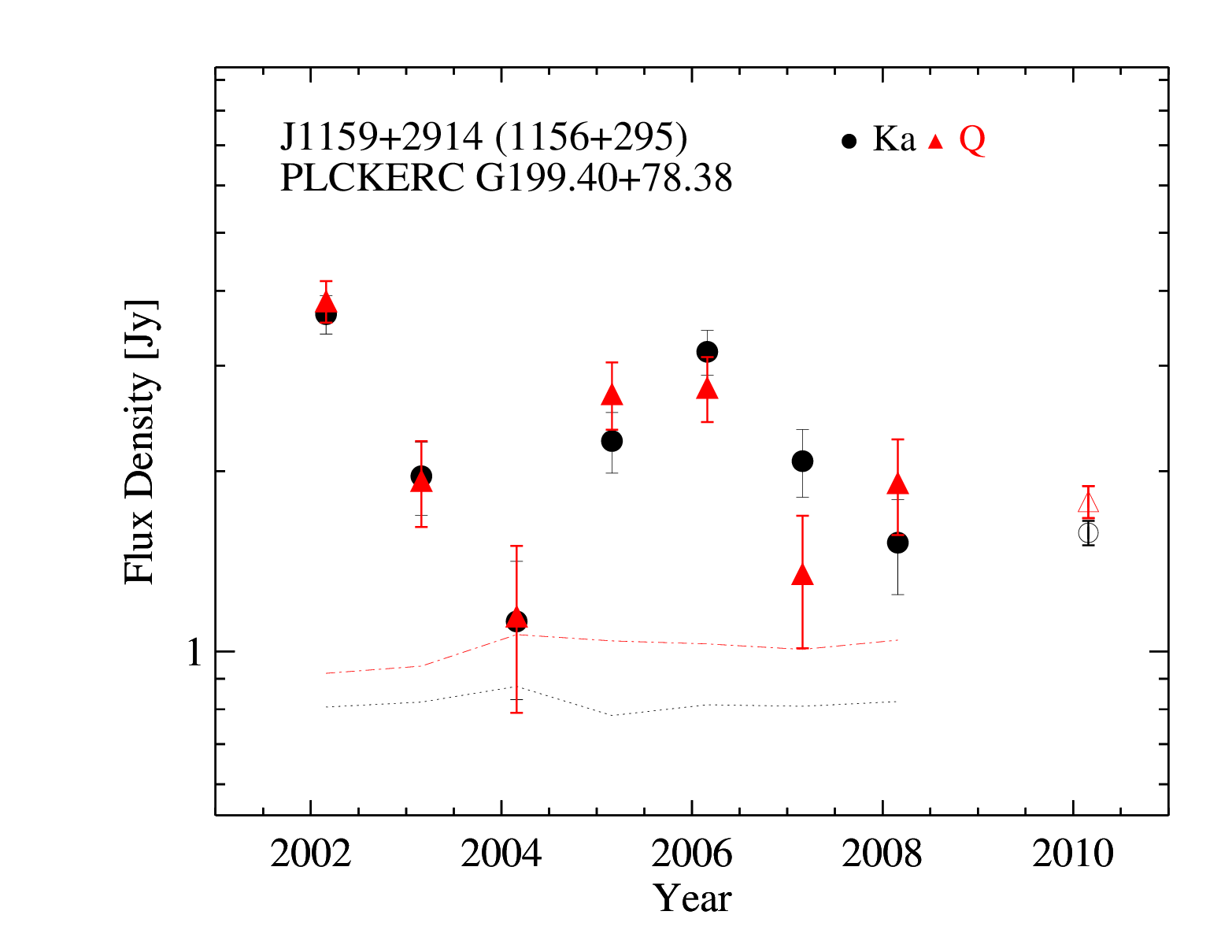} & \includegraphics[width=0.23\textwidth]{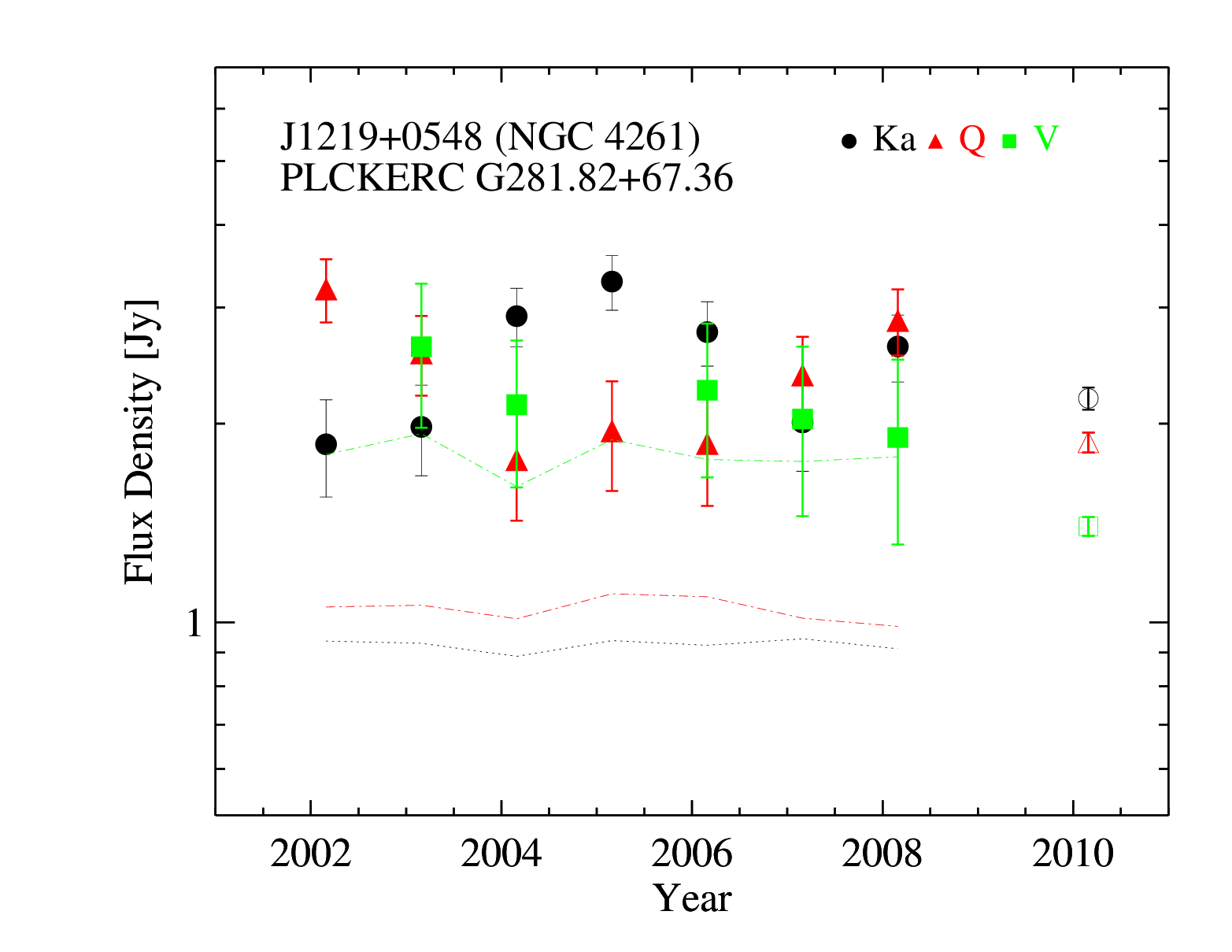}  \\
\end{tabular}
\end{figure*}

\begin{figure*}
\centering
\begin{tabular}{cccc}
\includegraphics[width=0.23\textwidth]{figures/lc/J1229+0202.eps} & \includegraphics[width=0.23\textwidth]{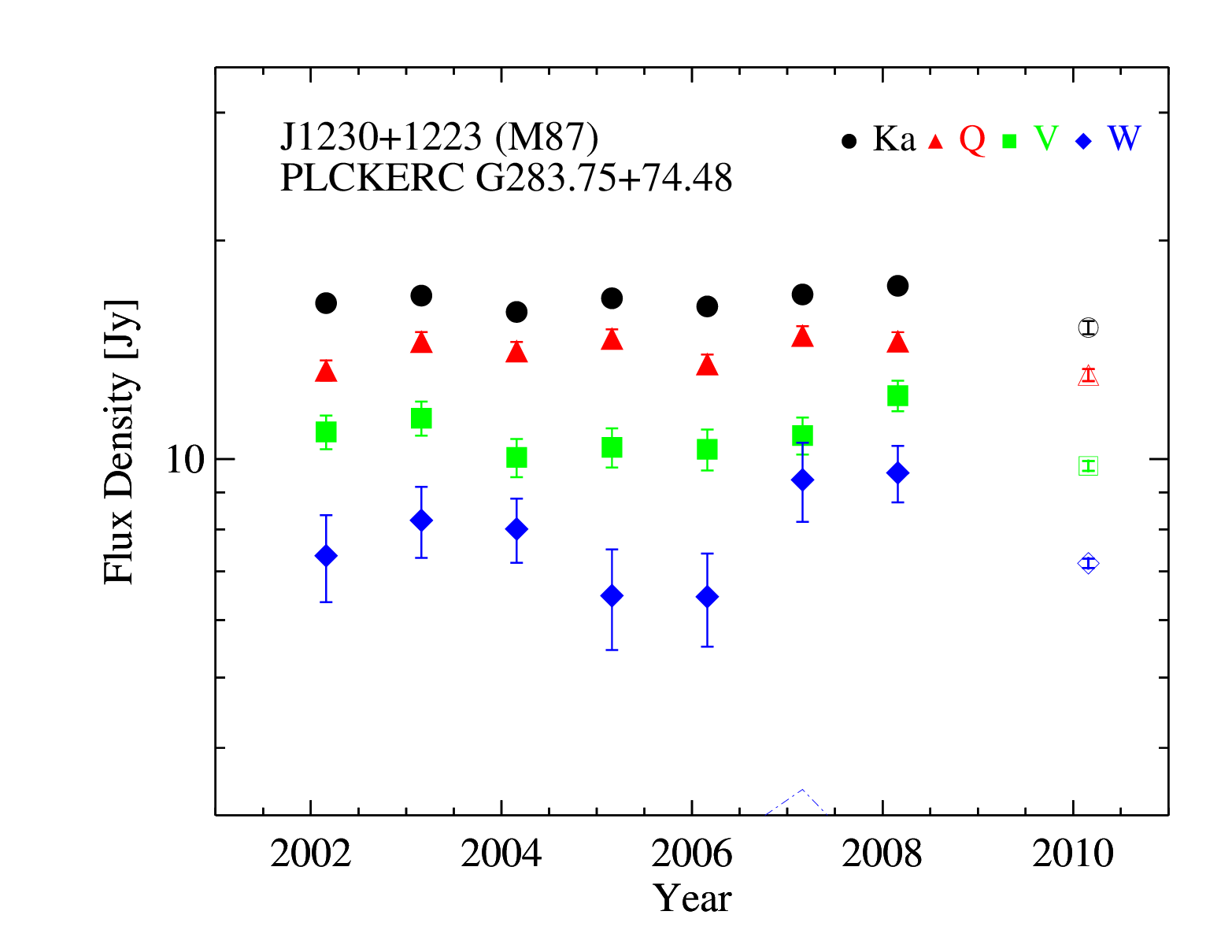}  & \includegraphics[width=0.23\textwidth]{figures/lc/J1256-0547.eps} & \includegraphics[width=0.23\textwidth]{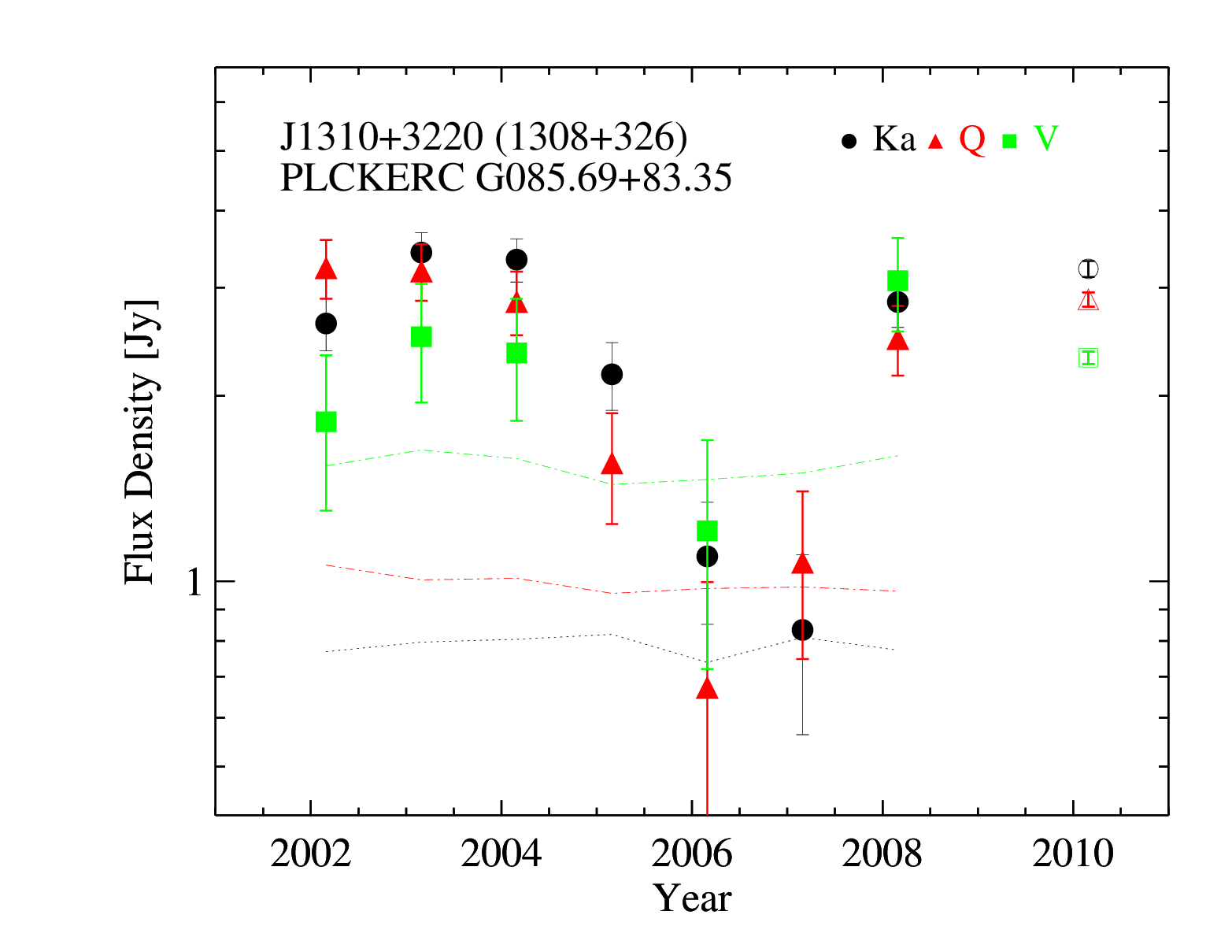}  \\
\includegraphics[width=0.23\textwidth]{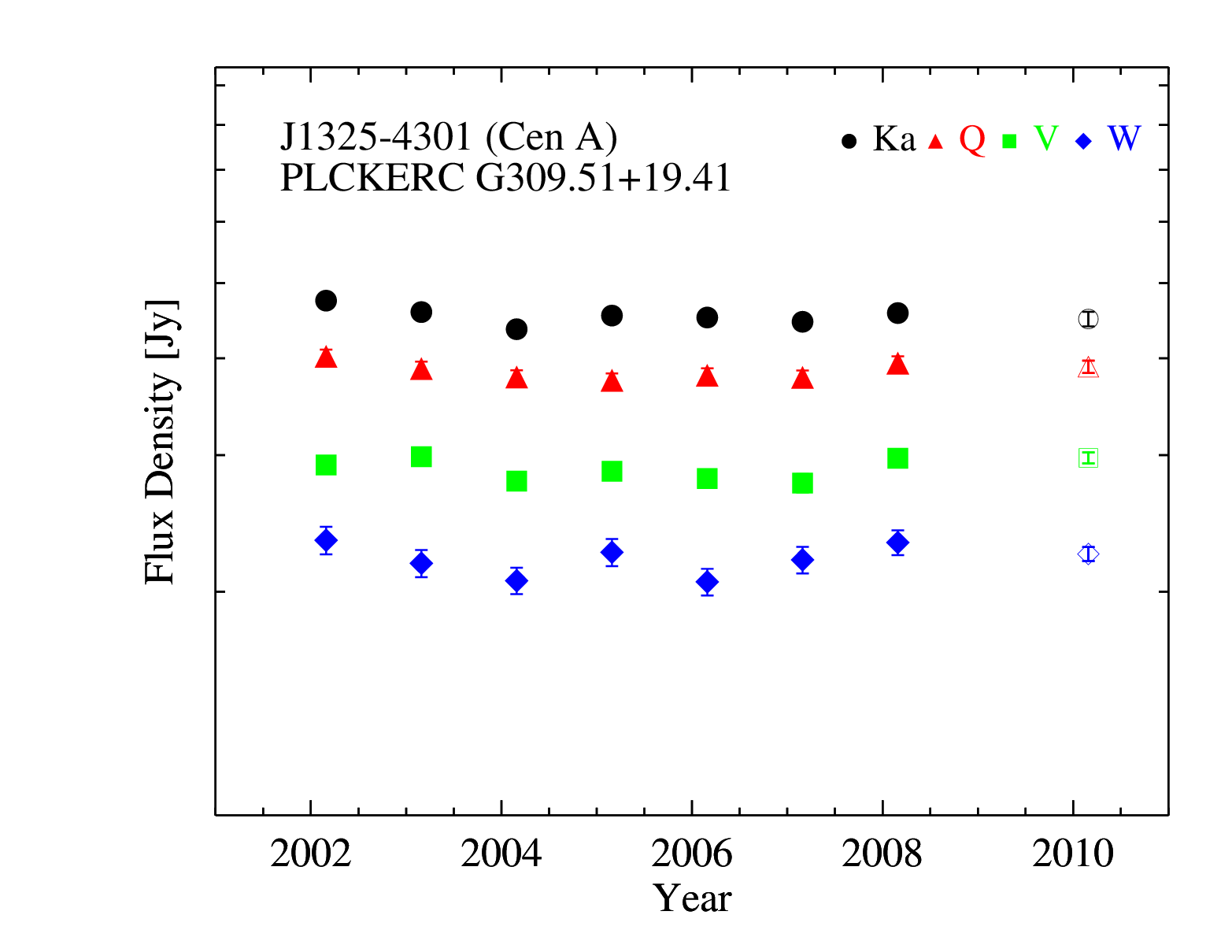} & \includegraphics[width=0.23\textwidth]{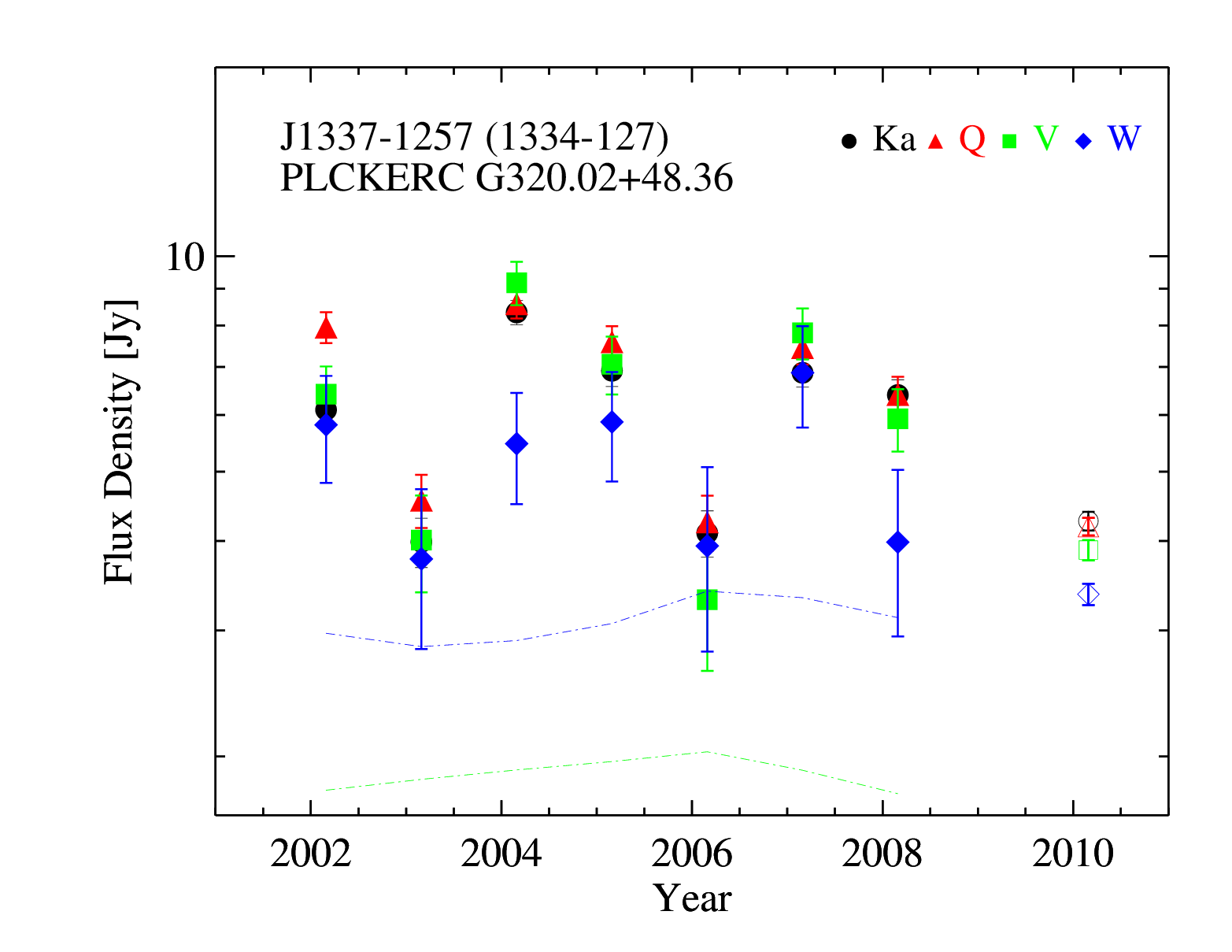}  & \includegraphics[width=0.23\textwidth]{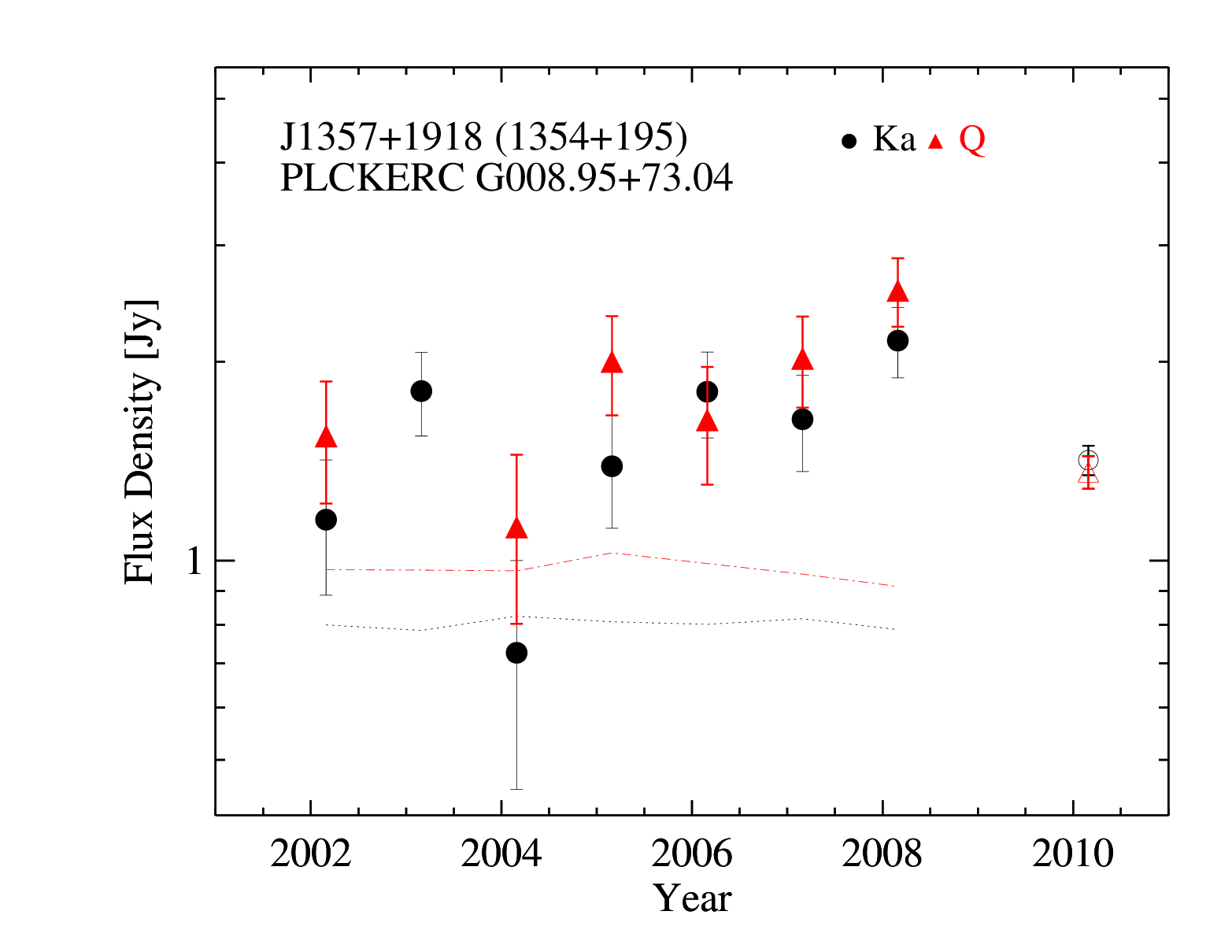} & \includegraphics[width=0.23\textwidth]{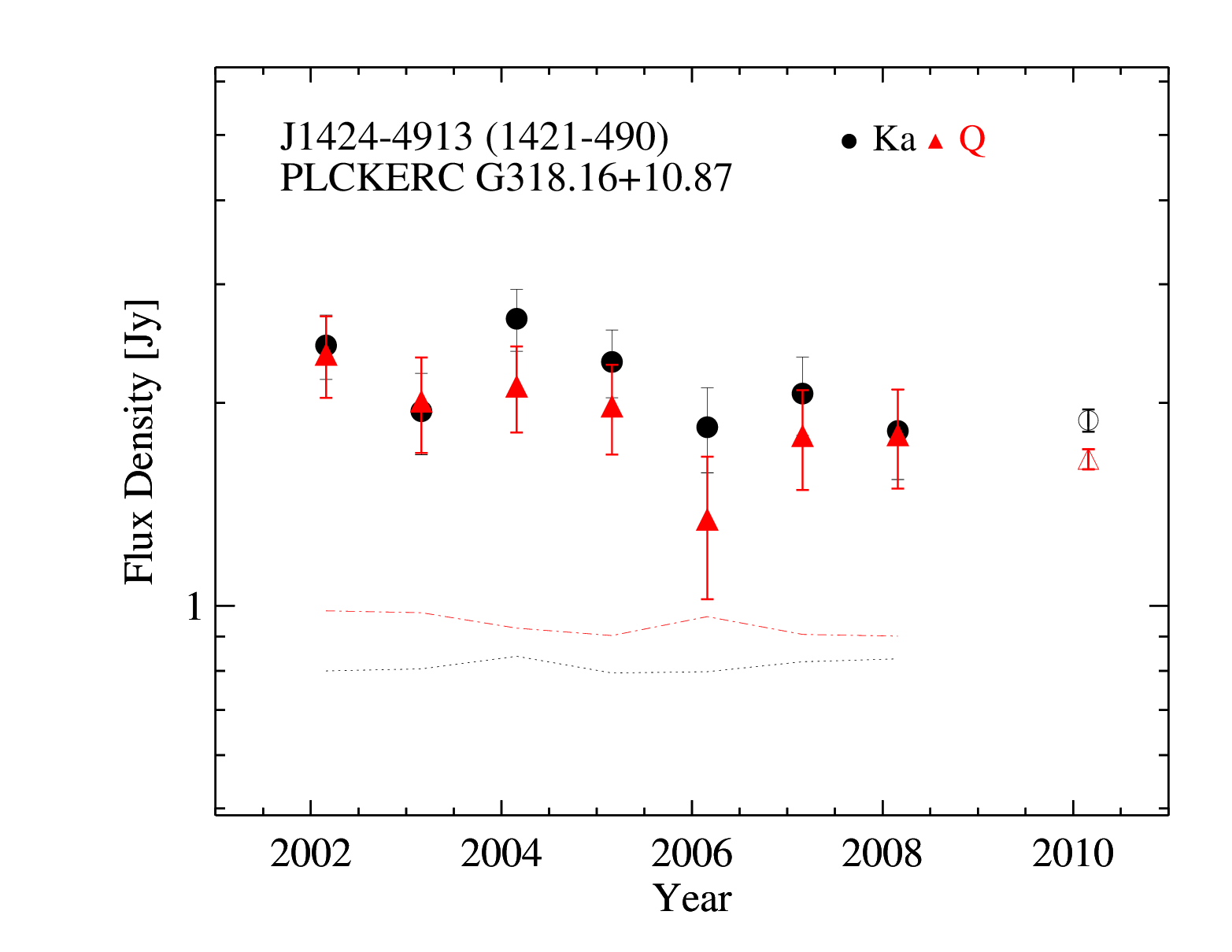}  \\
\includegraphics[width=0.23\textwidth]{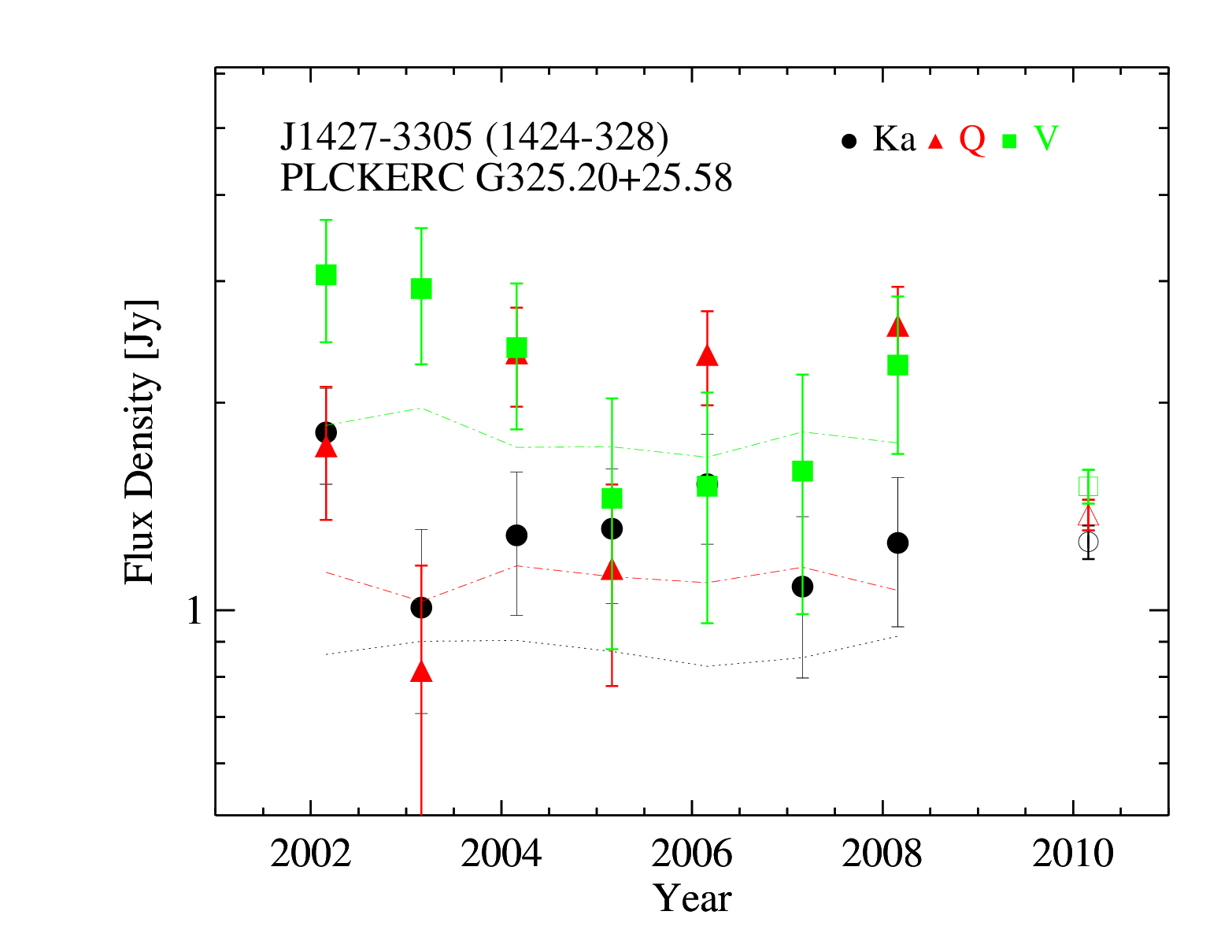} & \includegraphics[width=0.23\textwidth]{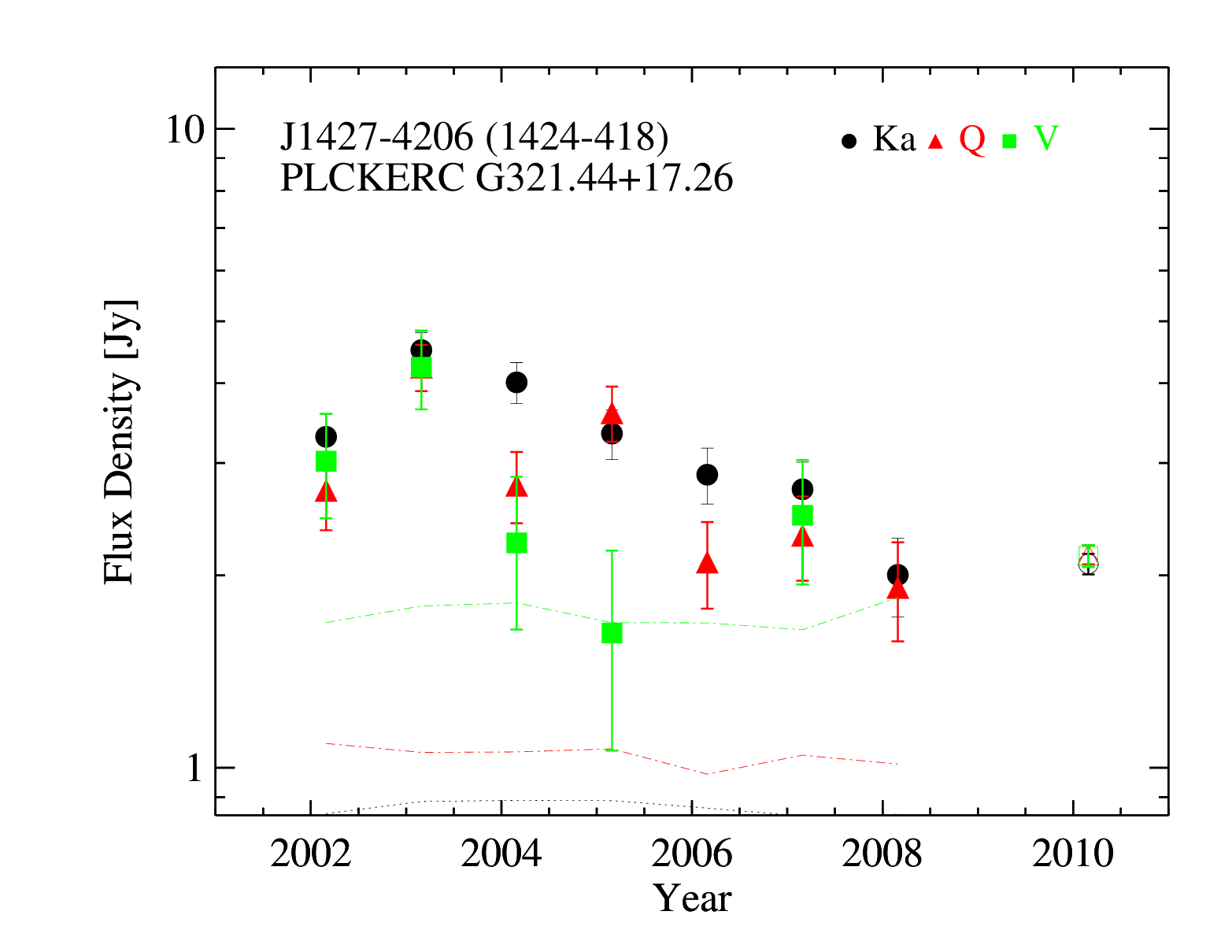}  & \includegraphics[width=0.23\textwidth]{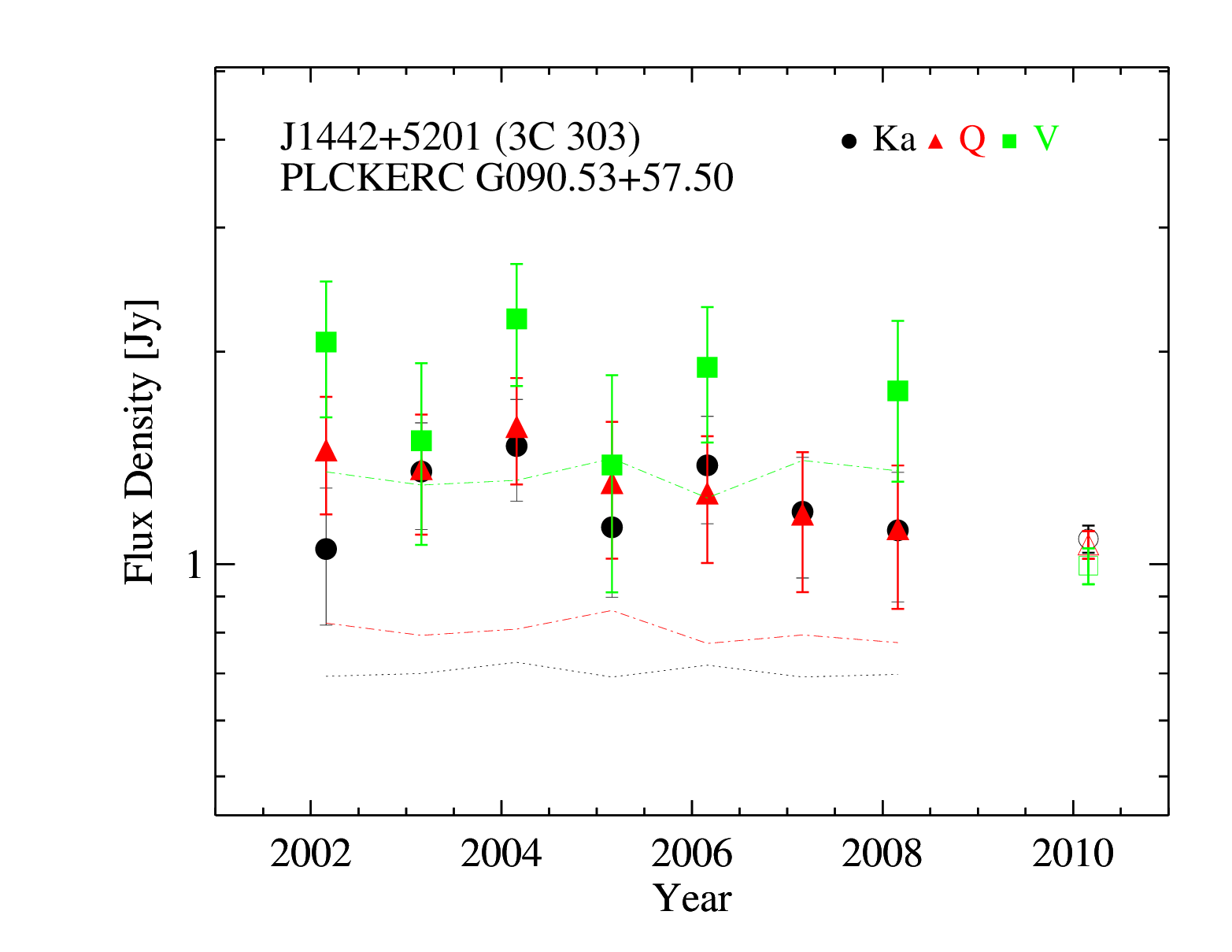} & \includegraphics[width=0.23\textwidth]{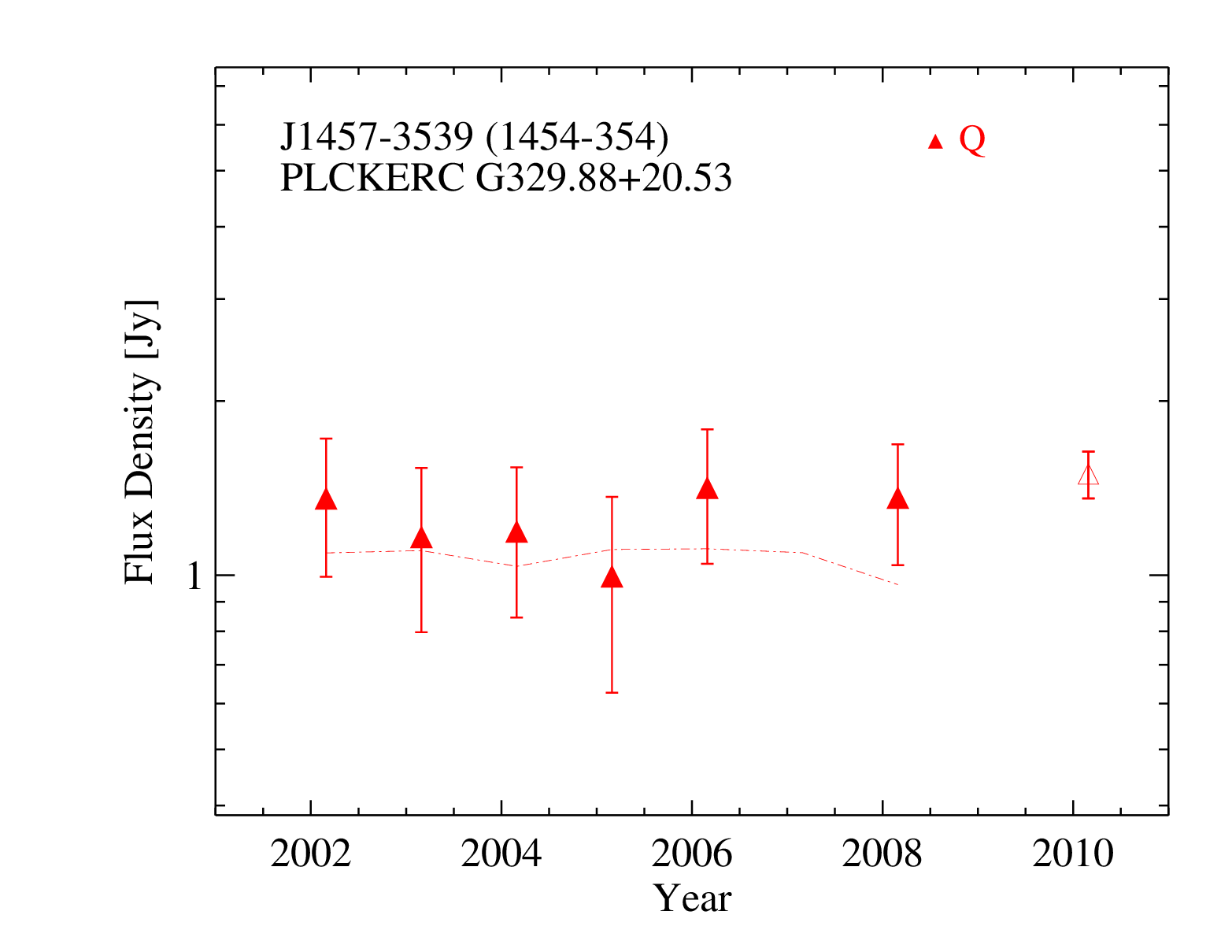}  \\
\includegraphics[width=0.23\textwidth]{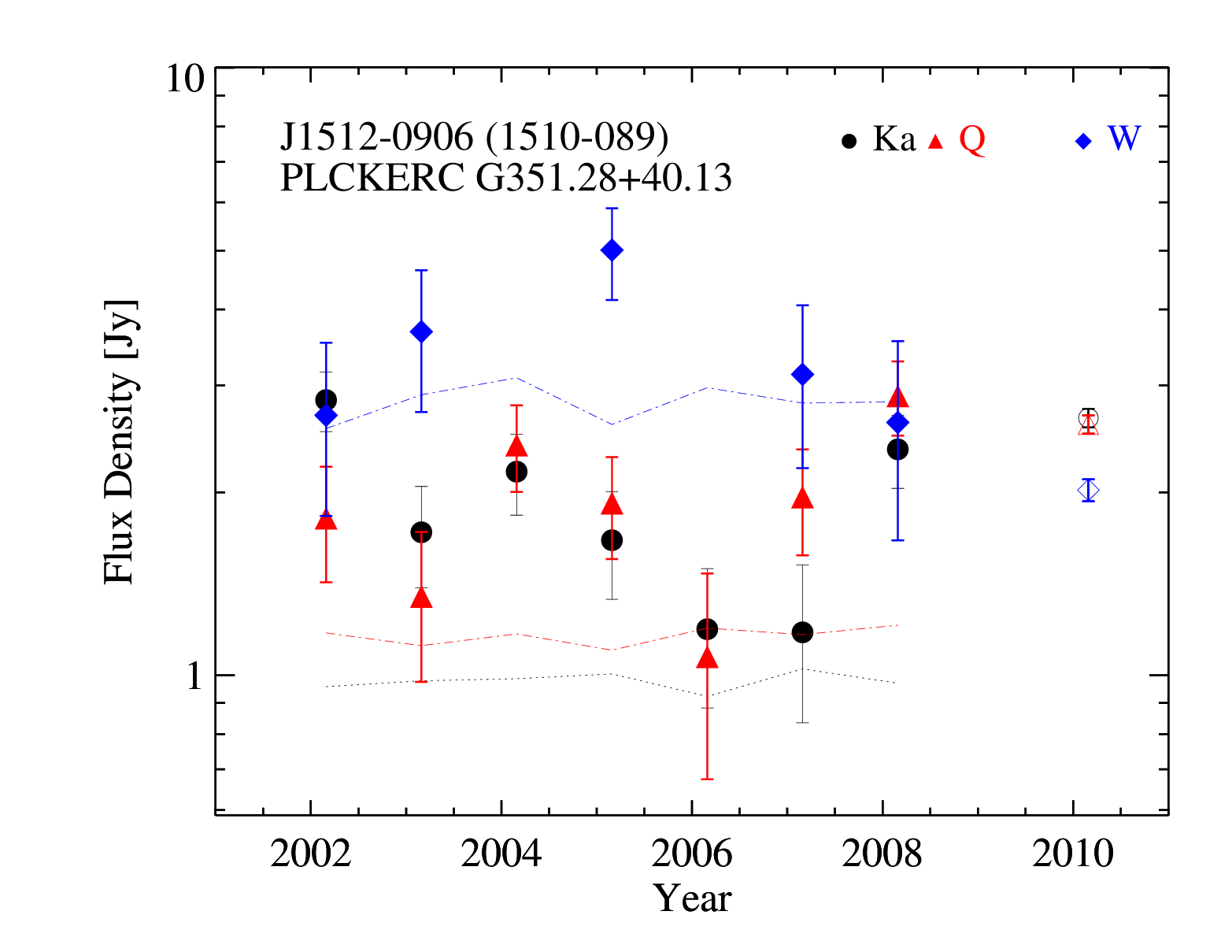} & \includegraphics[width=0.23\textwidth]{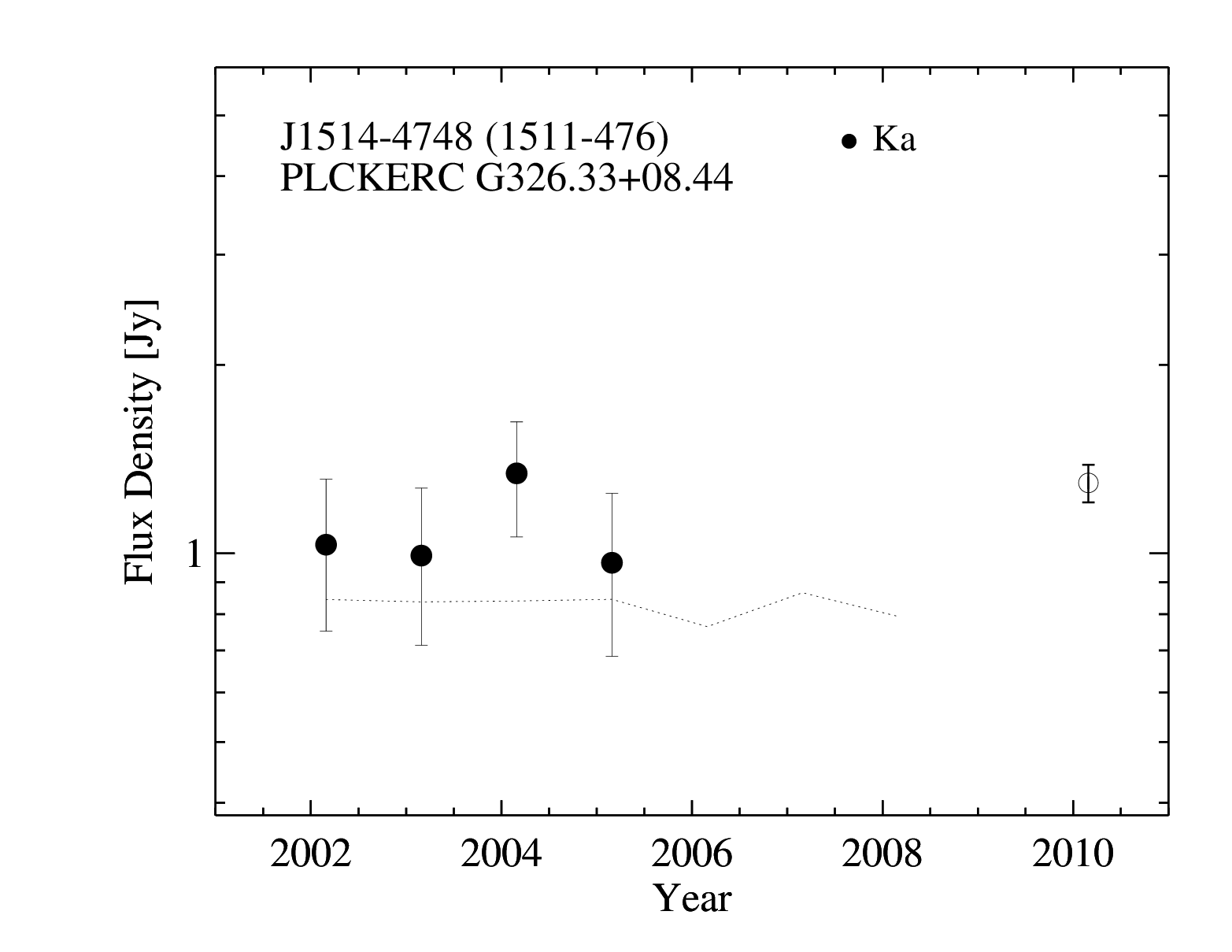}  & \includegraphics[width=0.23\textwidth]{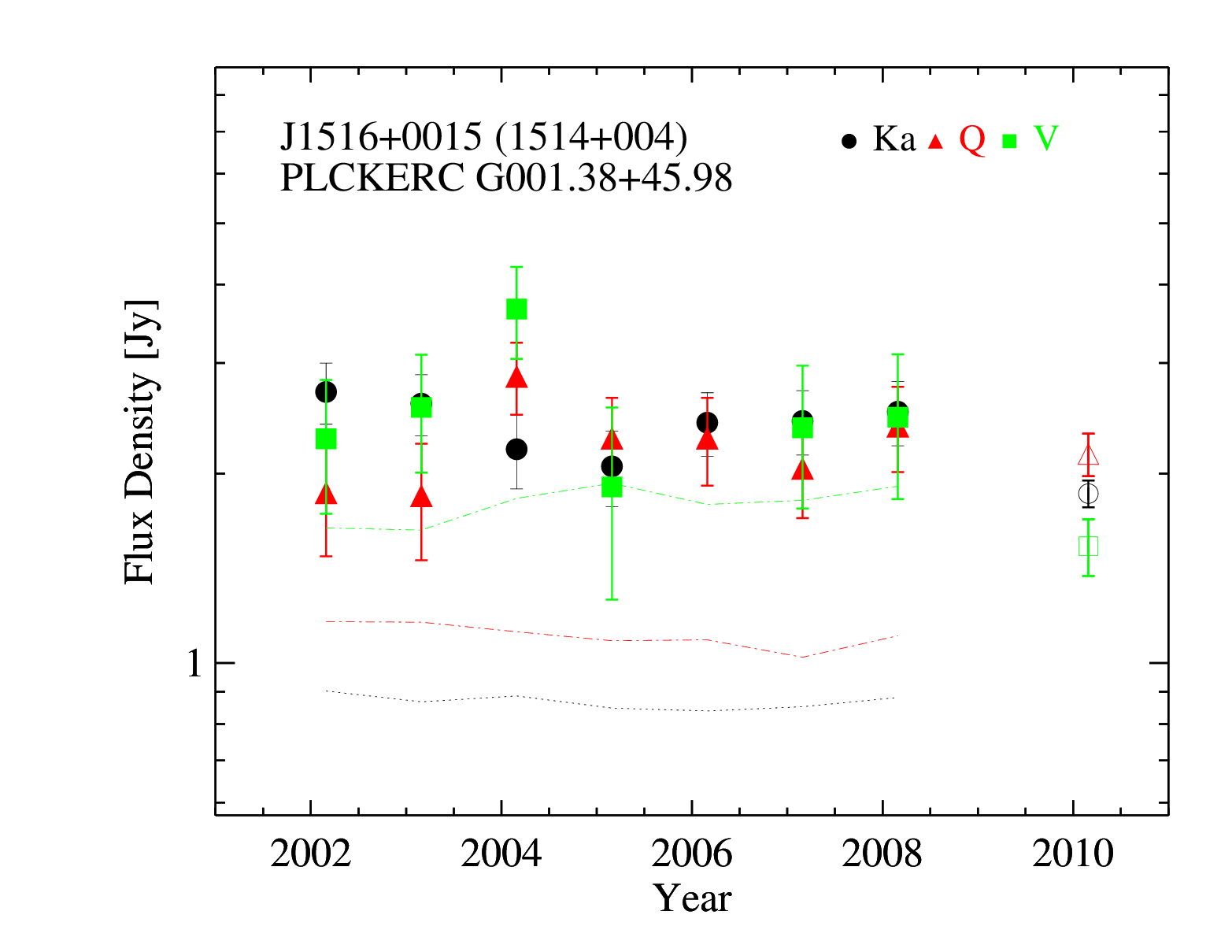} & \includegraphics[width=0.23\textwidth]{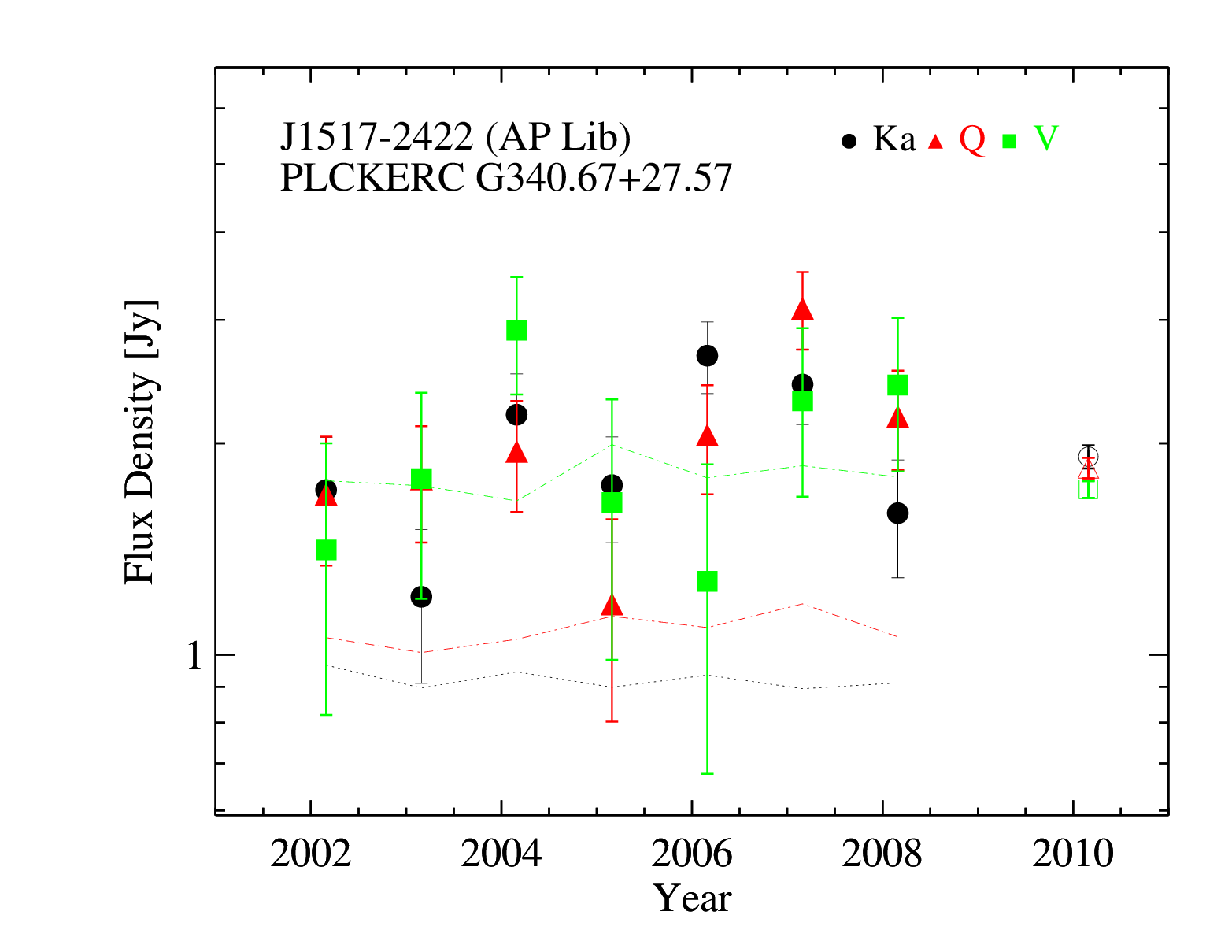}  \\
\includegraphics[width=0.23\textwidth]{figures/lc/J1549+0236.eps} & \includegraphics[width=0.23\textwidth]{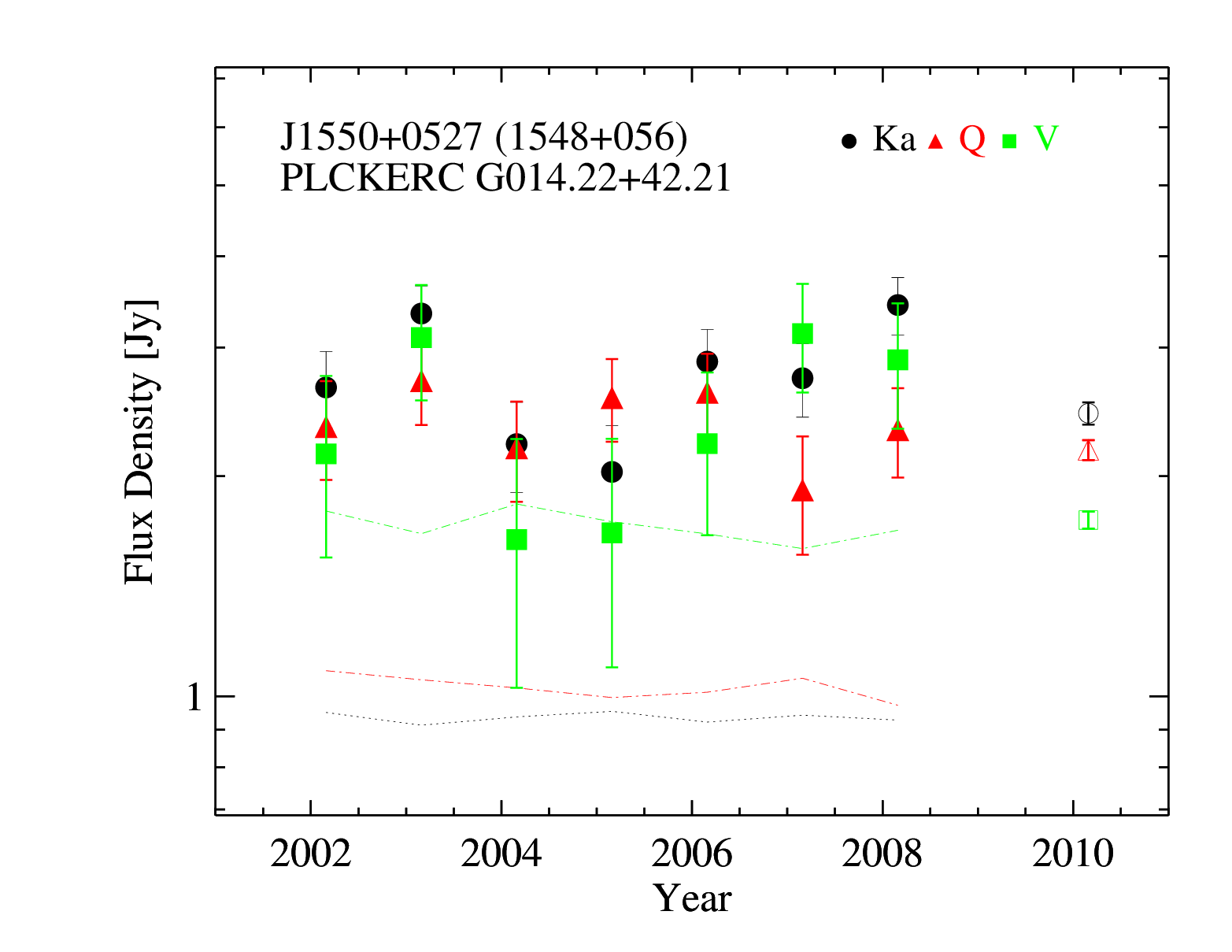}  & \includegraphics[width=0.23\textwidth]{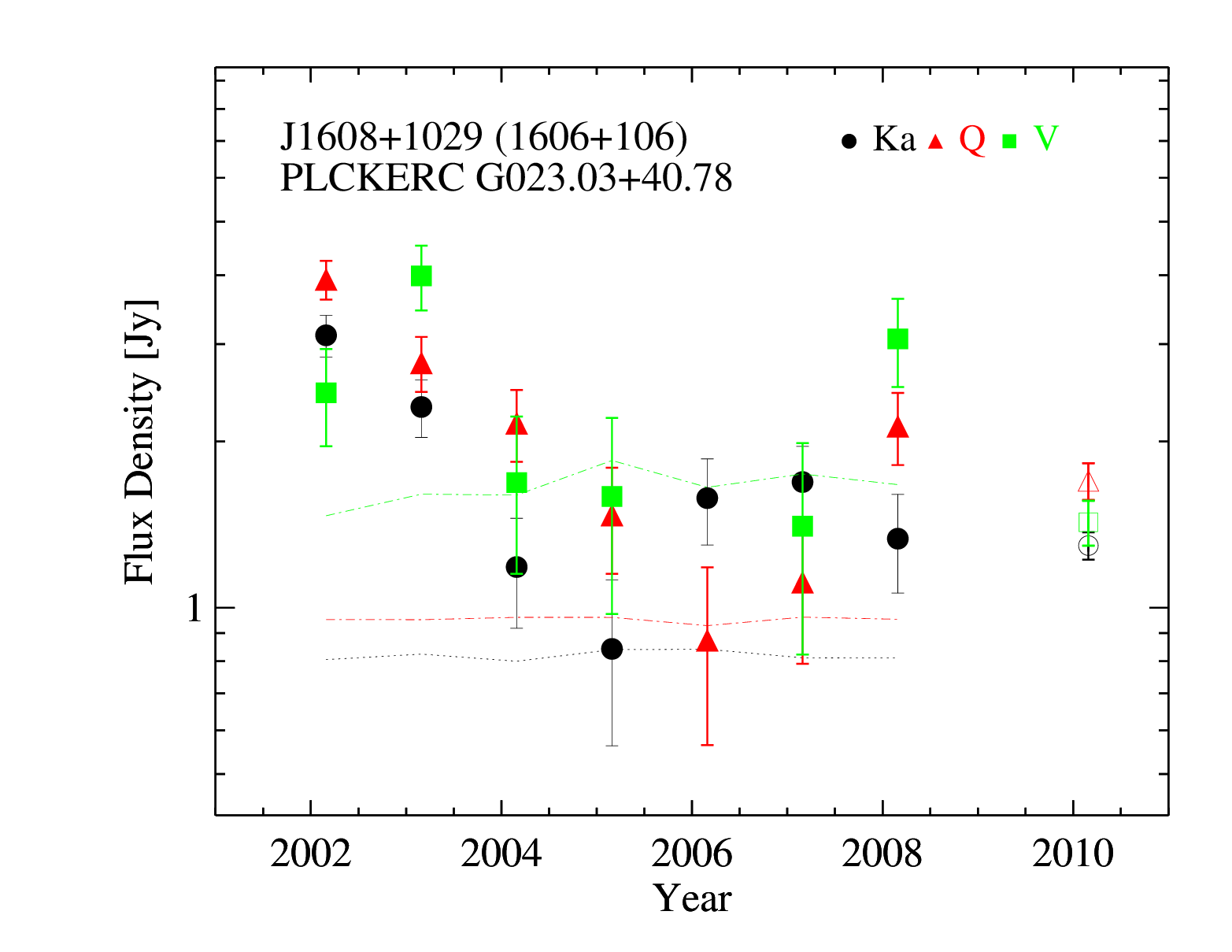} & \includegraphics[width=0.23\textwidth]{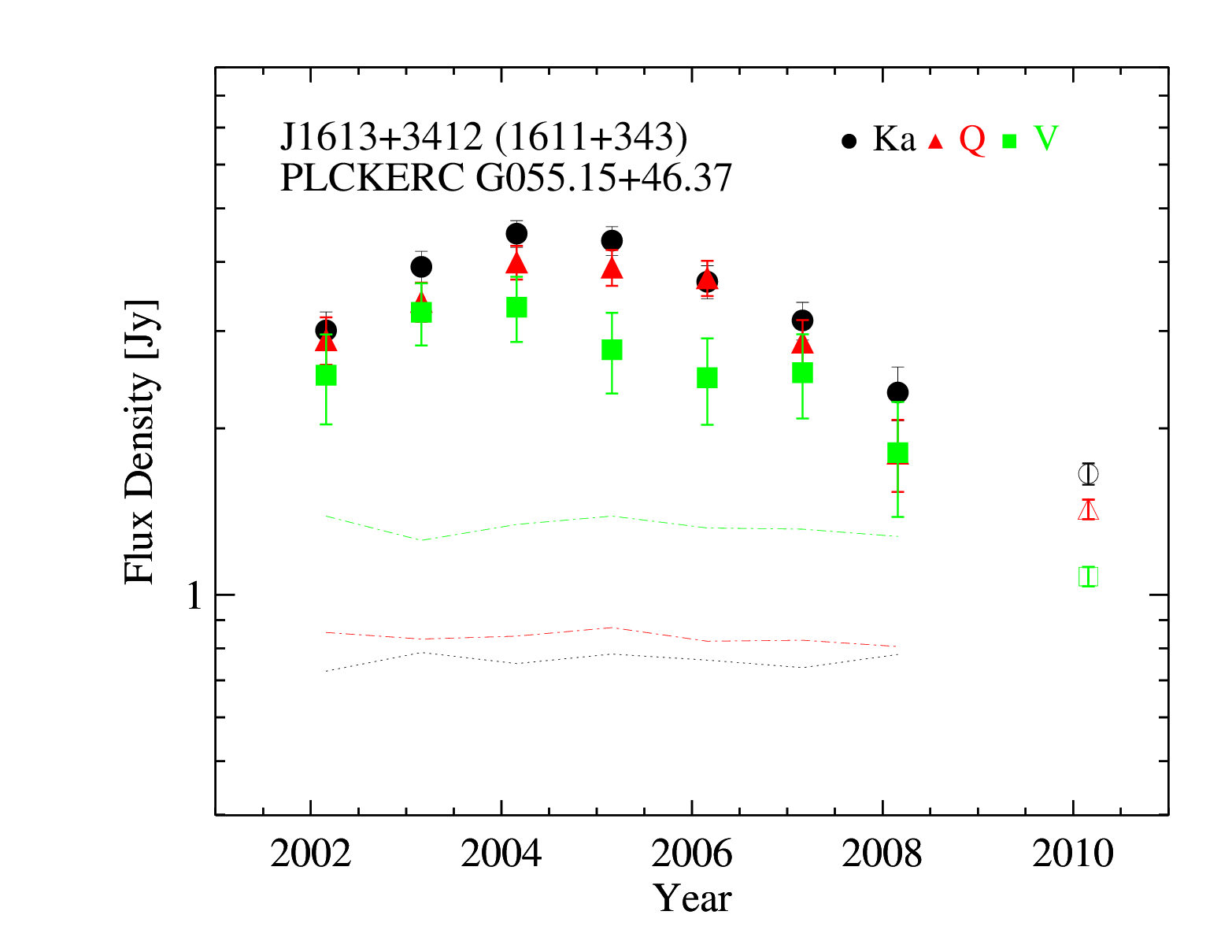}  \\
\includegraphics[width=0.23\textwidth]{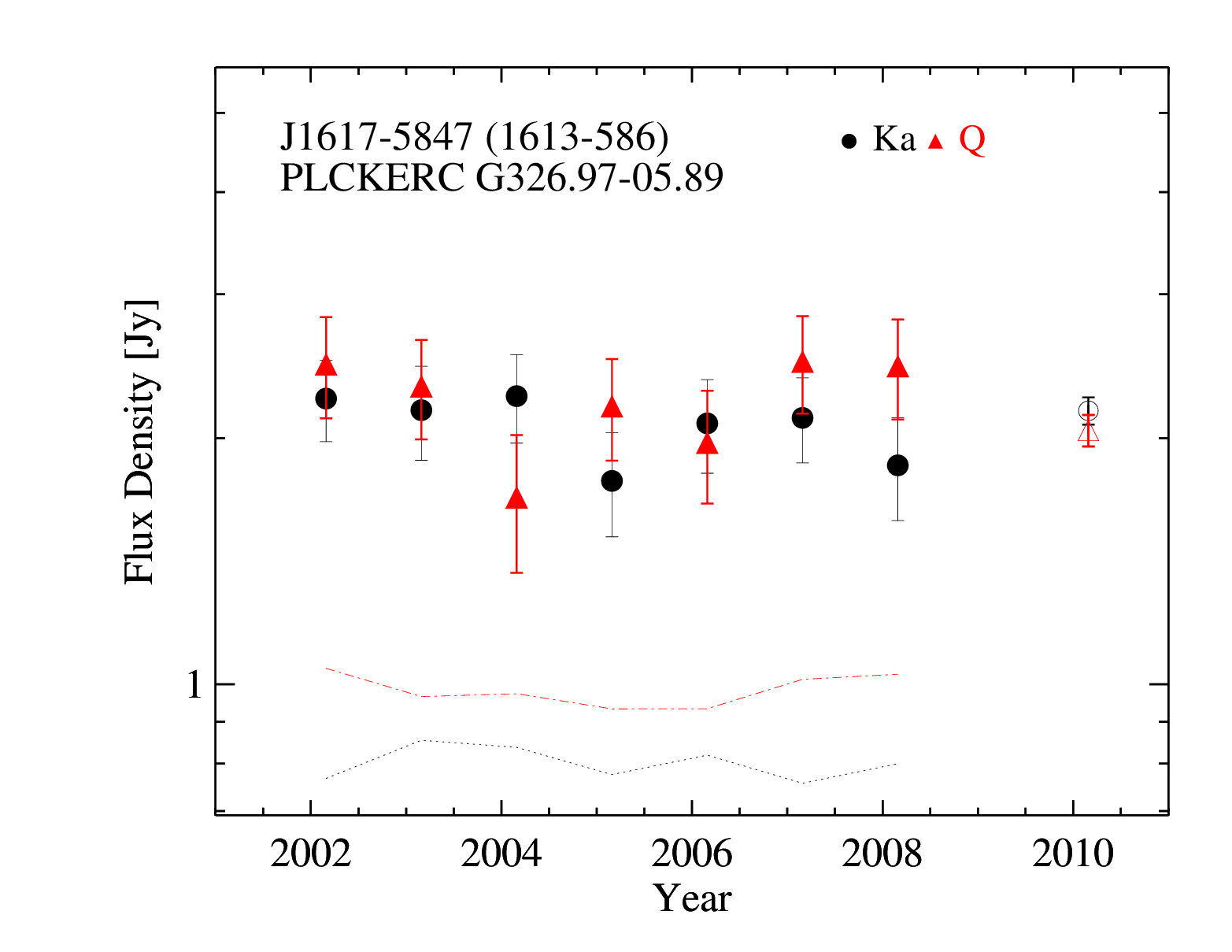} & \includegraphics[width=0.23\textwidth]{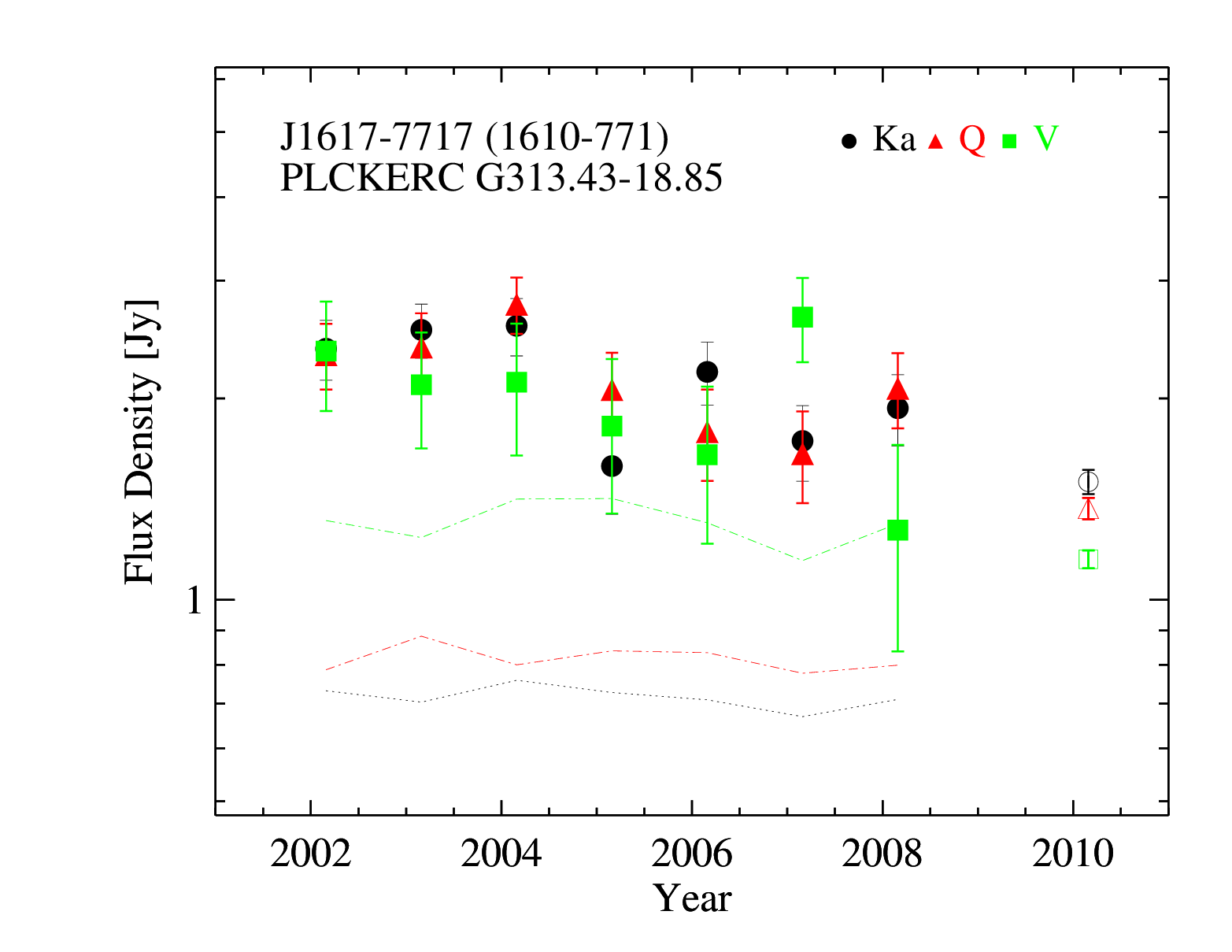}  & \includegraphics[width=0.23\textwidth]{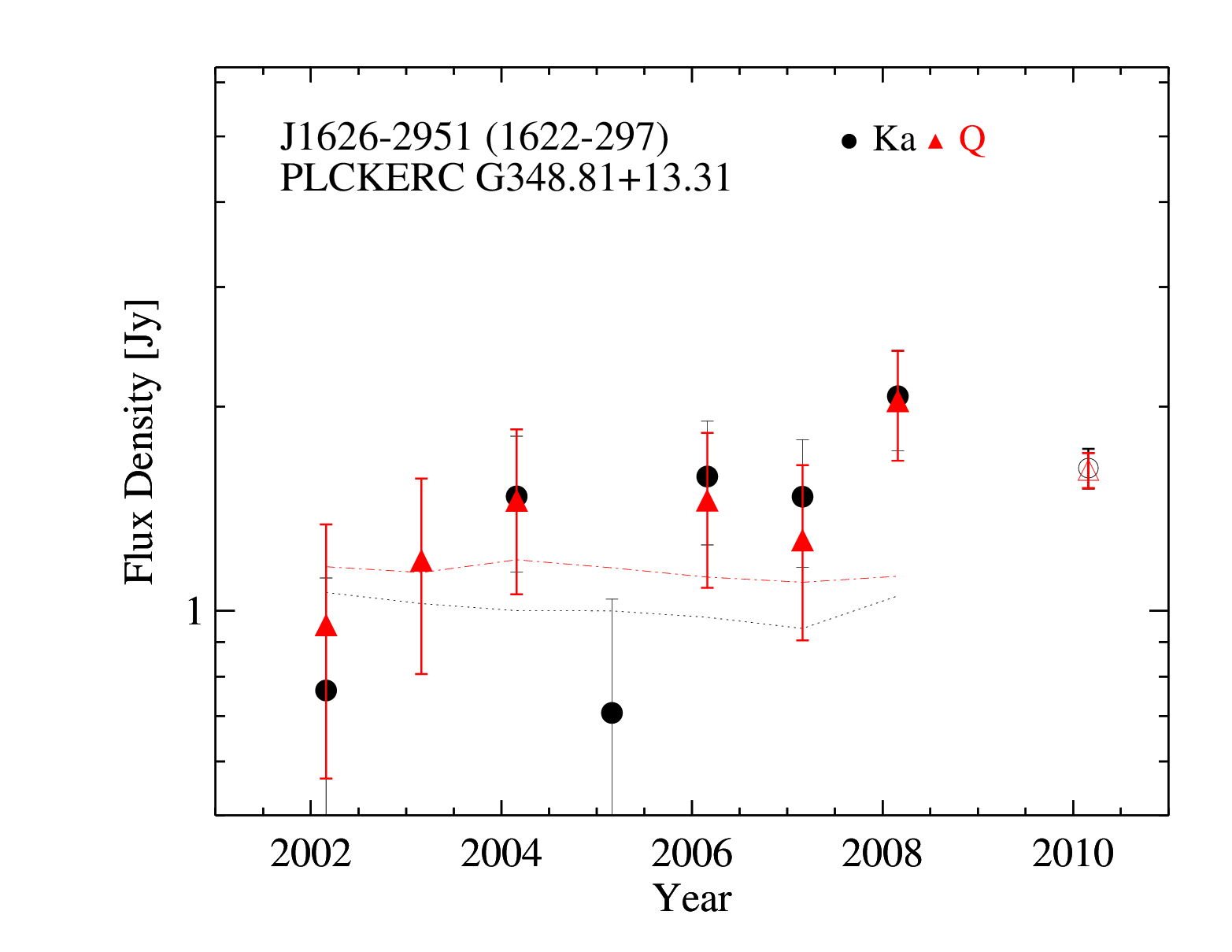} & \includegraphics[width=0.23\textwidth]{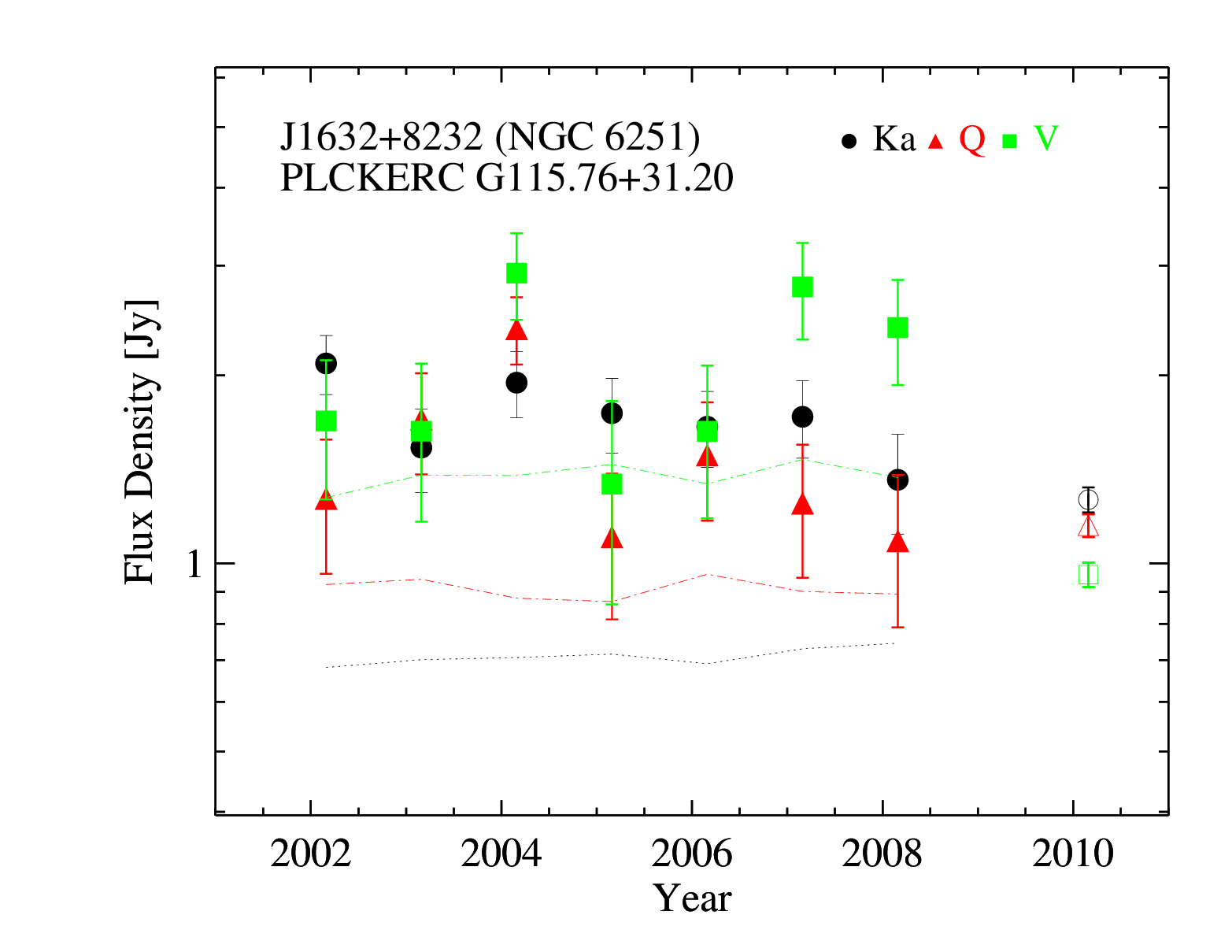}  \\
\includegraphics[width=0.23\textwidth]{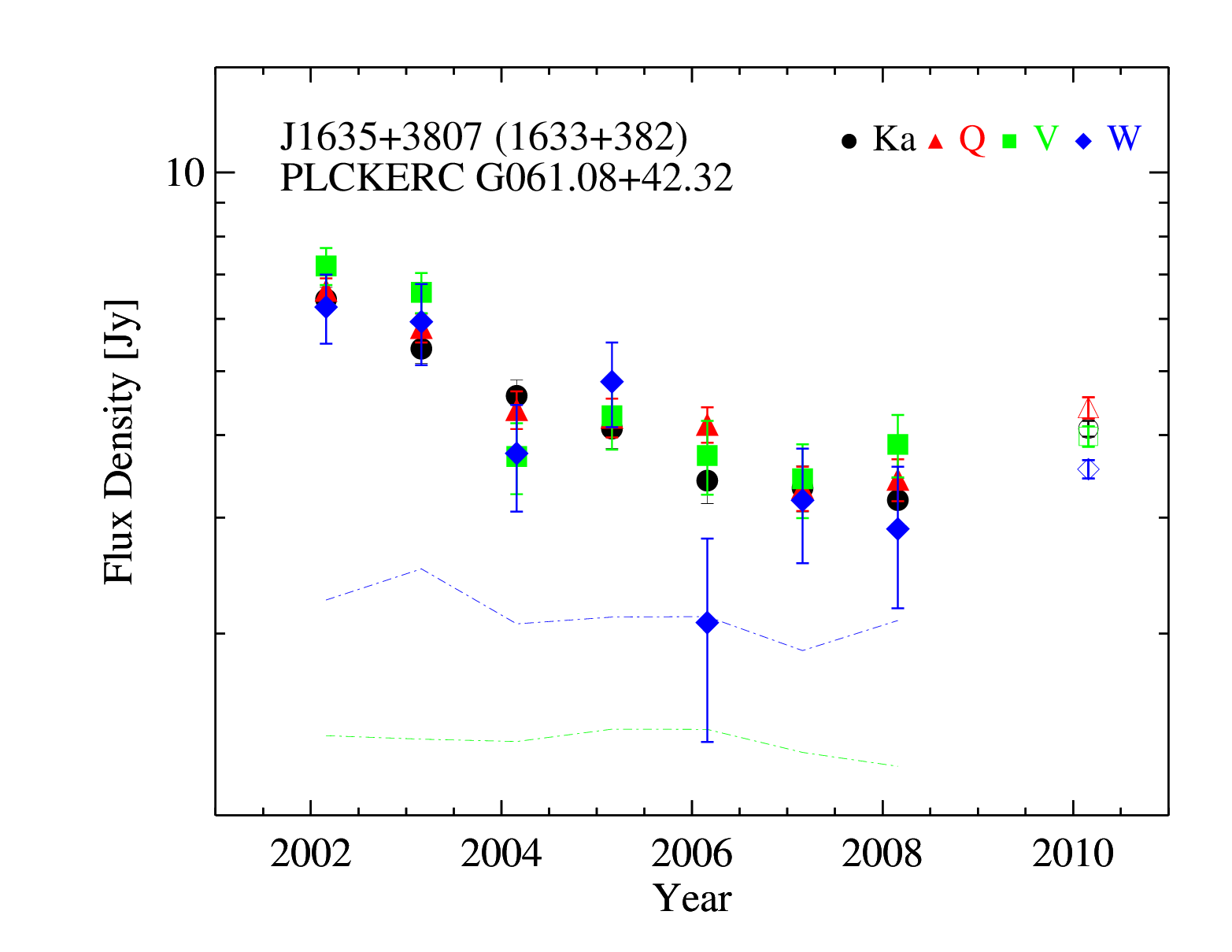} & \includegraphics[width=0.23\textwidth]{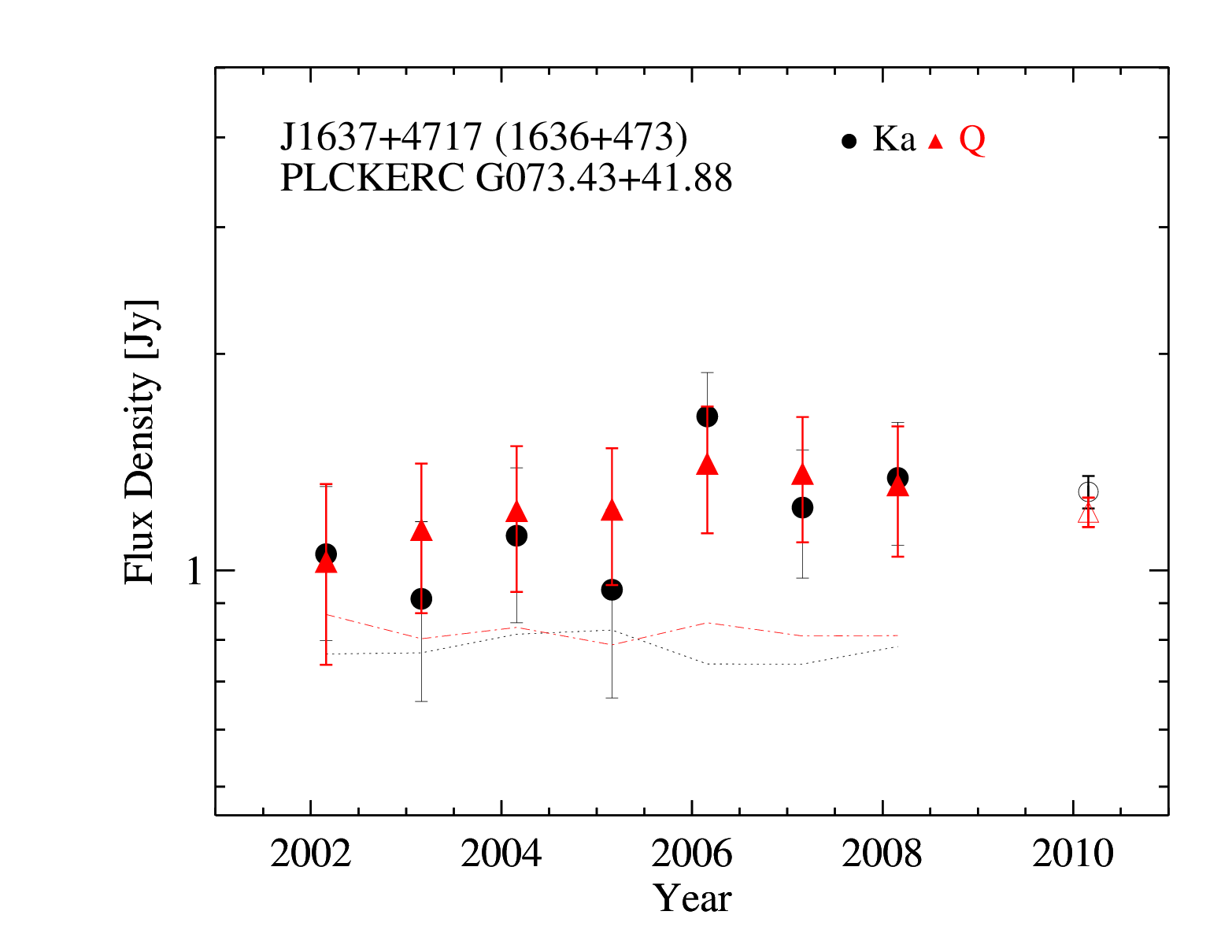}  & \includegraphics[width=0.23\textwidth]{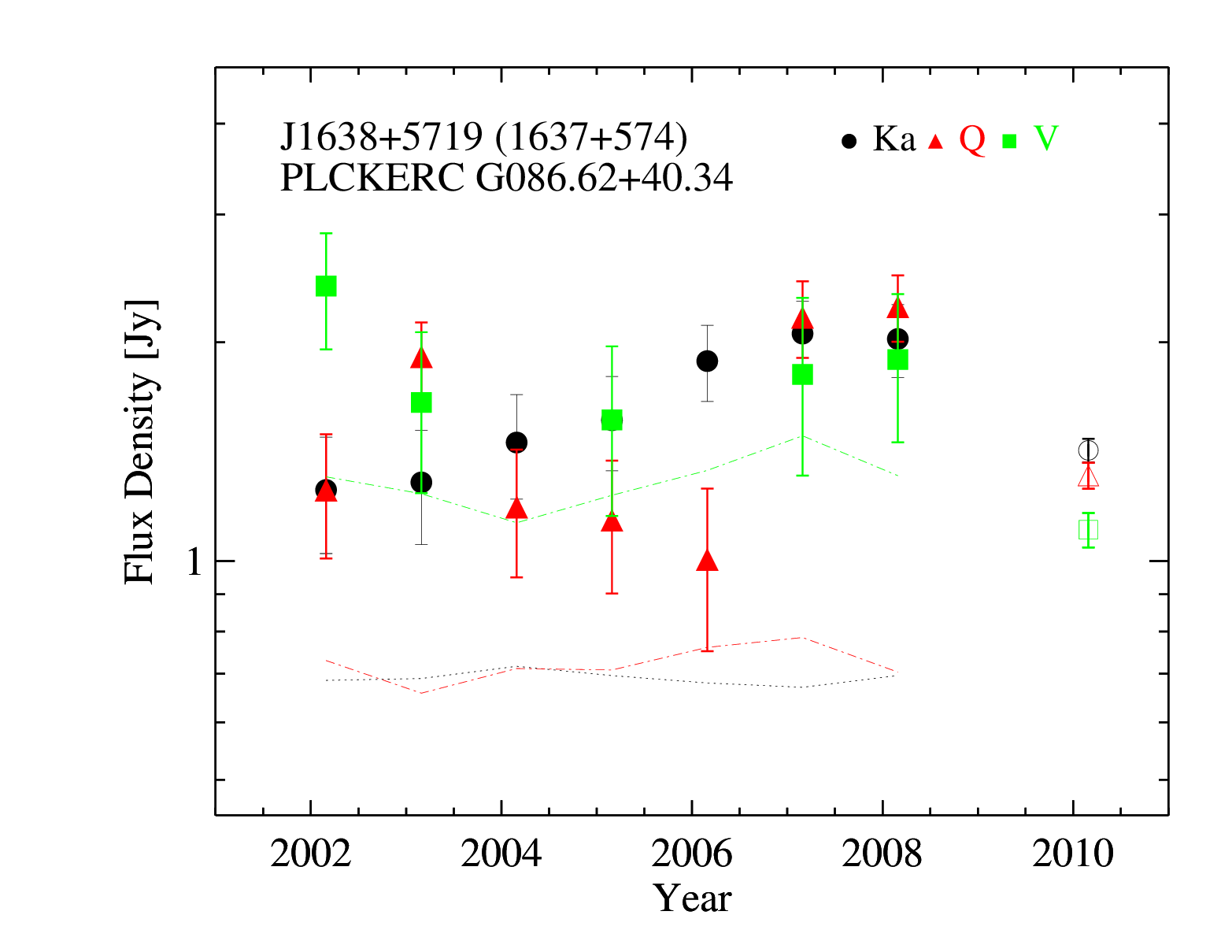} & \includegraphics[width=0.23\textwidth]{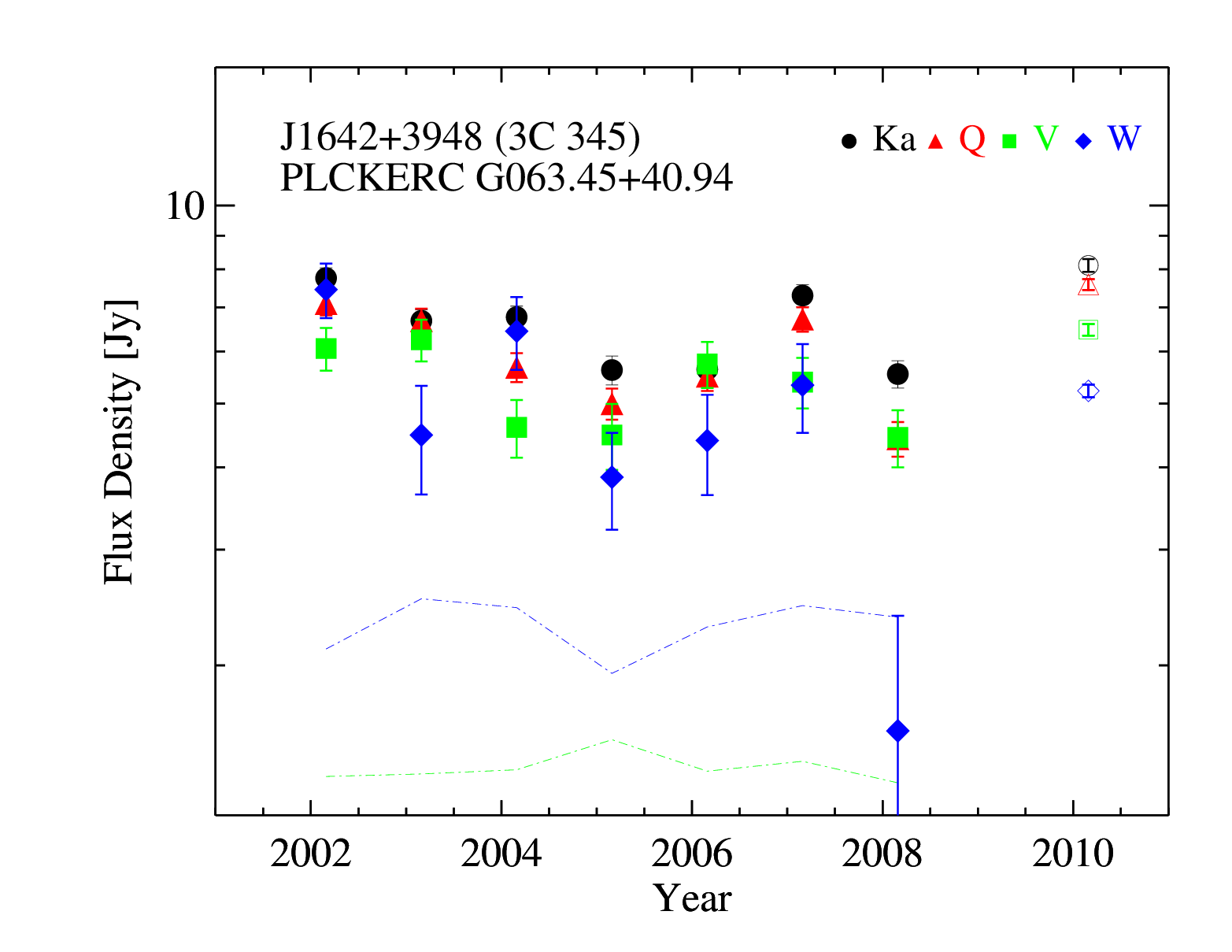}  \\
\end{tabular}
\end{figure*}

\begin{figure*}
\centering
\begin{tabular}{cccc}
\includegraphics[width=0.23\textwidth]{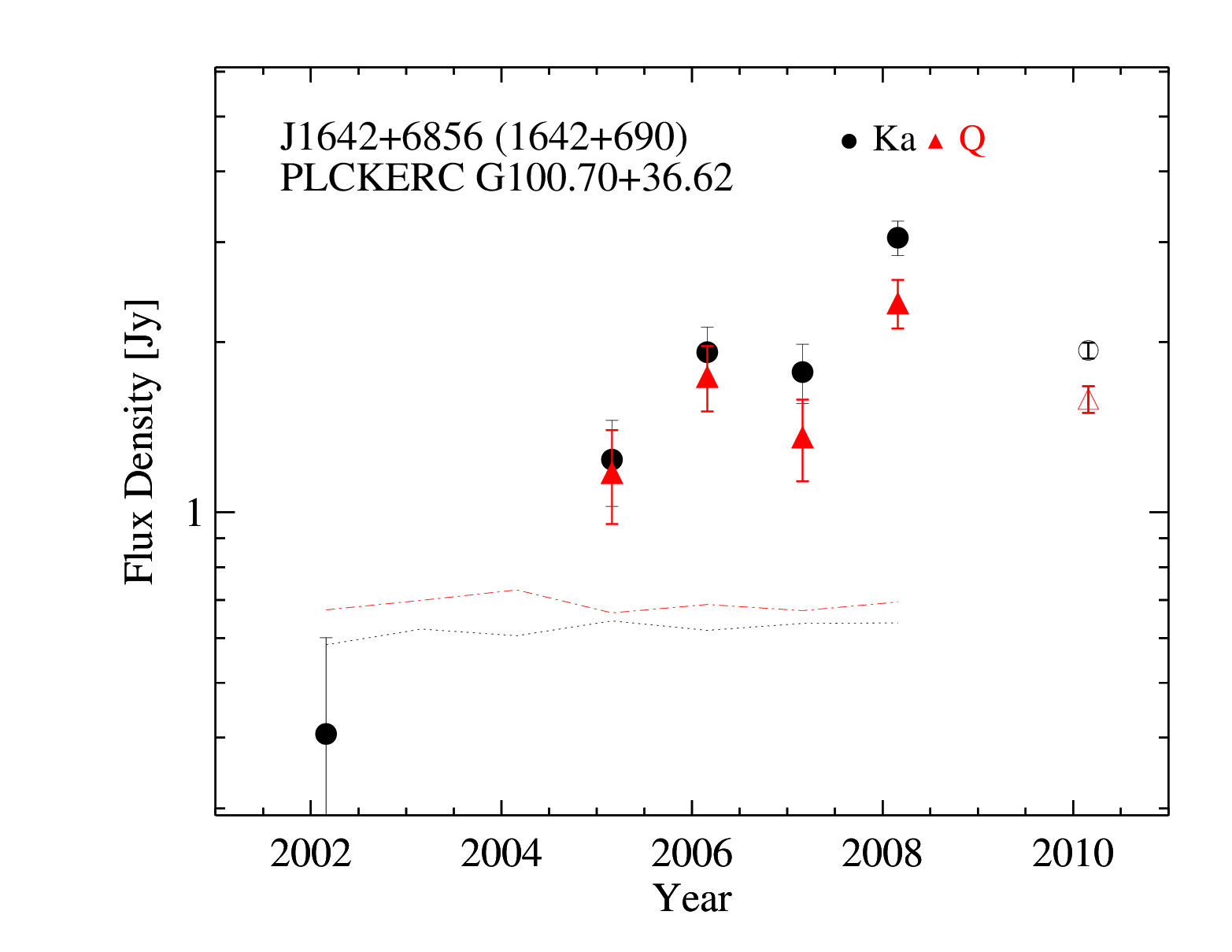} & \includegraphics[width=0.23\textwidth]{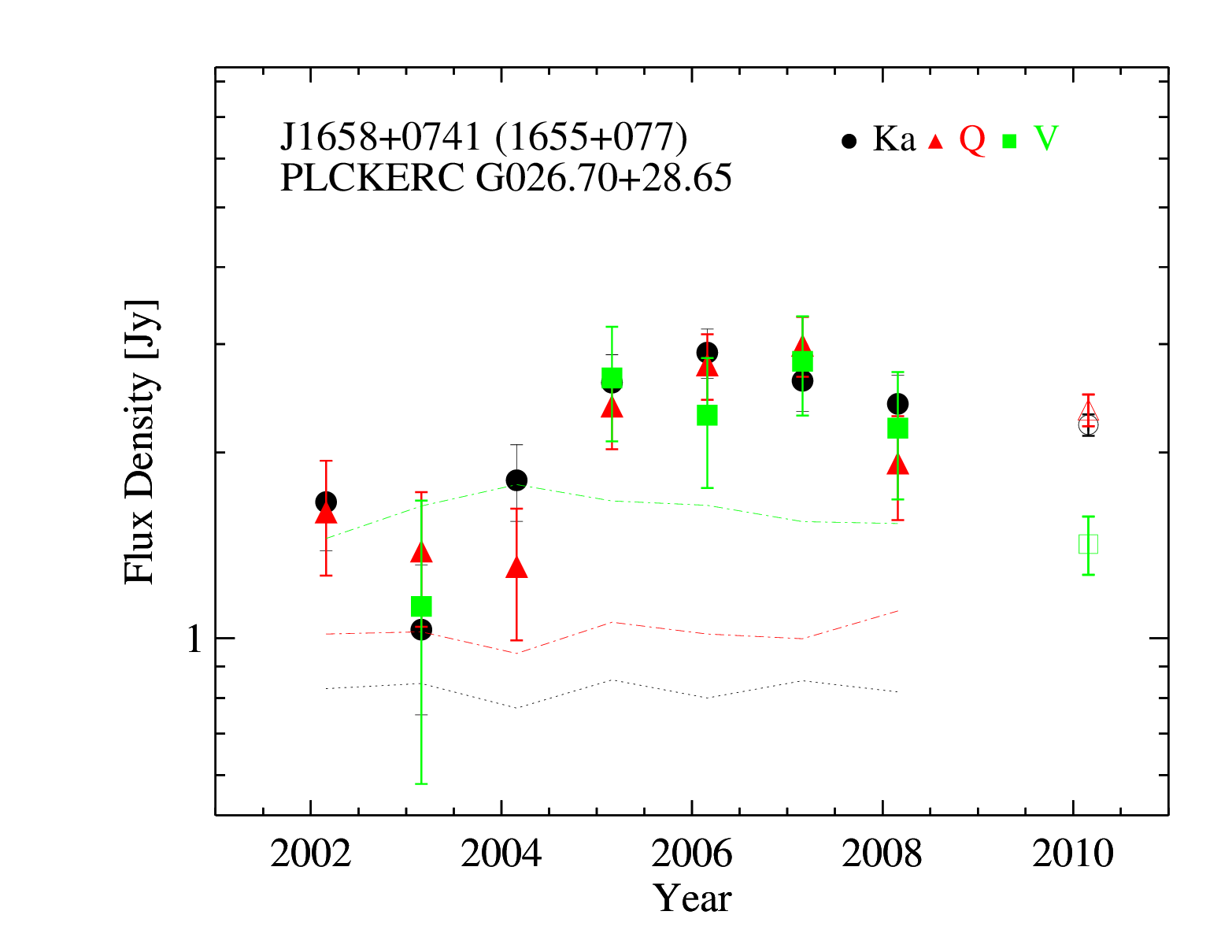}  & \includegraphics[width=0.23\textwidth]{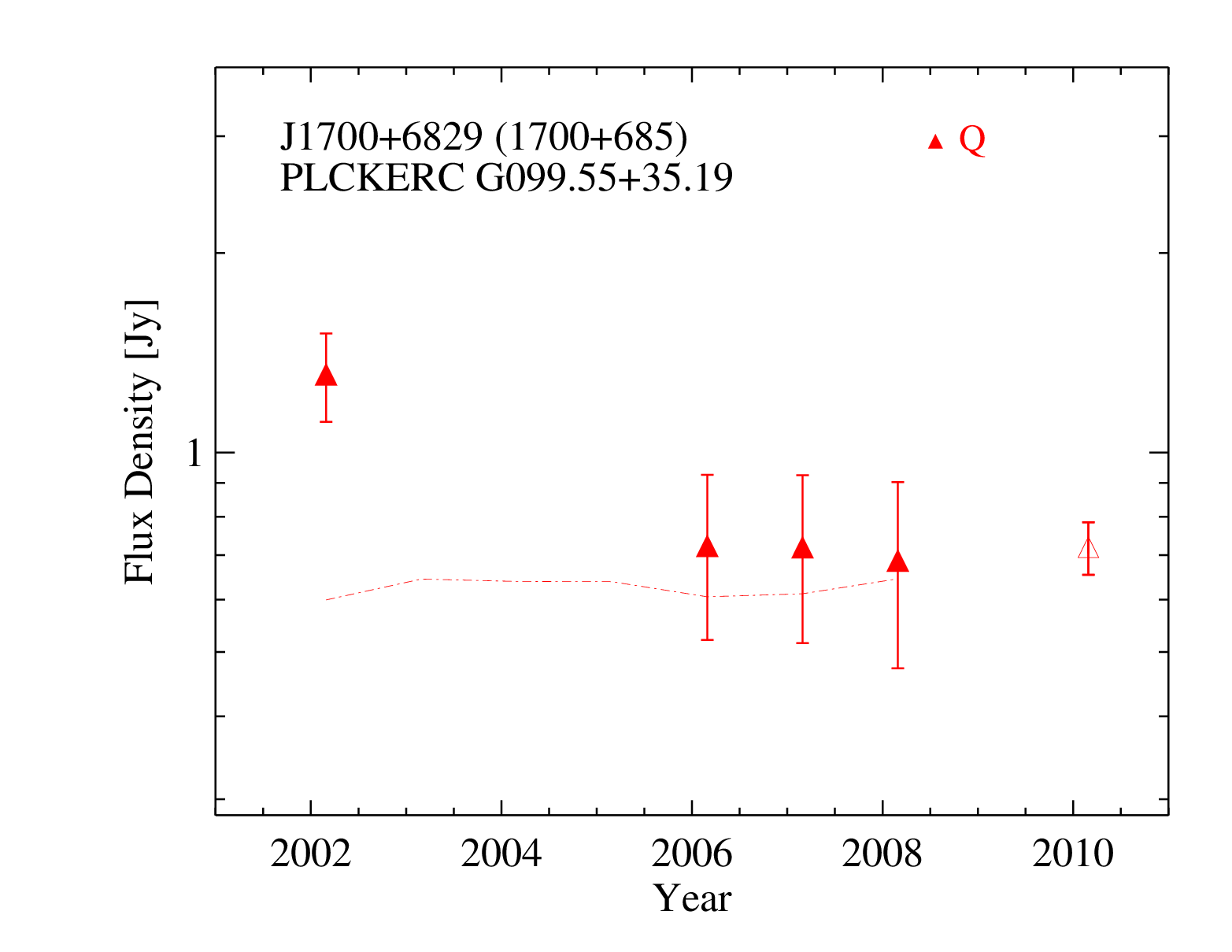} & \includegraphics[width=0.23\textwidth]{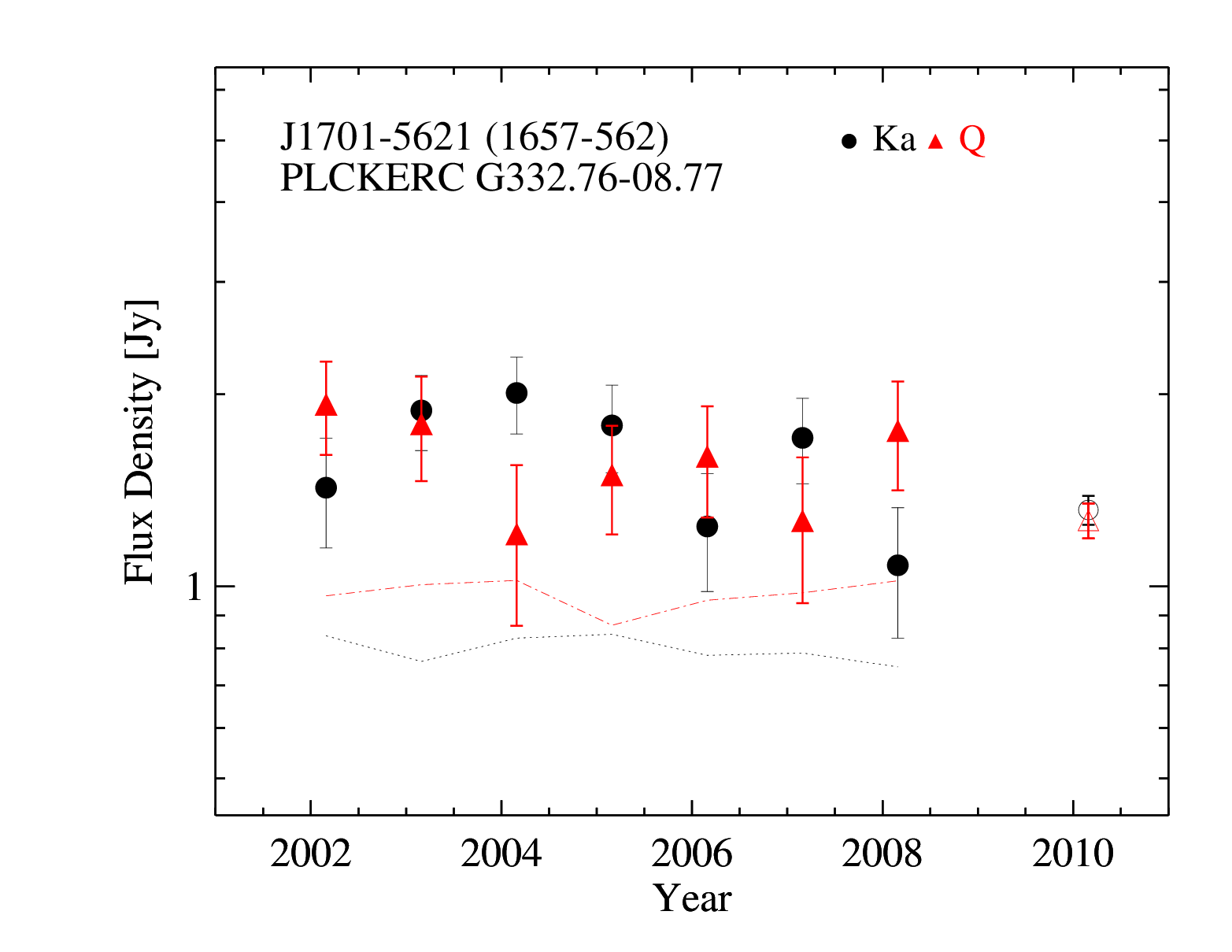}  \\
\includegraphics[width=0.23\textwidth]{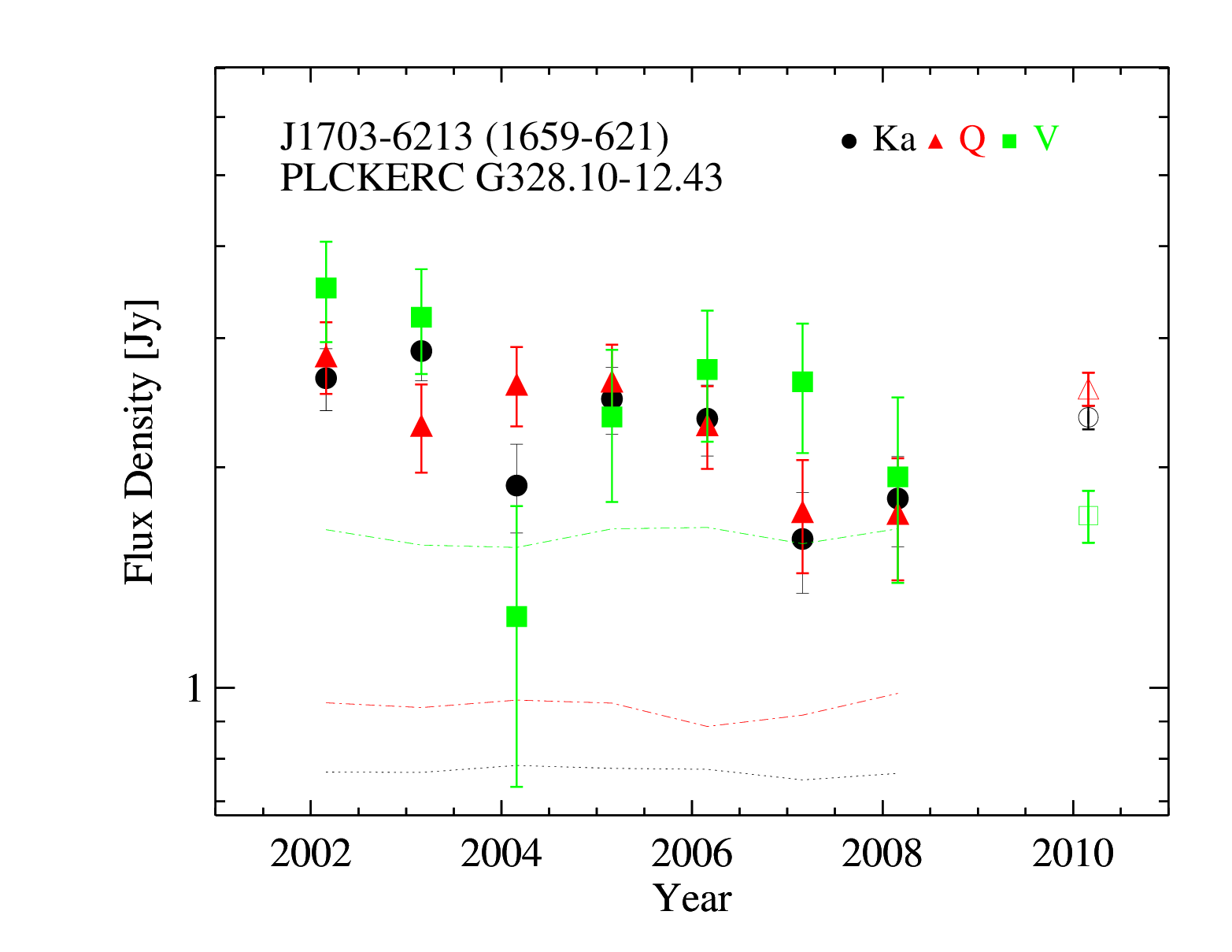} & \includegraphics[width=0.23\textwidth]{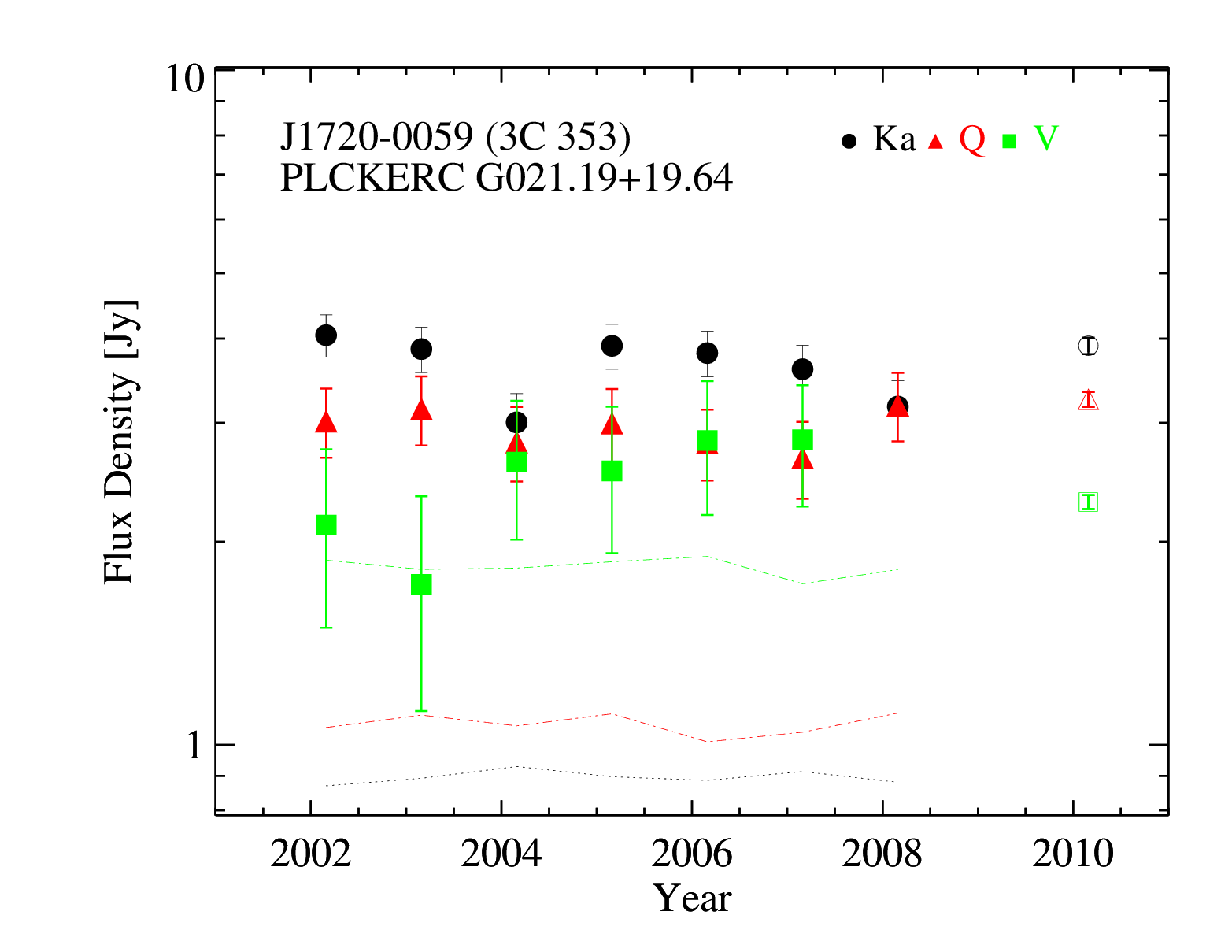}  & \includegraphics[width=0.23\textwidth]{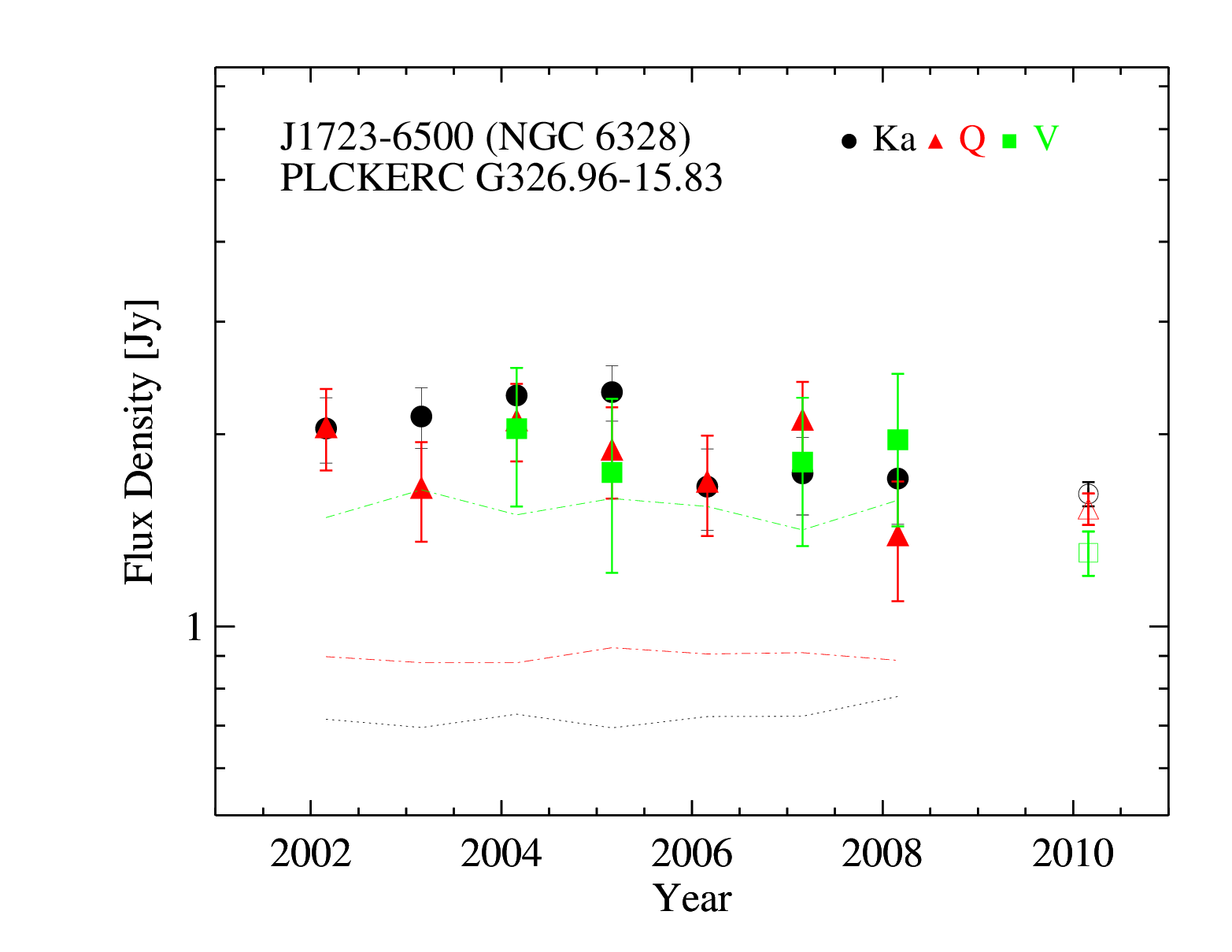} & \includegraphics[width=0.23\textwidth]{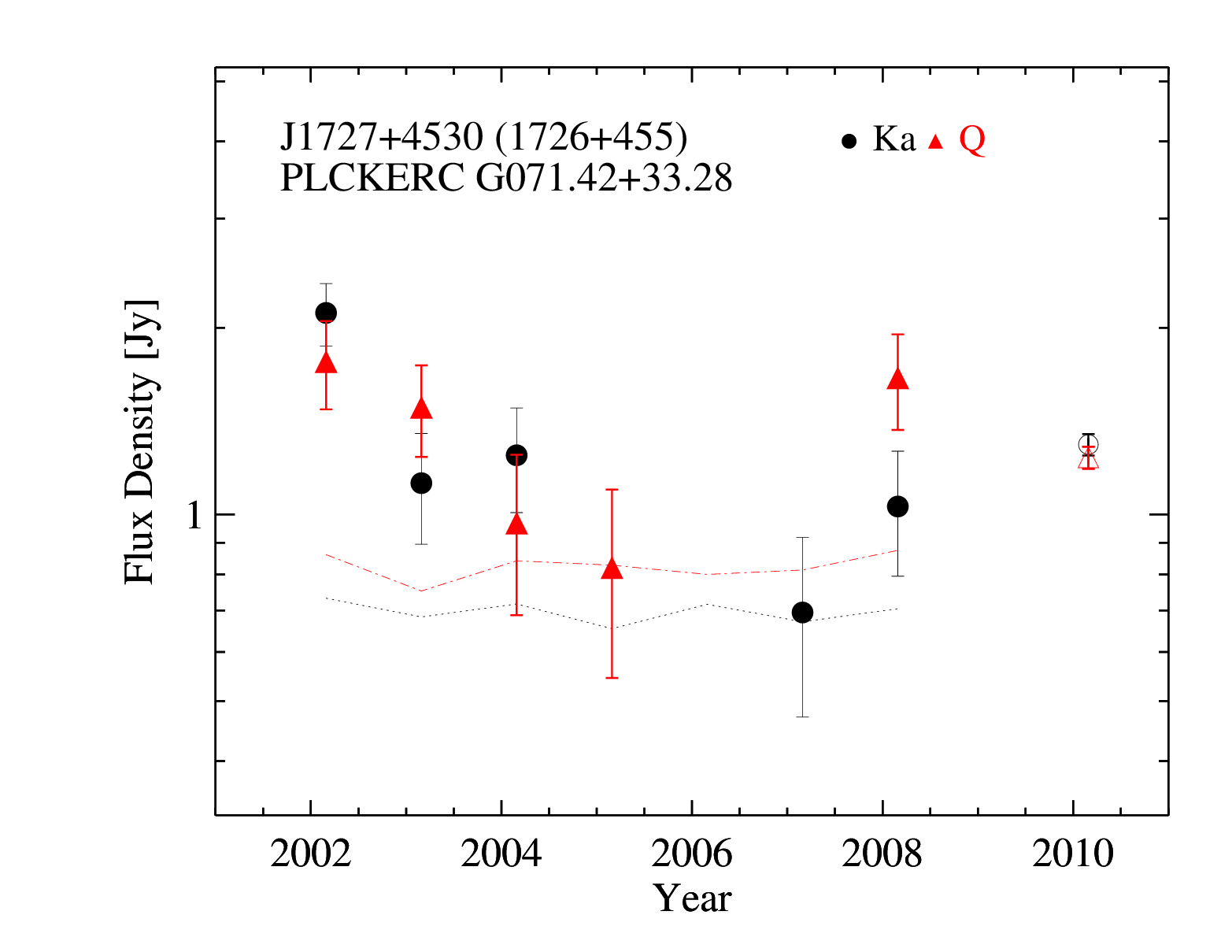}  \\
\includegraphics[width=0.23\textwidth]{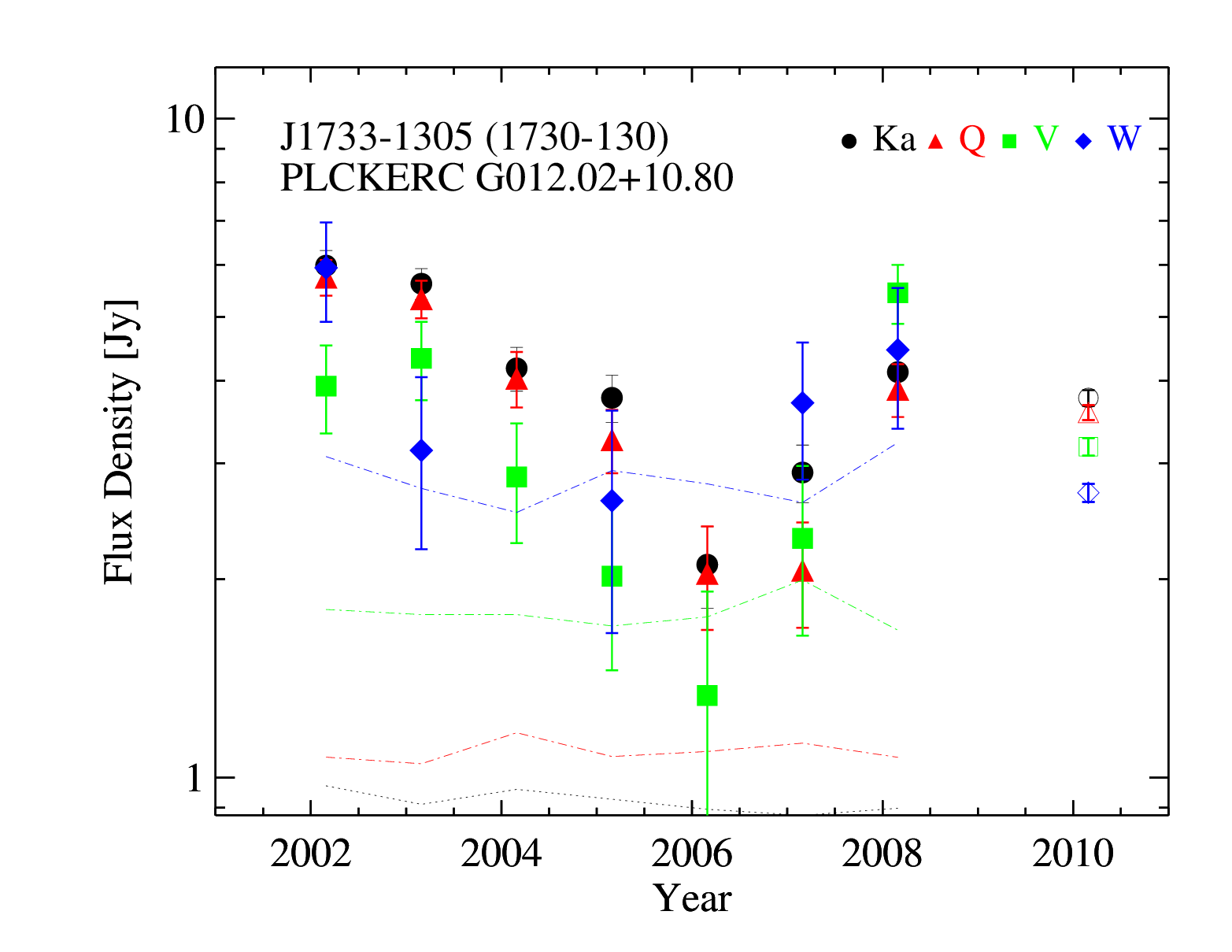} & \includegraphics[width=0.23\textwidth]{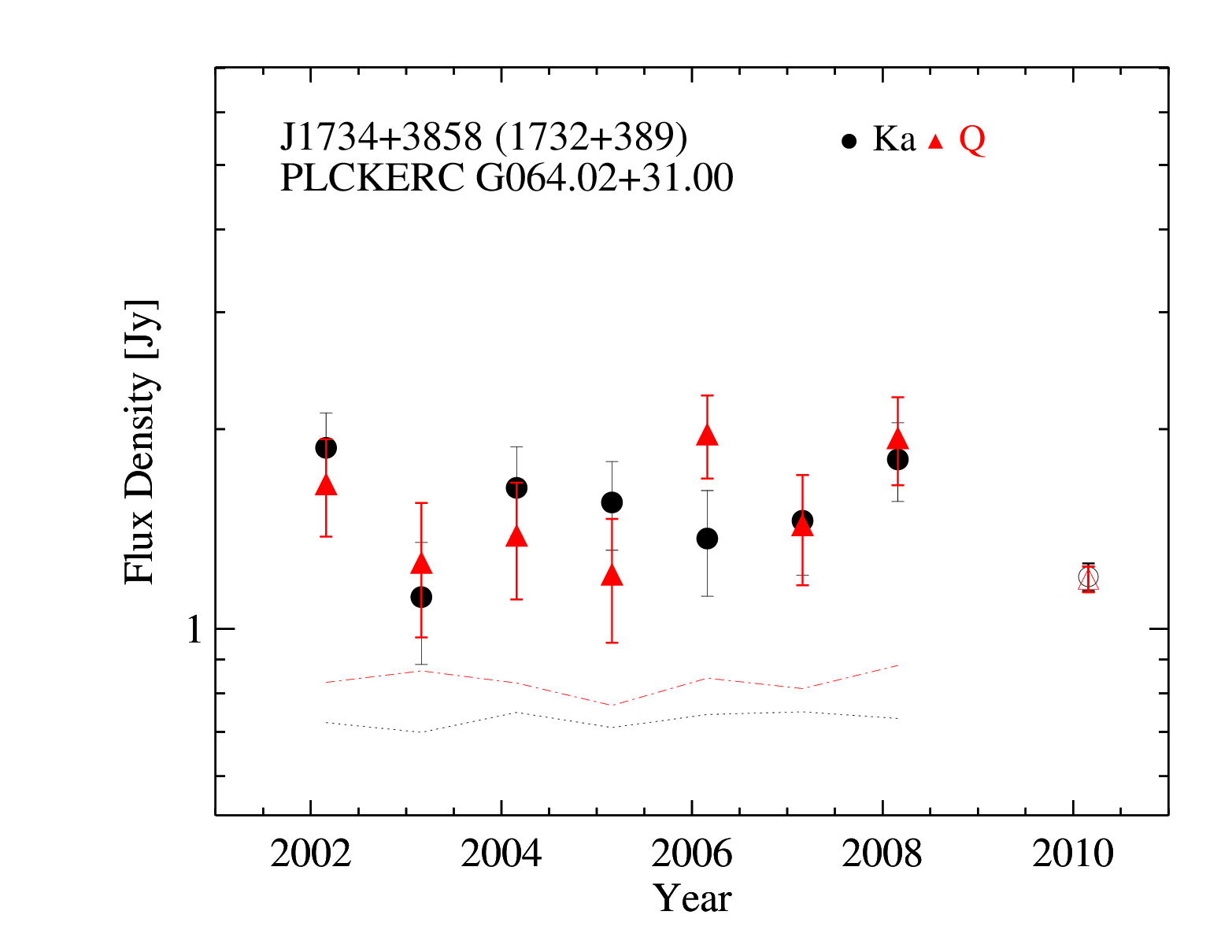}  & \includegraphics[width=0.23\textwidth]{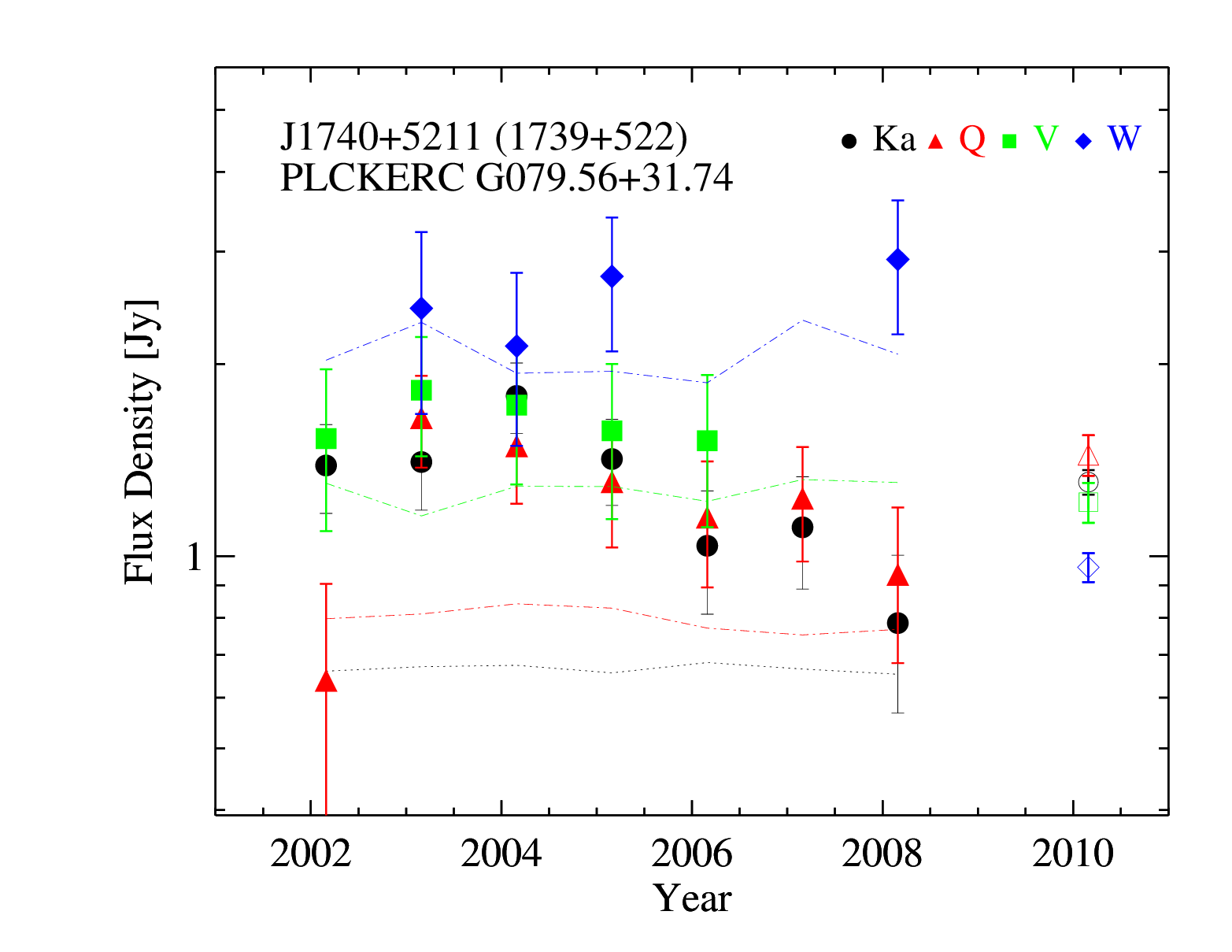} & \includegraphics[width=0.23\textwidth]{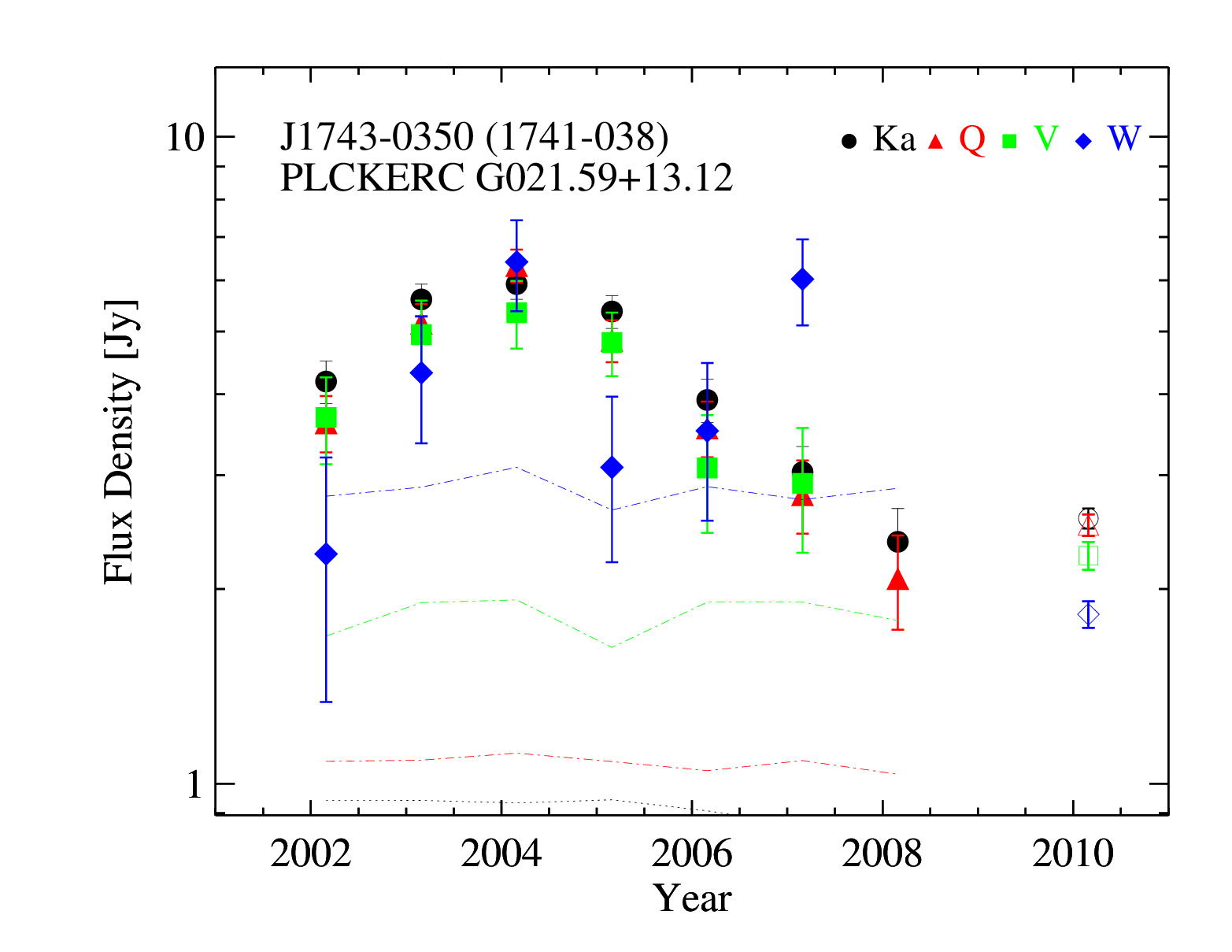}  \\
\includegraphics[width=0.23\textwidth]{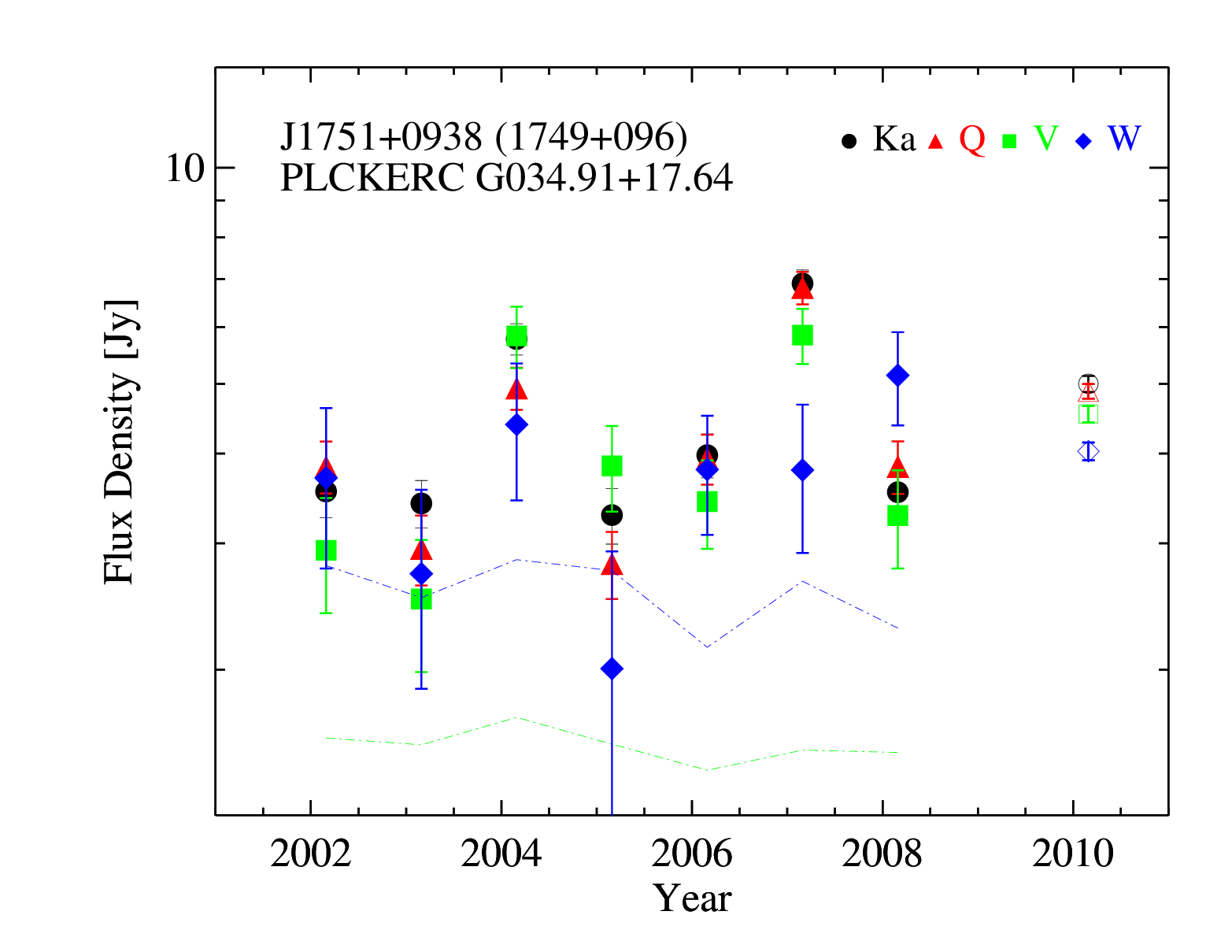} & \includegraphics[width=0.23\textwidth]{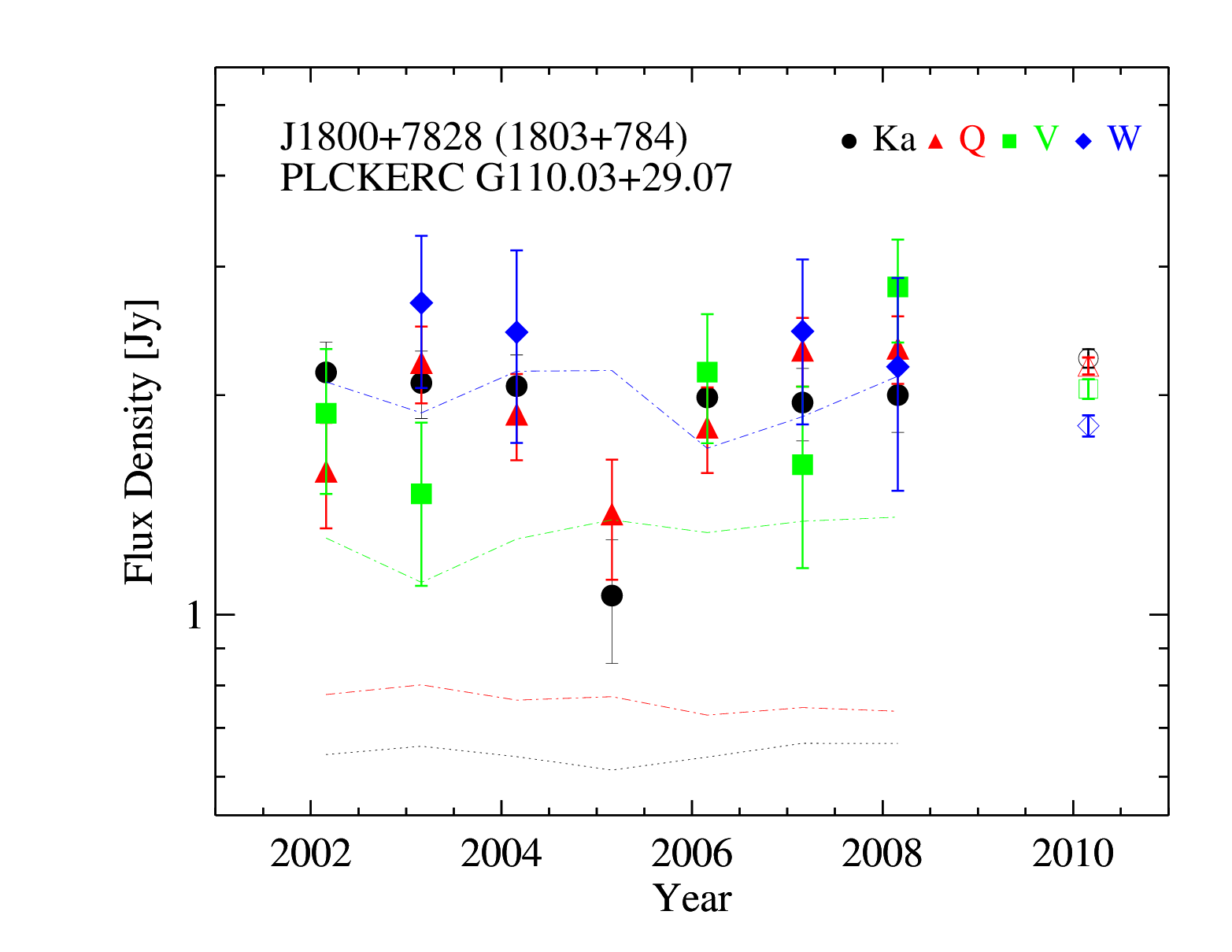}  & \includegraphics[width=0.23\textwidth]{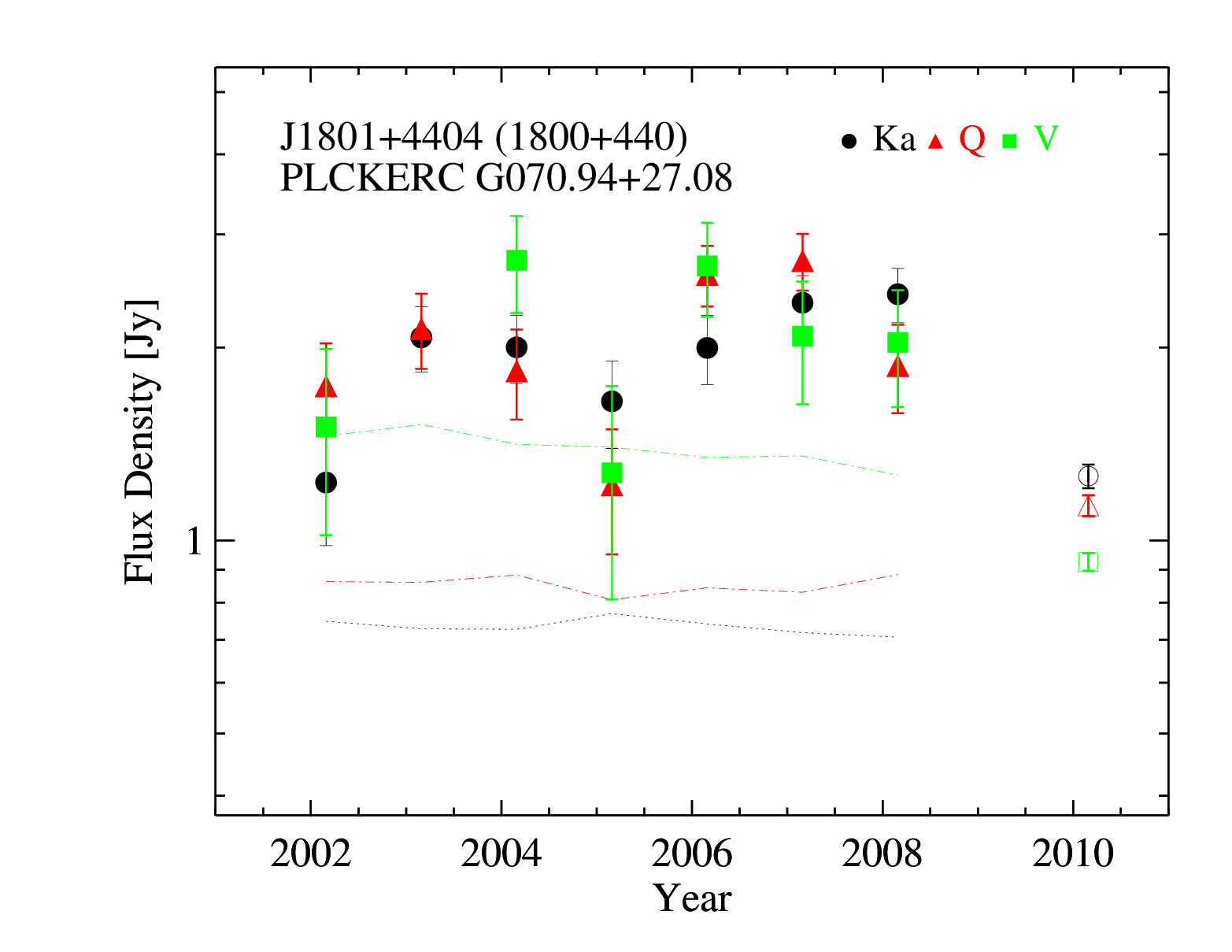} & \includegraphics[width=0.23\textwidth]{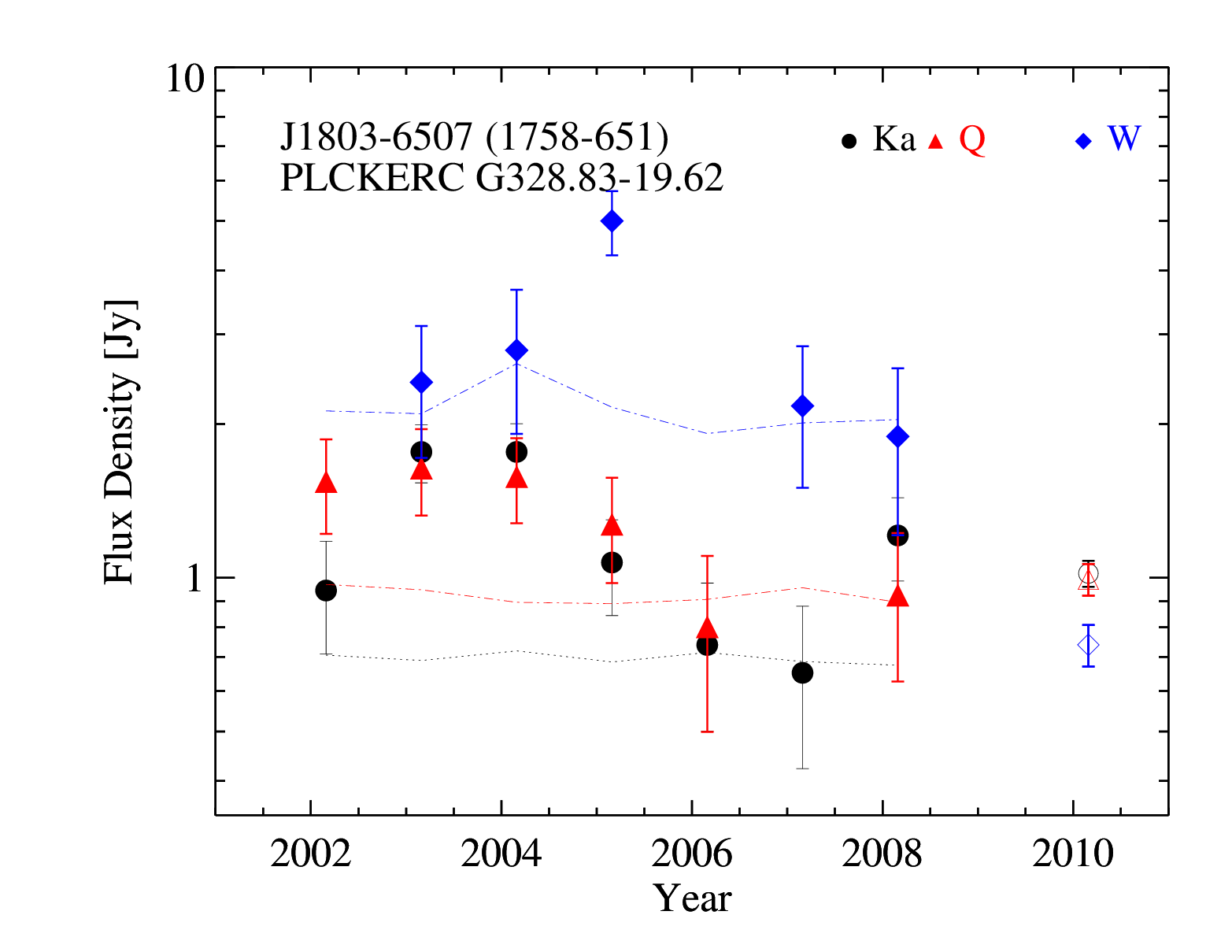}  \\
\includegraphics[width=0.23\textwidth]{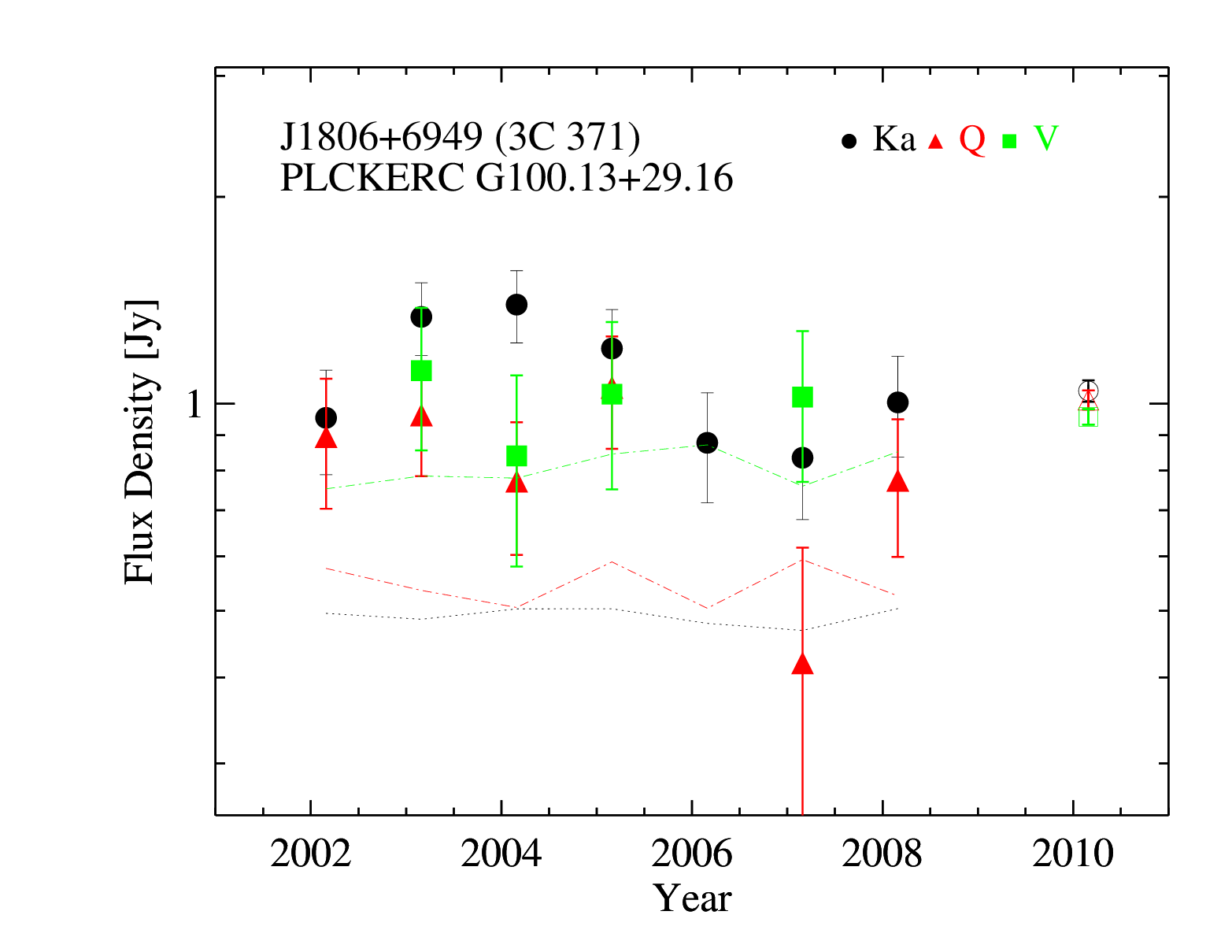} & \includegraphics[width=0.23\textwidth]{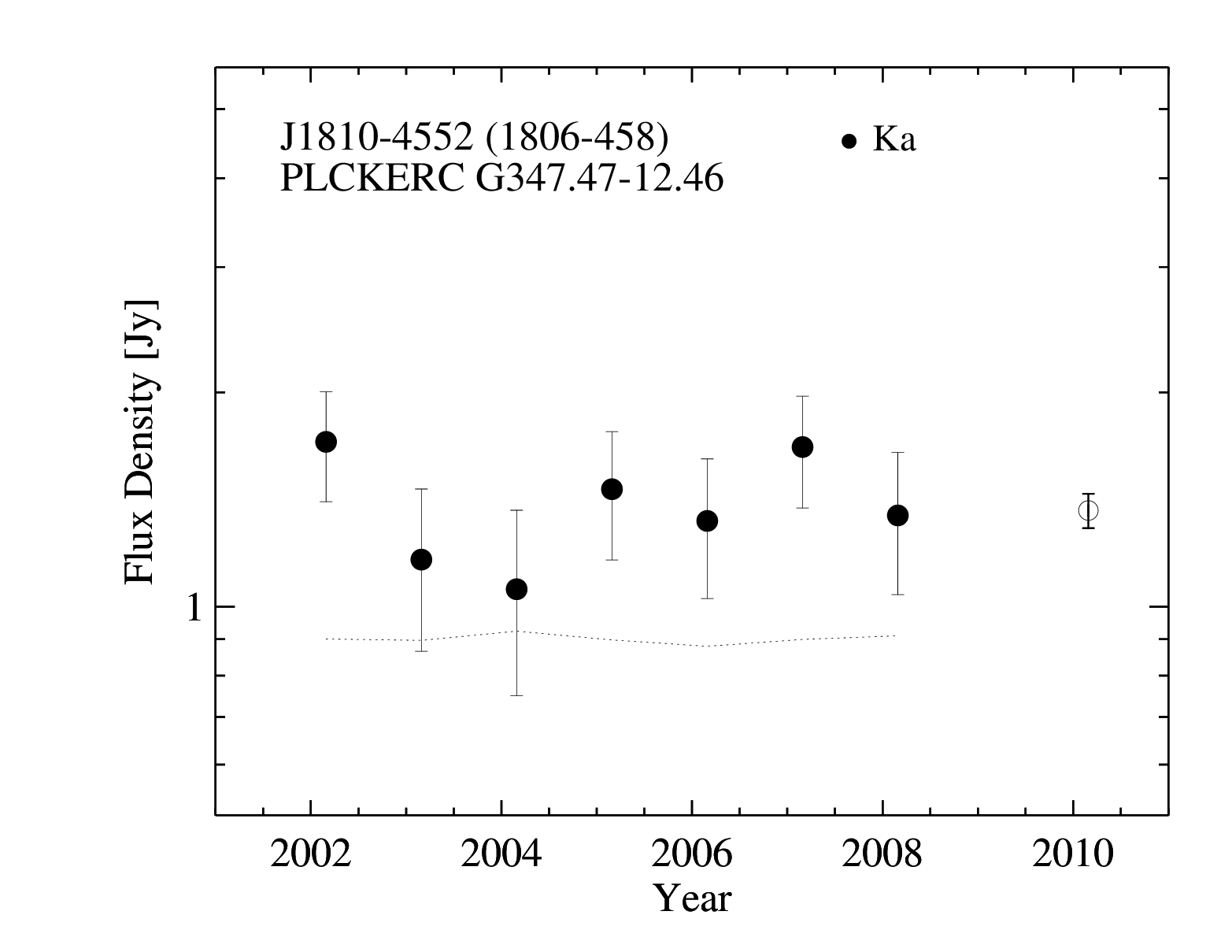}  & \includegraphics[width=0.23\textwidth]{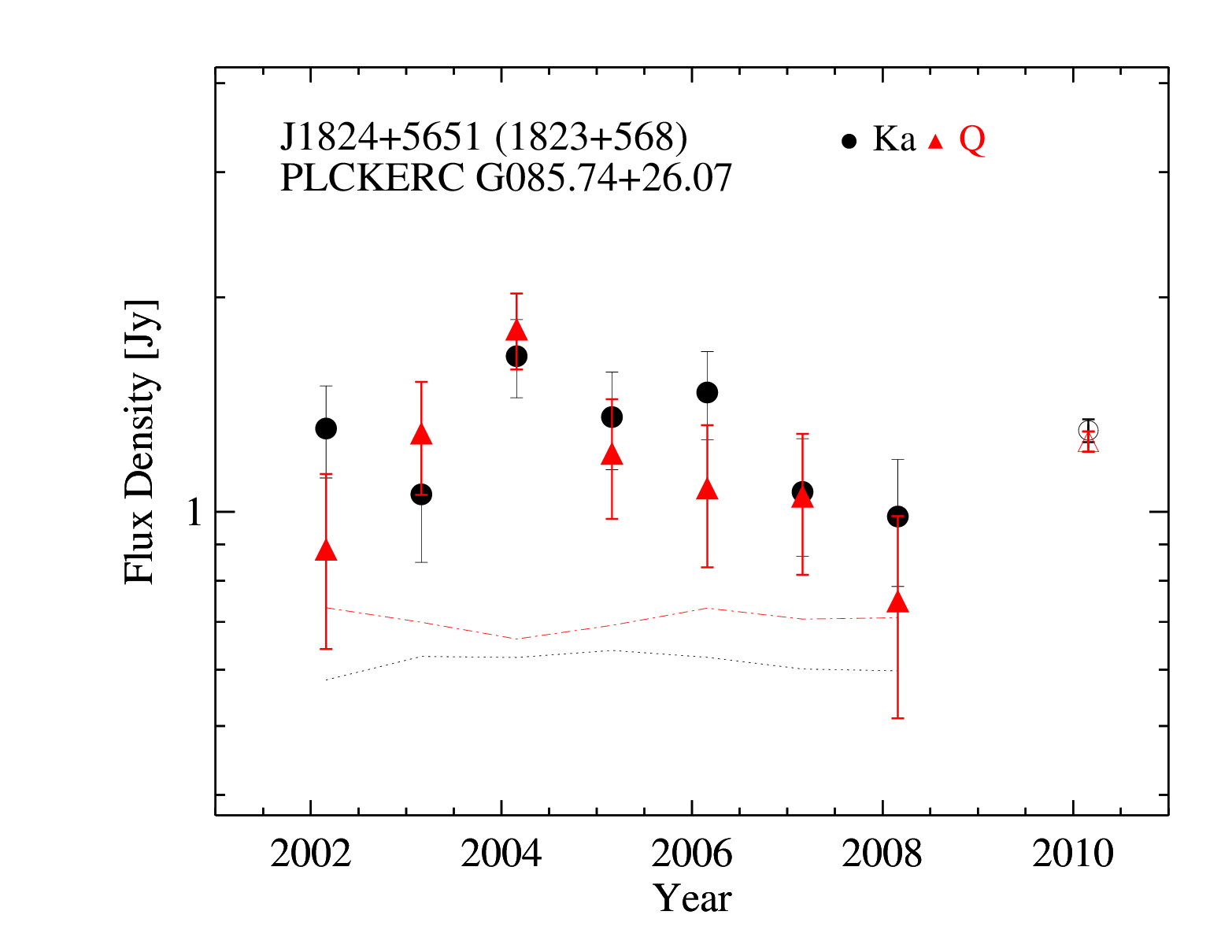} & \includegraphics[width=0.23\textwidth]{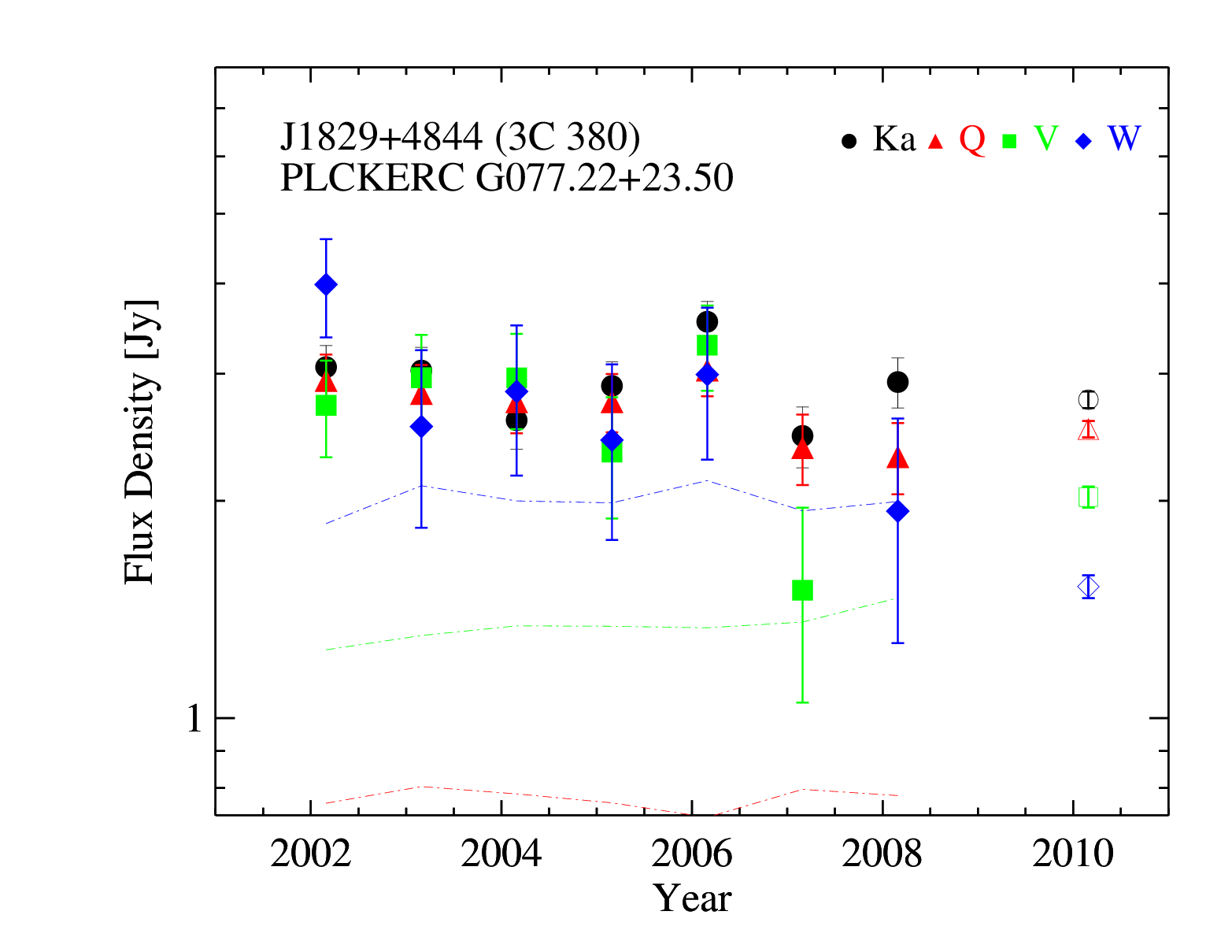}  \\
\includegraphics[width=0.23\textwidth]{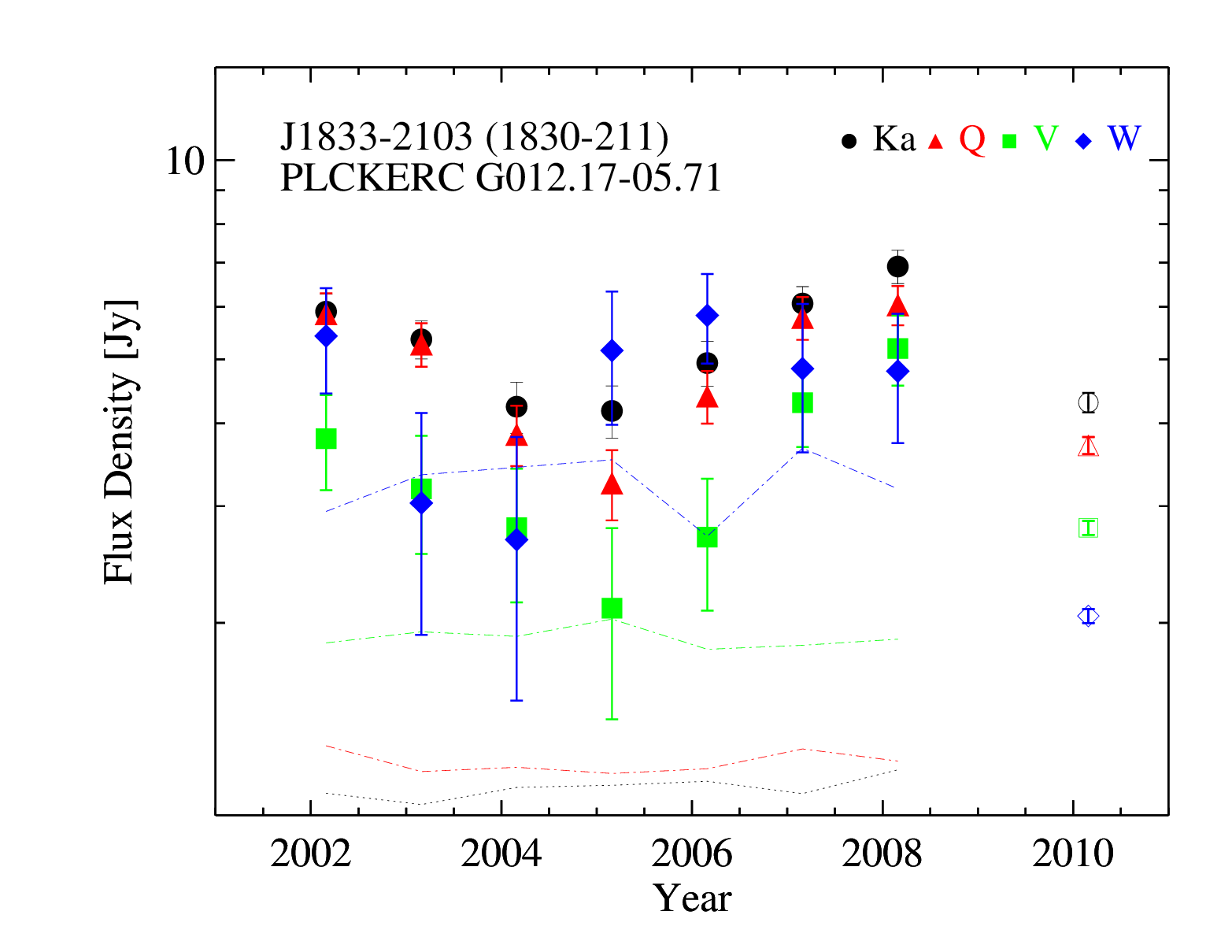} & \includegraphics[width=0.23\textwidth]{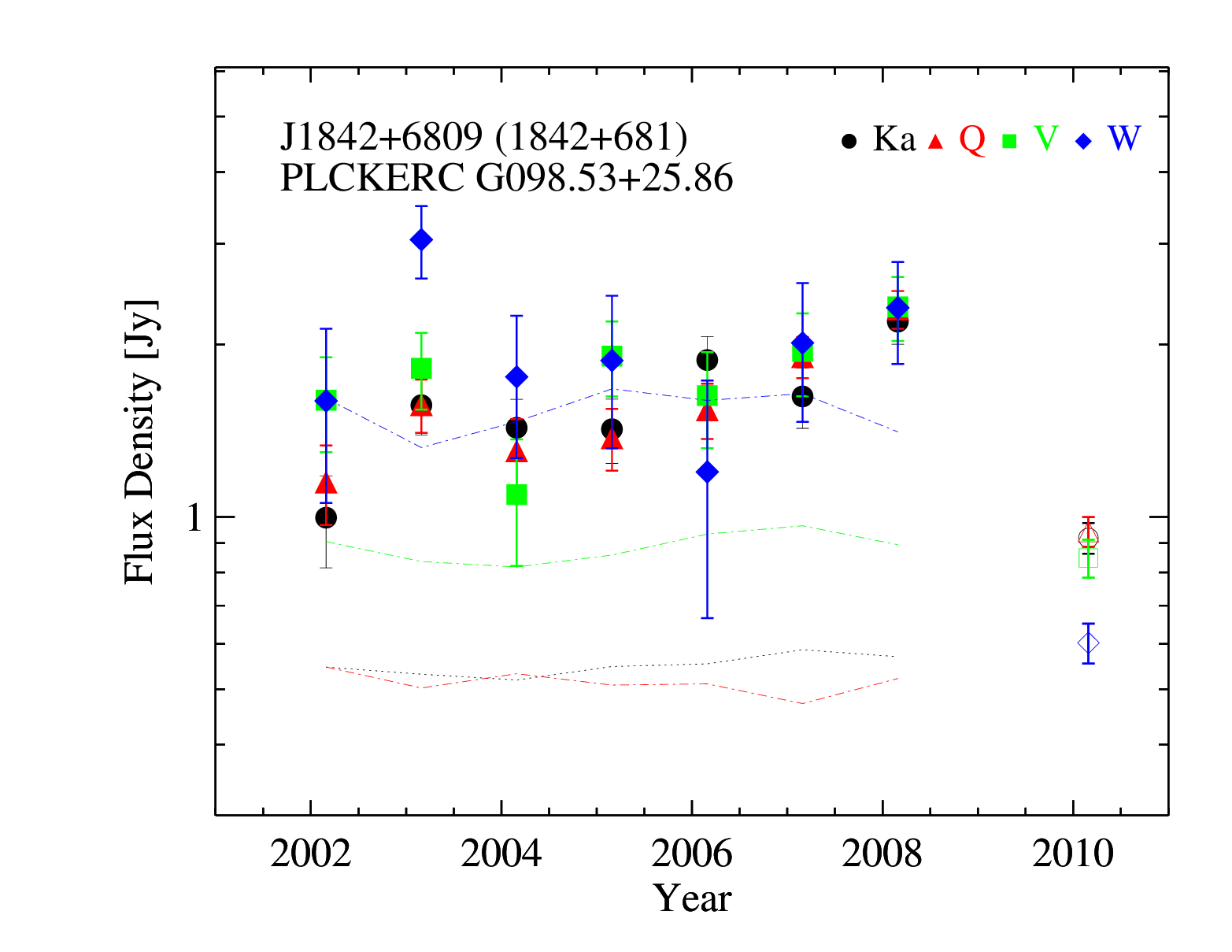}  & \includegraphics[width=0.23\textwidth]{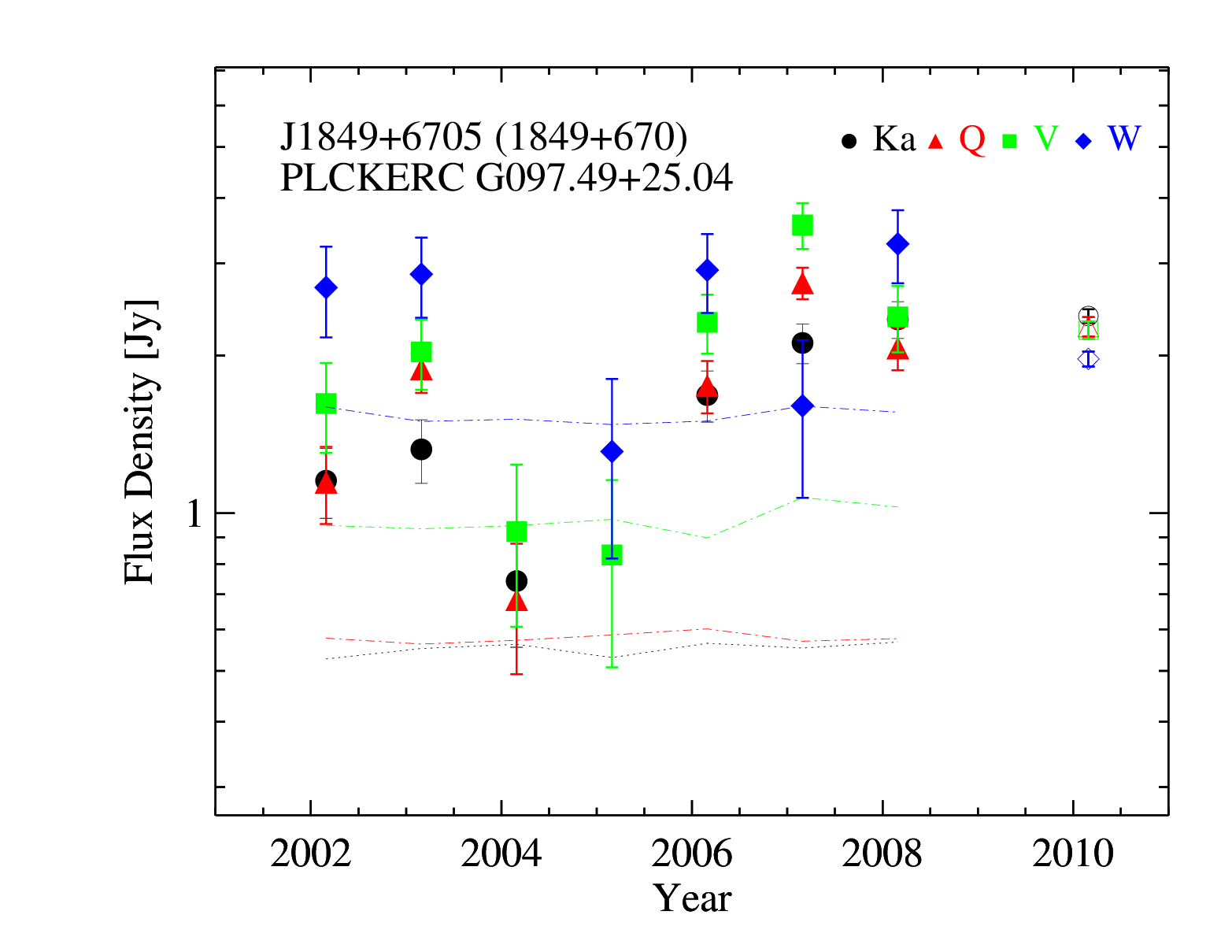} & \includegraphics[width=0.23\textwidth]{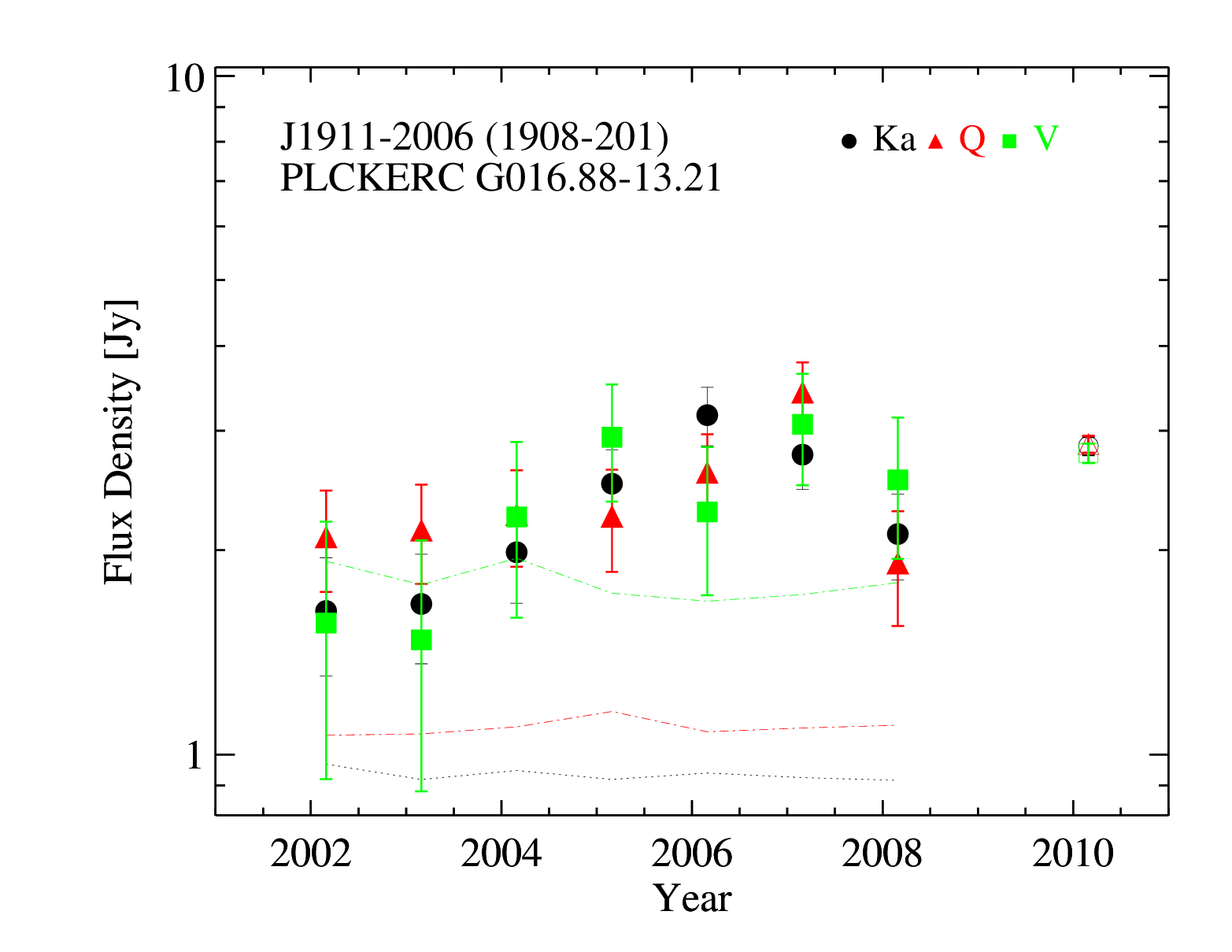}  \\
\includegraphics[width=0.23\textwidth]{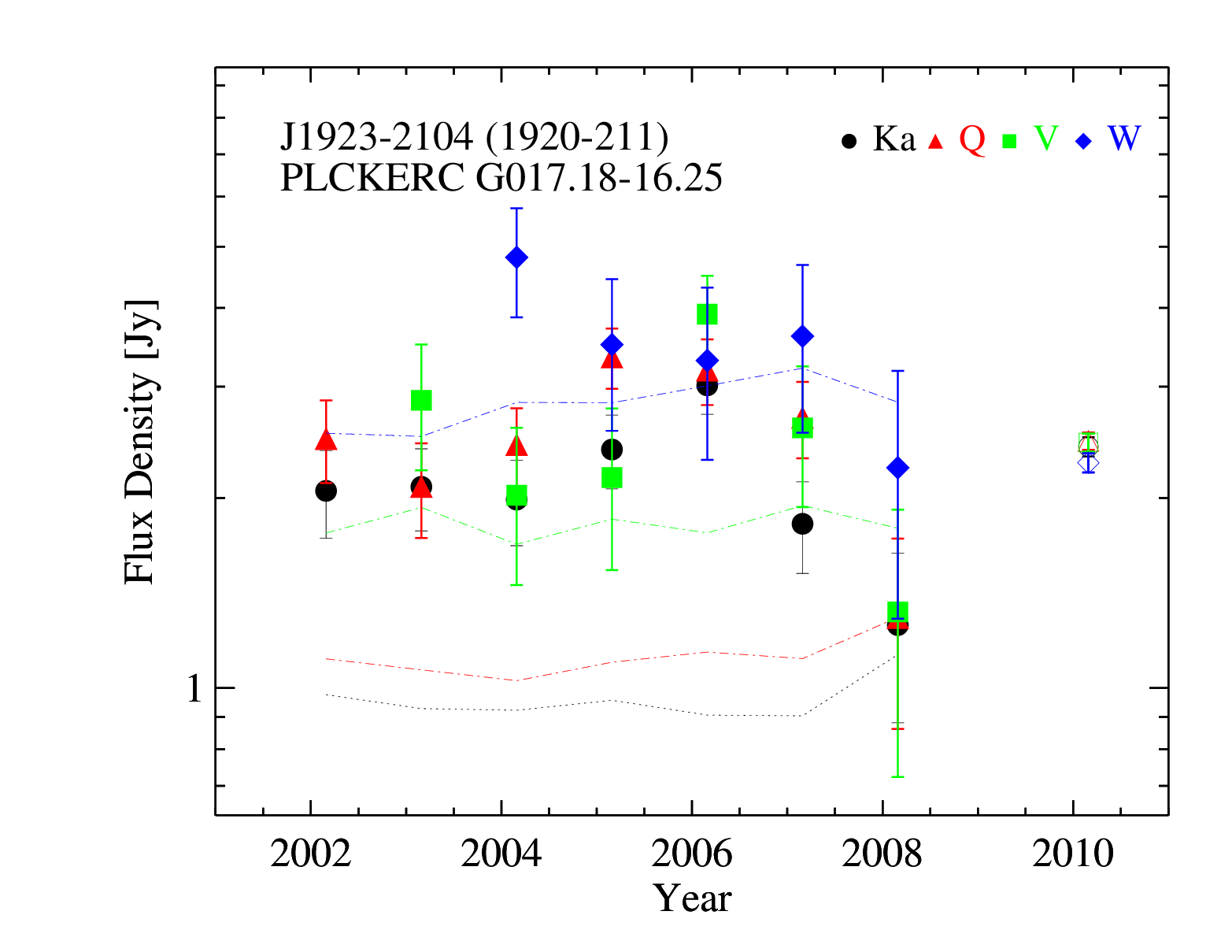} & \includegraphics[width=0.23\textwidth]{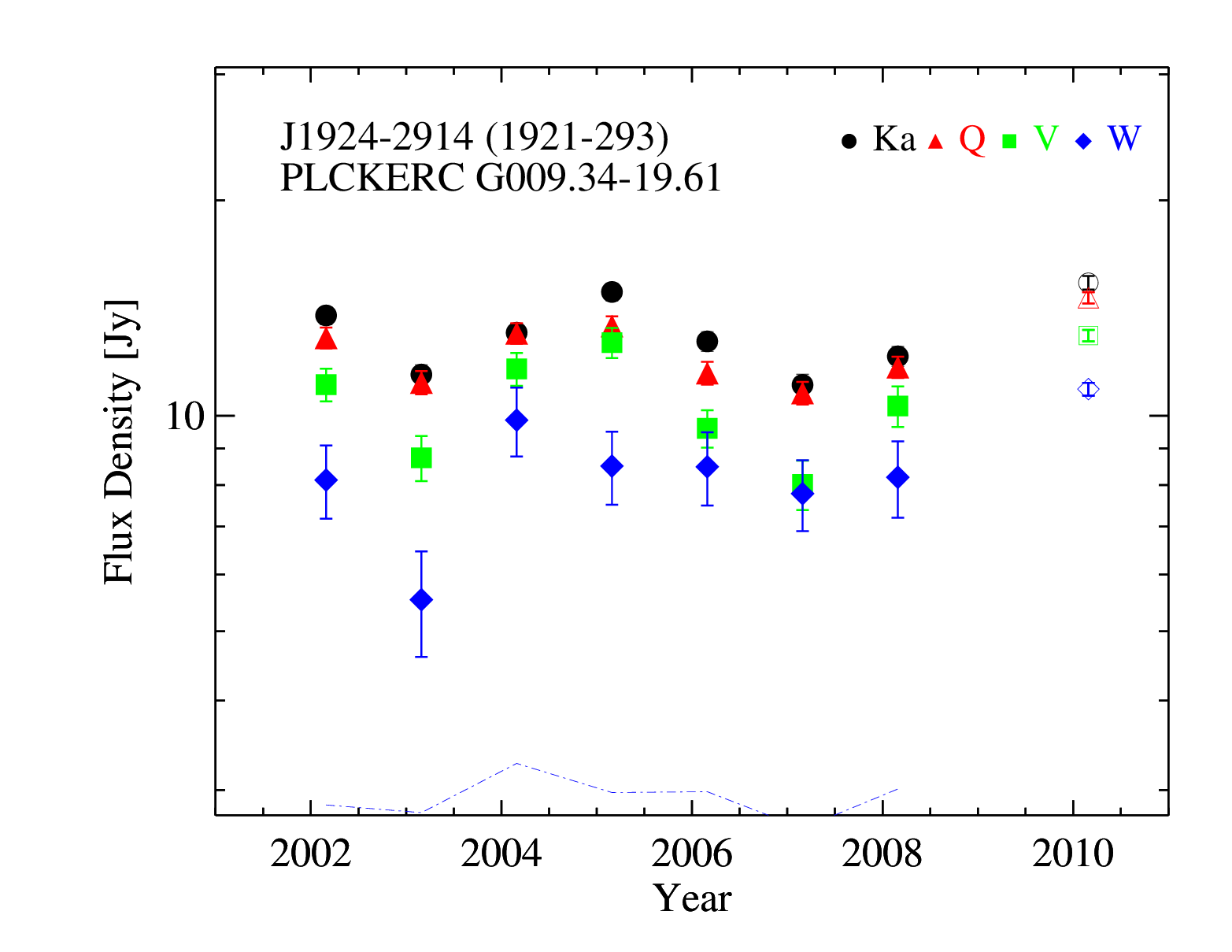}  & \includegraphics[width=0.23\textwidth]{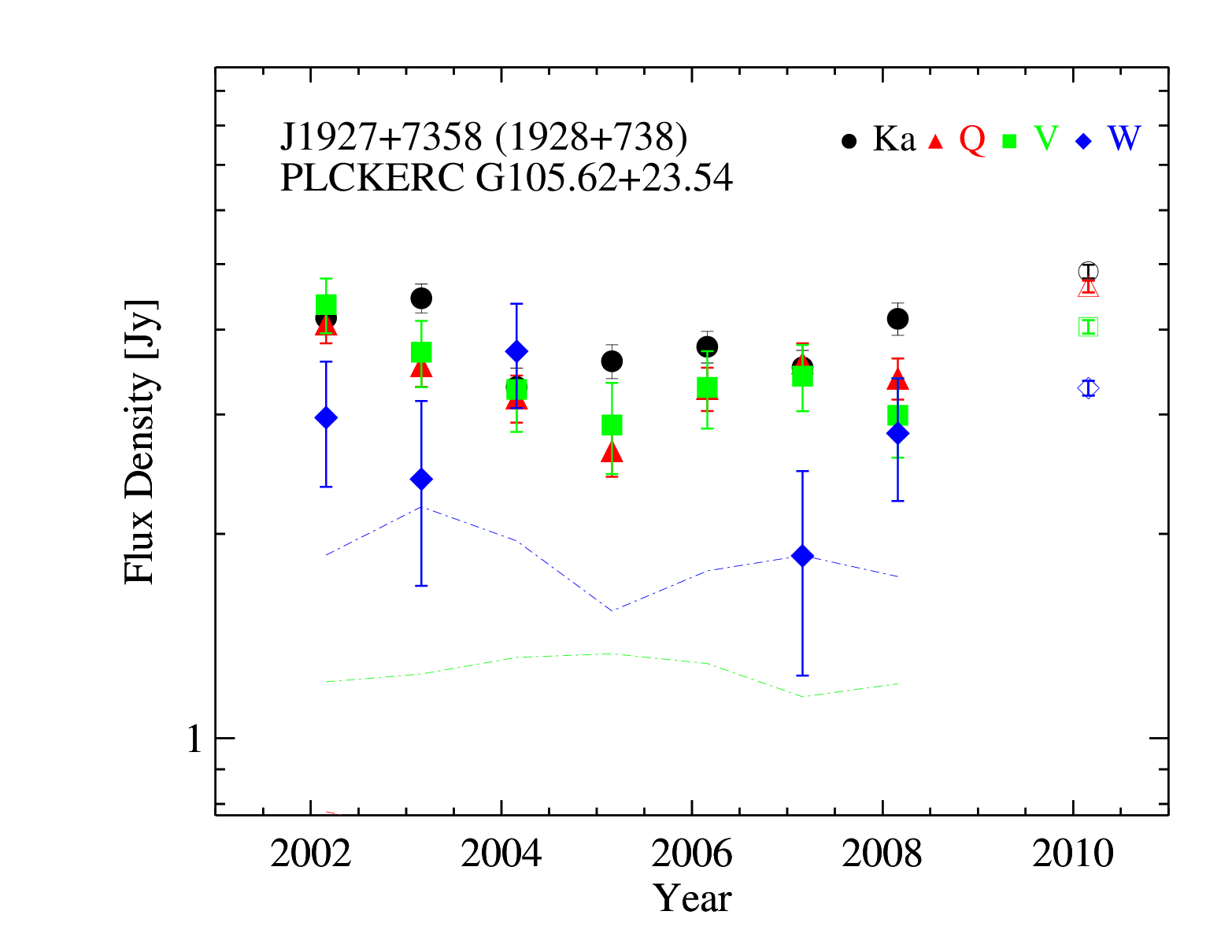} & \includegraphics[width=0.23\textwidth]{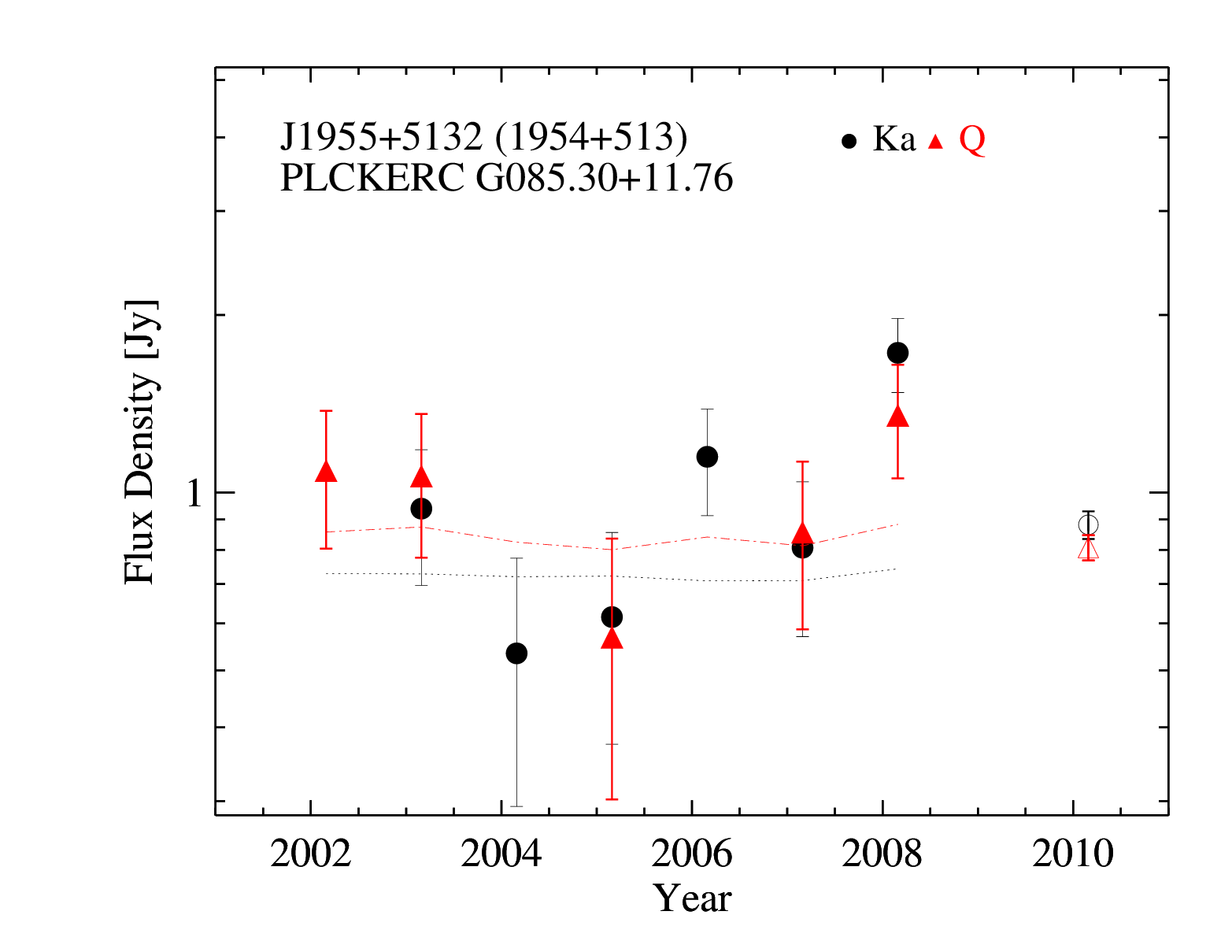}  \\
\end{tabular}
\end{figure*}

\begin{figure*}
\centering
\begin{tabular}{cccc}
\includegraphics[width=0.23\textwidth]{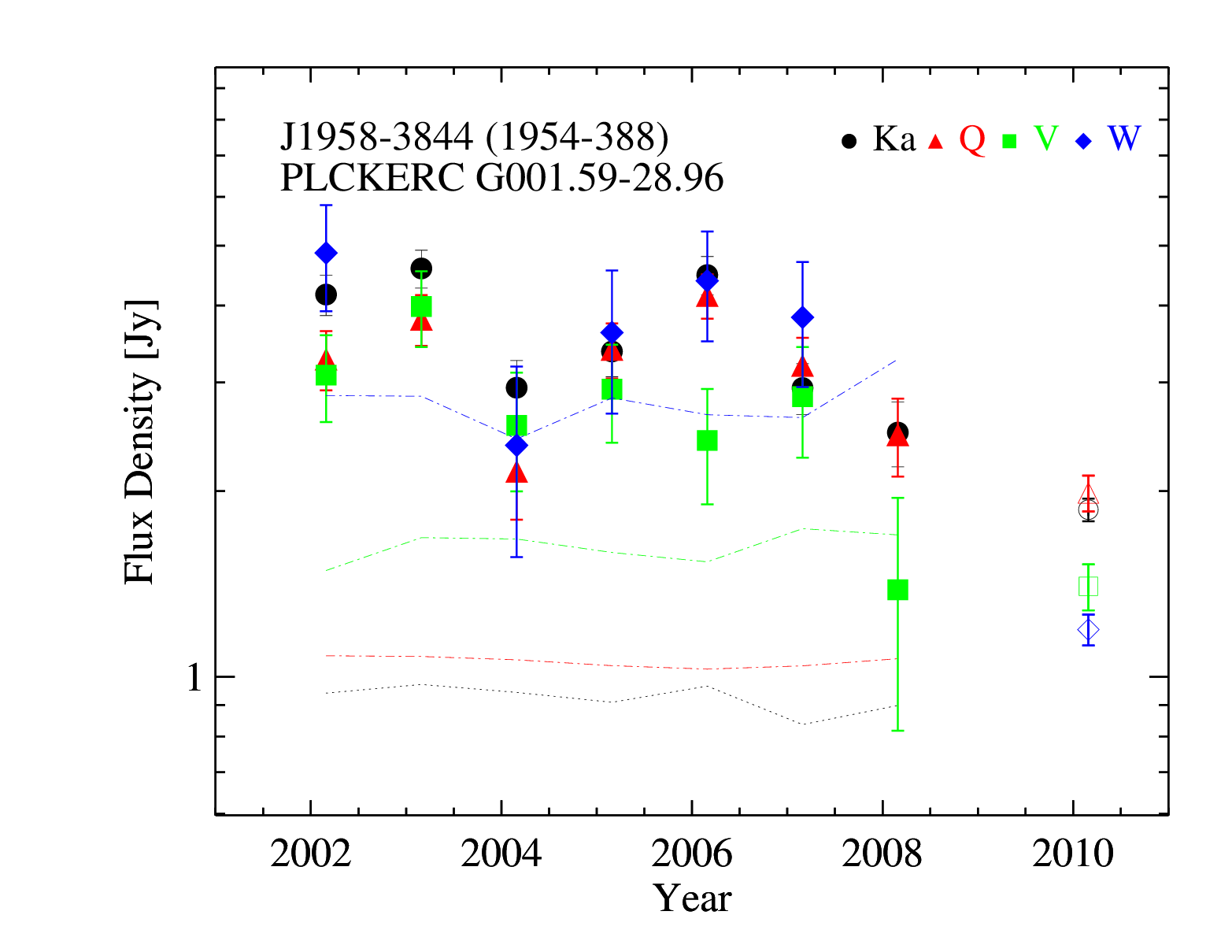} & \includegraphics[width=0.23\textwidth]{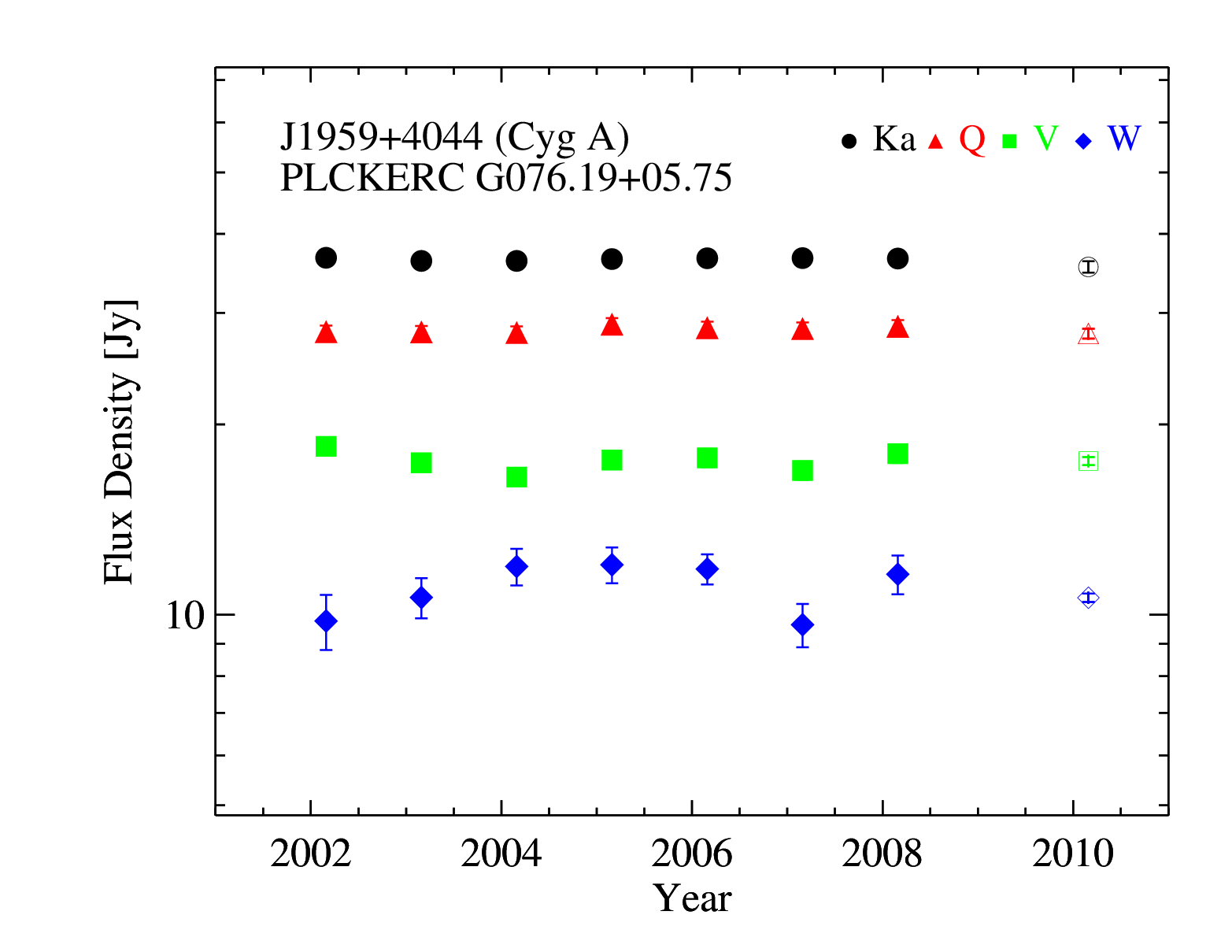}  & \includegraphics[width=0.23\textwidth]{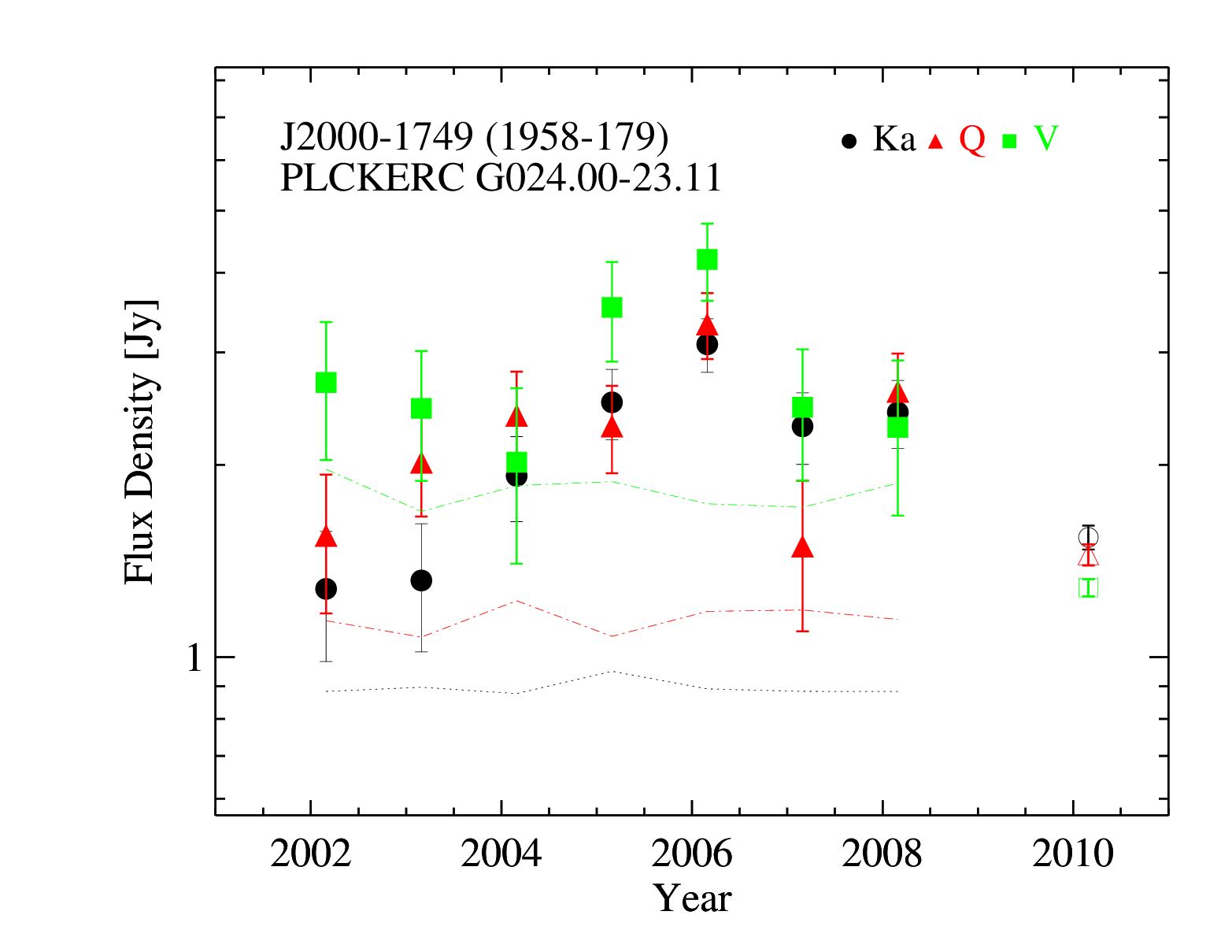} & \includegraphics[width=0.23\textwidth]{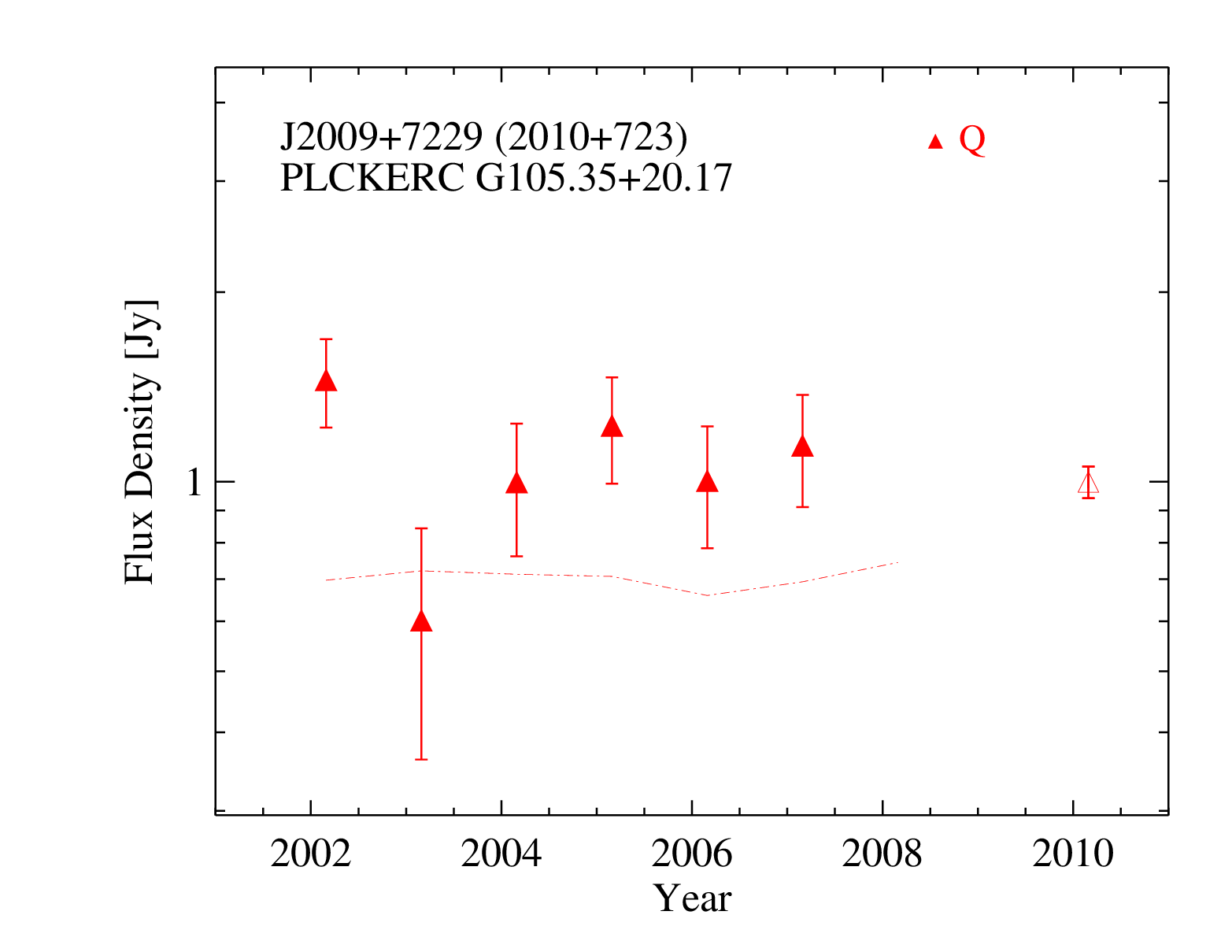}  \\
\includegraphics[width=0.23\textwidth]{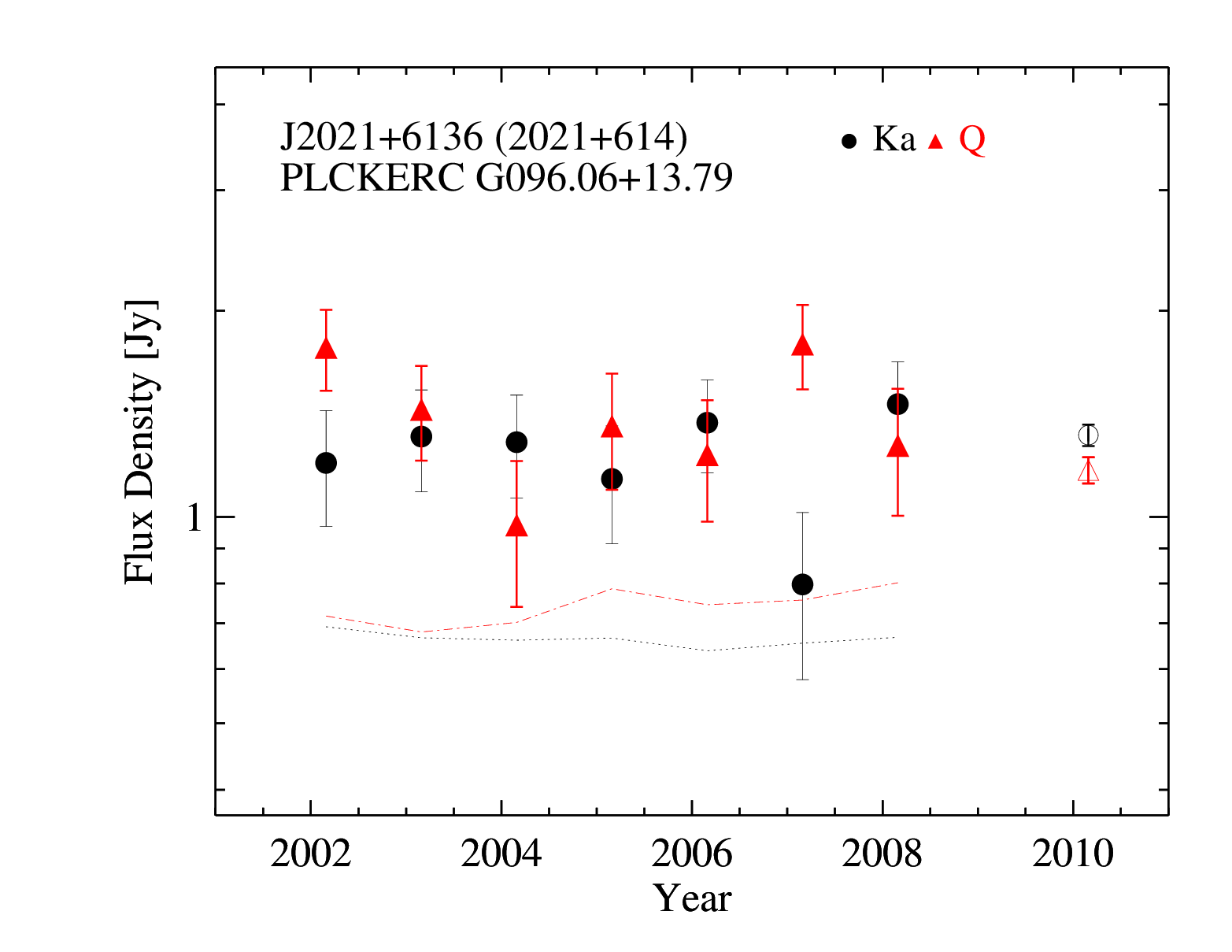} & \includegraphics[width=0.23\textwidth]{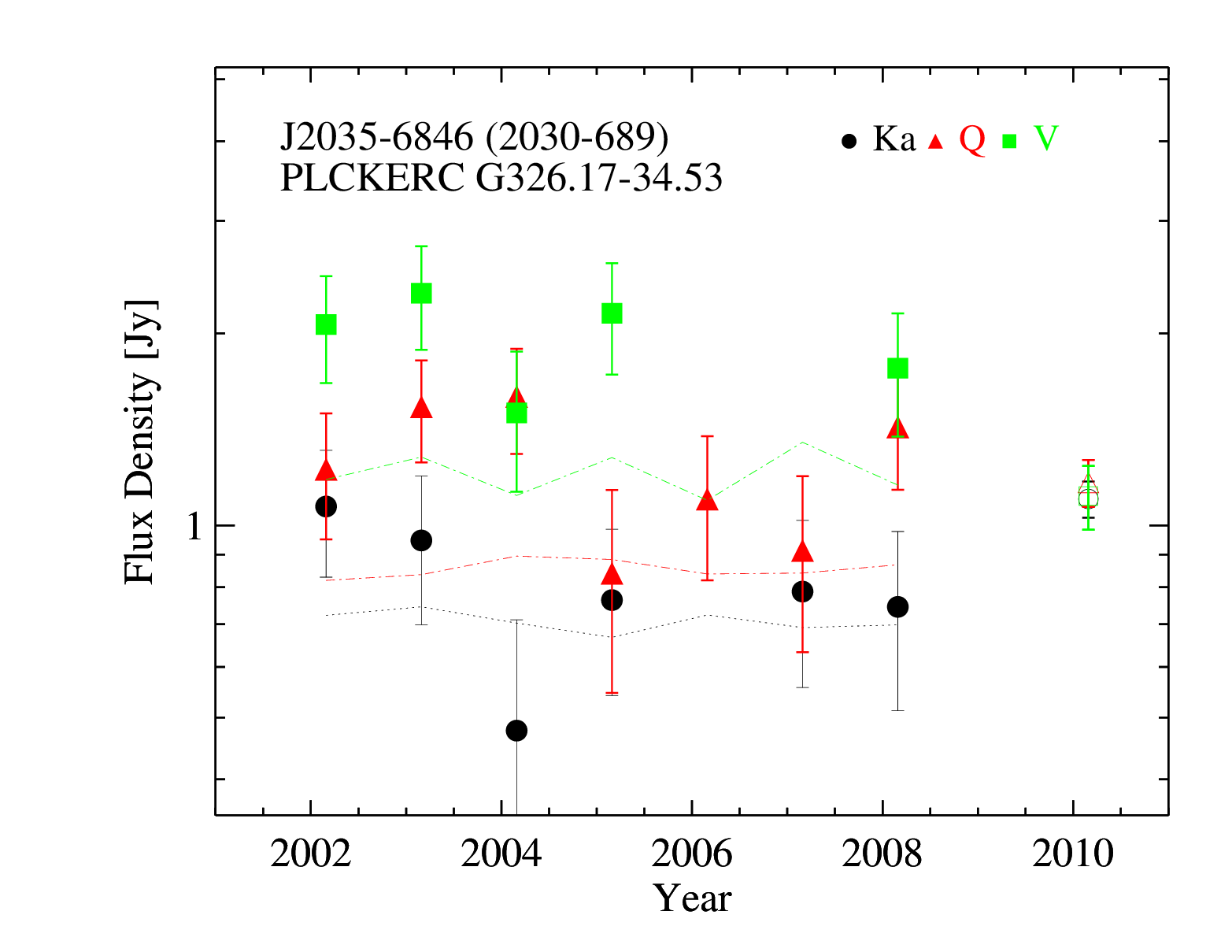}  & \includegraphics[width=0.23\textwidth]{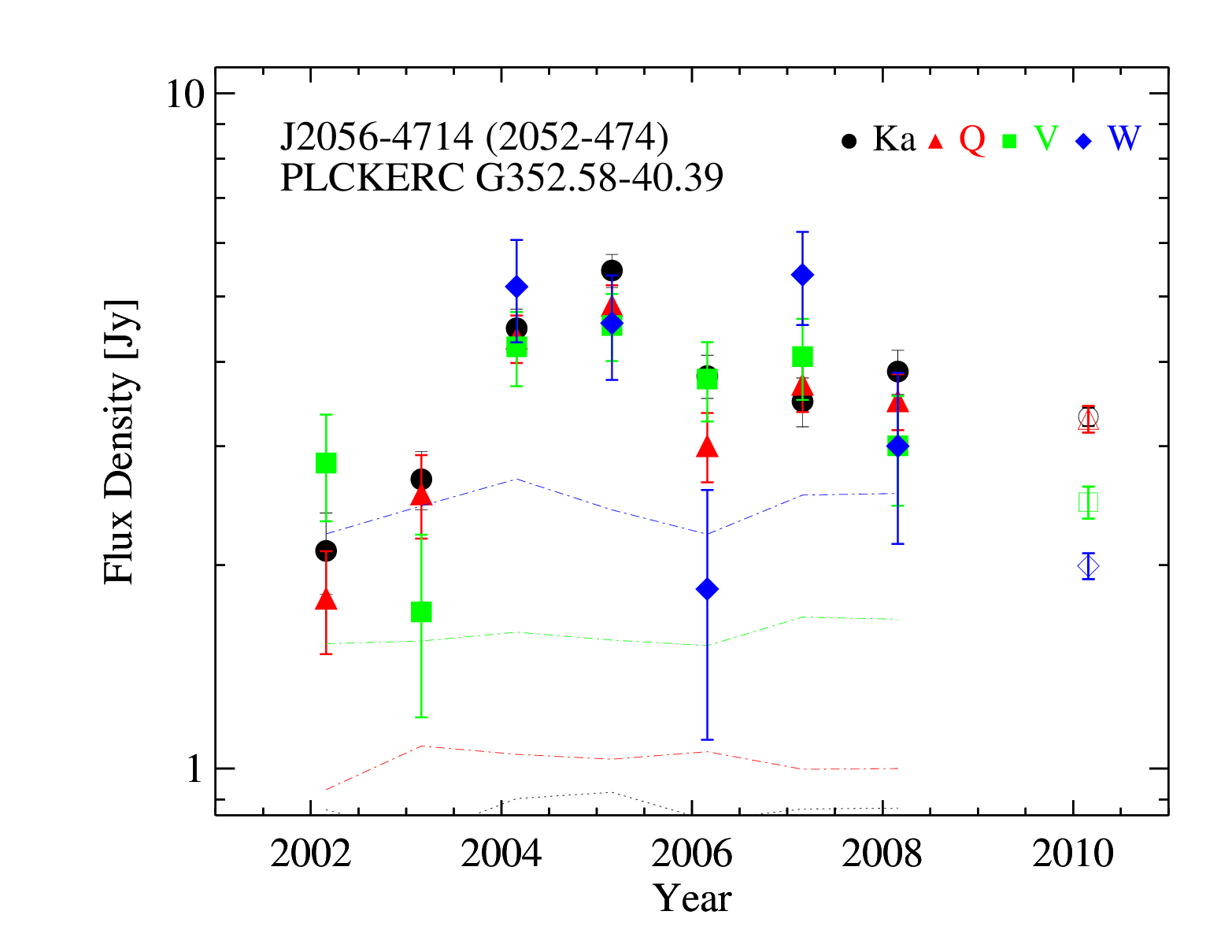} & \includegraphics[width=0.23\textwidth]{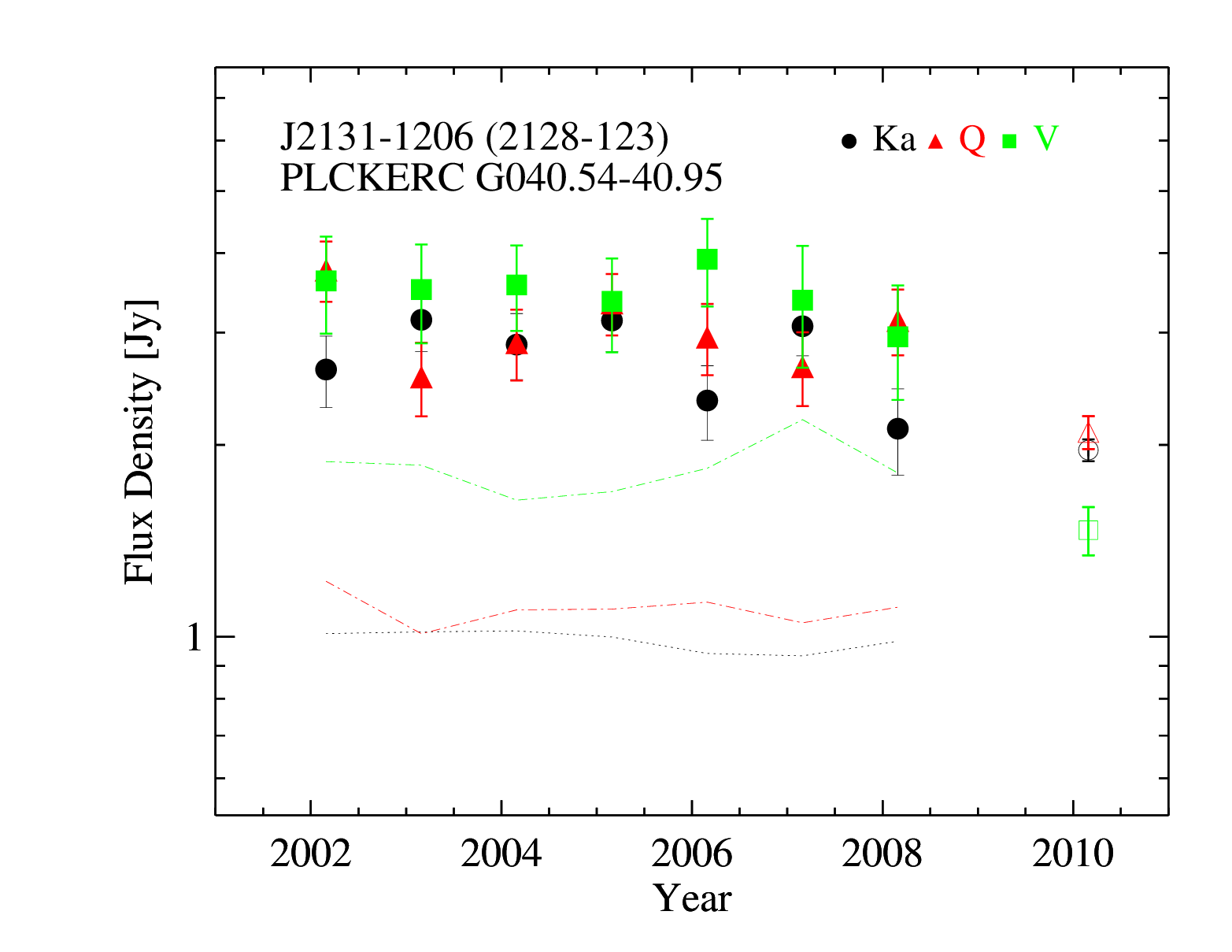}  \\
\includegraphics[width=0.23\textwidth]{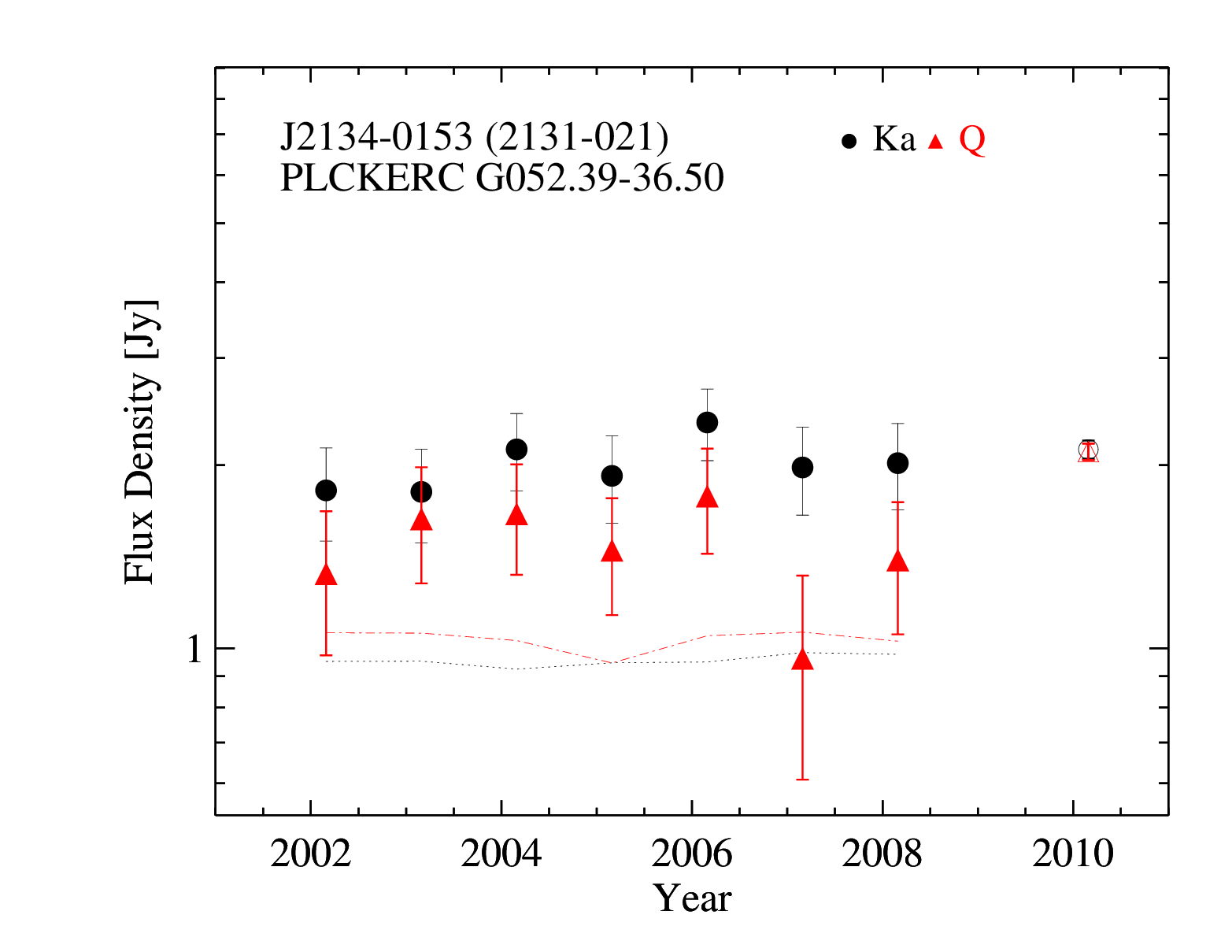} & \includegraphics[width=0.23\textwidth]{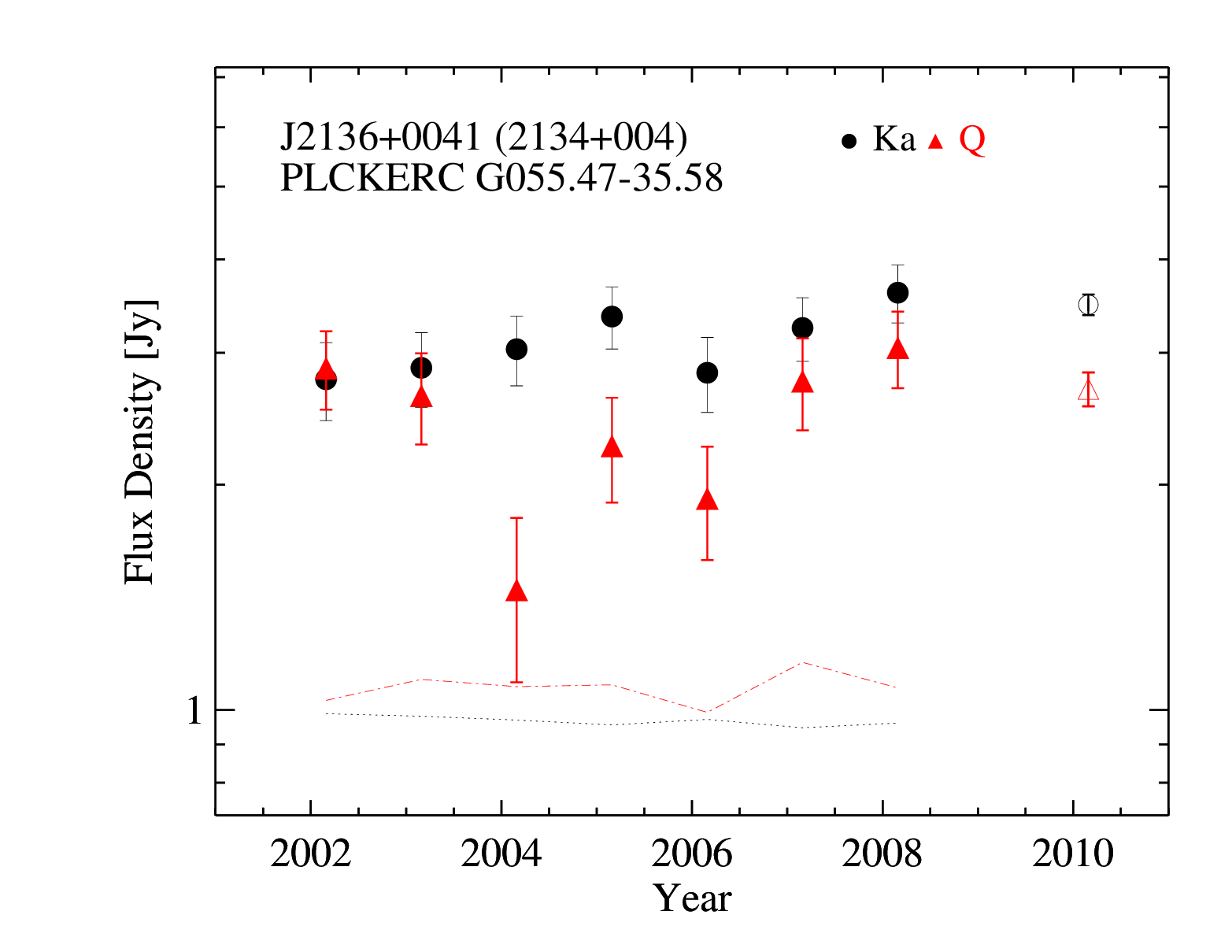}  & \includegraphics[width=0.23\textwidth]{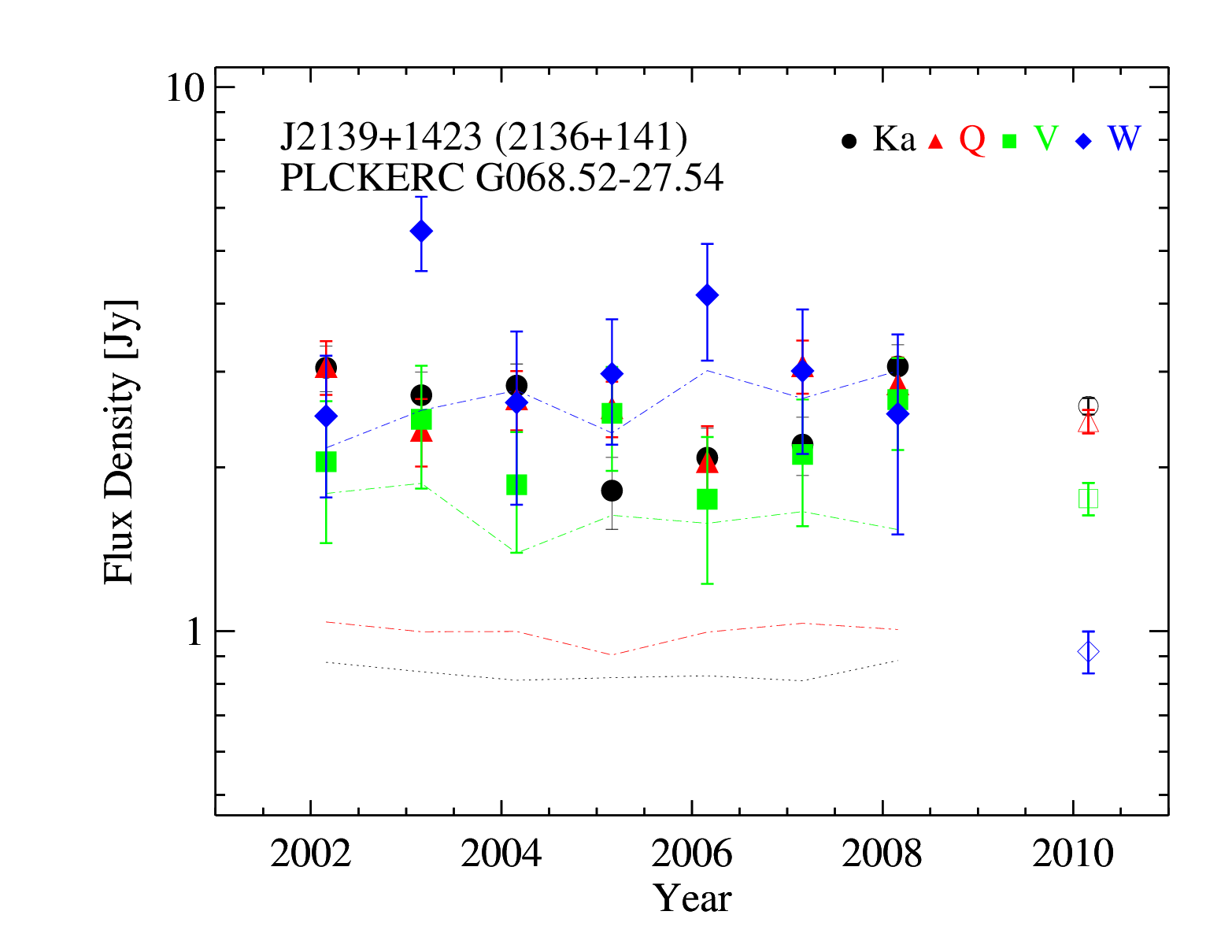} & \includegraphics[width=0.23\textwidth]{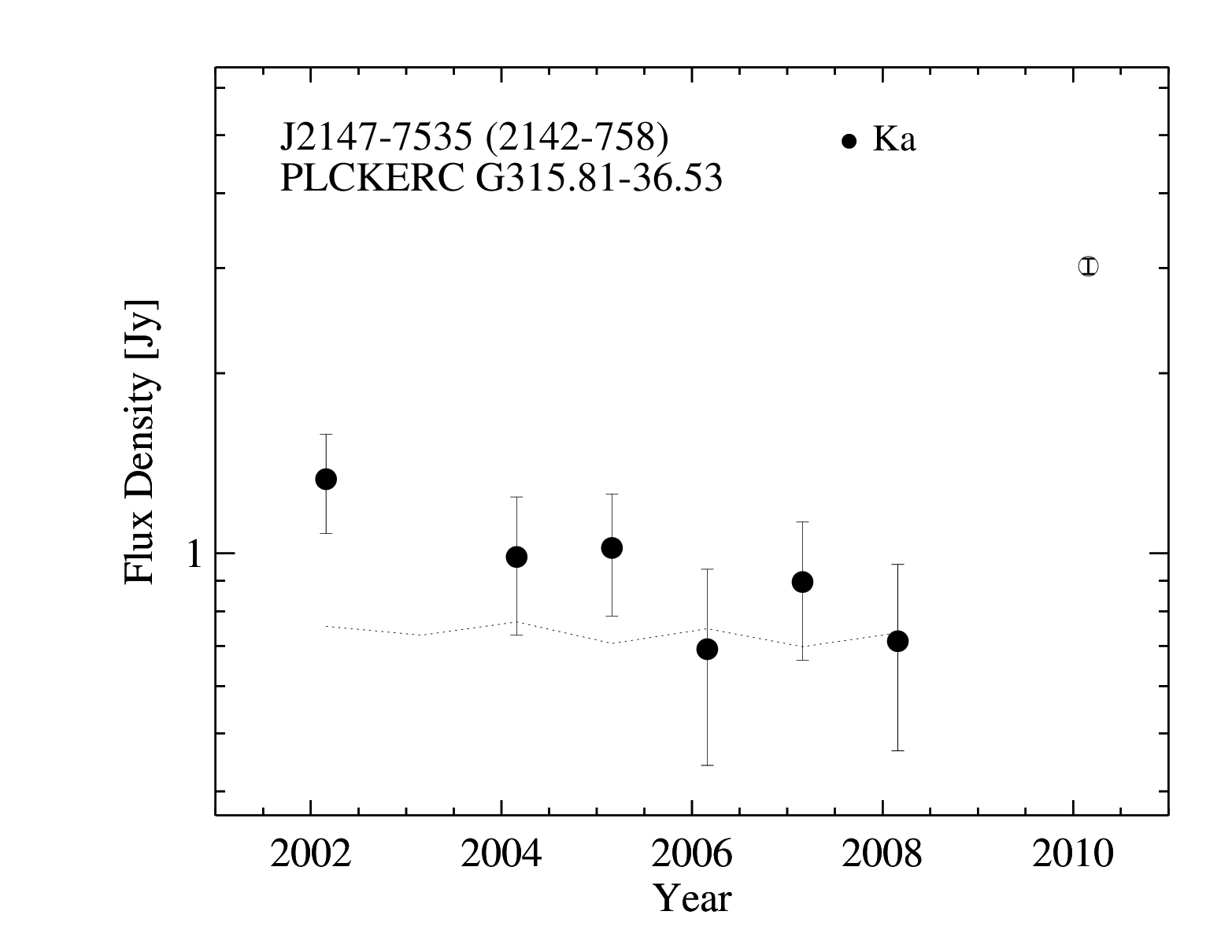}  \\
\includegraphics[width=0.23\textwidth]{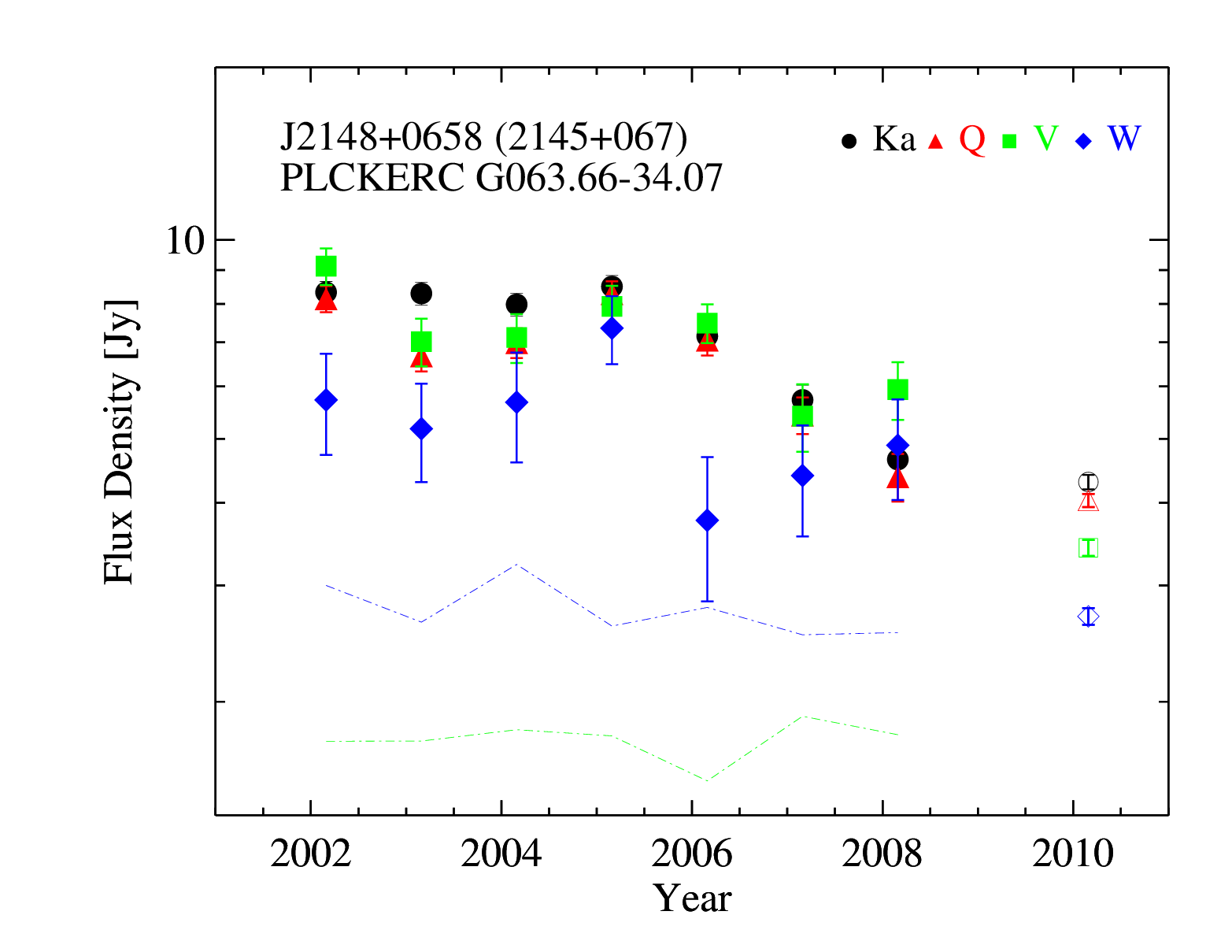} & \includegraphics[width=0.23\textwidth]{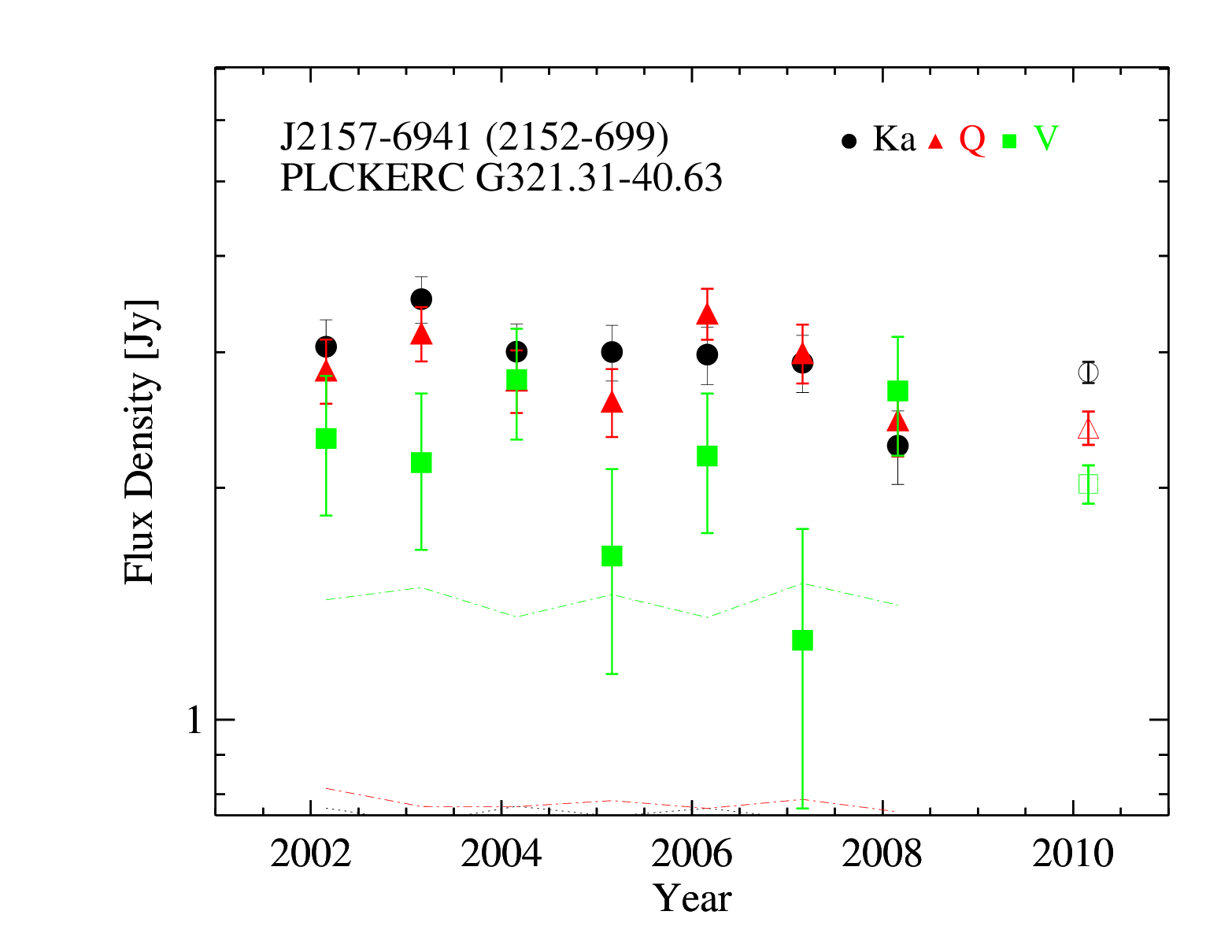}  & \includegraphics[width=0.23\textwidth]{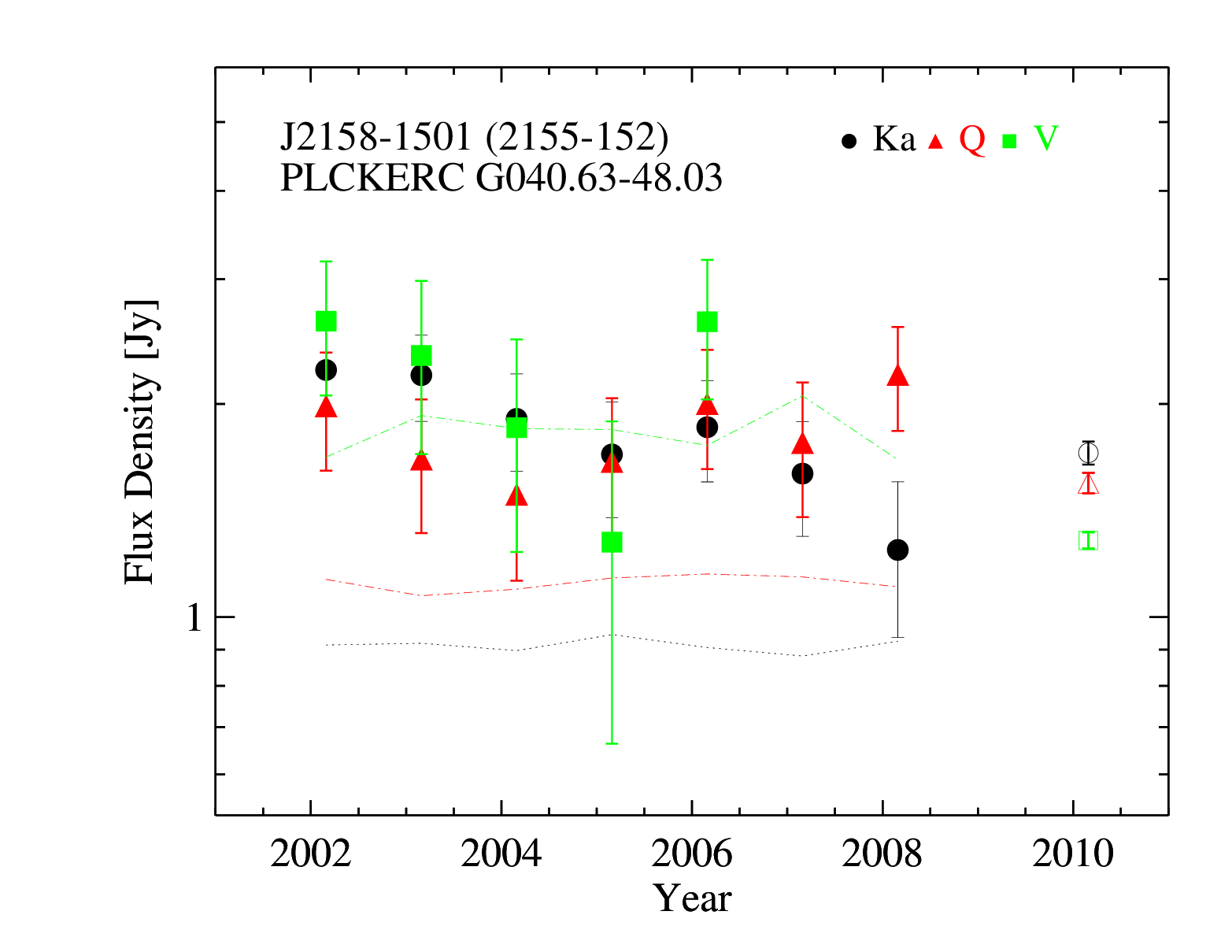} & \includegraphics[width=0.23\textwidth]{figures/lc/J2202+4217.eps}  \\
\includegraphics[width=0.23\textwidth]{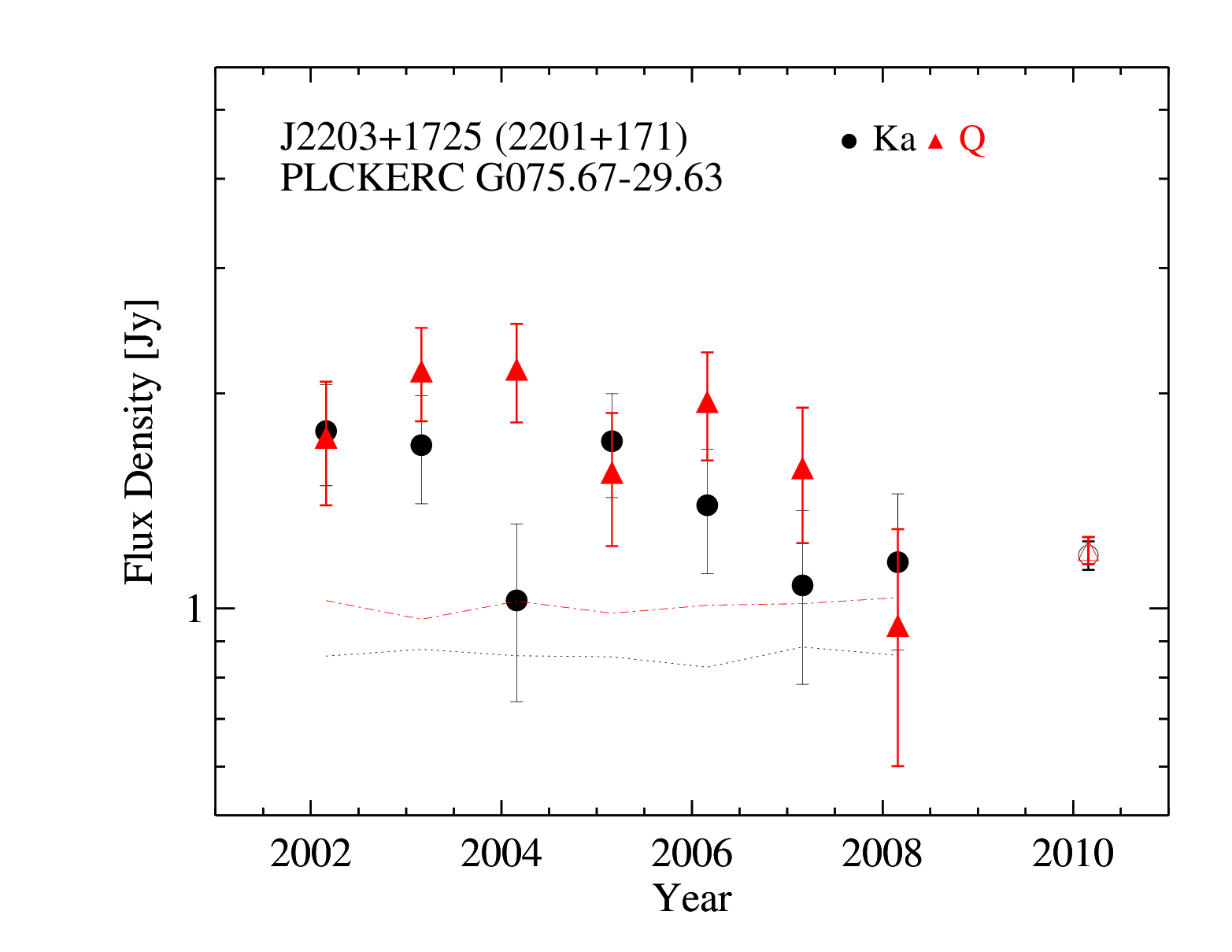} & \includegraphics[width=0.23\textwidth]{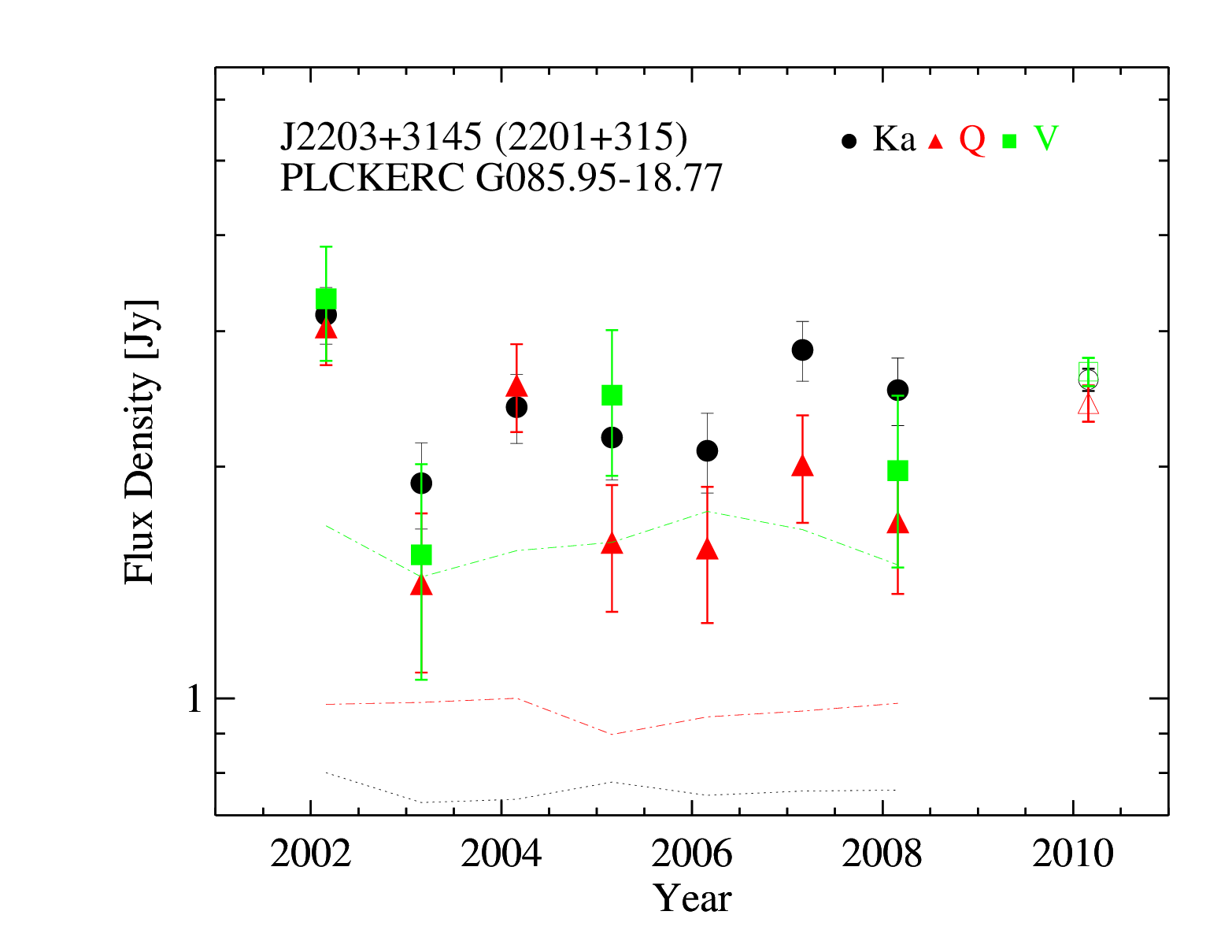}  & \includegraphics[width=0.23\textwidth]{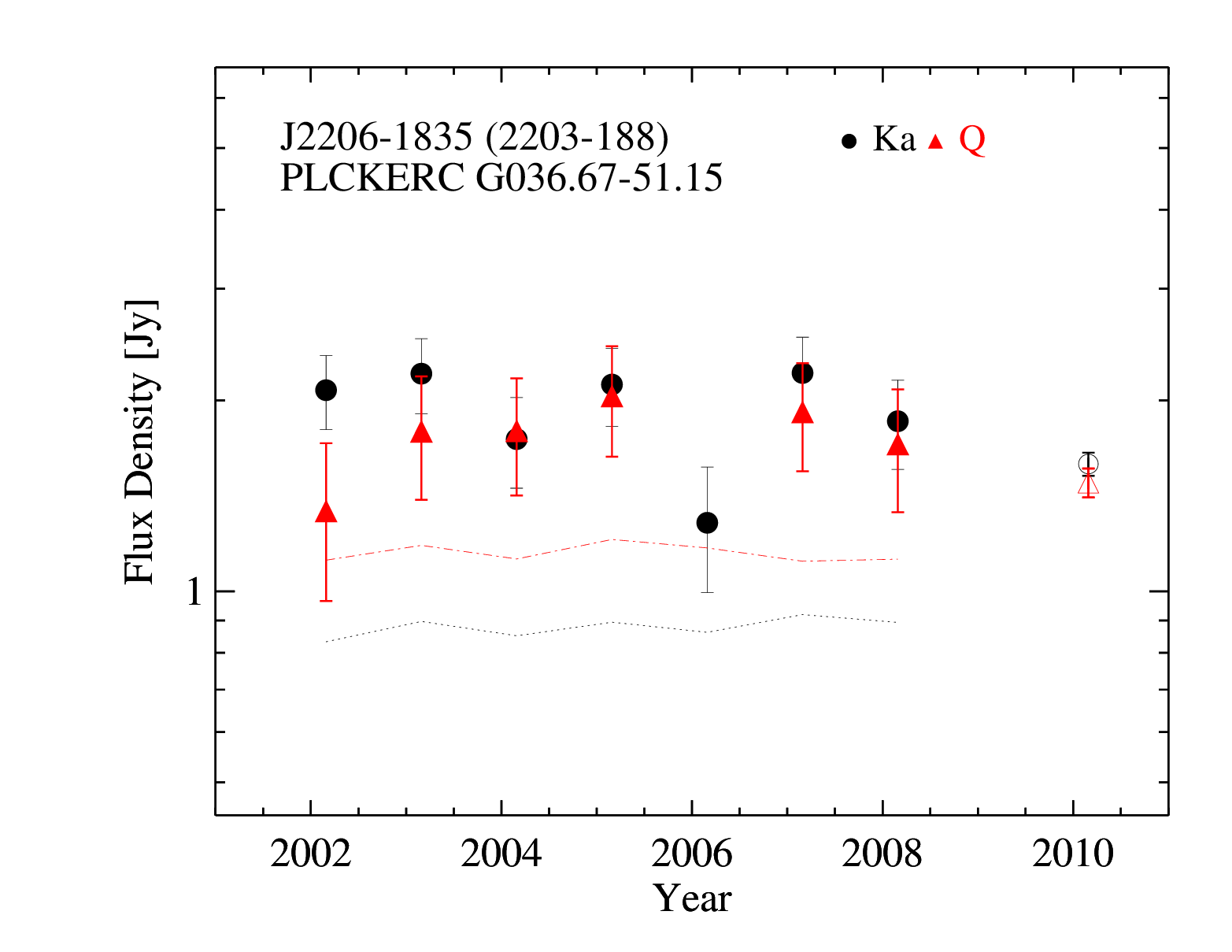} & \includegraphics[width=0.23\textwidth]{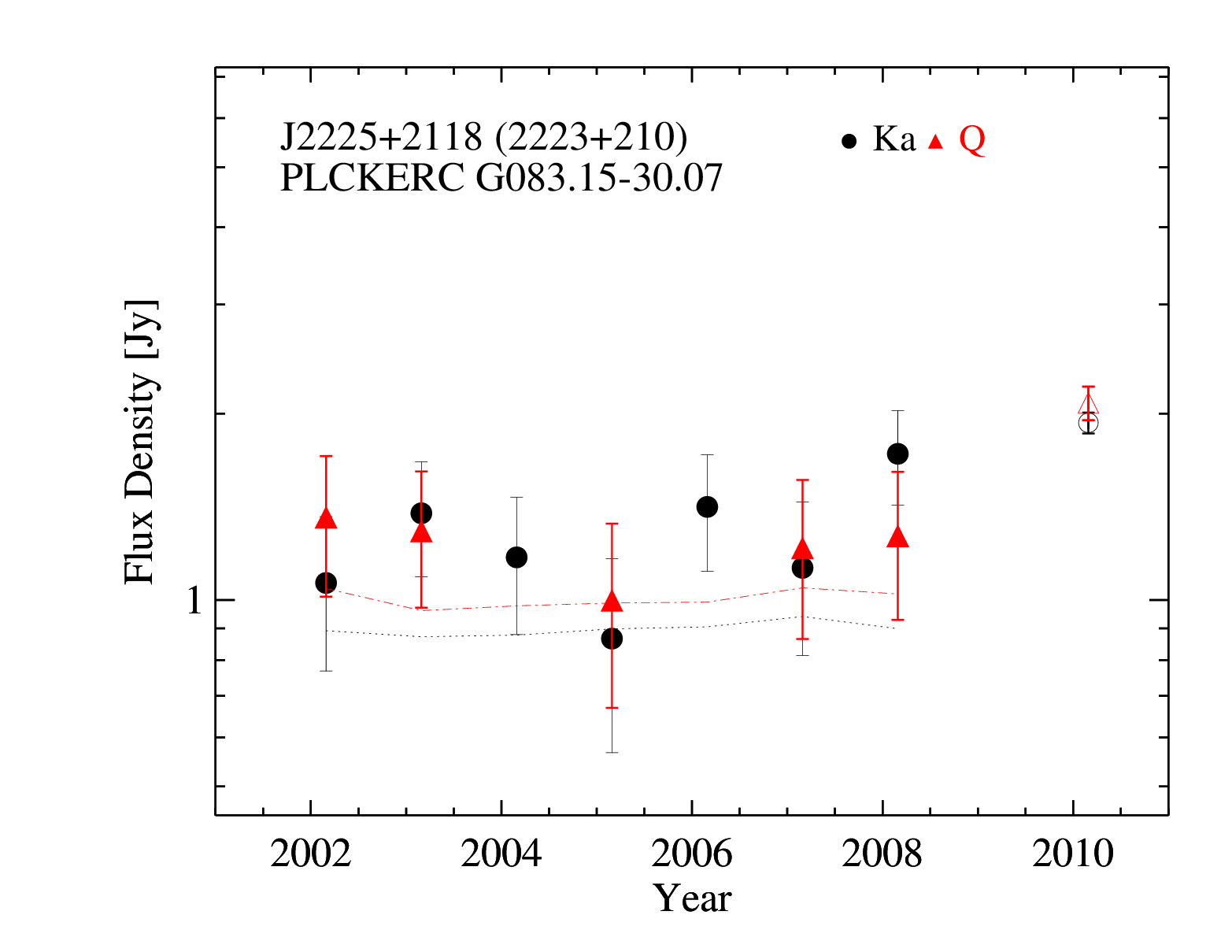}  \\
\includegraphics[width=0.23\textwidth]{figures/lc/J2225-0457.eps} & \includegraphics[width=0.23\textwidth]{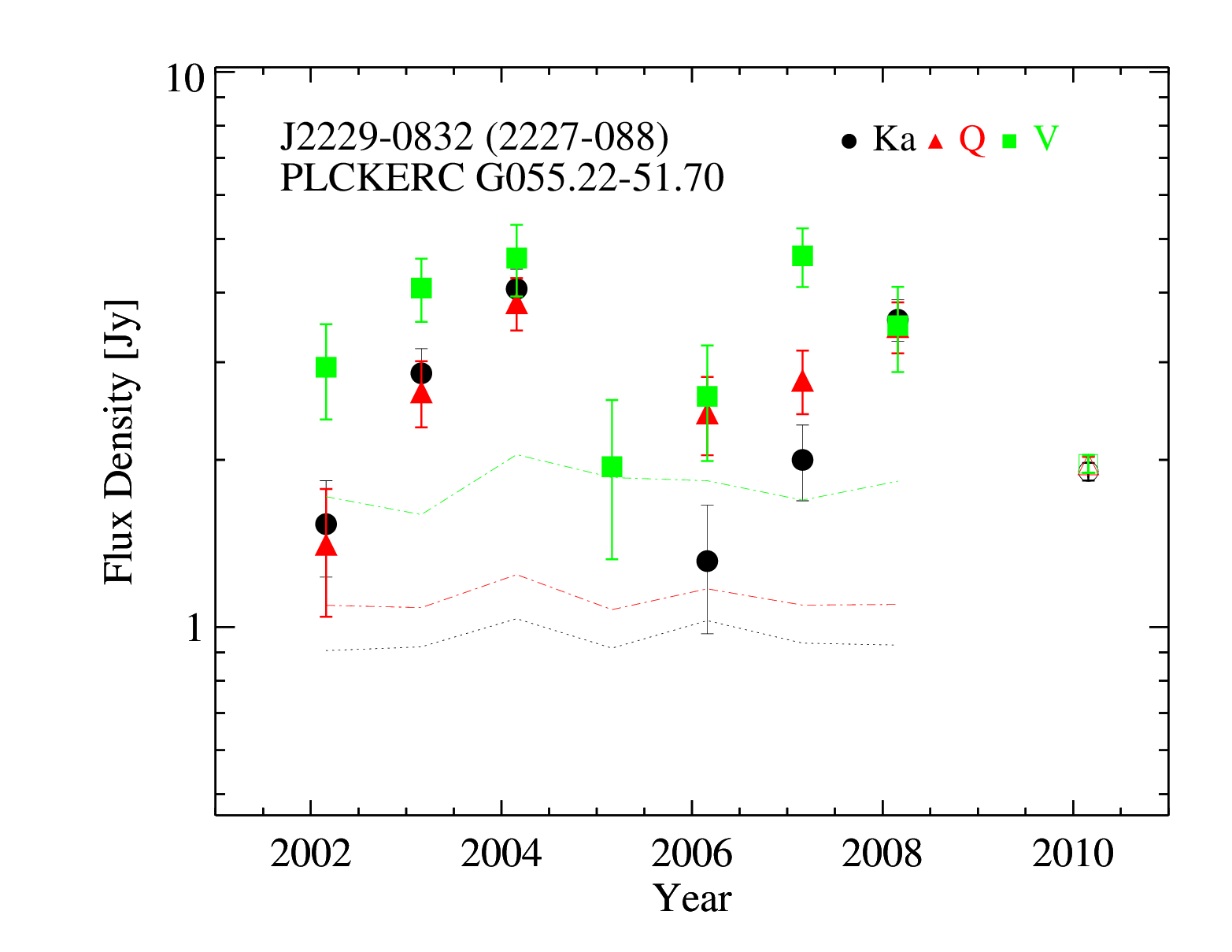}  & \includegraphics[width=0.23\textwidth]{figures/lc/J2232+1143.eps} & \includegraphics[width=0.23\textwidth]{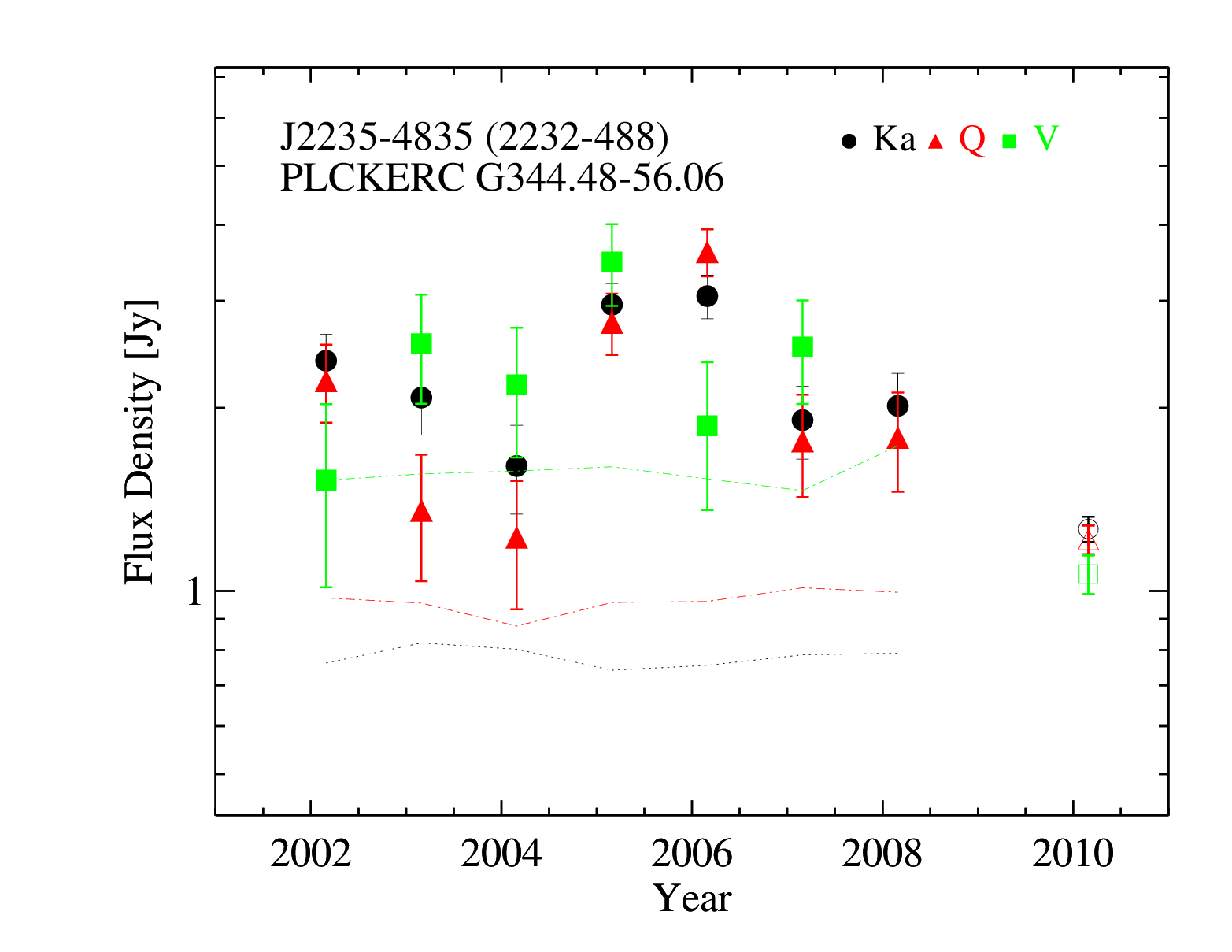}  \\
\includegraphics[width=0.23\textwidth]{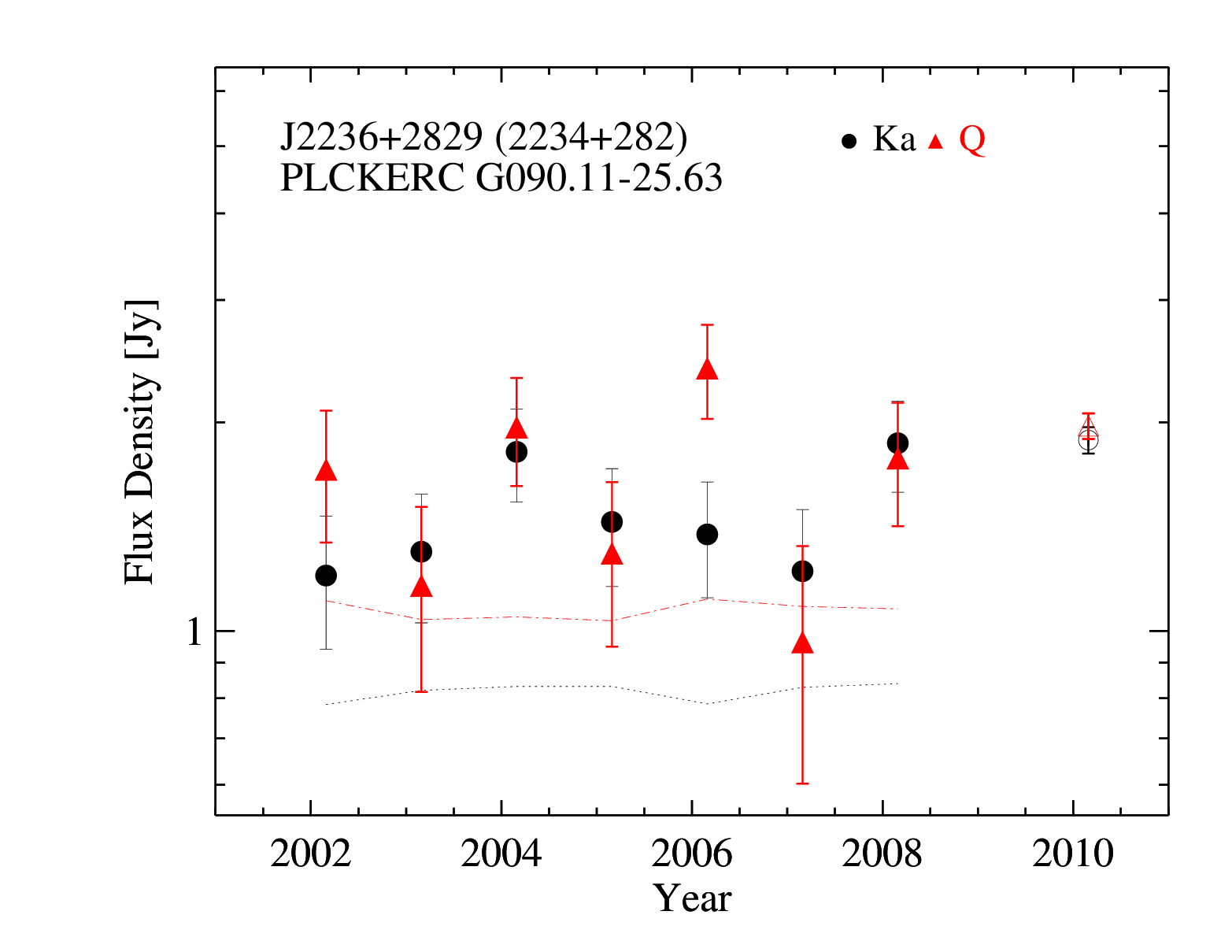} & \includegraphics[width=0.23\textwidth]{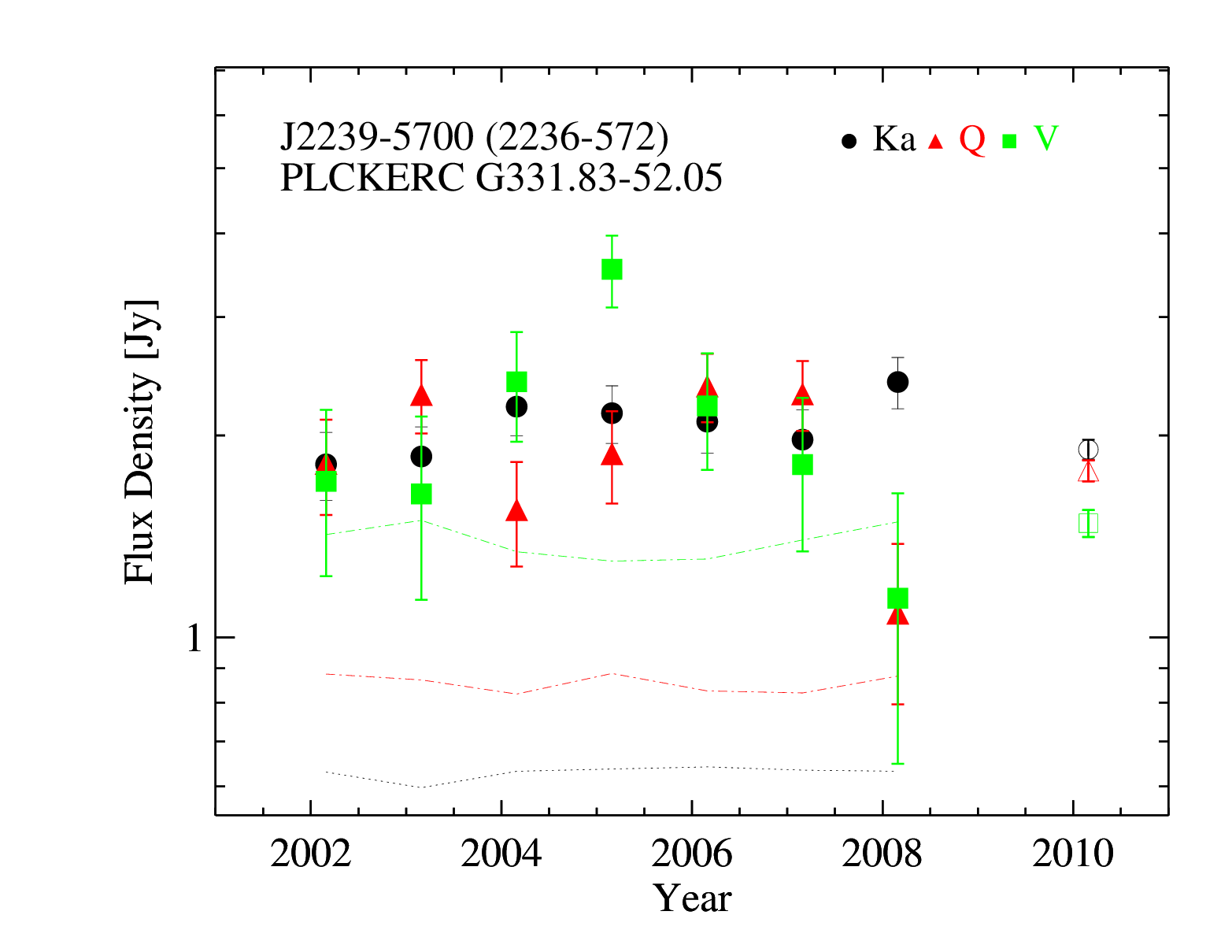}  & \includegraphics[width=0.23\textwidth]{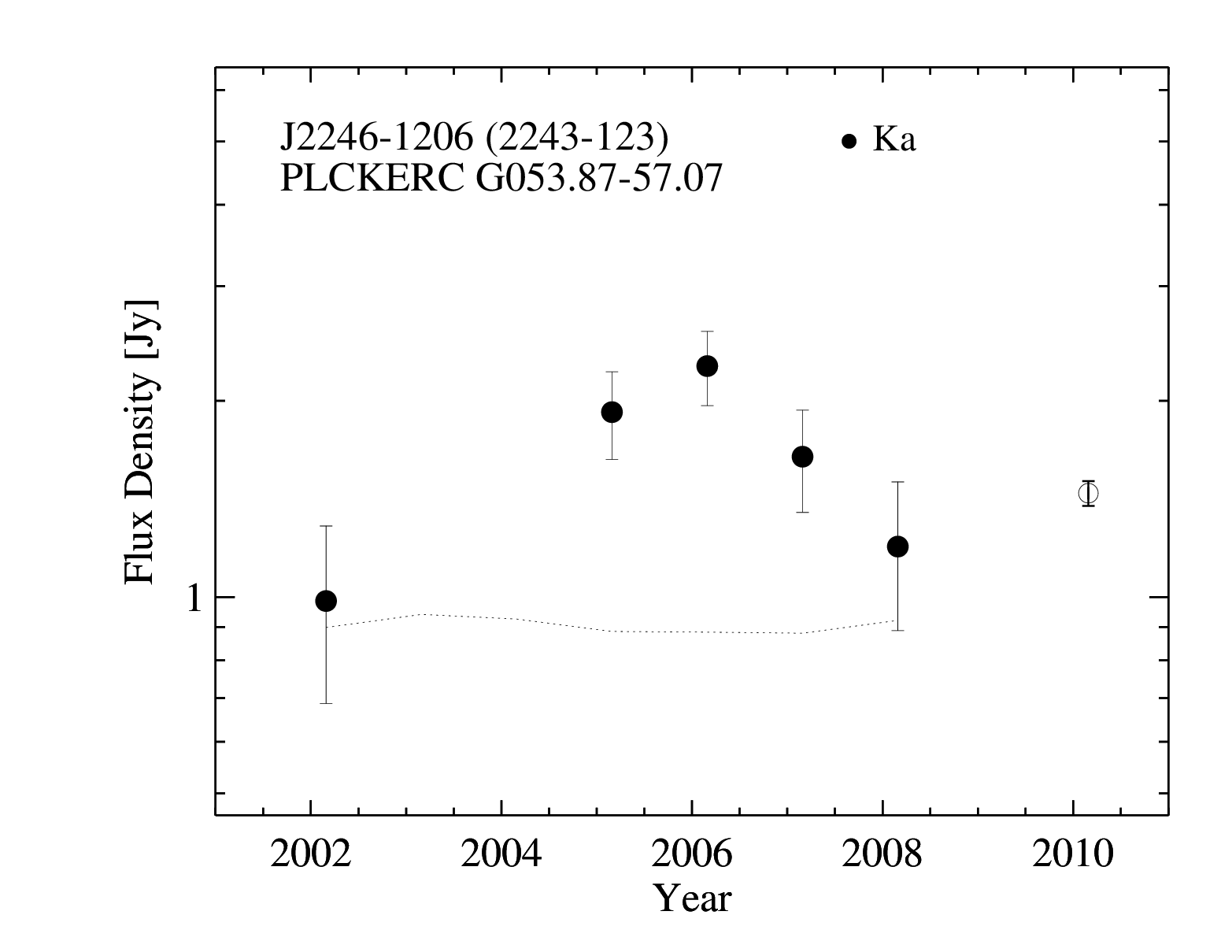} & \includegraphics[width=0.23\textwidth]{figures/lc/J2253+1609.eps}  \\
\end{tabular}
\end{figure*}

\begin{figure*}
\centering
\begin{tabular}{cccc}
\includegraphics[width=0.23\textwidth]{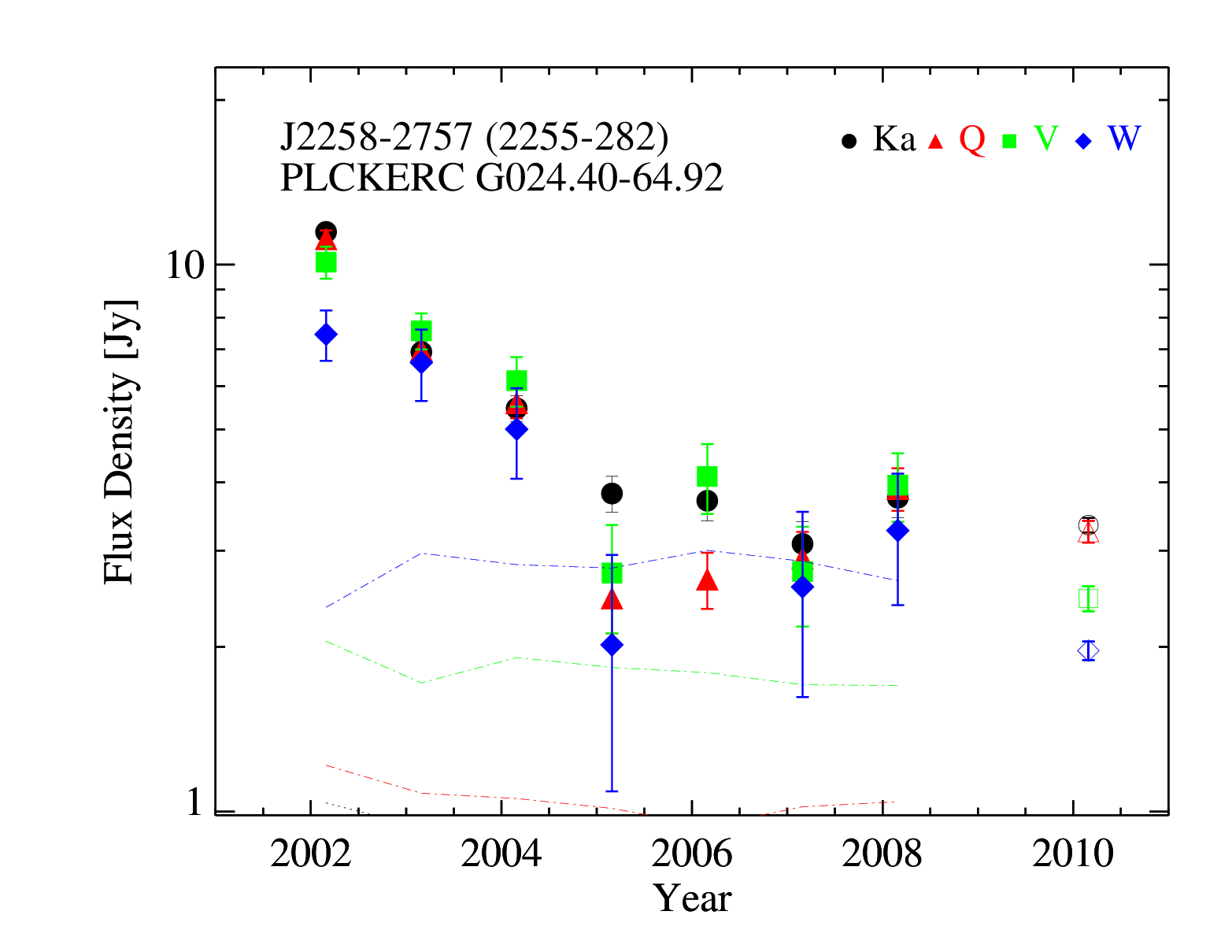} & \includegraphics[width=0.23\textwidth]{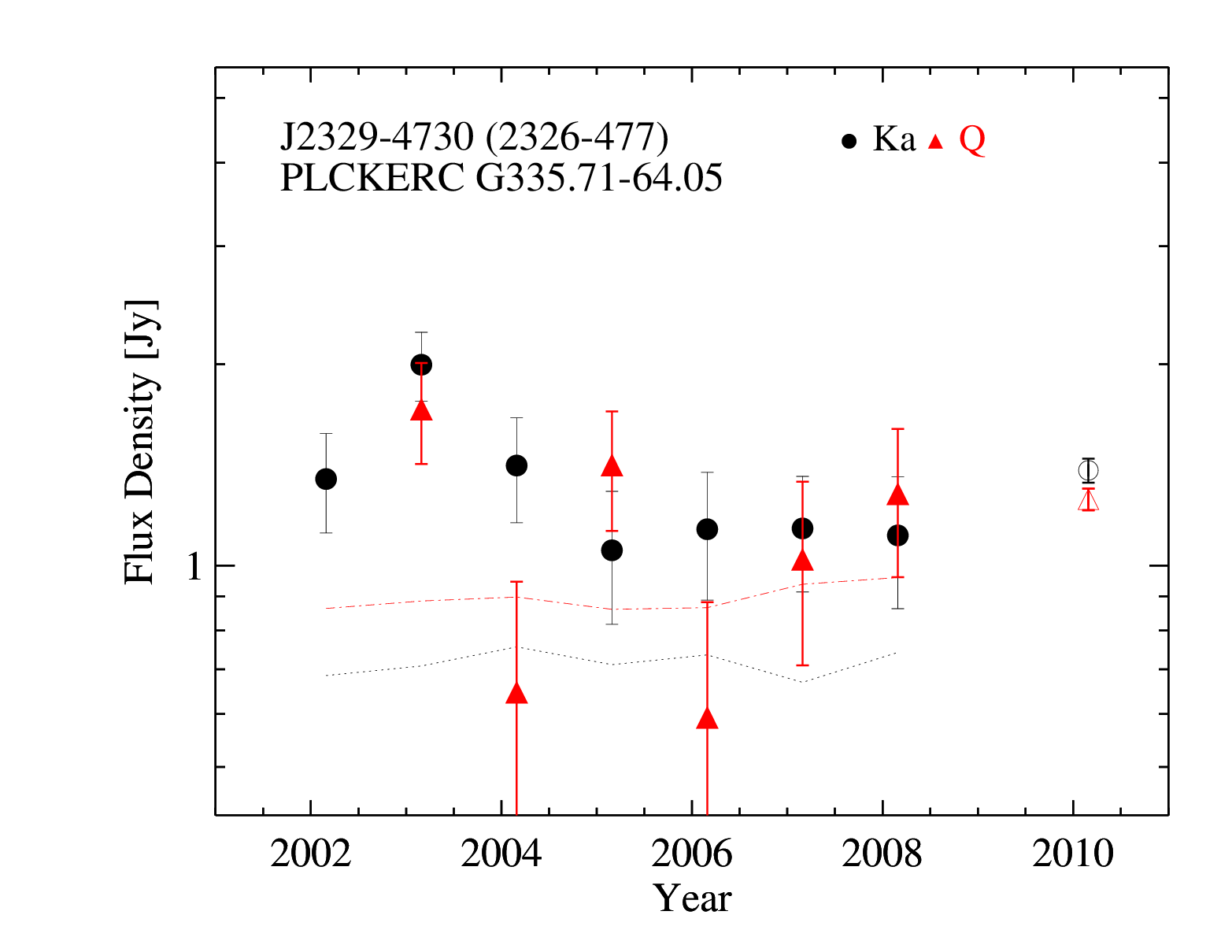}  & \includegraphics[width=0.23\textwidth]{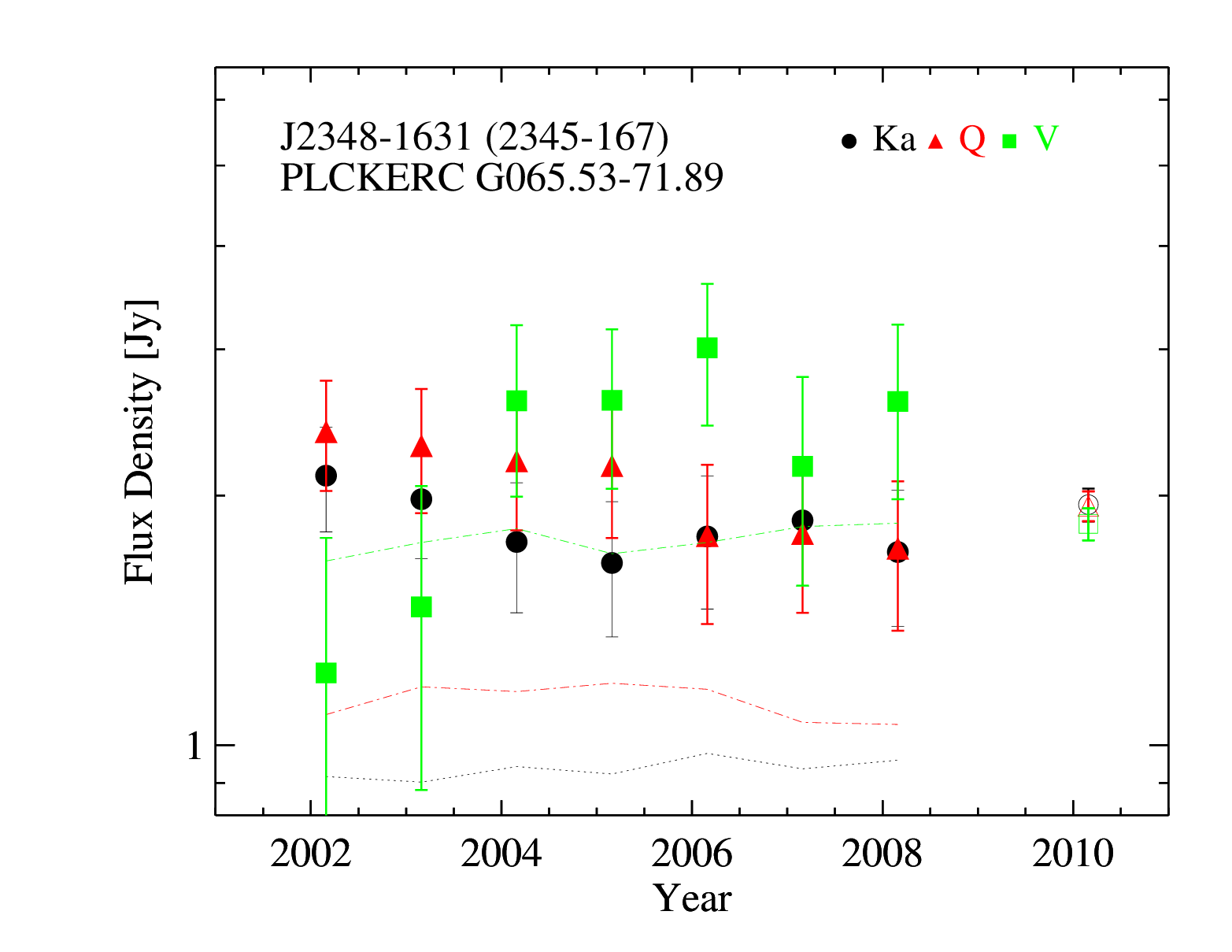} & \includegraphics[width=0.23\textwidth]{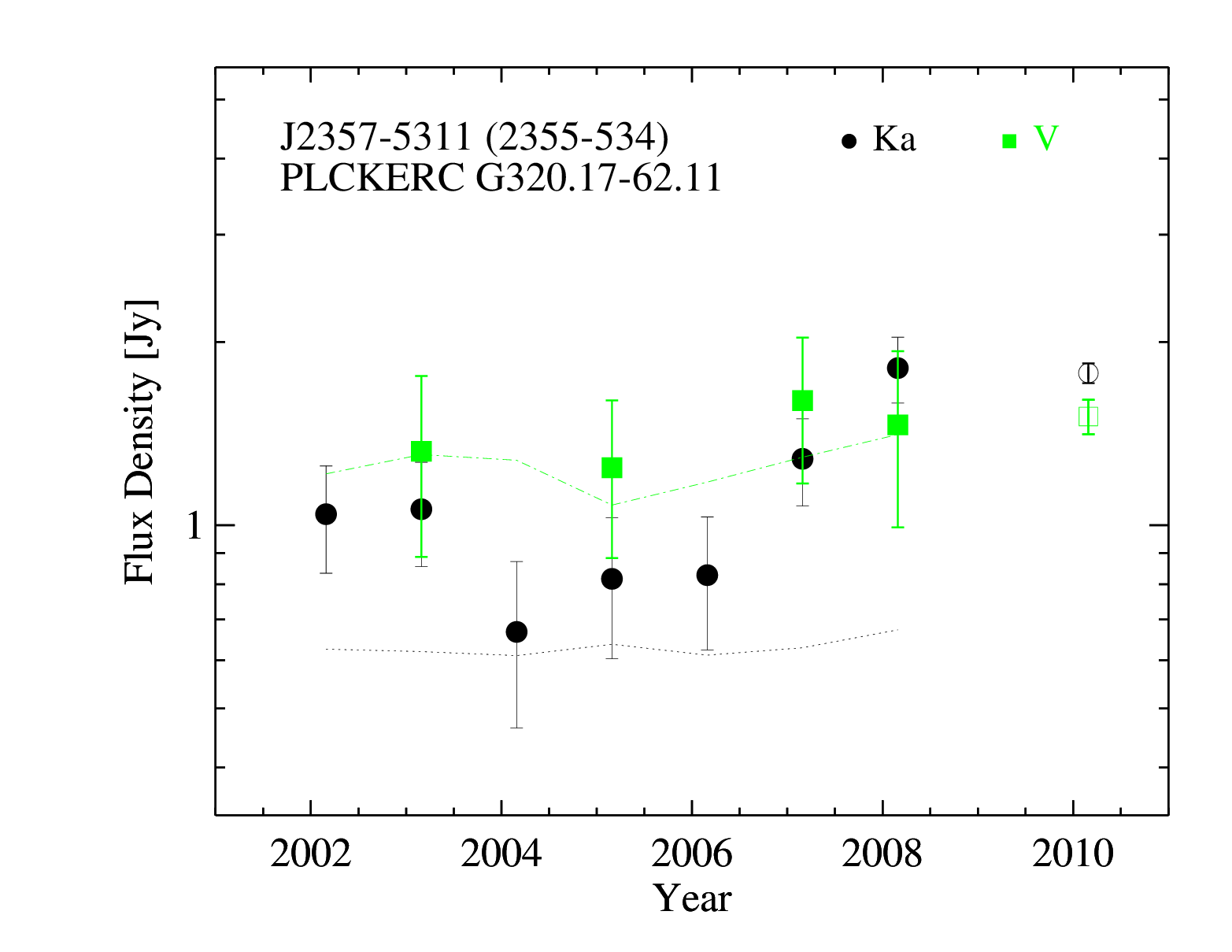}  \\
\end{tabular}
 \caption{\WMAP/\Planck\ light curves for sources with at least one good timeline (i.e., four or more $> 3\sigma$ data points in seven years of \WMAP\ observation) in our sample. For each good timeline, all $> 2\sigma$ data points are plotted,  with dash-dot lines indicating the 3$\sigma$ level. The \WMAP\ flux densities are shown  by filled symbols and  the \Planck\ flux densities are shown by  open symbols of the same shape. \label{fig:lightcurve}}
\end{figure*}
\end{appendix}

\end{document}